\documentclass[aps,prd,onecolumn,preprintnumbers,groupedaddress,showpacs,nofootinbib,amssymb]{revtex4}
\usepackage{graphicx}
\usepackage{amsmath}
\usepackage{amssymb}
\usepackage{amsfonts}
\usepackage{bm}

\begin{document}

%\makeatletter
%\renewcommand{\theequation}{\Roman{section}\,\Alph{subsection}.\arabic{equation}}
%\@addtoreset{equation}{subsection}{section}
%\makeatother
 
\makeatletter
\renewcommand{\theequation}{\Roman{section}.\arabic{equation}}
\@addtoreset{equation}{section}
\makeatother

\def\pp{{\, \mid \hskip -1.5mm =}}
\def\cL{\mathcal{L}}
\def\be{\begin{equation}}
\def\ee{\end{equation}}
\def\bea{\begin{eqnarray}}
\def\eea{\end{eqnarray}}
\def\tr{\mathrm{tr}\, }
\def\nn{\nonumber \\}
\def\e{\mathrm{e}}

\title{Unified cosmic history in modified gravity: from $F(R)$ theory to
Lorentz non-invariant models}

\author{Shin'ichi Nojiri$^{1,2}$
and Sergei D. Odintsov$^3$\footnote{also at Tomsk State Pedagogical University, Tomsk.}}

\affiliation{ $^1$ Department of Physics, Nagoya University, Nagoya 
464-8602,
Japan \\
$^2$ Kobayashi-Maskawa Institute for the Origin of Particles and
the Universe, Nagoya University, Nagoya 464-8602, Japan \\
$^3$Instituci\`{o} Catalana de Recerca i Estudis Avan\c{c}ats
(ICREA) and Institut de Ciencies de l'Espai (IEEC-CSIC), Barcelona, Spain}

\begin{abstract}

The classical generalization of general relativity is considered as 
the gravitational alternative for a unified description of the early-time 
inflation with late-time cosmic acceleration. 
The structure and cosmological properties of a 
number of modified theories, including traditional $F(R)$ and 
Ho\v{r}ava-Lifshitz $F(R)$ gravity,
scalar-tensor theory, string-inspired and Gauss-Bonnet theory,
non-local gravity, non-minimally coupled models, and power-counting
renormalizable covariant gravity are discussed. Different representations 
of and relations between such theories are investigated. It is shown that 
some versions of the above theories may be consistent with local tests 
and may provide a qualitatively reasonable unified description of inflation 
with the dark energy epoch.
The cosmological reconstruction of different modified gravities is 
provided in great detail. 
It is demonstrated that eventually any given universe evolution
may be reconstructed for the theories under consideration, and 
the explicit reconstruction is applied to an accelerating spatially flat 
Friedmann-Robertson-Walker (FRW) universe.
Special attention is paid to Lagrange multiplier constrained and
conventional $F(R)$ gravities, for latter $F(R)$ theory, the effective $\Lambda$CDM 
era and phantom divide crossing acceleration are obtained. 
The occurrences of the Big Rip and other finite-time future singularities in modified
gravity are reviewed along with their solutions via the addition of 
higher-derivative gravitational invariants.

\end{abstract}


\maketitle

\newpage

\tableofcontents

\newpage

\section{Introduction \label{Sec1}}

Current theoretical cosmology is faced with two fundamental problems: what
is inflation and what is dark energy? In other words, why and how did both 
the very early and the very late universe expand with acceleration?
What is the reason for the similarity of the universe evolution at small 
as well as at large curvature?

The scenarios to describe these early-time and late-time accelerations are
usually very similar, which is why it is quite natural to expect that the
same theory/principle lies behind
both accelerating cosmological epochs. Indeed, there are various proposals
to construct an acceptable dark energy model, such as: scalar, spinor, 
(non-)abelian vector theory, cosmological constant, fluid with a 
complicated equation of state, and higher dimensions. 
Remarkably, the same proposals are also intended to describe inflationary era. 
As a result, we have a number of competing scenarios with which 
to describe inflation and late-time acceleration. 
This situation is very reasonable due to the fact that the evolution
of the cosmological parameters is not defined with precise accuracy. Even
their current values are defined with at least 3-5\% error. Moreover,
working in framework of one of the above proposals, we are forced to introduce the
following extra cosmological components: inflaton, a dark component and dark matter.
Even if such a scenario seems to be partially successful,
it immediately introduce a new set of problems: such as the following: 
coupling with usual matter, (anti-)screening of
dark components during the evolution of the universe, compatibility with standard 
elementary particle theories, and consistency of formulation.
Thus, another natural proposal is to not introduce extra fields to resolve
the cosmological problems, which is the most economical solution in the spirit of
the Occam's razor principle.

The least understood and most fundamental force of nature is the gravitational
interaction.
It is expected that general relativity is just an approximation that is 
valid at small curvatures
and that is predicted by the Theory of Everything (strings/M-theory). In 
the very early universe, some unknown gravitational theory dictated
the evolution of the universe.
Thus, a very reasonable assumption is the gravitational alternative for
a unified description of the inflation and dark energy. In other words,
the modified gravity, which represents a classical generalization of
general relativity, should consistently describe the early-time inflation 
and late-time acceleration, without the introduction of any other dark 
component.
A sector of modified gravity that contains the gravitational terms
relevant at very high energies produced the inflationary epoch. In the
course of evolution, the curvature decreases and general relativity gives
a sufficient approximation in the intermediate universe. With a further
decrease of the curvature, when sub-dominant terms quickly grow, one sees
the change from deceleration to cosmic acceleration.
Thus, the early-time as well as the late-time cosmic speed-up is caused simply
by the fact that some sub-dominant terms of gravitational action may 
become essential at large or small curvatures.
Moreover, such an approach may also be considered as dynamic solution of 
the cosmological constant problem. Of course, the complete gravitational 
action should be defined by a fundamental theory, which remains to be the open 
problem of modern high-energy physics.
In the absence of fundamental quantum gravity, the modified gravity 
approach is a phenomenological model that is constructed by complying with
observational data and data from local tests.

In addition, just as by-product to be a cosmological model, 
the modified gravity may provide the
explanation for dark matter. It may resolve the coincidence problem simply
by the fact of the universe expansion. It also may describe the
transition from deceleration to acceleration of the universe, 
and it may be useful for
high-energy physics problems (i.e., unification of all interactions, 
and hierarchy problem resolution). 
Even if the current universe is entering the phantom phase,
modified gravity effectively describes the transition from the non-phantom to
phantom era without the need to
introduce the exotic matter (phantom) with extremely strange properties.

In this work, we present a detailed review of a number of popular models of
modified gravity. Their properties and different representations are
discussed. Assuming the spatially flat FRW cosmology, we investigate the
modified gravity background evolution with the aim of achieving 
a unified description of the universe's inflation with dark energy epoch. 
Using sufficient freedom in the choice
of the action as some function of curvature invariants, we develop the
cosmological reconstruction scheme, which is explicitly applied to obtain the
(effective quintessence/cosmological constant/phantom) accelerating
expansion for the models under discussion. Late-time dynamics is analyzed
and it is shown that the future universe may end in one of the four known
finite-time singularities. For instance, the effective phantom
superacceleration brings the universe to the famous Big Rip (or Type I)
singularity \cite{Caldwell:2003vq}. In order to avoid the future singularity
one has to add the extra higher-derivative gravitational terms, which are
relevant at the early universe and help in the unification of early-time
inflation with late-time acceleration because such terms induce inflation.

It should be noted that, as a rule, modified gravity equations of motion 
are higher-derivative differential equations. Thus, we consider only the 
background evolution of such theories in this report; we do not discuss 
the cosmological perturbation theory.
The reason for this is very simple. Actually, in almost all existing approaches to
cosmological perturbations in modified gravity, the corresponding 
equations are reduced to second-order differential equations.
Clearly, this is not a well-justified approach, and the results obtained in
this way are questionable.
The new higher-derivative covariant perturbation theory should be developed
to obtain correct answers to the related circle of questions.

The work is organized as follows. The second chapter gives the
introduction to several popular models of modified gravity:
$F(R)$-theory, modified Gauss-Bonnet and string-inspired gravity, 
non-local gravity, non-minimally coupled theories,
$F(R)$ Ho\v{r}ava-Lifshitz gravity, and power-counting renormalizable 
covariant gravity. 
The general properties of $F(R)$-theory are discussed in section \ref{IIA}.
Its different representations, namely Jordan frame, scalar-tensor form (Einstein
frame) and fluid representation, are given. The  thermodynamics of
Schwarzschild-(anti-)de Sitter space are briefly discussed. Some local 
tests, in particular, the Newtonian regime and matter instability, 
are investigated. Several specific models (i.e., theory with negative and 
positive curvature power terms, $\ln R$-model, and exponential theory), are 
presented.
The spatially flat FRW cosmology is investigated. Viable modified
gravities that allow the unification of the early-time inflation with
late-time acceleration are given (for the first work where such a 
unification was proposed, see \cite{Nojiri:2003ft}). 
Note that the literature on
$F(R)$-gravity is vast, and the earlier reviews on this theory are 
given in refs.~\cite{Nojiri:2006ri,review}. 
Different aspects of $F(R)$-gravity are
discussed in refs.~\cite{ASPECTS1,ASPECTS2,ASPECTS3}, and various exact
solutions are presented in refs.~\cite{SOLUTIONS}. 
The comparison with observational data and local
tests is made in refs.~\cite{Nojiri:2006ri,review,TESTS}. 
As is indicated in this work, there are a number of viable $F(R)$ 
models that successfully pass local tests and are in accordance with 
observational data.

Modified Gauss-Bonnet gravity is considered in section \ref{IIB}, and its 
spatially flat FRW equations are given. 
The de Sitter solution and its thermodynamics are also 
studied, and the possibility of the unification of inflation with dark energy is
mentioned. Section \ref{IIC} is devoted to a brief study of string-inspired
(scalar-Gauss-Bonnet) gravity. The emergence of an accelerating dark energy
epoch in such a model is proposed.
The non-local generalization of $F(R)$-gravity and Gauss-Bonnet gravity is
discussed in section \ref{IID}.
We find the scalar-tensor representation for such non-local gravities and
investigate their spatially flat FRW cosmology.
The possibility of early-time or late-time acceleration is again
demonstrated.
Non-minimally coupled theories are presented in section \ref{IIE}. We consider the
theories where scalar, vector or Yang-Mills Lagrangian is multiplied to
function $F(R)$. FRW equations are given and possibility of cosmic
acceleration is mentioned.
In section \ref{IIF} we propose modified $F(R)$ Ho\v{r}ava-Lifshitz gravity.
The original Ho\v{r}ava-Lifshitz theory, based on general relativity, has been
introduced as a candidate for renormalizable quantum gravity at the price of
the violation of the Lorentz invariance. The $F(R)$ generalization of
Ho\v{r}ava-Lifshitz gravity yields consistent FRW equations that coincide 
with equations associated with the conventional theory 
by choosing the parameters to be specific values. 
We study several explicit models that are introduced in section \ref{IIA} 
in this framework and show that they
may again predict the unified description of early-time inflation with 
dark energy. Motivated by the Ho\v{r}ava-Lifshitz proposal, the covariant 
power-counting renormalizable gravity is introduced in section \ref{IIG}. 
Two formulations of the theory: one utilizing an extra perfect fluid 
and the other using the Lagrange multiplier constraint, are given. 
Accelerating FRW cosmology is briefly discussed. 
In section \ref{IID},  it is shown that arbitrary modified gravity may be 
interpreted as an effective fluid with an inhomogeneous equation of state. 
The late-time evolution of a specific dark fluid with a power-law equation 
of state is investigated. 
It is demonstrated that an effective quintessence/phantom fluid
may drive the universe to a finite-time future singularity.

The background evolution of modified gravity is studied in the third chapter.
Usually, in theories like general relativity, we start from a theory 
that is defined by the action and solve the corresponding equations of
motion to define the background dynamics. However, an approach is 
already too complicated for simple models of modified gravity.
Thus, using the fact that modified gravity is defined up to some
arbitrary function of curvature invariants,
the inverse problem is analyzed. The corresponding cosmological
reconstruction scheme is proposed. Within the framework of this scheme, we show how 
the complicated background cosmology, which complies with observational data, 
may be reconstructed. This provides way to define (at least partially) 
the reconstructed form of modified gravity, which has a particular 
cosmological solution. Section \ref{IIIA} is devoted to a formulation of 
cosmological reconstruction (in terms of cosmological time or in terms 
of e-folding $N$) for scalar-tensor gravity. 
Several examples of a (super)accelerating epoch 
are used for the reconstruction of scalar potentials.
This formulation is extended for the theory with two scalars where the
reconstruction of the transition from matter dominance to late-time
acceleration as well as a reconstruction of the $\Lambda$CDM epoch is done. 
In section \ref{IIIB}, we study the reconstruction in the k-essence model, 
using both formulations. 
Cosmological reconstruction for the $F(R)$-model, using its 
presentation with an auxiliary scalar, is developed in section \ref{IIIC}. 
The differential equation that provides the form of function $F(R)$ 
for a given spatially flat FRW cosmology is obtained. 
The $\Lambda$CDM epoch and dark era with a phantom divide
crossing are reconstructed in $F(R)$-gravity with/without matter in great
detail. 
The general condition for the stability of the cosmological solution is derived.
$F(R)$-gravity with a Lagrange multiplier constraint is formulated, and its 
FRW cosmology is studied.
In sections \ref{IIID} and \ref{IIIE}, we propose the reconstruction 
of modified Gauss-Bonnet and scalar-Einstein-Gauss-Bonnet gravities, respectively. 
In section \ref{IIIF}, we show that the reconstruction is easily generalized 
for the case of $F(R)$ Ho\v{r}ava-Lifshitz gravity. 
The explicit reconstruction of the $\Lambda$CDM 
epoch and of phantom superacceleration is fulfilled. The cosmological
reconstruction of the power-law FRW solution for non-minimal Yang-Mills theory
is given in section \ref{IIIG}.

The fourth chapter is devoted to an investigation of late-time dynamics for
effective quintessence/phantom dark energy. It is demonstrated that, 
regardless of whether the dark energy model utilizes fluid, scalar or 
alternative gravity, the future universe may evolve to 
a finite-time singularity of one of four known types. 
The prescription to resolve such a future singularity via the addition of
a higher-derivative gravitational term is proposed.
In section \ref{IVA}, we consider dark fluid coupled with dark matter. 
For a power-law equation of state, it is shown that the future dark era may be
singular, depending on the parameter choice. The coupling with dark
matter may prevent some of the soft future singularities. However, such
coupling does not resolve the future singularity in a general case. 
Only the addition of the $R^2$-term may cure all types of future singularities 
in the model under consideration.
In section \ref{IVB}, using the reconstruction method, we show that $F(R)$ gravity
may have late-time accelerating cosmological solutions with all four types
of finite-time future singularities. To avoid this, one should
additionally modify the theory by the $R^2$ term. It is interesting that the
use of such a term may be motivated by conformal anomaly considerations.
The same problem with qualitatively similar results is studied for
modified Gauss-Bonnet and $F(R)$ Ho\v{r}ava-Lifshitz gravities in section
\ref{IVC} and \ref{IVD}. It is remarkable that, in the last case, one may need other
higher-derivative gravitational terms to resolve all types of future singularities.

A summary and outlook are given in the Discussion section.

%%%%%%%%%
%%%%%%%%%
%%%%%%%%%

\section{Modified gravity unifying the early-time inflation with late-time
acceleration \label{SecII}}

This chapter is devoted to the general review of many popular models
of modified gravity such as $F(R)$ gravity, modified Gauss-Bonnet gravity,
string-inspired theory, non-local gravity and non-minimal theories. Their
general properties, different representations and background 
spatially flat FRW
cosmologies are investigated. Special attention is paid to possible 
unification
of early-time inflation with late-time acceleration within such theories.
The extension of such a study to power-counting renormalizable gravity with
explicit or apparent Lorentz symmetry breaking is also made.

\subsection{$F(R)$ gravity \label{IIA}}

In this section, we give a review of general properties of $F(R)$ 
gravity. Its scalar-tensor description is presented, and some constant
curvature solutions are discussed. A number of popular models are introduced, 
and their applications for a unified description of inflation with dark energy 
are studied.

\subsubsection{General properties}

Let us start with the general introduction of the model.
In $F(R)$ gravity \cite{Nojiri:2006ri}, the scalar curvature $R$ in
the Einstein-Hilbert action\footnote{
We use the following convention for the curvatures:
\[
R=g^{\mu\nu}R_{\mu\nu} \, , \quad
R_{\mu\nu} = R^\lambda_{\ \mu\lambda\nu} \, , \quad
R^\lambda_{\ \mu\rho\nu} = -\Gamma^\lambda_{\mu\rho,\nu}
+ \Gamma^\lambda_{\mu\nu,\rho} - 
\Gamma^\eta_{\mu\rho}\Gamma^\lambda_{\nu\eta}
+ \Gamma^\eta_{\mu\nu}\Gamma^\lambda_{\rho\eta} \, ,\quad
\Gamma^\eta_{\mu\lambda} = \frac{1}{2}g^{\eta\nu}\left(
g_{\mu\nu,\lambda} + g_{\lambda\nu,\mu} - g_{\mu\lambda,\nu}
\right)\, .
\]
}
\be
\label{JGRG6}
S_\mathrm{EH}=\int d^4 x \sqrt{-g} \left( \frac{R}{2\kappa^2} +
\mathcal{L}_\mathrm{matter} \right)\, ,
\ee
is replaced by an appropriate function of the scalar curvature:
\be
\label{JGRG7}
S_{F(R)}= \int d^4 x \sqrt{-g} \left( \frac{F(R)}{2\kappa^2} +
\mathcal{L}_\mathrm{matter} \right)\, .
\ee
Let us review the general properties of $F(R)$ gravity.
For $F(R)$ theory, as for any other modified gravity theory, one can define an
effective EoS parameter using its fluid representation.
The FRW equations in the Einstein gravity coupled with perfect fluid are:
\be
\label{JGRG11}
\rho_\mathrm{matter}=\frac{3}{\kappa^2}H^2 \, ,\quad p_\mathrm{matter}= -
\frac{1}{\kappa^2}\left(3H^2 + 2\dot
H\right)\, .
\ee
For modified gravity, one may define an effective equation of state (EoS)
parameter as follows:
\be
\label{JGRG12}
w_\mathrm{eff}= - 1 - \frac{2\dot H}{3H^2} \, .
\ee
This relation is very useful for a generalized fluid
description of modified gravity.

The equation of motion for modified gravity is given by
\be
\label{JGRG13}
\frac{1}{2}g_{\mu\nu} F(R) - R_{\mu\nu} F'(R) - g_{\mu\nu} \Box F'(R)
+ \nabla_\mu \nabla_\nu F'(R) = - \frac{\kappa^2}{2}T_{\mathrm{matter}\,
\mu\nu}\, .
\ee
Assuming a spatially flat FRW universe,
\be
\label{JGRG14}
ds^2 = - dt^2 + a(t)^2 \sum_{i=1,2,3} \left(dx^i\right)^2\, ,
\ee
the FRW like equations are as follows:
\bea
\label{JGRG15}
0 &=& -\frac{F(R)}{2} + 3\left(H^2 + \dot H\right) F'(R)
 - 18 \left( 4H^2 \dot H + H \ddot H\right) F''(R)
+ \kappa^2 \rho_\mathrm{matter}\, ,\\
\label{Cr4b}
0 &=& \frac{F(R)}{2} - \left(\dot H + 3H^2\right)F'(R)
+ 6 \left( 8H^2 \dot H + 4 {\dot H}^2 + 6 H \ddot H + \dddot H\right) F''(R)
+ 36\left( 4H\dot H + \ddot H\right)^2 F'''(R) \nn 
&& + \kappa^2 p_\mathrm{matter}\, .
\eea
Here, the Hubble rate $H$ is defined by $H=\dot a/a$ and
the scalar curvature $R$ is given by $R=12H^2 + 6\dot H$.

Several (often exact) solutions of (\ref{JGRG15}) maybe found.
However, due to the complicated nature of field equations, the number of exact
solutions is much less than, for instance, that in general relativity.
Without any matter, assuming that the Ricci tensor is covariantly
constant, that is, $R_{\mu\nu}\propto g_{\mu\nu}$, Eq.~(\ref{JGRG13}) 
reduces to the algebraic equation:
\be
\label{JGRG16}
0 = 2 F(R) - R F'(R)\, .
\ee
If Eq.~(\ref{JGRG16}) has a solution, the (anti-)de Sitter and/or 
Schwarzschild- (anti-)de Sitter space 
\be
\label{SdS}
ds^2 = - \left( 1 - \frac{2MG}{r} \mp \frac{r^2}{L^2} \right) dt^2
+ \left( 1 - \frac{2MG}{r} \mp \frac{r^2}{L^2}
\right)^{-1} dr^2
+ r^2 d\Omega^2\, ,
\ee
or Kerr - (anti-)de Sitter space is an exact vacuum solution.
In (\ref{SdS}), the minus and plus signs in $\pm$ correspond to the de Sitter and 
anti-de Sitter space, respectively. 
$M$ is the mass of the black hole, $G = \frac{\kappa^2}{8\pi}$,
and $L$ is the length parameter of (anti-)de Sitter space, which is related to 
the curvature $R=\pm \frac{12}{L^2}$
(the plus sign for the de Sitter space and the minus sign for the anti-de 
Sitter space).
For example, if we consider the model 
\be
\label{FR1B}
F(R) = R - \frac{\mu^{2m+2}}{R^m}\, ,
\ee
with a constant $\mu$ with a mass dimension,
the solution of (\ref{JGRG16}) is given by
\be
\label{FR2}
R = \left( 2+m \right)^{\frac{1}{m+1}} \mu^2 \, .
\ee
As another example, for a logarithmic model
\be
\label{FR3}
F(R) = \mu^2 \ln \frac{R}{R_0}\, ,
\ee
with a constant $R_0$, one gets
\be
\label{FR4}
R = R_0 \e^{\frac{1}{2}}\, .
\ee

The entropy of the Schwarzschild - (anti-)de Sitter space can be evaluated 
by the WKB approximation for the Euclidean partition function $Z$ 
\cite{Cognola:2005de}: 
\be
\label{FR5}
Z \sim \e^{S_\mathrm{SdS}}\, .
\ee
Here, $S_\mathrm{SdS}$ is the value of the classical action calculated in
the Schwarzschild - (anti-)de Sitter space. Because the value of
$S_\mathrm{SdS}$ diverges, we regularize the action by subtracting the 
value when the solution of the (anti-)de Sitter space without a black hole 
is substituted \cite{Cognola:2005de}.
The free energy $\mathcal{F}$ is given by
\be
\label{FR6}
\mathcal{F} = - \frac{1}{\beta} \ln Z\, .
\ee
And, using the standard thermodynamics relations,
\be
\label{FR7B}
\mathcal{S} = - \frac{dF}{dT_H}\, ,
\ee
one can evaluate the entropy $\mathcal{S}$.
Here, $T_H$ is the Hawking temperature. Then, it follows that 
\be
\label{FR8}
\mathcal{S} = \frac{F'(R) A_\mathrm{H} }{4G}\, .
\ee
Here, $A_H$ expresses the area of the horizon.

Let us now assume that $F(R)$ behaves as $F(R) \propto f_0 R^m$.
Eq.~(\ref{JGRG15}) gives 
\be
\label{M7}
0 = f_0 \left\{ - \frac{1}{2} \left(6\dot H + 12 H^2\right)^m
+ 3 m \left(\dot H + H^2\right)\left(6\dot H + 12 H^2\right)^{m - 1}
 -3 m H \frac{d}{dt} \left\{\left(6\dot H + 12 H^2
\right)^{m -1}\right\}\right\} + \kappa^2 \rho_0 a^{-3(1+w)}\, .
\ee
Eq.~(\ref{Cr4b}) is irrelevant because it can be derived from (\ref{M7}).
When the contribution from the matter can be neglected ($\rho_0=0$), 
the following solution appears:
\be
\label{JGRG17}
H \sim \frac{-\frac{(m-1)(2m-1)}{m-2}}{t}\, ,
\ee
with a corresponding EoS parameter (\ref{JGRG12}):
\be
\label{JGRG18}
w_\mathrm{eff}=-\frac{6m^2 - 7m - 1}{3(m-1)(2m -1)}\, .
\ee
On the other hand, when the matter contribution is include with a 
constant EoS parameter $w$, an exact solution of (\ref{M7}) is given by
\bea
\label{M8}
&& a=a_0 t^{h_0} \, ,\quad h_0\equiv \frac{2m}{3(1+w)} \, ,\nn
&& a_0\equiv \left[-\frac{3f_0h_0}{\kappa^2 \rho_0}\left(-6h_0 + 12
h_0^2\right)^{m-1}
\left\{\left(1-2m\right)\left(1-m\right) -
(2-m)h_0\right\}\right]^{-\frac{1}{3(1+w)}}\, ,
\eea
and the effective EoS parameter (\ref{JGRG12}) is
\be
\label{JGRG20}
w_\mathrm{eff}= -1 + \frac{w+1}{m}\, .
\ee
These solutions (\ref{JGRG17}) and (\ref{M8}) show that modified gravity 
may describe early/late-time universe acceleration. Moreover, it is very 
natural to propose that a more complicated modified gravity from the above 
class may give the unified description for inflation with 
late-time acceleration.

\subsubsection{Scalar-tensor description}

Let us now introduce perfect fluid and scalar-tensor representations 
for the modified gravity under consideration. 
These mathematically equivalent representations may simplify 
the investigation of a number of problems.

Eqs.~(\ref{JGRG15}) and (\ref{Cr4b}) show that
the effective energy density $\rho_\mathrm{eff}$ and the effective 
pressure
$p_\mathrm{eff}$ including the contribution from $f(R)$ are given by
(see, for instance, \cite{Nojiri:2009xw})
\bea
\label{Cr4}
\rho_\mathrm{eff} &=& \frac{1}{\kappa^2}\left(-\frac{1}{2}f(R) + 
3\left(H^2 + \dot H\right) f'(R)  - 18 \left(4H^2 \dot H 
+ H \ddot H\right)f''(R)\right) + \rho_\mathrm{matter}\, ,\\
\label{Cr4bb}
p_\mathrm{eff} &=& \frac{1}{\kappa^2}\left(\frac{1}{2}f(R) - \left(3H^2 
+ \dot H \right)f'(R) + 6 \left(8H^2 \dot H + 4{\dot H}^2
+ 6 H \ddot H + \dddot H \right)f''(R) + 36\left(4H\dot H 
+ \ddot H\right)^2f'''(R) \right) \nn
&& + p_\mathrm{matter}\, .
\eea
The equations (\ref{JGRG15}) and (\ref{Cr4b}) can be rewritten as in the
Einstein gravity case (\ref{JGRG11}):
\be
\label{JGRG11B}
\rho_\mathrm{eff}=\frac{3}{\kappa^2}H^2 \, ,
\quad p_\mathrm{eff}= - \frac{1}{\kappa^2}\left(3H^2 + 2\dot H\right)\, .
\ee
Such a mathematically equivalent fluid representation for the FRW equations 
in modified gravity may lead to a number of assumptions that are not well 
justified. Indeed, the generalized gravitational fluid contains
higher-derivative curvature invariants. This fact is often ignored in the 
study of cosmological perturbations in modified gravity when such a theory 
is developed in strict similarity with general relativity, such that 
corresponding differential equations become of second order (see,
for instance, \cite{DeFelice:2010aj}). Indeed, first of all, the field
equations of the theory under investigation are of fourth order. 
It is evident that cosmological perturbation equations should be of 
fourth order as well.
Moreover, such equations should be covariant. The corresponding covariant 
and higher-derivative cosmological perturbation theory (for review, see 
\cite{Carloni:2007yv,Carloni:2009gp}) is developed, but it is 
extremely complicated.

One can also rewrite $F(R)$ gravity in the scalar-tensor form.
By introducing the auxiliary field $A$, the action (\ref{JGRG7}) of
the $F(R)$ gravity is rewritten in the following form:
\be
\label{JGRG21}
S=\frac{1}{2\kappa^2}\int d^4 x \sqrt{-g} \left\{F'(A)\left(R-A\right) +
F(A)\right\}\, .
\ee
By the variation of $A$, one obtains $A=R$. Substituting $A=R$ into
the action (\ref{JGRG21}), one can reproduce the action in (\ref{JGRG7}). 
Furthermore, we rescale the
metric in the following way (conformal transformation):
\be
\label{JGRG22}
g_{\mu\nu}\to \e^\sigma g_{\mu\nu}\, ,\quad \sigma = -\ln F'(A)\, .
\ee
Thus, the Einstein frame action is obtained:
\bea
\label{JGRG23}
S_E &=& \frac{1}{2\kappa^2}\int d^4 x \sqrt{-g} \left( R -
\frac{3}{2}g^{\rho\sigma}
\partial_\rho \sigma \partial_\sigma \sigma - V(\sigma)\right) \, ,\nn
V(\sigma) &=& \e^\sigma g\left(\e^{-\sigma}\right)
 - \e^{2\sigma} f\left(g\left(\e^{-\sigma}\right)\right) = 
\frac{A}{F'(A)}
 - \frac{F(A)}{F'(A)^2}\, .
\eea
Here $g\left(\e^{-\sigma}\right)$ is given by solving the equation
$\sigma = -\ln\left( 1 + f'(A)\right)=- \ln F'(A)$ as
$A=g\left(\e^{-\sigma}\right)$.
Due to the scale transformation (\ref{JGRG22}), a coupling
of the scalar field $\sigma$ with usual matter arises. .
The mass of $\sigma$ is given by
\be
\label{JGRG24}
m_\sigma^2 \equiv \frac{3}{2}\frac{d^2 V(\sigma)}{d\sigma^2}
=\frac{3}{2}\left\{\frac{A}{F'(A)} - \frac{4F(A)}{\left(F'(A)\right)^2} +
\frac{1}{F''(A)}\right\}\, .
\ee
Unless $m_\sigma$ is very large, the large correction to Newton's law 
appears.
Naively, one expects the order of the mass $m_\sigma$ to be that of the
Hubble rate, that is, $m_\sigma \sim H \sim 10^{-33}\,\mathrm{eV}$, 
which is very light and could make the correction very large.

In \cite{Hu:2007nk}, a ``realistic'' $F(R)$ model was proposed. 
It has been found, however, that the model has an instability where 
the large curvature can be easily produced (manifestation of a 
possible future singularity). 
In the model of \cite{Hu:2007nk}, a parameter 
$m\sim 10^{-33}\, \mathrm{eV}$ with a mass dimension is included. 
The parameter $m$ plays a role of the effective cosmological constant.
When the curvature $R$ is large enough compared with $m^2$, $R\gg m^2$,
$F(R)$ \cite{Hu:2007nk} looks as follows:
\be
\label{HS1}
F(R) = R - c_1 m^2 + \frac{c_2 m^{2n+2}}{R^n} +
\mathcal{O}\left(R^{-2n}\right)\, .
\ee
Here $c_1$, $c_2$, and $n$ are positive dimensionless constants. 
The potential $V(\sigma)$ (\ref{JGRG23}) has the following asymptotic 
form:
\be
\label{HS2}
V(\sigma) \sim \frac{c_1 m^2}{A^2}\, .
\ee
Then, the infinite curvature $R=A\to \infty$ corresponds to a small 
value of the potential and, therefore, the large curvature can be 
easily produced.

%%%%%%%%%%%%
%%%%%%%%%%%%

Let us assume that when $R$ is large, $F(R)$ behaves as
\be
\label{HS3}
F(R) \sim F_0 R^{\epsilon}\, .
\ee
Here, $F_0$ and $\epsilon$ are positive constants. 
We also assume $\epsilon>1$ so that this term
dominates compared with general relativity. Then, the potential
$V(\sigma)$(\ref{JGRG23}) behaves as
\be
\label{HS4}
V(\sigma) \sim \frac{\epsilon -1}{\epsilon^2 F_0 R^{\epsilon -2}}\, .
\ee
Therefore, if $1<\epsilon<2$, the potential $V(\sigma)$ diverges when $R\to
\infty$ and, therefore, the large curvature is not realized so easily.
When $\epsilon=2$, $V(\sigma)$ takes a finite value $1/F_0$ when 
$R\to \infty$.
As long as $1/F_0$ is large enough, the large curvature can be prevented.

Note that the anti-gravity regime appears when $F'(R)$ is negative, 
which follows from Eq.~(\ref{JGRG21}) \cite{Nojiri:2003ft}.
Then, we need to require
\be
\label{FR1}
F'(R) > 0 \, .
\ee

We should also note that
\be
\label{FRV1}
\frac{dV(\sigma)}{dA} = \frac{F''(A)}{F'(A)^3}
\left( - AF'(A) + 2F(A) \right)\, .
\ee
Therefore, if
\be
\label{FRV2}
0 = - AF'(A) + 2F(A) \, ,
\ee
the scalar field $\sigma$ is on the local maximum or local minimum of the
potential and, therefore, $\sigma$ can be a constant. 
Note that the condition (\ref{FRV2}) is nothing but the
condition (\ref{JGRG16}) for the existence of the de Sitter solution.
When the condition (\ref{FRV2}) is satisfied, the mass (\ref{JGRG24}) can 
be rewritten as
\be
\label{FRV3}
m_\sigma^2 = \frac{3}{2 F'(A)} \left( - A 
+ \frac{F'(A)}{F''(A)} \right)\, .
\ee
Then, when the condition (\ref{FR1}) for the non-existence of the 
anti-gravity is satisfied, the mass squared $m_\sigma^2$ is positive and, 
therefore the, scalar field is on the local minimum if
\be
\label{FRV4}
 - A + \frac{F'(A)}{F''(A)} > 0\, .
\ee
On the other hand, if
\be
\label{FRV5}
 - A + \frac{F'(A)}{F''(A)} < 0\, ,
\ee
the scalar field is on the local maximum of the potential and
the mass squared $m_\sigma^2$ is negative.
As we will see later, the condition (\ref{FRV4}) is nothing but the 
condition for stability of the de Sitter space.

We have rewritten the action (\ref{JGRG7}) of $F(R)$ gravity
in a scalar-tensor form (\ref{JGRG23}).
Inversely, it is always possible to rewrite the action of the scalar-tensor
theory
\be
\label{f6}
S=\int d^4 x
\sqrt{-g}\left\{\frac{1}{2\kappa^2}R - \frac{1}{2}\partial_\mu
\varphi \partial^\mu \varphi - \tilde V(\varphi)\right\}\, ,
\ee
as the action of $F(R)$ gravity \cite{Capozziello:2005mj}.
By comparing (\ref{f6}) with (\ref{JGRG23}),
we find
\be
\label{fff1}
\varphi = \frac{\sqrt{3}}{\kappa} \sigma\, , \quad
\tilde V(\varphi) = V\left(\frac{\kappa}{\sqrt{3}} \varphi \right)\, .
\ee
One now uses the conformal transformation
\be
\label{f7}
g_{\mu\nu}\to
\e^{\pm\kappa \varphi\sqrt{\frac{2}{3}}}g_{\mu\nu}\, ,
\ee
to make the kinetic term of $\varphi$ vanish.
Then, we obtain
\be
\label{f8}
S=\int d^4 x \sqrt{-g}\left\{\frac{\e^{\pm\kappa
\varphi\sqrt{\frac{2}{3}}}}{2\kappa^2}
R - {\e^{\pm 2\kappa \varphi\sqrt{\frac{2}{3}}}}
\tilde V(\varphi)\right\}\, .
\ee
The action (\ref{f8}) is often called the ``Jordan frame action'', 
whereas the action (\ref{f6}) is the ``Einstein frame action'' due to
either the non-minimal coupling or the standard coupling in front
of the scalar curvature.
Because $\varphi$ now becomes an auxiliary field, which does not have a 
kinetic term, one may delete $\varphi$
by using the corresponding equation of motion:
\be
\label{f9}
R=\e^{\pm\kappa \varphi\sqrt{\frac{2}{3}}}
\left(4\kappa^2 \tilde V(\varphi) \pm 2\kappa \sqrt{\frac{3}{2}}
\tilde V'(\varphi)\right)\, ,
\ee
which can be solved with respect
to $\varphi$ as a function of $R$ as $\varphi=\varphi(R)$.
Thus, we can rewrite the action (\ref{f8}) in the form of $F(R)$ gravity:
\be
\label{f10}
S=\int d^4 x \sqrt{-g}F(R)\, , \quad
F(R) \equiv
\frac{\e^{\pm \kappa \varphi(R)\sqrt{\frac{2}{3}}}}{2\kappa^2}
R - {\e^{\pm 2\kappa \varphi(R)\sqrt{\frac{2}{3}}}}\tilde
V\left(\varphi(R)\right)\, .
\ee
Note that one can rewrite the scalar-tensor theory (\ref{f6}) 
only when the sign in front of the kinetic term is negative, that is, 
when the scalar field is canonical.
One cannot rewrite the action if the sign is positive, as
\be
\label{f6p}
S=\int d^4 x
\sqrt{-g}\left\{\frac{1}{2\kappa^2}R + \frac{1}{2}\partial_\mu
\varphi \partial^\mu \varphi - \tilde V(\varphi)\right\}\, ,
\ee
which corresponds to the phantom scalar \cite{Caldwell:1999ew}.

Note that, for the metric transformed as in (\ref{f7}),
even if the Einstein frame universe is in a non-phantom phase, 
where the effective EoS $w_\mathrm{eff}$ in (\ref{JGRG12}) is
larger than $-1$, the Jordan frame universe can be, in general, in a
phantom phase. In other words, despite the mathematical equivalence 
between two frames there occurs some kind of physical 
non-equivalence \cite{Capozziello:2006dj}.
More precisely, if physics in one frame is matched with observations 
and describes the observable accelerating universe, then another frame 
of physics becomes extremely unconventional. 
For instance, instead of the acceleration, the observer sees 
the deceleration, and matter couples with scalar, etc.

\subsubsection{Matter instability}

Let us discuss the matter instability issue \cite{Dolgov:2003px} in 
modified gravity. It is related to the fact that the spherical body solution in 
general relativity may not be the solution in modified gravity theory. 
The matter instability may appear when the energy density or the curvature 
is large compared with the average density or curvature in the universe, 
as is the case inside of a planet.
Multiplying $g^{\mu\nu}$ with Eq.~(\ref{JGRG13}), one obtains
\be
\label{JGRG27}
\Box R + \frac{F^{(3)}(R)}{F^{(2)}(R)}\nabla_\rho R \nabla^\rho R
+ \frac{F'(R) R}{3F^{(2)}(R)} - \frac{2F(R)}{3 F^{(2)}(R)}
= \frac{\kappa^2}{6F^{(2)}(R)}T_\mathrm{matter}\, .
\ee
Here,  $T_\mathrm{matter}$ is the trace of the matter energy-momentum 
tensor: $T_\mathrm{matter} \equiv {T_\mathrm{matter}}_\rho^{\ \rho}$. 
We also denote $d^nF(R)/dR^n$ by $F^{(n)}(R)$.
Let us now consider the perturbation from the solution of general 
relativity.
We denote the scalar curvature solution given by the matter density in the
Einstein gravity by $R_b\sim (\kappa^2/2)\rho_\mathrm{matter}>0$ and 
separate the scalar curvature $R$ into the sum of $R_b$
and the perturbed part $R_p$ as 
$R=R_b + R_p$ $\left(\left|R_p\right|\ll \left|R_b\right|\right)$.
Then, Eq.~(\ref{JGRG27}) leads to the following perturbed equation:
\bea
\label{JGRG28}
0 &=& \Box R_b + \frac{F^{(3)}(R_b)}{F^{(2)}(R_b)}\nabla_\rho R_b 
\nabla^\rho R_b + \frac{F'(R_b) R_b}{3F^{(2)}(R_b)} \nn
&& - \frac{2F(R_b)}{3 F^{(2)}(R_b)} - \frac{R_b}{3F^{(2)}(R_b)} + \Box R_p
+ 2\frac{F^{(3)}(R_b)}{F^{(2)}(R_b)}\nabla_\rho R_b \nabla^\rho R_p 
+ U(R_b) R_p\, .
\eea
Here, $U(R_b)$ is given by
\bea
\label{JGRG29}
U(R_b) &\equiv& \left(\frac{F^{(4)}(R_b)}{F^{(2)}(R_b)} -
\frac{F^{(3)}(R_b)^2}{F^{(2)}(R_b)^2}\right)
\nabla_\rho R_b \nabla^\rho R_b + \frac{R_b}{3} \nn
&& - \frac{F^{(1)}(R_b) F^{(3)}(R_b) R_b}{3 F^{(2)}(R_b)^2} -
\frac{F^{(1)}(R_b)}{3F^{(2)}(R_b)}
+ \frac{2 F(R_b) F^{(3)}(R_b)}{3 F^{(2)}(R_b)^2}
 - \frac{F^{(3)}(R_b) R_b}{3 F^{(2)}(R_b)^2} \, .
\eea
It is convenient to consider the case that $R_b$ and $R_p$ are uniform;
that is, they do not depend on the spatial coordinates. 
Thus, the d'Alembertian can be replaced with the
second derivative with respect to the time coordinate:
$\Box R_p \to - \partial_t^2 R_p$ and Eq.~(\ref{JGRG29}) has
the following structure:
\be
\label{JGRG30}
0=-\partial_t^2 R_p + U(R_b) R_p + \mathrm{const.}\, .
\ee
Then, if $U(R_b)>0$, $R_p$ becomes exponentially large with time $t$:
$R_p\sim \e^{\sqrt{U(R_b)} t}$ and the system is unstable.
In the $1/R$ model 
\cite{Capozziello:2002rd,Carroll:2003wy,Capozziello:2003tk},
because the order of the mass parameter $m_\mu$ is
\be
\label{JGRG31}
\mu^{-1}\sim 10^{18} \mbox{sec} \sim \left( 10^{-33} \mbox{eV} 
\right)^{-1}\, ,
\ee
one finds
\be
\label{JGRG32}
U(R_b) = - R_b + \frac{R_b^3}{6\mu^4} \sim \frac{R_0^3}{\mu^4}
\sim \left(10^{-26} \mbox{sec}\right)^{-2}
\left(\frac{\rho_\mathrm{matter}}{\mbox{g\,cm}^{-3}}\right)^3\, ,\quad
R_b \sim \left(10^3 \mbox{sec}\right)^{-2}
\left(\frac{\rho_\mathrm{matter}}{\mbox{g\,cm}^{-3}}\right)\, .
\ee
Thus, the model is unstable and it presumably decays in $10^{-26}$ sec 
(for a planet-sized object).
On the other hand, in a $1/R + R^2$ model \cite{Nojiri:2003ft}, one gets
\be
\label{JGRG33}
U(R_0)\sim \frac{R_0}{3}>0\, .
\ee
Then, the system could be unstable again, but the decay time is $\sim$
$1,000$ sec, that is, macroscopic.
For the model \cite{Hu:2007nk}, $U(R_b)$ is negative 
\be
\label{JGRG34}
U(R_0) \sim - \frac{(n+2)m^2 c_2^2}{c_1 n(n+1)} < 0 \, .
\ee
Therefore, there is no matter instability.
Note that the models considered in the following sections successfully pass the 
matter instability test.

\subsubsection{Unified theories without a flat space solution}

We now review the $F(R)$ models which do not have a flat space solution because
of the non-analytical behavior at zero curvature. Some of these 
such models may induce large corrections to Newton's law.

\

\noindent
{\it Modified gravity with negative and positive powers of the curvature}

In \cite{Nojiri:2003ft}, the following action was proposed:
\be
\label{RR12}
F(R)= R - \frac{c}{\left( R - \Lambda_1 \right)^n}
+ b \left( R - \Lambda_2\right)^m \, .
\ee
Here $n$, $m$, $c$, and $b$ are constants. In general, $n$ and $m$ may be
fractional. Let us show that the above model leads to an acceptable cosmic
speed-up and is consistent with some of the solar system tests.

For the action (\ref{RR12}), Eq.~(\ref{JGRG16}) has the following form:
\be
\label{RR17}
0 = R - \frac{c\left\{(n+2) R - 2 \Lambda_1
\right\}}{\left(R-\Lambda_1\right)^{n+1}}
+ b \left\{(2 - m) R - 2\Lambda_2\right\} \left(R-\Lambda_2\right)^m\, .
\ee
Especially when $n=1$, $m=2$, and $\Lambda_1=0$, one gets
\be
\label{RR18}
R=R_\pm = \frac{b \Lambda_2 \pm \sqrt{b^2 \Lambda_2^2
+ 3 c \left( 1 + 2 b \Lambda_2 \right)}}{ 1 + 2 b \Lambda_2 }\, .
\ee
If $b\Lambda_2>$, $c<0$, and $b^2 \Lambda_2^2
+ 3 c \left( 1 + 2 b \Lambda_2 \right) >0$,
both of the solutions express the de Sitter space.
Thus, the natural possibility for the unification of
early-time inflation with late-time acceleration appears.
Here, higher-derivative terms act in favor of early-time inflation, and 
the $1/R$ term supports a cosmic speed-up.

We now neglect the contribution from matter. When $n=1$, $m=2$,
and $\Lambda_1=\Lambda_2=0$ in (\ref{RR12})
and the curvature is small, we obtain the solution of (\ref{JGRG15}):
$\hat a \propto t^2$, which is consistent with the result
in \cite{Capozziello:2002rd,Carroll:2003wy,Capozziello:2003tk}.

If the present universe expands with the above power-law, the
curvature of the present universe should be small compared with that
of the de Sitter universe solution in (\ref{RR18}).
As the Hubble rate in the present universe is
$10^{-33}\, \mathrm{eV}$, the magnitude of the parameter
$c$, which corresponds to $\mu^4$ in 
\cite{Capozziello:2002rd,Carroll:2003wy,Capozziello:2003tk},
is on the order of $\left(10^{-33}\mathrm{eV}\right)^4$.

Let us consider the more general case that $F(R)$ is given by (\ref{RR12})
when the curvature is small and $\Lambda_1=0$.
Neglecting the contribution from the matter again, solving (\ref{JGRG15}),
we obtain $\hat a \propto t^{\frac{(n+1)(2n+1)}{n+2}}$ (see (\ref{JGRG17})).
It is quite remarkable that actually any negative power of the curvature
supports the cosmic acceleration. This gives the freedom in modification
of the model to achieve the better consistency with experimental tests of
Newtonian gravity.

On the other hand, when the scalar curvature $R$ is large, one obtains
$\hat a \propto t^{-\frac{(m-1)(2m-1)}{m-2}}$ (see (\ref{JGRG17})).
When $\frac{(m-1)(2m-1)}{m-2}>0$, the universe is shrinking but with a  
change in the arrow of time by $t\to t_s - t$, the inflation occurs with 
the inverse power-law.
At $t=t_s$, the size of the universe diverges. It is remarkable that
when $m$ is fractional (or irrational) and $1<m<2$,
the expression of $\hat a$ is still valid and the power
becomes positive, the universe evolves with the (fractional) power-law
expansion.

It is interesting that the model in (\ref{RR12}) is more
consistent that 
refs.~\cite{Capozziello:2002rd,Carroll:2003wy,Capozziello:2003tk}
(for a discussion of the Newtonian limit in $1/R$ theory, see 
\cite{Capozziello:2004sm})
as it may pass (at least, some) solar system tests. 
An example of the tests is matter instability as discusses 
in the previous subsection.
Indeed, in \cite{Dolgov:2003px}, small gravitational object like
the earth or the sun in the model 
\cite{Capozziello:2002rd,Carroll:2003wy,Capozziello:2003tk} is considered. 
It has been shown that the system becomes unstable.
This may cause an unacceptable force between distant 
galaxies \cite{Soussa:2003re,Woodard:2006nt}.
As shown in the previous subsection (see ref.~\cite{Nojiri:2003ft}), however, 
by adding the positive power (higher than $1$)
of the scalar curvature to the action, the instability could be 
significantly improved.
Furthermore, the account of quantum effects in modified gravity also acts
against instability \cite{Nojiri:2003wx}.

It has also been mentioned in \cite{Chiba:2003ir} that the $1/R$
model \cite{Capozziello:2002rd,Carroll:2003wy,Capozziello:2003tk}, which is
equivalent to scalar-tensor gravity, is ruled out as a realistic theory
due to the constraints to such Brans-Dicke-type theories.
In \cite{Nojiri:2003ft}, it was shown that by
adding the scalar curvature squared term to the action, the mass
of the scalar field can be adjusted to be very large
and the scalar field can decouple.
Thus, the modified gravity theory (\ref{RR12}) (after some
fine-tuning \cite{Nojiri:2003ft}) passes most of the solar system tests.
In addition, at precisely the same parameter values, the above modified
gravity has Newtonian limit that does not deviate significantly from
that in general relativity. Nevertheless, as has been mentioned at 
the very beginning, such theory has non-analytical behavior at zero (or some
constant) curvature which may be phenomenologically unacceptable.

%%%%%%%%%%
%%%%%%%%%%

\

\noindent
{\it $\ln R$ gravity}

Other gravitational alternatives for dark energy may be suggested along 
this same line.
As an extension of the theory (\ref{RR12}), one may consider the model 
\cite{Nojiri:2003ni} containing the logarithm of the scalar curvature
$R$:
\be
\label{RD15}
F(R)=R + \alpha' \ln \frac{R}{\mu^2} + \beta R^m\, .
\ee
We should note that the $m=2$ choice simplifies the model.
Assuming the scalar curvature is constant and the Ricci tensor is also
covariantly constant, the equations (\ref{JGRG16}) are as follows:
\be
\label{RD19}
0=2F(R) - RF'(R) = \tilde F(R) \equiv R
+ 2\alpha' \ln \frac{R}{\mu^2} - \alpha'\, .
\ee
If $\alpha'>0$, $\tilde F(R)$ is a monotonically increasing function and
$\lim_{R\to 0}\tilde F(R)=-\infty$ and $\lim_{R\to +\infty}\tilde
F(R)=+\infty$. There is one and only one solution of (\ref{RD19}). 
This solution may correspond to the inflation. 
On the other hand, if $\alpha'<0$, $\lim_{R\to 0}\tilde F(R)=\infty$
and $\lim_{R\to +\infty}=+\infty$. 
Because $\tilde F'(R) = 1 + \frac{\alpha'}{R}$, the minimum of $\tilde F(R)$, 
where $\tilde F'(R)=0$, is given by $R=-2\alpha'$. If $\tilde f(-2\alpha')>0$, 
there is no solution of (\ref{RD19}).
If $\tilde f(-2\alpha')=0$, there is only one solution and if $\tilde
f(-2\alpha')<0$, there are two solutions.
Because $\tilde f(-2\alpha')=-2\alpha'\left(1 - \ln \frac{-2\alpha'}{\mu^2}\right)$,
there are two solutions if $- \frac{2\alpha'}{\mu^2}>\e$.
Because the square root of the curvature $R$ corresponds to the rate of the
expansion of the universe, the larger of the two solutions might correspond to 
the inflation in the early universe and the smaller one to the present 
accelerating universe.

We can consider the late FRW cosmology when the scalar curvature $R$ is small.
Solving (\ref{JGRG15}), it follows that the power-law inflation could 
occur: $\hat a \propto t^{\frac{1}{2}}$.

One may discuss further generalizations like \cite{Nojiri:2003ni} (see 
also \cite{Meng:2003en})
\be
\label{GR1}
F(R)=R + \gamma R^{-n}\left(\ln \frac{R}{\mu^2}\right)^m\, .
\ee
Here, $n$ is restricted by $n>-1$ ($m$ is an arbitrary integer) in order that
the second term could be more dominant than the Einstein term
when $R$ is small.
For this model, we find
\be
\label{GR5}
\hat a\sim t^{\frac{(n+1)(2n+1)}{n+2}}\, .
\ee
This does not depend on $m$. The logarithmic factor is almost irrelevant.
The effective $w_\mathrm{eff}$ (\ref{JGRG12}) is given by
\be
\label{GR6}
w_\mathrm{eff}=-\frac{6n^2 + 7n - 1}{3(n+1)(2n+1)}\, .
\ee
Then, $w_\mathrm{eff}$ can be negative if
\be
\label{GR7}
 -1<n<-\frac{1}{2}\ \mbox{or}\
n>\frac{-7 + \sqrt{73}}{12}=0.1287\cdots \, .
\ee
 From (\ref{GR5}), the condition that the universe
could accelerate is given by $\frac{(n+1)(2n+1)}{n+2}>1$, that is:
\be
\label{GR8}
n> \frac{-1 + \sqrt{3}}{2}=0.366\cdots \, .
\ee
Clearly, the effective EoS parameter $w$ may be within the existing
bounds. Thus, the logarithmic model may also propose the unified 
description of inflation with dark energy era.

\subsubsection{Viable modified gravities}

Using the previous arguments, let us review the viable models of gravity that could
unify the accelerating expansion of the present universe and the inflation of 
the early universe.

\

\noindent
{\it The realistic $F(R)$ models unifying inflation with dark
energy}

In refs.~\cite{Nojiri:2007as,Nojiri:2007cq,Cognola:2007zu},
viable models of $F(R)$ gravity unifying the late-time acceleration 
and the inflation were proposed.
To construct these models, we have required several conditions:
\begin{enumerate}
\item To generate the inflation, one requires
\be
\label{Uf1}
\lim_{R\to\infty} f (R) = - \Lambda_i\, .
\ee
Here, $\Lambda_i$ is an effective cosmological constant at the early 
universe; therefore, it is natural to assume $\Lambda_i \gg 
\left(10^{-33}\mathrm{eV}\right)^2$.
For instance, it is natural to have $\Lambda_i\sim 10^{20 \sim 38}\left(
\mathrm{eV}\right)^2$.
\item To enable the current cosmic acceleration to be generated,
the current $f(R)$ gravity is considered to be a small constant, that is,
\be
\label{Uf3}
f(R_0)= - 2\tilde R_0\, ,\quad f'(R_0)\sim 0\, .
\ee
Here, $R_0$ is the current curvature 
$R_0\sim \left(10^{-33}\mathrm{eV}\right)^2$.
Note that $R_0> \tilde R_0$ due to the contribution from matter. 
In fact, if one can regard $f(R_0)$ as an effective cosmological constant, 
the effective Einstein equation gives 
$R_0=\tilde R_0 - \kappa^2 T_\mathrm{matter}$.
Here, $T_\mathrm{matter}$ is the trace of the matter energy-momentum tensor.
Note that $f'(R_0)$ need not vanish completely. 
Because we consider the time scale from one billion years to ten 
billion years, we only require
$\left| f'(R_0) \right| \ll \left(10^{-33}\,\mathrm{eV}\right)^4$.

Instead of the model corresponding to (\ref{Uf1}), one may consider a 
model that satisfies
\be
\label{UU2}
\lim_{R\to\infty} f (R) = \alpha R^m \, ,
\ee
with a positive integer $m>1$ and a constant $\alpha$.
To avoid the anti-gravity $f'(R)>-1$, $\alpha>0$ and, 
therefore, $f(R)$ should be positive at the early universe.
On the other hand, Eq.~(\ref{Uf3}) shows that $f(R)$ is negative at
the present universe. Therefore, $f(R)$ should cross zero in the past.
\item The last condition is
\be
\label{Uf4}
\lim_{R\to 0} f(R) = 0\, ,
\ee
which means that there is a flat space-time solution.
\end{enumerate}

One typical model proposed in ref.~\cite{Nojiri:2007as} that satisfies the
above conditions is
\be
\label{UU2d}
f(R)= \frac{\alpha R^{2n} - \beta R^n}{1 + \gamma R^n}\, .
\ee
Here, $\alpha$, $\beta$, and $\gamma$ are positive constants and $n$ is a
positive integer.
Eq.~(\ref{UU2d}) gives \cite{Nojiri:2007as}
\be
\label{UUU7}
R_0=\left\{ \left(\frac{1}{\gamma}\right)
\left(1+ \sqrt{ 1 + \frac{\beta\gamma}{\alpha} }\right)\right\}^{1/n}\, ,
\ee
and, therefore,
\be
\label{UU6}
f(R_0) \sim -2 \tilde R_0 = \frac{\alpha}{\gamma^2}
\left( 1 + \frac{\left(1 - \frac{\beta\gamma}{\alpha} \right)
\sqrt{ 1 + \frac{\beta\gamma}{\alpha}}}{2 + \sqrt{ 1 +
\frac{\beta\gamma}{\alpha}}} \right) \, .
\ee
Then, it follows that 
\be
\label{UU9}
\alpha \sim 2 \tilde R_0 R_0^{-2n}\, ,\quad \beta \sim 4 {\tilde R_0}^2
R_0^{-2n} R_I^{n-1}\, ,\quad
\gamma \sim 2 \tilde R_0 R_0^{-2n} R_I^{n-1}\, .
\ee
In the model (\ref{UU2d}), the correction to Newton's law is small
because the mass $m_\sigma$ (\ref{JGRG24}) is large and is given by
$m_\sigma^2 \sim 10^{-160 + 109 n}\,\mathrm{eV}^2$ in the solar system and
$m_\sigma^2 \sim 10^{-144 + 98 n}\,\mathrm{eV}^2$ in the air
on the earth \cite{Nojiri:2007as}.
In both cases, the mass $m_\sigma$ is very large if $n\geq 2$.

Another model corresponding to (\ref{Uf1}) was proposed in 
\cite{Cognola:2007zu}
\be
\label{tan7}
f(R) = -\alpha_0 \left( \tanh \left(\frac{b_0\left(R-R_0\right)}{2}\right)
+ \tanh \left(\frac{b_0 R_0}{2}\right)\right) %\nn %&&
 -\alpha_I \left( \tanh \left(\frac{b_I\left(R-R_I\right)}{2}\right)
+ \tanh \left(\frac{b_I R_I}{2}\right)\right)\, .
\ee
One now assumes
\be
\label{tan8}
R_I\gg R_0\, ,\quad \alpha_I \gg \alpha_0\, ,\quad b_I \ll b_0\, ,
\ee
and
\be
\label{tan8b}
b_I R_I \gg 1\, .
\ee
When $R\to 0$ or $R\ll R_0,\, R_I$, $f(R)$ behaves as
\be
\label{tan9}
f(R) \to - \left(\frac{\alpha_0 b_0 }{2\cosh^2 
\left(\frac{b_0 R_0}{2}\right) }
+ \frac{\alpha_I b_I }{2\cosh^2 \left(\frac{b_I R_I}{2}\right) }
\right)R\, .
\ee
and $f(0)=0$ again. When $R\gg R_I$, it follows that 
\be
\label{tan10}
f(R) \to - 2\Lambda_I \equiv -\alpha_0
\left( 1 + \tanh \left(\frac{b_0 R_0}{2}\right)\right) -\alpha_I
\left( 1 + \tanh \left(\frac{b_I R_I}{2}\right)\right)
\sim -\alpha_I \left( 1 + \tanh \left(\frac{b_I R_I}{2}\right)\right)\, .
\ee
On the other hand, when $R_0\ll R \ll R_I$, one gets
\be
\label{tan11}
f(R) \to -\alpha_0 \left[ 1 + \tanh \left(\frac{b_0 R_0}{2}\right)\right]
 - \frac{\alpha_I b_I R}{2\cosh^2 \left(\frac{b_I R_I}{2}\right) }
\sim -2\Lambda_0 \equiv -\alpha_0 \left[ 1 
+ \tanh \left(\frac{b_0 R_0}{2}\right)\right] \, .
\ee
Here, we have assumed the condition (\ref{tan8b}). One also finds
\be
\label{tan12}
f'(R)= - \frac{\alpha_0 b_0 }{2\cosh^2 \left(\frac{b_0 \left(R -
R_0\right)}{2}\right) }- \frac{\alpha_I b_I }{2\cosh^2 
\left(\frac{b_I \left(R - R_I\right)}{2}\right) }\, ,
\ee
which has two valleys when $R\sim R_0$ or $R\sim R_I$. When $R= R_0$, 
\be
\label{tan13}
f'(R_0)= - \alpha_0 b_0 - \frac{\alpha_I b_I }{2\cosh^2 
\left(\frac{b_I \left(R_0 - R_I\right)}{2}\right) }
> - \alpha_I b_I - \alpha_0 b_0 \, .
\ee
On the other hand, when $R=R_I$, it follows that 
\be
\label{tan14}
f'(R_I)= - \alpha_I b_I - \frac{\alpha_0 b_0 }{2\cosh^2
\left(\frac{b_0 \left(R_0 - R_I\right)}{2}\right) }
> - \alpha_I b_I - \alpha_0 b_0 \, .
\ee
Due to the condition (\ref{FR1}) required to avoid the anti-gravity regime, 
one obtains
\be
\label{tan15}
\alpha_I b_I + \alpha_0 b_0 < 2\, .
\ee
In the solar system domain, on or inside the earth, where $R\gg R_0$,
$f(R)$ can be approximated by
\be
\label{tan16}
f(R) \sim -2 \Lambda_\mathrm{eff} + 2\alpha \e^{-b(R-R_0)}\, .
\ee
On the other hand, because $R_0\ll R \ll R_I$, by assuming 
Eq.~(\ref{tan8b}), $f(R)$ (\ref{tan7}) could also be approximated by
\be
\label{tan17}
f(R) \sim -2 \Lambda_0 + 2\alpha \e^{-b_0(R-R_0)}\, ,
\ee
which has the same expression, after having identified 
$\Lambda_0 = \Lambda_\mathrm{eff}$ and $b_0=b$.
Thus, one may check the case (\ref{tan16}) only.
The effective mass has the following form
\be
\label{tan18}
m_\sigma^2 \sim \frac{\e^{b(R-R_0)}}{4\alpha b^2}\, ,
\ee
which could be very large, that is, $m_\sigma^2 \sim 
10^{1,000}\,\mathrm{eV}^2$
in the solar system and 
$m_\sigma^2 \sim 10^{10,000,000,000}\,\mathrm{eV}^2$ in the air
surrounding the earth. Thus, the correction to Newton's law becomes
negligible.

One may consider another model \cite{Nojiri:2007as}:
\be
\label{VI}
f(R) = - \frac{\left(R - R_0\right)^{2k+1} + R_0^{2k+1}}{f_0 + f_1 
\left\{\left(R - R_0\right)^{2k+1} + R_0^{2k+1} \right\}}\, .
\ee
It has been shown \cite{Nojiri:2007as} that for $k\geq 10$, such modified
gravity passes the local tests. 
It also unifies the early-time inflation with the dark energy epoch.
In (\ref{VI}), $R_0$ is the current curvature 
$R_0\sim \left(10^{-33}\,\mathrm{eV}\right)^2$.
We also require
\be
\label{Uf7}
f_0 \sim \frac{R_0^{2n}}{2} \, ,\quad
f_1=\frac{1}{\Lambda_i}\, .
\ee
Here $\Lambda_i$ is the effective cosmological constant in the inflation 
epoch.
When $R\gg \Lambda_i$, $f(R)$ (\ref{VI}) behaves as
\be
\label{VIII}
f(R) \sim - \frac{1}{f_1} + \frac{f_0}{f_1^2 R^{2n+1}}\, .
\ee
The trace equation, which is the trace part of (\ref{JGRG13}), 
is as follows:
\be
\label{Scalaron}
3\Box f'(R)= R+2f(R)-Rf'(R)-\kappa^2 T_\mathrm{matter}\, .
\ee
Here, $T_\mathrm{matter}$ is the trace part of the matter energy-momentum 
tensor.
For the FRW metric (\ref{JGRG14}), one finds
\be
\label{R}
R \sim \left(t_s - t\right)^{-2/\left(2n+3\right)}\, ,
\ee
which diverges at a finite future time $t=t_s$.
By a similar analysis, we can show that if $f(R)$ behaves as $f(R) \sim
R^\epsilon$ for large $R$ with a constant $\epsilon$, a future singularity 
appears if $\epsilon>2$ or $\epsilon<0$, which is consistent with 
the analysis (\ref{HS4}) via the scalar-tensor form of the action.
Conversely if $2\geq \epsilon \geq 0$, the singularity does not appear.
Thus, adding the term $R^2 \tilde f(R)$, where 
$\lim_{R\to 0} \tilde f(R) = c_1$, $\lim_{R\to \infty} \tilde f(R) = c_2$, 
to $f(R)$ (\ref{VI}), the future singularity (\ref{R}) disappears. 
More details about future singularities in modified gravity will be given 
in the fourth chapter.

Let us investigate the above situation in more detail. Assume
\be
\label{X}
f(R) \sim F_0 + F_1 R^\epsilon\, ,
\ee
when $R$ is large. Here $F_0$ and $F_1$ are constants where $F_0$ may
vanish but $F_1\neq 0$.
In the case of (\ref{VIII})
\be
\label{XI}
F_0 = - \frac{1}{f_1} \, ,\quad
F_1 = \frac{f_0}{f_1^2} \, ,\quad
\epsilon = -\left(2n+1\right) \, .
\ee
Under the assumption (\ref{X}), the trace equation (\ref{Scalaron}) gives
\be
\label{XII}
3 F_1 \Box R^{\epsilon -1} = \left\{
\begin{array}{ll} R & \ \mbox{when $\epsilon<0$ or $\epsilon=2$} \\
\left(2-\epsilon\right) F_1 R^\epsilon & \ \mbox{when $\epsilon>1$ or
$\epsilon\neq 2$}
\end{array} \right. \, .
\ee
In the FRW background (\ref{JGRG14}), 
when the Hubble rate has a singularity as
\be
\label{XIII}
H \sim \frac{h_0}{\left(t_s - t\right)^\beta}\, ,
\ee
with constants $h_0$ and $\beta$, the scalar curvature 
$R=6\dot H + 12 H^2$ behaves as
\be
\label{XIV}
R \sim \left\{ \begin{array}{ll}
\frac{12h_0^2}{\left(t_s - t\right)^{2\beta}} & \ \mbox{when $\beta>1$} \\
\frac{6 h_0 + 12 h_0^2}{\left(t_s - t\right)^2} & \ \mbox{when $\beta=1$} \\
\frac{6\beta h_0}{\left(t_s - t\right)^{\beta + 1}} & \ \mbox{when 
$\beta<1$}
\end{array} \right. \, .
\ee
In (\ref{XIII}) or (\ref{XIV}), the $\beta\geq 1$ case corresponds to
Type I (Big Rip) singularity in \cite{Caldwell:2003vq,ref5},
$1>\beta>0$ to Type III, $0>\beta>-1$ to Type II, and $\beta<-1$
(but $\beta\neq\mbox{integer}$) to Type IV.

The classification of the finite-time future singularities used above is
given in ref.~\cite{Nojiri:2005sx}:
\begin{itemize}
\item Type I (``Big Rip'') : For $t \to t_s$, $a \to \infty$,
$\rho_\mathrm{eff} \to \infty$ and $\left|p_\mathrm{eff}\right| \to 
\infty$.
This also includes the case of $\rho_\mathrm{eff}$, $p_\mathrm{eff}$ being
finite at $t_s$.
This singularity has been discovered in ref.~\cite{Caldwell:2003vq}. Its
manifestations in different models have been studied in 
refs.~\cite{Nojiri:2005sx}.
\item Type II (``sudden'') \cite{barrow}: For $t \to t_s$, $a \to a_s$,
$\rho_\mathrm{eff} \to \rho_s$ and 
$\left|p_\mathrm{eff}\right| \to \infty$.
\item Type III : For $t \to t_s$, $a \to a_s$,
$\rho_\mathrm{eff} \to \infty$ and 
$\left|p_\mathrm{eff}\right| \to \infty$.
\item Type IV : For $t \to t_s$, $a \to a_s$,
$\rho_\mathrm{eff} \to 0$, $\left|p_\mathrm{eff}\right| \to 0$ and higher
derivatives of $H$ diverge.
This also includes the case in which $p_\mathrm{eff}$ 
($\rho_\mathrm{eff}$) or both of $p_\mathrm{eff}$ and $\rho_\mathrm{eff}$
tend to some finite values, whereas higher derivatives of $H$ diverge.
\end{itemize}
Here, $\rho_\mathrm{eff} $ and $p_\mathrm{eff}$ are defined by
\be
\label{IV}
\rho_\mathrm{eff} \equiv \frac{3}{\kappa^2} H^2 \, , \quad
p_\mathrm{eff} \equiv - \frac{1}{\kappa^2} \left( 2\dot H + 3 H^2 
\right)\, .
\ee
Substituting (\ref{XIV}) into (\ref{XII}), one finds that there are two
classes of consistent solutions.
The first solution is specified by $\beta =1$ and $\epsilon>1$ (but
$\epsilon\neq 2$) case, which corresponds to the Big Rip
($h_0>0$ and $t<t_s$) or Big Bang ($h_0<0$ and $t>t_s$) singularity at 
$t=t_s$.
Another solution is $\epsilon<1$, and
$\beta = - \epsilon/\left(\epsilon - 2\right)$ ($-1<\beta<1$) case, which
corresponds to (\ref{R}) and to the Type II future singularity.
In fact, we find that $\epsilon = - 2n -1$ and, therefore, 
$\beta + 1 = - 2/(2n+3)$.
We should note that when $\epsilon=2$, that is, $f(R) \sim R^2$,
there is no singular solution.
Therefore, if the above term $R^2 \tilde f(R)$, where 
$\lim_{R\to 0} \tilde f(R) = c_1$,
$\lim_{R\to \infty} \tilde f(R) = c_2$, is added to $f(R)$ in (\ref{VI}),
the added term dominates, and the 
modified $f(R)$ behaves as $f(R)\sim R^2$.
The future singularity (\ref{R}) disappears.
We also note that if the $R^n$-term with $n=3,4,5,\cdots$ is added, the
singularity becomes (in some sense)
worse because this case corresponds to $\epsilon=n>1$, that is the Big Rip case.
Using the potential that appears when we transform $F(R)$ gravity to
scalar-tensor theory \cite{Nojiri:2008fk}, it has been found that
the future singularity may not appear in the case where $0<\epsilon<2$.

Thus, to avoid the finite-time future singularity, one has to add the 
$R^2$-term to the above viable unification models. 
\bea
\label{UU2dR2}
f(R) &=& \frac{\alpha R^{2n} - \beta R^n}{1 + \gamma R^n} + c R^2\, ,\\
\label{tan7R2}
f(R) &=& -\alpha_0 \left( \tanh 
\left(\frac{b_0\left(R-R_0\right)}{2}\right)
+ \tanh \left(\frac{b_0 R_0}{2}\right)\right) \nn
&& -\alpha_I \left( \tanh \left(\frac{b_I\left(R-R_I\right)}{2}\right)
+ \tanh \left(\frac{b_I R_I}{2}\right)\right) + c R^2\, , \\
\label{VIR2}
f(R) &=& - \frac{\left(R - R_0\right)^{2k+1} + R_0^{2k+1}}{f_0 + f_1
\left\{\left(R - R_0\right)^{2k+1} + R_0^{2k+1} \right\}} + c R^2\, .
\eea
The addition of this term was proposed first in ref.~\cite{Abdalla:2004sw}, 
where it was shown that, in this case, the Big Rip singularity disappears.
Moreover, such a term, which effectively eliminates future singularity also 
supports the early-time inflation.
In other words, adding such a $R^2$ term to the gravitational dark energy 
model (for instance, to viable dark energy models \cite{Hu:2007nk,others}) may
lead to the emergence of an inflationary phase in the model as was first 
observed in ref.~\cite{Nojiri:2003ft}. 
In the case when a model already contains the inflationary era, 
its dynamics will be changed by the $R^2$-term.
The investigation showing that the $R^2$-term cures all types of future 
singularities was carried out in
refs.~\cite{Nojiri:2008fk,Bamba:2008ut,Capozziello:2009hc,Bamba:2009uf}.
In fact, it has been recently realized that some phenomenological problems
\cite{Kobayashi:2008wc,Sami:2009jx,Thongkool:2009js,Babichev:2009fi,Appleby:2010dx}
of $F(R)$ dark energy (like achieving a consistent description of
neutron stars) may be resolved in the presence of the $R^2$ term.
Moreover, the traditional phantom/quintessence (fluid/scalar) dark energy
models often bring the universe to a finite-time future singularity.
It was demonstrated in ref.~\cite{Nojiri:2009pf}
that the natural method for eliminating the singularity in these models is again
the addition of the $R^2$ term. In other words, resolving future 
singularities in specific dark energy
models requires (at least partly) modified gravity. 
We discuss these questions in more detail in the fourth chapter. Note also
that in the above models, there is no matter instability (Eq.~(\ref{JGRG30})).

It should be noted that the exponential-type model may represent a very
interesting proposal for the resolution of the dark energy problem.
Some versions of such a theory are
\be
\label{exp1}
F(R) = R + \alpha \left(\e^{-bR} - 1 \right)
\ \mbox{or}\ F(R)
= R - \alpha \left(\frac{\e^{bR} - 1}{\e^{bR} + \e^{b R_0}} \right)\, ,
\ee
with constants $\alpha$, $b$, and $R_0$. This model, which was introduced 
in ref.~\cite{Cognola:2007zu} provides the accelerating cosmological 
solutions without a future singularity. As a rule, the accelerating late-time 
cosmology leads asymptotically to a de Sitter universe.
It was shown \cite{Yang} that such a model describes a realistic dark energy 
epoch that is compatible with local and observational tests. 
Moreover, the number of free parameters is not big. 
Of course, one can consider different modifications of the above model.

\

\noindent
{\it A more complicated but viable $F(R)$ model}

\begin{figure}%[b]

\begin{center}

\unitlength=1mm

\begin{picture}(120,100)

\put(7,7){\makebox(0,0){$0$}}
\put(10,85){\makebox(0,0){$\frac{F(R)}{R^2}$}}
\put(115,10){\makebox(0,0){$R$}}
\put(5,25){\makebox(0,0){$f_\infty$}}
\put(20,5){\makebox(0,0){$R_e$}}
\put(40,5){\makebox(0,0){$R_L$}}
\put(70,5){\makebox(0,0){$R_I$}}

\thicklines

\put(10,10){\vector(1,0){100}}
\put(10,10){\vector(0,1){70}}

\qbezier(11,80)(11,50)(20,50)
\qbezier(20,50)(25,50)(30,55)
\qbezier(30,55)(35,60)(40,60)
\qbezier(40,60)(45,60)(50,55)
\qbezier(50,55)(60,45)(70,45)
\qbezier(70,45)(80,45)(90,35)
\qbezier(90,35)(99,26)(110,26)

\thinlines

\put(10,25){\line(1,0){100}}
\put(20,10){\line(0,1){40}}
\put(40,10){\line(0,1){50}}
\put(70,10){\line(0,1){35}}

\end{picture}

\end{center}

\caption{\label{fig1} The qualitative behavior of $\frac{F(R)}{R^2}$
versus $R$ in a viable model.}
\end{figure}
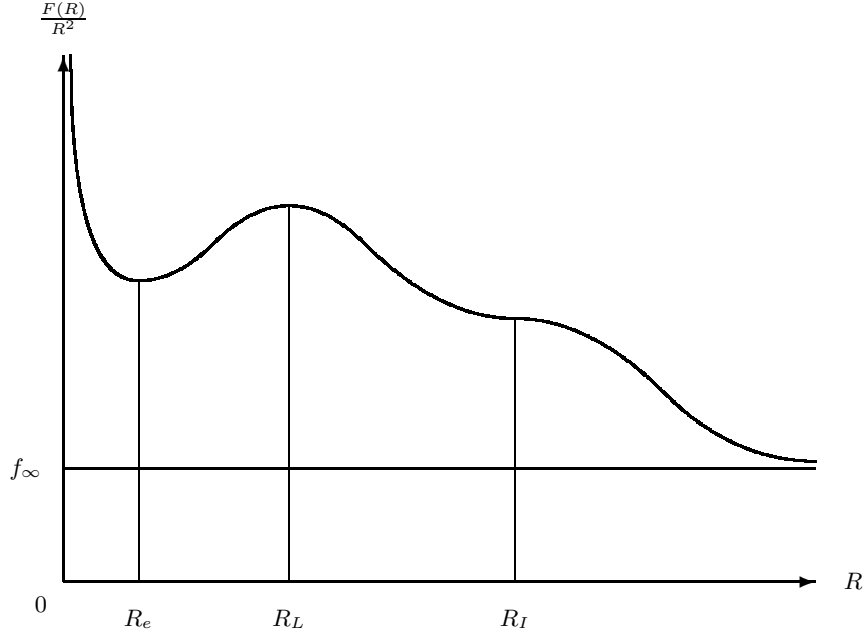

The above analysis shows that, to obtain a realistic and viable model,
$F(R)$ gravity should satisfy the following conditions:
\begin{enumerate}
\item\label{req1} When $R\to 0$,
the Einstein gravity is recovered, that is,
\be
\label{E1}
F(R) \to R \quad \mbox{that is,} \quad \frac{F(R)}{R^2} 
\to \frac{1}{R}\, .
\ee
This also means that there is a flat space solution as in (\ref{Uf4}).
\item\label{req2} As will be discussed after Eq.~(\ref{dS6}),
there appears a stable de Sitter solution,
which corresponds to the late-time acceleration and, therefore, the 
curvature is small
$R\sim R_L \sim \left( 10^{-33}\, \mathrm{eV}\right)^2$.
This requires, when $R\sim R_L$,
\be
\label{E2}
\frac{F(R)}{R^2} = f_{0L} - f_{1L} \left( R - R_L \right)^{2n+2}
+ o \left( \left( R - R_L \right)^{2n+2} \right)\, .
\ee
Here, $f_{0L}$ and $f_{1L}$ are positive constants and $n$ is a positive
integer. Of course, in some cases this condition may not be strictly 
necessary.
One can use it, for instance, to avoid a future singularity.
\item\label{req3} As will also be discussed after Eq.~(\ref{dS6}),
there appears a quasi-stable de Sitter solution that corresponds 
to the inflation of the early universe and, therefore, the curvature is large
$R\sim R_I \sim \left( 10^{16 \sim 19}\, \mathrm{GeV}\right)^2$. 
The de Sitter space should not be exactly stable so that the curvature decreases very slowly. It requires
\be
\label{E3}
\frac{F(R)}{R^2} = f_{0I} - f_{1I} \left( R - R_I \right)^{2m+1}
+ o \left( \left( R - R_I \right)^{2m+1} \right)\, .
\ee
Here, $f_{0I}$ and $f_{1I}$ are positive constants and $m$ is a positive
integer.
\item\label{req4} Following the discussion after (\ref{HS3}), when $R\to
\infty$, to avoid the curvature singularity, it is proposed that 
\be
\label{E4}
F(R) \to f_\infty R^2 \quad \mbox{that is} \quad \frac{F(R)}{R^2} \to 
f_\infty
\, .
\ee
Here, $f_\infty$ is a positive and sufficiently small constant.
Instead of (\ref{E4}), we may take
\be
\label{E5}
F(R) \to f_{\tilde \infty} R^{2 - \epsilon} \quad \mbox{that is}
\quad \frac{F(R)}{R^2} \to \frac{f_{\tilde\infty}}{R^\epsilon} \, .
\ee
Here, $f_{\tilde\infty}$ is a positive constant and $0< \epsilon <1$.
The above condition (\ref{E4}) or (\ref{E5}) prevents both the future
singularity and the singularity due to large density of matter.
\item\label{req5} As in (\ref{FR1}), to avoid the anti-gravity, 
we require
%%%%
%%%%%
%%%%%
\be
\label{E6}
F'(R)>0\, ,
\ee
which is rewritten as
\be
\label{E7}
\frac{d}{dR} \left( \ln \left( \frac{F(R)}{R^2} \right)\right)
> - \frac{2}{R}\, .
\ee
\item\label{req6}
Combining conditions (\ref{E1}) and (\ref{E6}), one finds
\be
\label{E8}
F(R)>0\, .
\ee
\item To avoid the matter instability in (\ref{JGRG30})
\cite{Dolgov:2003px},
we require 
\be
\label{E8B}
U(R_b) \equiv \frac{R_b}{3} - \frac{F^{(1)}(R_b) F^{(3)}(R_b) R_b}
{3 F^{(2)}(R_b)^2} - \frac{F^{(1)}(R_b)}{3F^{(2)}(R_b)}
+ \frac{2 F(R_b) F^{(3)}(R_b)}{3 F^{(2)}(R_b)^2} - \frac{F^{(3)}(R_b) R_b}{3
F^{(2)}(R_b)^2}< 0\, .
\ee
\end{enumerate}
The conditions \ref{req1} and \ref{req2} show that an extra, unstable de Sitter 
solution must appear at $R=R_e$ $\left( 0< R_e < R_L \right)$ 
(see Fig.\ref{fig1}).
The universe evolution will stop at $R=R_L$ because the de Sitter
solution $R=R_L$ is stable; the curvature never becomes smaller than $R_L$ 
and, therefore, the extra de Sitter solution is not realized.
The behavior of $\frac{F(R)}{R^2}$, which satisfies the conditions 
(\ref{E1}), (\ref{E2}), (\ref{E3}), (\ref{E4}), and (\ref{E8}) 
is given in Fig.\ref{fig1}.

An example of such $F(R)$ gravity is given in \cite{Nojiri:2010ny}
\bea
\label{EE1}
\frac{F(R)}{R^2} &=& 
\left\{ \left(X_m \left(R_I;R\right) - X_m\left(R_I;R_1\right) \right)
\left(X_m\left(R_I;R\right) - X_m\left(R_I;R_L\right) \right)^{2n+2} 
\right.
\nn
&& \left. + X_m\left(R_I;R_1\right) X_m\left(R_I;R_L\right)^{2n+2}
+ f_\infty^{2n+3} \right\}^{\frac{1}{2n+3}} \, , \nn
X_m\left(R_I;R\right) &\equiv & 
\frac{\left(2m +1\right) R_I^{2m}}{\left( R - R_I \right)^{2m+1} 
+ R_I^{2m+1}} \, .
\eea
Here, $n$ and $m$ are integers greater or equal to unity, and $n,m\geq 1$ 
and $R_1$ is a parameter related with $R_e$ by
\be
\label{EE2}
X\left(R_I;R_e\right) = \frac{\left(2n+2\right)
X\left(R_I;R_1\right)X\left(R_I;R_1\right) + X\left(R_I;R_L\right)}{2n+3}\, .
\ee
It is assumed that
%%%%%%%%%
%%%%%%%%%
%%%%%%%%%
%%%%%%%%%
\be
\label{EEE1}
0<R_1<R_L \ll R_I \, .
\ee
$X\left(R_I;R\right)$ is a monotonically decreasing function of $R$
and, in the limit $R\to 0$, $X\left(R_I;R\right)$ behaves as
\be
\label{EE3}
X\left(R_I;R\right) \to \frac{1}{R}\, ,
\ee
which shows that in the limit $R\to 0$, $F(R)$ (\ref{EE1}) reproduces 
Eq.~(\ref{E1}) and, therefore, the condition \ref{req1} is satisfied.
On the other hand, when $R\to \infty$,
\be
\label{EE4}
X\left(R_I;R\right) \to \frac{\left(2m+1\right) R_I^{2m}}{R^{2m+1}} \to 
0\, ,
\ee
and, therefore, $F(R)$ behaves as (\ref{E4}) and the condition \ref{req4}
is satisfied.

When $R\sim R_L$, we find the behavior of (\ref{E2}), where
\bea
\label{EE5}
f_{0L} &=& \left\{ X_m\left(R_I;R_1\right) X_m\left(R_I;R_L\right)^{2n+2}
+ f_\infty^{2n+3} \right\}^{\frac{1}{2n+3}} \, , \nn
f_{1L} &=& \frac{1}{2m+3} \left\{ X_m\left(R_I;R_1\right)
X_m\left(R_I;R_L\right)^{2n+2}
+ f_\infty^{2n+3} \right\}^{- \frac{2\left(n+1\right)}{2n+3}}
\left(X_m \left(R_I;R_1\right) - X_m\left(R_I;R_L\right) \right) \nn
&& \times \frac{\left( 2m + 1 \right)^{4\left(m+1\right)}
\left\{ R_I \left(R_L - R_I\right)\right\}^{4m\left(m+1\right)} }
{\left\{ \left( R_L - R_I \right)^{2m+1} + R_I^{2m+1}
\right\}^{4\left(m+1\right)}} \, .
\eea
Then, the condition \ref{req2} is satisfied.

On the other hand, when $R\sim R_I$, we also find the behavior of 
(\ref{E3}), where
\bea
\label{EE6}
f_{0I} &=& \left\{ 
\left(X_m \left(R_I;R_I\right) - X_m\left(R_I;R_1\right) \right) 
\left(X_m\left(R_I;R_I\right) - X_m\left(R_I;R_L\right) \right)^{2n+2} 
\right. \nn
&& \left. + X_m\left(R_I;R_1\right) X_m\left(R_I;R_L\right) ^{2n+2}
+ f_\infty^{2n+3} \right\}^{\frac{1}{2n+3}} \nn
f_{1I} &=& \frac{2m+1}{R_I^{2m+2}}
\left\{ \left(X_m \left(R_I;R_I\right) - X_m\left(R_I;R_1\right) \right)
\left(X_m\left(R_I;R_I\right) - X_m\left(R_I;R_L\right) \right)^{2n+2} 
\right.
\nn
&& \left. + X_m\left(R_I;R_1\right) X_m\left(R_I;R_L\right)^{2n+2}
+ f_\infty^{2n+3} \right\}^{\frac{2\left(n+1\right)}{2n+3}} \left\{
\left(X_m\left(R_I;R_I\right) - X_m\left(R_I;R_L\right) \right)^{2n+2} 
\right. \nn
&& \left. + \left(2n+2\right)\left(X_m \left(R_I;R_I\right) -
X_m\left(R_I;R_1\right) \right)
\left(X_m\left(R_I;R_I\right) - X_m\left(R_I;R_L\right) \right)^{2n+1}
\right\}\, .
\eea
Thus, the condition \ref{req3} is satisfied.

Let us now investigate the mass of the scalar field $\sigma$ to 
determine whether or not the Chameleon mechanism 
\cite{Khoury:2003rn,Brax:2004qh} works or not. 
For this purpose, one investigates the region
$R_1<R_L \ll R \ll R_I$. In this region, $X_m\left(R_I;R\right)$ can be
approximated as
\be
\label{EE7}
X_m\left(R_I;R\right) \sim \frac{1}{R}\, ,\quad
X_m\left(R_I;R_1\right) \sim \frac{1}{R_1}\, ,\quad
X_m\left(R_I;R_L\right) \sim \frac{1}{R_L}\, .
\ee
Then< $F(R)$ can be approximated as
\be
\label{EE8}
\frac{F(R)}{R^2} \sim f_\infty + \frac{f_n}{R}\, ,\quad
f_n \equiv \frac{R_1 + \left(2n+2\right) R_L f_\infty^{-2n -2}}
{\left(2n+3\right) R_1 R_L^{2n+2}}
\sim \frac{f_\infty^{-2n -2}}{R_L^{2n+2}}\, .
\ee
In the last equation, it was assumed that $R_1\sim R_L$.
If $f_\infty R \ll f_n$ or
\be
\label{EE9}
\frac{1}{f_\infty^{2n+3}} \gg R R_L^{2n+2}\, ,
\ee
one gets
\be
\label{EE10}
m_\sigma^2 \sim \frac{3}{4f_\infty}\, .
\ee
If we consider the region inside the earth, because 
$1\,{g}\sim 6\times 10^{32}\,\mathrm{eV}$
and $1\,\mathrm{cm}\sim \left(2\times 10^{-5}\,\mathrm{eV}\right)^{-1}$,
the density is about
$\rho_\mathrm{matter}\sim 1 \mathrm{g/cm^3} \sim 5\times 10^{18}\,
\mathrm{eV}^4$.
This shows that the magnitude of the curvature is
$R \sim \kappa^2 \rho_\mathrm{matter} \sim
\left(10^{-19}\,\mathrm{eV}\right)^2$.
In the air on the earth, one finds $\rho_\mathrm{matter}\sim 10^{-6}
\mathrm{g/cm^3} \sim 10^{12}\, \mathrm{eV}^4$,
which gives $R_0 \sim \kappa^2 \rho_\mathrm{matter} \sim
\left(10^{-25}\,\mathrm{eV}\right)^2$.
In the solar system, there could be interstellar gas. Typically, in the
interstellar gas, there is one proton (or hydrogen atom) per
$1\,\mathrm{cm}^3$, which shows 
$\rho_\mathrm{matter}\sim 10^{-5}\,\mathrm{eV}^4$, $R_0\sim 
10^{-61}\, \mathrm{eV}^2$.
Then, the condition (\ref{EE9}) can be easily satisfied; for example, we 
can choose $\frac{1}{f_\infty}\sim \mathrm{MeV}^2$. Then, the Compton length
of the scalar field becomes very small, and the correction to Newton's 
law is negligible.

One can now check the matter (in)stability issue \cite{Dolgov:2003px}.
In the earth or sun, the condition $R_1<R_L \ll R \ll R_I$ is satisfied 
and, therefore, $F(R)$ behaves as (\ref{EE8}). Then, $U(R_b)$ 
in (\ref{E8B}) is given by
\be
\label{URb}
U(R_b) \sim - \frac{f_n}{6f_\infty} < 0\, .
\ee
Because $U(R_b)$ is negative, the matter instability does not occur in the 
model under consideration.

The condition \ref{req5} remain to be checked. 
Because $F(R)$ (\ref{EE1}) satisfies Eqs.~(\ref{E1}) and (\ref{E4}), it is 
clear that Eq.~(\ref{E6}) or (\ref{E7}) and, therefore, 
the condition \ref{req5} are satisfied when $R\to 0$ or
$R\to \infty$. Because $\frac{F(R)}{R^2}$ is a monotonically increasing 
function of $R$ in the region $R_e<R<R_L$, Eq.~(\ref{E6}) or (\ref{E7}) 
is trivially satisfied in the region.
In the region $R_1<R_L \ll R \ll R_I$, because $\frac{F(R)}{R^2}$ behaves as
(\ref{EE8}), Eq.~(\ref{E6}) or (\ref{E7}) is again satisfied.
Then, the condition \ref{req5} seems to be satisfied throughout the whole region.

After the inflation at $R=R_I$, the radiation and the matter are 
generated.
If the matter and radiation energy densities dominate compared with
the contribution from $f(R)=F(R) - R$, that is, only the first term in
(\ref{Cr4}) dominates, a radiation/matter dominated-universe will be realized.
The radiation and matter densities decrease rapidly compared with
the contribution from the $f(R)$ term at late-time universe. 
When curvature arrives at $R\sim R_L$, the late-time acceleration occurs.

The condition (\ref{E5}) indicates that the $R^2$ term appears in the 
very high curvature region.
The $R^2$ term will generate an inflation besides the inflation at 
$R=R_I$ if the universe started with a very high curvature $R\gg R_I$. 
Because the inflation due to the $R^2$-term is unstable, the inflation 
at the very early stage would stop; however when the curvature 
decreases and reaches $R\sim R_I$, the inflationary phase 
will occur again.

Thus, we demonstrated that a number of viable $F(R)$ gravity models may 
explain the early-time inflation in addition to the dark energy epoch 
in a unified way. It will be shown in the third chapter that when such 
a unification is problematic, one can use
cosmological (partial/complete) reconstruction to ensure it.

\subsection{$f(\mathcal{G})$ gravity \label{IIB}}

\subsubsection{General properties}

Let us consider another class of modified gravity where the arbitrary 
function added to the action of the general relativity depends on 
topological Gauss-Bonnet invariant. 
It has been shown recently that such a class of modified gravity is
closely related with (super)string theory.
The starting action is given by 
\cite{Nojiri:2005jg,Nojiri:2005am,Cognola:2006eg}:
\be
\label{GB1b}
S=\int d^4x\sqrt{-g} \left(\frac{1}{2\kappa^2}R + f(\mathcal{G}) 
+ \mathcal{L}_\mathrm{matter}\right)\, .
\ee
Here, $\mathcal{L}_\mathrm{matter}$ is the Lagrangian density of matter and
$\mathcal{G}$ is the Gauss-Bonnet invariant:
\be
\label{GB}
\mathcal{G}=R^2 -4 R_{\mu\nu} R^{\mu\nu} + R_{\mu\nu\xi\sigma}
R^{\mu\nu\xi\sigma}\, .
\ee
By the variation of the metric $g_{\mu\nu}$, we obtain
\bea
\label{GB4b} && 0= \frac{1}{2\kappa^2}\left(- R^{\mu\nu} 
+ \frac{1}{2} g^{\mu\nu} R\right) + T_\mathrm{matter}^{\mu\nu} 
+ \frac{1}{2}g^{\mu\nu} f(\mathcal{G}) -2 f'(\mathcal{G}) R R^{\mu\nu} \nn
&& + 4f'(\mathcal{G})R^\mu_{\ \rho}
R^{\nu\rho} -2 f'(\mathcal{G}) R^{\mu\rho\sigma\tau}
R^\nu_{\ \rho\sigma\tau} - 4 f'(\mathcal{G}) 
R^{\mu\rho\sigma\nu}R_{\rho\sigma}
+ 2 \left( \nabla^\mu \nabla^\nu f'(\mathcal{G})\right)R \nn
&& - 2 g^{\mu\nu} \left( \nabla^2 f'(\mathcal{G})\right)
R - 4 \left( \nabla_\rho \nabla^\mu f'(\mathcal{G})\right)
R^{\nu\rho} - 4 \left( \nabla_\rho \nabla^\nu 
f'(\mathcal{G})\right)R^{\mu\rho} \nn
&& + 4 \left( \nabla^2 f'(\mathcal{G}) \right)R^{\mu\nu} + 4g^{\mu\nu} 
\left( \nabla_{\rho} \nabla_\sigma f'(\mathcal{G}) \right)
R^{\rho\sigma} - 4 \left(\nabla_\rho \nabla_\sigma f'(\mathcal{G}) \right)
R^{\mu\rho\nu\sigma} \, .
\eea
It is interesting to note that some memory of the fact that the Gauss-Bonnet 
term is topological invariant is contained in the above equation of motion.
Indeed, it will be seen that such an equation of motion does not contain the
higher-derivative terms as, for instance, the theory in the previous section.

Choosing the spatially flat FRW universe metric (\ref{JGRG14}),
the equation corresponding to the first FRW equation has the following 
form:
\be
\label{GB7b}
0=-\frac{3}{\kappa^2}H^2 - f(\mathcal{G})
+ \mathcal{G}f'(\mathcal{G}) - 24 \dot{\mathcal{G}}f''(\mathcal{G}) H^3
+ \rho_\mathrm{matter}\, .
\ee
In the FRW universe (\ref{JGRG14}), $\mathcal{G}$ looks like:
\be
\label{mGB000}
\mathcal{G} = 24 \left( H^2 \dot H + H^4 \right)\, .
\ee
Then, from Eq.~(\ref{GB7b}), the FRW like equations, as in the
Einstein gravity case (\ref{JGRG11}), we find 
\be
\label{mGB1BB}
\rho^\mathcal{G}_\mathrm{eff}=\frac{3}{\kappa^2}H^2 \, ,
\quad p^\mathcal{G}_\mathrm{eff}= - \frac{1}{\kappa^2}\left(3H^2 
+ 2\dot H\right)\, .
\ee
Here, 
\bea
\label{mGB2BB}
\rho^\mathcal{G}_\mathrm{eff} &\equiv& - f(\mathcal{G}) +
\mathcal{G}f'(\mathcal{G}) - 24 \dot{\mathcal{G}}f''(\mathcal{G}) H^3
+ \rho_\mathrm{matter}\, , \nn
p^\mathcal{G}_\mathrm{eff} &\equiv& 
f(\mathcal{G}) - \mathcal{G}f'(\mathcal{G})
+ \frac{2 \mathcal{G} \dot{\mathcal{G}}}{3H} f''(\mathcal{G})
+ 8 H^2 \ddot{\mathcal{G}} f''(\mathcal{G})
+ 8 H^2 {\dot{\mathcal{G}}}^2 f'''(\mathcal{G}) + p_\mathrm{matter}\, .
\eea
When $\rho_\mathrm{matter}=0$, Eq.~(\ref{GB7b}) has a de Sitter universe
solution where $H$, and therefore $\mathcal{G}$, are constant. For 
$H=H_0$, with a constant $H_0$, Eq.~(\ref{GB7b}) turns into
\be
\label{GB7bb}
0=-\frac{3}{\kappa^2}H_0^2 + 24H_0^4
f'\left( 24H_0^4 \right) - f\left( 24H_0^4\right) \, .
\ee
As an example, we consider the model
\be
\label{mGB1}
f(\mathcal{G})=f_0\left|\mathcal{G}\right|^\beta\, ,
\ee
with constants $f_0$ and $\beta$. Then, the solution of Eq.~(\ref{GB7bb}) is
given by
\be
\label{GBGB2}
H_0^4 = \frac{1}{24 \left( 8 \left(n-1\right) \kappa^2 f_0
\right)^{\frac{1}{\beta - 1}}}\, .
\ee
For a large number of choices of the function $f(\mathcal{G})$, 
Eq.~(\ref{GB7bb}) has a non-trivial ($H_0\neq 0$) real solution 
for $H_0$ (de Sitter universe). 
The late-time cosmology for above theory without matter has been 
discussed for a number of examples
in refs.~\cite{Nojiri:2005jg,Nojiri:2005am,Cognola:2006eg}. 
We will refer to these studies in the following. 

The de Sitter universe can be expressed as an accelerating, expanding
universe,
\be
\label{dS1BB}
ds^2 = - dt^2 + \e^{2H_0 t} \sum_{i=1,2,3} \left(dx^i\right)^2\, .
\ee
Changing the coordinates, the de Sitter universe can be described by the
static metric, which can be obtained by putting $M=0$ in the metric 
(\ref{SdS}) (and choosing the minus sign in $\mp$):
\be
\label{dS2BB}
ds^2 = - \left( 1 - \frac{r^2}{L^2} \right) dt^2
+ \left( 1 - \frac{r^2}{L^2} \right)^{-1} dr^2
+ r^2 d\Omega^2\, .
\ee
The length parameter $L$ can be identified with $1/H_0$.
Note that there is a cosmological horizon at $r=L$.
When $r\sim L$, by the Wick-rotating time coordinate $t=i\tau$, the metric has 
the following form:
\be
\label{dS3BB}
ds^2\sim 2\left( 1 - \frac{r}{L} \right) d\tau^2 + \frac{dr^2}{2\left( 
1 - \frac{r}{L} \right)} + l^2 d\Omega^2\, .
\ee
Changing the radial coordinate $r$ to $\rho$ as
\be
\label{dS4BB}
r = L \left( 1 - \frac{\rho^2}{2L^2} \right)\, ,
\ee
the metric (\ref{dS3BB}) can be rewritten as
\be
\label{dS5BB}
ds^2\sim \frac{\rho^2}{L^2} d\tau^2 + d\rho^2+ l^2 d\Omega^2\, .
\ee
To avoid the conical singularity at $\rho=0$, the
periodicity condition for the
coordinate $\tau$ should be imposed as $\tau \sim \tau + 2\pi L$.
Thus, the Hawking temperature $T_\mathrm{H}$ is defined as
\be
\label{dS6BB}
T_\mathrm{H} = \frac{1}{2\pi L}\, .
\ee
Let us now calculate the entropy like in Eq.~(\ref{FR8}).
The partition function $Z$ (\ref{FR6}) can be evaluated from the action
(\ref{GB1b}) by neglecting the matter Lagrangian $\mathcal{L}_\mathrm{matter}$ and
substituting the metric of the de Sitter universe (\ref{dS2BB}):
\bea
\label{dS7BB}
 - \ln Z \sim S &=& 4\pi \int_0^{2\pi L} dt \int_0^L dr r^2
\left(\frac{6}{\kappa^2 L^2}
+ f\left(\frac{24}{L^4}\right) \right) \nn
&=& \frac{8\pi^2 L^4}{3} \left(\frac{6}{\kappa^2 L^2}
+ f\left(\frac{24}{L^4}\right) \right) \nn
&=& \frac{1}{6\pi^2 T_\mathrm{H}^4}
\left(\frac{24\pi^2 T_\mathrm{H}^2}{\kappa^2 } + f\left(384 \pi^4
T_\mathrm{H}^4\right) \right) \, .
\eea
Note that $R=\frac{12}{L^2}$ and $\mathcal{G} = \frac{24}{L^4}$.
Using (\ref{FR6}) and (\ref{FR7}), the free energy $\mathcal{F}$
and the entropy $\mathcal{S}$ are
\bea
\label{dS8BB}
\mathcal{F} &=& \frac{1}{6\pi^2 T_\mathrm{H}^3}
\left(\frac{24\pi^2 T_\mathrm{H}^2}{\kappa^2} + f\left(384 \pi^4
T_\mathrm{H}^4\right) \right) \, , \\
\label{dS9BB}
\mathcal{S} &=& - \frac{4}{\kappa^2
T_\mathrm{H}^2} - \frac{f\left(384 \pi^4 T_\mathrm{H}^4 \right)}
{2\pi^2 T_\mathrm{H}^4}
256 \pi^2 f'\left(384 \pi^4 T_\mathrm{H}^4 \right) \nn
&=& - \frac{16\pi^2 L^2}{\kappa^2} - 8\pi^2 L^4 f\left(\frac{24}{L^4} 
\right) + 256 \pi^2 f'\left(\frac{24}{L^4} \right) \, .
\eea
Because $H_0 = \frac{1}{L}$, Eq.~(\ref{GB7bb}) can be rewritten as
\be
\label{dS10BB}
0=-\frac{3}{\kappa^2 L^2} + \frac{24}{L^4}
f'\left( \frac{24}{L^4} \right) - f\left( \frac{24}{L^4} \right) \, .
\ee
By using (\ref{dS10BB}), the entropy (\ref{dS9BB}) has the following
expression:
\be
\label{dS11BB}
\mathcal{S} = \frac{A_\mathrm{H}}{4G} + 64\pi^2 
f'\left(\frac{24}{L^4} \right)\, .
\ee
Here, $A_\mathrm{H}$ is the area of the cosmological horizon: 
$A_\mathrm{H} = 4\pi L^2$ and $G=\frac{\kappa^2}{8\pi}$.
When $f(\mathcal{G})$ is a constant, which corresponds to the cosmological
constant, Eq.~(\ref{dS11BB}) reproduces the standard result.

One now considers the case $\rho_\mathrm{matter}\neq 0$. Assuming that the
EoS parameter $w\equiv p_\mathrm{matter}/\rho_\mathrm{matter}$ 
is a constant, using the conservation of energy:
\be
\label{CEm}
\dot \rho_\mathrm{matter} + 3H\left(\rho_\mathrm{matter} 
+ p_\mathrm{matter}\right)=0\, ,
\ee
we find $\rho=\rho_0 a^{-3(1+w)}$. When the function $f(\mathcal{G})$ is 
chosen as in (\ref{mGB1}), again, if $\beta<1/2$, the $f(\mathcal{G})$ 
term becomes dominant compared with the Einstein term when the curvature 
is small. 
If we neglect the contribution from the Einstein term in (\ref{GB7b}),
the following solution may be found
\bea
\label{mGB3}
&& a= a_0 t^{h_0} \quad \mbox{when}\ h_0>0\, , \quad
a = a_0 \left(t_s - t\right)^{h_0} \quad \mbox{when}\ h_0<0\, , \nn
&& h_0=\frac{4\beta}{3(1+w)}\, ,\quad
a_0=\Bigl[ -\frac{f_0(\beta - 1)}
{\left(h_0 - 1\right)\rho_0}\left\{24 \left|h_0^3 \left(- 1
+ h_0\right) \right|\right\}^\beta \left( h_0 - 1
+ 4\beta\right)\Bigr]^{-\frac{1}{3(1+w)}}\, .
\eea
Thus, the effective EoS parameter $w_\mathrm{eff}$ (\ref{JGRG12})
is less than $-1$ if $\beta<0$, and for $w>-1$, it is
\be
\label{mGB3b}
w_\mathrm{eff}=-1 + \frac{2}{3h_0}=-1 + \frac{1+w}{2\beta}\, ,
\ee
which is again less than $-1$ for $\beta<0$. 
Thus, if $\beta<0$, we obtain an effective phantom with
negative $h_0$ even when $w>-1$. In the phantom phase,
the Big Rip might seem to occur at $t=t_s$.
Near this Big Rip, however, the curvature becomes
dominant and then the Einstein term dominates, so the
$f(\mathcal{G})$ term can be neglected. Therefore, the universe behaves as
$a=a_0t^{2/3(w+1)}$ and as a consequence, the Big Rip does not
occur. The phantom era is thus transient.

The case when $0<\beta<1/2$ may also be considered. As $\beta$ is
positive, the universe does not reach the phantom phase. When
the curvature is large, the $f(\mathcal{G})$ term in the action (\ref{GB1b})
can be neglected and one can work with the Einstein gravity. Then, if
$w$ is positive, the matter energy density $\rho_\mathrm{matter}$ should 
behave as $\rho_\mathrm{matter}\sim t^{-2}$, but $f(\mathcal{G})$ goes as
$f(\mathcal{G})\sim t^{-4\beta}$.
For late times (large $t$), the $f(\mathcal{G})$ term may become
dominant as compared with the Lagrangian density of matter. Neglecting the
contribution from matter, Eq.~(\ref{GB7b}) has a de Sitter universe
solution where $H$, and, therefore, $\mathcal{G}$, are constant. If $H=H_0$ 
with a constant $H_0$, Eq.~(\ref{GB7b}) looks as (\ref{GB7bb}). As a
consequence, even if we start from the deceleration phase with
$w>-1/3$, an asymptotically de Sitter universe emerges.
Correspondingly, a transition also occurs here from deceleration 
to acceleration of the universe.

Now, we consider the case in which the contributions from the Einstein and matter
terms can be neglected. Eq.~(\ref{GB7b}) reduces to
\be
\label{mGB9}
0=\mathcal{G}f'(\mathcal{G}) - f(\mathcal{G}) - 24 \dot{\mathcal{G}}
f''(\mathcal{G}) H^3 \, .
\ee
If $f(\mathcal{G})$ behaves as (\ref{mGB1}), assuming
\be
\label{mGB2}
a=\left\{\begin{array}{ll} a_0t^{h_0}\ &\mbox{when}\ h_0>0\
\mbox{(quintessence)} \\
a_0\left(t_s - t\right)^{h_0}\ &\mbox{when}\ h_0<0\ \mbox{(phantom)} \\
\end{array} \right. \, ,
\ee
one obtains
\be
\label{mGB10}
0=\left(\beta - 1\right)h_0^6\left(h_0 - 1\right)
\left(h_0 - 1 + 4\beta \right)\, .
\ee
As $h_0=1$ implies $\mathcal{G}=0$, one may choose
\be
\label{mGB11}
h_0 = 1 - 4\beta\, ,
\ee
and Eq.~(\ref{JGRG12}) gives
\be
\label{mGB12}
w_\mathrm{eff}=-1 + \frac{2}{3(1-4\beta)}\, .
\ee
Therefore, if $\beta>0$, the universe is accelerating 
($w_\mathrm{eff}<-1/3$), and if $\beta>1/4$, the universe is in 
a phantom phase ($w_\mathrm{eff}<-1$).
Thus, we are led to consider the following model:
\be
\label{mGB13}
f(\mathcal{G})=f_i\left|\mathcal{G}\right|^{\beta_i}
+ f_l\left|\mathcal{G}\right|^{\beta_l} \, ,
\ee
where it is assumed that
\be
\label{mGB14}
\beta_i>\frac{1}{2}\, ,\quad \frac{1}{2}>\beta_l>\frac{1}{4}\, .
\ee
Then, when the curvature is large, as in the primordial universe, the first term
dominates, compared with the second term and the Einstein term, and it gives
\be
\label{mGB15}
 -1>w_\mathrm{eff}=-1
+ \frac{2}{3(1-4\beta_i)}>- \frac{5}{3}\, .
\ee
On the other hand, when the curvature is small, as is the case in 
the present universe, the second term in (\ref{mGB13}) dominates 
compared with the first term and the Einstein term and yields
\be
\label{mGB16} w_\mathrm{eff}= -1
+ \frac{2}{3(1-4\beta_l)}< - \frac{5}{3}\, .
\ee
Therefore, theory (\ref{mGB13}) can produce a model that is 
able to describe inflation and the late-time acceleration 
of the universe in a unified manner.

Instead of (\ref{mGB14}), one may also choose $\beta_l$ as
\be
\label{mGB17}
\frac{1}{4}>\beta_l>0\, ,
\ee
which gives
\be
\label{mGB18} -\frac{1}{3}>w_\mathrm{eff}>-1\, .
\ee
Then, we obtain an effective quintessence epoch. Moreover, by
properly adjusting the couplings $f_i$ and $f_l$ in (\ref{mGB13}), 
one can obtain a period where the Einstein term dominates and 
the universe is in a deceleration phase. After that, a transition 
occurs from deceleration to acceleration when the Gauss-Bonnet term 
becomes the dominant one. 
More choices of $f(\mathcal{G})$ may be studied for the purpose of
the construction of the current accelerating universe. Nevertheless, many
non-linear choices for this function may be approximated by the above model.
For instance, one can mention some realistic examples of $f(\mathcal{G})$ 
gravity: 
\be
f_{1}(\mathcal{G}) =
\frac{a_{1}\mathcal{G}^{n}+b_{1}}{a_{2}\mathcal{G}^{n}+b_{2}}\, ,
\quad f_{2}(\mathcal{G}) =
\frac{a_{1}\mathcal{G}^{n+N}+b_{1}}{a_{2}\mathcal{G}^{n}+b_{2}}\,.
\label{due}
\ee

Let us address the issue of the correction to Newton's law. Let
$g_{(0)}$ be a solution of (\ref{GB4b}) and represent the
perturbation of the metric as 
$g_{\mu\nu}=g_{(0)\mu\nu} + h_{\mu\nu}$. 
First, we consider the perturbation around the de Sitter
background. The de Sitter space metric is taken as $g_{(0)\mu\nu}$, 
which gives the following Riemann tensor:
\be
\label{GB35}
R_{(0)\mu\nu\rho\sigma}=H_0^2\left(g_{(0)\mu\rho}
g_{(0)\nu\sigma} - g_{(0)\mu\sigma}g_{(0)\nu\rho}\right)\, .
\ee
The flat background corresponds to the limit of $H_0\to 0$. For
simplicity, the following gauge condition is chosen:
$g_{(0)}^{\mu\nu} h_{\mu\nu}=\nabla_{(0)}^\mu h_{\mu\nu}=0$. Then
Eq.~(\ref{GB4b}) gives
\be
\label{GB38b}
0=\frac{1}{4\kappa^2} \left(\nabla^2 h_{\mu\nu} - 2H_0^2 h_{\mu\nu}\right)
+ T_{\mathrm{matter}\, \mu\nu}\, .
\ee
The Gauss-Bonnet term contribution does not appear except
in the length parameter $1/H_0$ of the de Sitter space, which is
determined by taking into account the Gauss-Bonnet term.
This may occur due to the special structure of the Gauss-Bonnet invariant.
Eq.~(\ref{GB38b}) shows that there is no correction to Newton's law in 
de Sitter space and even in the flat background corresponding to 
$H_0\to 0$, regardless of the form of $f$ (at least, with the above 
gauge condition). In contrast, the study of the Newtonian limit 
in $1/R$ gravity indicates that significant
corrections to Newton's law may appear
as is shown in refs.~\cite{Capozziello:2004sm}.
For most $1/R$ models, the corrections to Newton's law do not comply 
with solar system tests.

By introducing two auxiliary fields, $A$ and $B$, one can rewrite
the action (\ref{GB1b}) as
\bea
\label{GB3}
S&=&\int d^4 x\sqrt{-g}\Bigl(\frac{1}{2\kappa^2}R 
+ B\left(\mathcal{G}-A\right) \nn
&& + f(A) + \mathcal{L}_\mathrm{matter} \Bigr)\, .
\eea
By the variation of the action (\ref{GB3}) with respect to $B$, it follows that
$A=\mathcal{G}$. Using this in (\ref{GB3}), the action (\ref{GB1b}) is
recovered. 
Alternatively, varying the action (\ref{GB3}) with respect to $A$ 
in (\ref{GB3}), one gets $B=f'(A)$, and, thus, 
\be
\label{GB6}
S=\int d^4
x\sqrt{-g}\Bigl(\frac{1}{2\kappa^2}R
+ f'(A)\mathcal{G} - Af'(A) + f(A)\Bigr)\, .
\ee
By varying the action (\ref{GB6}) with respect to $A$, 
the relation $A=\mathcal{G}$ is obtained again. 
The scalar is not dynamic as it has no kinetic
term. One may add, however, a kinetic term to the action by hand
\be
\label{GB6b}
S=\int d^4 x\sqrt{-g}\Bigl(
\frac{1}{2\kappa^2}R - \frac{\epsilon}{2}\partial_\mu A \partial^\mu A 
+ f'(A)\mathcal{G} - Af'(A) + f(A)\Bigr)\, .
\ee
Then, one obtains a dynamical scalar theory
coupled with the Gauss-Bonnet invariant and with a potential. It is
known that, in general,  a theory of this kind has no ghosts and is stable. 
Actually, it is related to string-inspired dilaton gravity, which was 
proposed as an alternative for dark energy 
\cite{Nojiri:2005vv,Sami:2005zc,Calcagni:2005im}.
Then, when the limit $\epsilon\to 0$ can be
obtained smoothly, the corresponding $f(\mathcal{G})$ theory has no
ghost and could actually be stable. It may be of interest to study the
cosmology of such a theory in the limit $\epsilon\to 0$.

Thus, we investigated the modified Gauss-Bonnet gravity and demonstrated that 
it may naturally lead to the unified cosmic history.
Further study of cosmological properties of such a theory as well as 
discussion of particular models of the modified Gauss-Bonnet gravity 
may be found in refs.~\cite{1GB}.

%%%%%%%%
%%%%%%%%
%%%%%%%%
%%%%%%%%

\subsubsection{$F(\mathcal{G},R)$ gravity \label{S52}}

More general models include the dependence from
curvature as well as from the Gauss-Bonnet term \cite{Cognola:2006eg}:
\be
\label{GR1b}
S=\int d^4 x\sqrt{-g}\left(F(\mathcal{G},R) 
+ \mathcal{L}_\mathrm{matter}\right)\, .
\ee
An example is given by
\be
\label{GR7b}
F(\mathcal{G},R)=R \tilde f\left(\frac{\mathcal{G}}{R^2}\right)\, ,
\quad \tilde f\left(\frac{\mathcal{G}}{R^2}\right)=\frac{1}{2\kappa^2}
+ f_0 \left(\frac{\mathcal{G}}{R^2}\right)\, ,
\ee
which gives a solution describing the FRW universe:
\be
\label{GR11}
H=\frac{h_0}{t}\, ,\quad
h_0 = \frac{\frac{3}{\kappa^2} - 2f_0 
\pm \sqrt{8f_0\left(f_0 - \frac{3}{8\kappa^2}\right)}}
{\frac{6}{\kappa^2} + 2f_0}\, .
\ee
Then, for example, if $\kappa^2 f_0<-3$, there is a solution
describing a phantom with $h_0<-1-\sqrt{2}$ and a solution
describing the effective matter with $h_0>-1+\sqrt{2}$.
Late-time cosmology in other versions of such theory can be constructed.
The unification with inflation in such model may be achieved via 
cosmological reconstruction, which is developed in the third chapter.

\subsection{String-inspired model and scalar-Einstein-Gauss-Bonnet 
gravity \label{IIC}}

In string theories, the compactification from higher dimensions to 
four dimensions induces many scalar fields, such as moduli or 
dilaton fields. These scalars couple with curvature invariants.
Neglecting the moduli fields associated with the radii of the internal
space, we may consider the following action of the low-energy 
effective string theories \cite{Sami:2005zc,Calcagni:2005im}:
\be
\label{eq:action}
S = \int d^4 x
\sqrt{-g}\left[\frac{R}{2}+\mathcal{L}_{\phi}
+ \mathcal{L}_{c}+\ldots\right]\, ,
\ee
where $\phi$ is the dilaton field, which is related to the
string coupling, $\mathcal{L}_{\phi}$ is the Lagrangian of $\phi$,
and $\mathcal{L}_c$ expresses the string curvature correction terms
to the Einstein-Hilbert action,
\be
\label{stcor}
\mathcal{L}_{\phi} = -\partial_{\mu}\phi\partial^{\mu}\phi-V(\phi) \, , 
\quad
\mathcal{L}_c = c_1 {\alpha'} \e^{2\frac{\phi}{\phi_0}}\mathcal{L}_c^{(1)}
+ c_2{\alpha'}^2\e^{4\frac{\phi}{\phi_0}}\mathcal{L}_c^{(2)}
+ c_3{\alpha'}^3\e^{6\frac{\phi}{\phi_0}}\mathcal{L}_c^{(3)}\, ,
\ee
where $1/\alpha'$ is the string tension, $\mathcal{L}_c^{(1)}$,
$\mathcal{L}_c^{(2)}$, and $\mathcal{L}_c^{(3)}$ express the 
leading-order (Gauss-Bonnet term $\mathcal{G}$ in (\ref{GB})),
the second-order, and the third-order curvature corrections, respectively.
The terms $\mathcal{L}_c^{(1)}$, $\mathcal{L}_c^{(2)}$ and
$\mathcal{L}_c^{(3)}$ in the Lagrangian have the following form
\be
\label{ccc}
\mathcal{L}_c^{(1)} = \Omega_2\, , \quad
\mathcal{L}_c^{(2)} = 2 \Omega_3 + R^{\mu\nu}_{\alpha \beta}
R^{\alpha\beta}_{\lambda\rho}
R^{\lambda\rho}_{\mu\nu}\, , \quad
\mathcal{L}_c^{(3)}
= \mathcal{L}_{31} - \delta_{H} \mathcal{L}_{32} -\frac{\delta_{B}}{2}
\mathcal{L}_{33}\, .
\ee
Here, $\delta_B$ and $\delta_H$ take the value of $0$ or $1$ and
\bea
\Omega_2 &=& \mathcal{G} \, , \nn
\Omega_3 &\propto & \epsilon^{\mu\nu\rho\sigma\tau\eta}
\epsilon_{\mu'\nu'\rho'\sigma'\tau'\eta'}
R_{\mu\nu}^{\ \ \mu'\nu'} R_{\rho\sigma}^{\ \ \rho'\sigma'}
R_{\tau\eta}^{\ \ \tau'\eta'} \, , \nn
\mathcal{L}_{31} &=& \zeta(3) R_{\mu\nu\rho\sigma}R^{\alpha\nu\rho\beta}
\left( R^{\mu\gamma}_{\ \ \delta\beta}
R_{\alpha\gamma}^{\ \ \delta\sigma} - 2 R^{\mu\gamma}_{\ \ \delta\alpha}
R_{\beta\gamma}^{\ \ \delta\sigma} \right)\, , \nn
\mathcal{L}_{32} &=& \frac{1}{8} \left( R_{\mu\nu\alpha\beta}
R^{\mu\nu\alpha\beta}\right)^2
+ \frac{1}{4} R_{\mu\nu}^{\ \ \gamma\delta}
R_{\gamma\delta}^{\ \ \rho\sigma} R_{\rho\sigma}^{\ \ \alpha\beta}
R_{\alpha\beta}^{\ \ \mu\nu} - \frac{1}{2} R_{\mu\nu}^{\ \ \alpha\beta}
R_{\alpha\beta}^{\ \ \rho\sigma}
R^\mu_{\ \sigma\gamma\delta}
R_\rho^{\ \nu\gamma\delta} - R_{\mu\nu}^{\ \ \alpha\beta}
R_{\alpha\beta}^{\ \ \rho\nu} R_{\rho\sigma}^{\ \ \gamma\delta}
R_{\gamma\delta}^{\ \ \mu\sigma}\, , \nn
\mathcal{L}_{33} &=& \left( R_{\mu\nu\alpha\beta}
R^{\mu\nu\alpha\beta}\right)^2 - 10 R_{\mu\nu\alpha\beta}
R^{\mu\nu\alpha\sigma}
R_{\sigma\gamma\delta\rho}
R^{\beta\gamma\delta\rho} - R_{\mu\nu\alpha\beta} 
R^{\mu\nu\rho}_{\ \ \ \sigma}
R^{\beta\sigma\gamma\delta}
R_{\delta\gamma\rho}^{\ \ \ \alpha} \, .
\eea
The correction terms are different depending on the type of string theory; 
the dependence is encoded in the curvature invariants and in
the coefficients $(c_1,c_2,c_3)$ and $\delta_H$, $\delta_B$, as follows,
\begin{itemize}
\item For the Type II superstring theory: $(c_1,c_2,c_3) = (0,0,1/8)$
and $\delta_H=\delta_B=0$.
\item For the heterotic superstring theory: $(c_1,c_2,c_3) = (1/8,0,1/8)$
and $\delta_H=1,\delta_B=0$.
\item For the bosonic superstring theory: $(c_1,c_2,c_3) = (1/4,1/48,1/8)$
and $\delta_H=0,\delta_B=1$.
\end{itemize}

Motivated by the string considerations, we consider the
scalar-Einstein-Gauss-Bonnet gravity\footnote{For pioneering work
on the scalar-Einstein-Gauss-Bonnet gravity, see \cite{Boulware:1986dr}. }
based on \cite{Nojiri:2005vv,Nojiri:2006je}. It was first proposed in
ref.~\cite{Nojiri:2005vv} to consider such a theory as a gravitational 
alternative for dark energy, so-called Gauss-Bonnet dark energy.

The starting action is:
\be
\label{ma22}
S=\int d^4 x \sqrt{-g}\left[ \frac{R}{2\kappa^2} - \frac{1}{2}
\partial_\mu \phi
\partial^\mu \phi - V(\phi) - \xi(\phi) \mathcal{G}\right]\, .
\ee
Here, we do not restrict the forms of $V(\phi)$ and $\xi(\phi)$ which 
should be given by the non-perturbative string theory (\ref{stcor}).
Note also that the action (\ref{ma22}) is given by adding the kinetic
term for the scalar field $\phi$ to the action of the 
$F(\mathcal{G})$ gravity in the scalar-tensor form that appeared 
in the previous section.

Making the variation of the action (\ref{ma22}) over the metric 
$g_{\mu\nu}$, the field equations follow as:
\bea
\label{ma23}
&& 0= \frac{1}{\kappa^2}\left(- R^{\mu\nu} 
+ \frac{1}{2} g^{\mu\nu} R\right)
+ \frac{1}{2}\partial^\mu \phi \partial^\nu \phi - \frac{1}{4}g^{\mu\nu}
\partial_\rho \phi \partial^\rho \phi
+ \frac{1}{2}g^{\mu\nu}\left( - V(\phi) + \xi(\phi) \mathcal{G} \right) 
\nn
&& -2 \xi(\phi) R R^{\mu\nu} - 4\xi(\phi)R^\mu_{\ \rho} 
R^{\nu\rho} -2 \xi(\phi) R^{\mu\rho\sigma\tau}R^\nu_{\ \rho\sigma\tau}
+4 \xi(\phi) R^{\mu\rho\nu\sigma}R_{\rho\sigma} \nn
&& + 2 \left( \nabla^\mu \nabla^\nu \xi(\phi)\right)R - 2 g^{\mu\nu}
\left( \nabla^2\xi(\phi)\right)R - 4 \left(
\nabla_\rho \nabla^\mu \xi(\phi)\right)R^{\nu\rho} - 4 \left(
\nabla_\rho \nabla^\nu \xi(\phi)\right)R^{\mu\rho} \nn
&& + 4 \left( \nabla^2 \xi(\phi) \right)R^{\mu\nu}
+ 4g^{\mu\nu} \left( \nabla_{\rho} \nabla_\sigma \xi(\phi) \right)
R^{\rho\sigma}
+ 4 \left(\nabla_\rho \nabla_\sigma \xi(\phi) \right) R^{\mu\rho\nu\sigma} 
\, .
\eea
In Eq.~(\ref{ma23}), the derivatives of curvature such as $\nabla R$, 
do not appear. Therefore, the derivatives higher than two do not appear, 
which can be contrasted with a general
$\alpha R^2 + \beta R_{\mu\nu}R^{\mu\nu} + \gamma
R_{\mu\nu\rho\sigma}R^{\mu\nu\rho\sigma}$ gravity,
where fourth derivatives of $g_{\mu\nu}$ appear.
For the classical theory, if we specify the values of
$g_{\mu\nu}$ and $\dot g_{\mu\nu}$
on a spatial surface as an initial condition, the time 
development is uniquely determined.
This situation is similar to the case in classical mechanics, 
in which one only needs to specify the values of
position and velocity of particle as initial conditions.
In general $\alpha R^2 + \beta R_{\mu\nu}R^{\mu\nu} + \gamma
R_{\mu\nu\rho\sigma}R^{\mu\nu\rho\sigma}$ gravity,
we need to specify the values of $\ddot g_{\mu\nu}$ and $\dddot g_{\mu\nu}$
in addition to $g_{\mu\nu}$, $\dot g_{\mu\nu}$
so that a unique time development will follow.
In Einstein gravity, only the specific value of $g_{\mu\nu}$,
$\dot g_{\mu\nu}$, should be given as an initial condition.
Thus, the scalar-Gauss-Bonnet gravity is a natural extension of the 
Einstein gravity.

In the FRW universe (\ref{JGRG14}), Eq.~(\ref{ma23}) becomes 
the following:
\bea
\label{ma24}
0&=& - \frac{3}{\kappa^2}H^2 + \frac{1}{2}{\dot\phi}^2 + V(\phi)
+ 24 H^3 \frac{d \xi(\phi(t))}{dt}\, ,\\
\label{GBany5}
0&=& \frac{1}{\kappa^2}\left(2\dot H + 3 H^2 \right)
+ \frac{1}{2}{\dot\phi}^2 - V(\phi) - 8H^2 \frac{d^2 \xi(\phi(t))}
{dt^2} - 16H \dot H \frac{d\xi(\phi(t))}{dt} - 16 H^3 
\frac{d \xi(\phi(t))}{dt} \, .
\eea
On the other hand, by the variation of the action (\ref{ma22}) 
with respect to the scalar field, the scalar equation
of motion follows as
\be
\label{ma24b}
0=\ddot \phi + 3H\dot \phi + V'(\phi) + \xi'(\phi) \mathcal{G}\, .
\ee

In particular when we consider the following string-inspired model 
\cite{Nojiri:2005vv},
\be
\label{NOS1}
V=V_0\e^{-\frac{2\phi}{\phi_0}}\, , \quad \xi(\phi)=\xi_0
\e^{\frac{2\phi}{\phi_0}}\, ,
\ee
the de Sitter space solution follows: 
\be
\label{NOS2}
H^2 = H_0^2 \equiv - \frac{\e^{-\frac{2\varphi_0}{\phi_0}}}
{8\xi_0 \kappa^2} \, , \quad \phi = \varphi_0 \, .
\ee
Here, $\varphi_0$ is an arbitrary constant.
If $\varphi_0$ is chosen to be larger, the Hubble rate $H=H_0$ becomes
smaller.
Then, if $\xi_0\sim \mathcal{O}(1)$, by choosing 
$\varphi_0/\phi_0\sim 140$,
the value of the Hubble rate $H=H_0$ is consistent with the observations.
The model (\ref{NOS1}) also has another solution:
\bea
\label{NOS3}
& H=\frac{h_0}{t}\, ,\quad \phi=\phi_0 \ln \frac{t}{t_1}\
& \mbox{when}\ h_0>0\, ,\nn
& H=-\frac{h_0}{t_s - t}\, ,\quad \phi=\phi_0 \ln \frac{t_s - t}{t_1}\ &
\mbox{when}\ h_0<0\, .
\eea
Here, $h_0$ is obtained by solving the following algebraic equations:
\be
\label{NOS4}
0 = -\frac{3h_0^2}{\kappa^2} + \frac{\phi_0^2}{2}
+ V_0 t_1^2 - \frac{48 \xi_0 h_0^3}{t_1^2}\, ,\quad
0 = \left( 1 - 3h_0 \right)\phi_0^2 + 2V_0 t_1^2
+ \frac{48 \xi_0 h_0^3}{t_1^2}\left(h_0 - 1\right)\, .
\ee
Eqs.~(\ref{NOS4}) can be rewritten as
\bea
\label{NOS5}
V_0 t_1^2 &=& - \frac{1}{\kappa^2\left(1 + h_0\right)}\left\{3h_0^2 \left( 
1 - h_0\right)
+ \frac{\phi_0^2 \kappa^2 \left( 1 - 5 h_0\right)}{2}\right\}\, ,\\
\label{NOS6}
\frac{48 \xi_0 h_0^2}{t_1^2} &=& - \frac{6}{\kappa^2\left( 1 
+ h_0\right)}\left(h_0 - \frac{\phi_0^2 \kappa^2}{2}\right)\, .
\eea
The arbitrary value of $h_0$ can be realized by properly 
choosing $V_0$ and $\xi_0$.
With the appropriate choice of $V_0$ and $\xi_0$, we can obtain 
a negative $h_0$ and, therefore, the effective EoS parameter 
(\ref{JGRG12}) is less than $-1$, $w_\mathrm{eff} < -1$,
which corresponds to the effective phantom.
In usual (canonical) scalar-tensor theory without the Gauss-Bonnet term, 
the phantom cannot be realized by the canonical scalar.

For example, if $h_0=-80/3<-1$ and, therefore, $w= - 1.025$, which is 
consistent with the observed value, we find
\be
\label{NOS6b}
V_0t_1^2 = \frac{1}{\kappa^2}\left( \frac{531200}{231}
+ \frac{403}{154}\gamma \phi_0^2 \kappa^2 \right)>0\, , \quad
\frac{f_0}{t_1^2} = -\frac{1}{\kappa^2}\left( \frac{9}{49280}
+ \frac{27}{7884800}\gamma \phi_0^2 \kappa^2 \right)\, .
\ee
For other choices of scalar potentials one can realize other types of dark
energy universes, for instance, the effective quintessence. Moreover, one 
can propose the potentials in such a way that the unification of 
the inflation with dark energy naturally occurs. Different 
cosmological aspects of scalar-Gauss-Bonnet dark energy have been 
studied in refs.~\cite{sGB}. An extension of these studies taking 
into account the Euler invariant and other higher-derivative
stringy terms has also been done \cite{Elizalde:2007pi}.

\subsection{Non-local gravity \label{IID}}

\subsubsection{Non-local $F(R)$ gravity}

In this section, we review non-local gravity 
\cite{Deser:2007jk,Nojiri:2007uq,Jhingan:2008ym} and
present its local (scalar-tensor) formulation. The explicit example of 
such a theory, which naturally leads to the unification of inflation 
with late-time cosmic acceleration, is worked out.

The starting action of the simplest non-local gravity is given by
\be
\label{nl1}
S=\int d^4 x \sqrt{-g}\left\{
\frac{1}{2\kappa^2}\left\{ R\left(1 + f(\Box^{-1}R )\right) -2 \Lambda 
\right\} + \mathcal{L}_\mathrm{matter} \left(Q; g\right)
\right\}\, .
\ee
Here, $f$ is some function, $\Box$ is the d'Alembertian for the 
scalar field, $\Lambda$ is a cosmological constant
and $Q$ represents the matter fields.
The above action can be rewritten by introducing two scalar fields, 
$\eta$ and $\xi$, in the following form:
\bea
\label{nl2}
S&=&\int d^4 x \sqrt{-g}\left[
\frac{1}{2\kappa^2}\left\{R\left(1 + f(\eta)\right)
+ \xi\left(\Box\eta - R\right) - 2 \Lambda \right\}
+ \mathcal{L}_\mathrm{matter} \right] \nn
&=&\int d^4 x \sqrt{-g}\left[
\frac{1}{2\kappa^2}\left\{R\left(1
+ f(\eta)\right) - \partial_\mu \xi \partial^\mu \eta - \xi R - 2 \Lambda
\right\} + \mathcal{L}_\mathrm{matter}
\right] \, .
\eea
By the variation of the action (\ref{nl2}) over $\xi$, we obtain
$\Box\eta=R$ or $\eta=\Box^{-1}R$.
Substituting the above equation into (\ref{nl2}), one re-obtains
(\ref{nl1}).

The variation of (\ref{nl2}) with respect to the metric 
tensor $g_{\mu\nu}$ gives
\bea
\label{nl4}
0 &=& \frac{1}{2}g_{\mu\nu} \left\{R
\left(1 + f(\eta) - \xi\right) - \partial_\rho \xi
\partial^\rho \eta - 2 \Lambda \right\} - R_{\mu\nu}
\left(1 + f(\eta) - \xi\right) \nn
&& + \frac{1}{2}\left(\partial_\mu \xi \partial_\nu \eta
+ \partial_\mu \eta \partial_\nu \xi \right) -\left(
g_{\mu\nu}\Box - \nabla_\mu \nabla_\nu\right)
\left( f(\eta) - \xi\right) + \kappa^2T_{\mu\nu}\, .
\eea
On the other hand, the variation with respect to $\eta$ gives
\be
\label{nl5}
0=\Box\xi+ f'(\eta) R\, .
\ee
Now, we assume that the spatially flat FRW metric
and the scalar fields $\eta$ and $\xi$ only depend on time.
Then, Eq.~(\ref{nl4}) has the following form:
\bea
\label{nl7a}
0 &=& - 3 H^2\left(1 + f(\eta) - \xi\right)
+ \frac{1}{2}\dot\xi \dot\eta - 3H\left(f'(\eta)\dot\eta - \dot\xi\right)
+ \Lambda + \kappa^2 \rho_\mathrm{matter}\, ,\\
\label{nl7b}
0 &=& \left(2\dot H + 3H^2\right) \left(1 + f(\eta) - \xi\right) +
\frac{1}{2}\dot\xi \dot\eta
+ \left(\frac{d^2}{dt^2} + 2H \frac{d}{dt} \right)
\left( f(\eta) - \xi \right) - \Lambda + \kappa^2 p\, .
\eea
On the other hand, the scalar equations are:
\bea
\label{nl8a}
0 &=& \ddot \eta + 3H \dot \eta + 6 \dot H + 12 H^2 \, , \\
\label{nl8b}
0 &=& \ddot \xi + 3H \dot \xi - \left( 6 \dot H + 12 H^2\right)f'(\eta) 
\, .
\eea

For the de Sitter solution $H=H_0$, Eq.~(\ref{nl8a}) can be solved as
\be
\label{NLdS1}
\eta= - 4H_0 t - \eta_0 \e^{-3H_0 t} + \eta_1\, ,
\ee
with constants of integration $\eta_0$ and $\eta_1$.
For simplicity, let $\eta_0=\eta_1=0$ and $f(\eta)$ be given by
\be
\label{NLdS2}
f(\eta)=f_0 \e^{\frac{\eta}{\beta}}= f_0 \e^{-\frac{4H_0 t}{\beta}}\, .
\ee
Here, $\beta$ is a constant.
Then, Eq.~(\ref{nl8b}) can be solved as follows:
\be
\label{NLdS3}
\xi= - \frac{3f_0 \beta}{3\beta - 4} \e^{-\frac{4H_0 t}{\beta}} +
\frac{\xi_0}{3H_0}\e^{-3H_0 t} - \xi_1\, .
\ee
Here, $\xi_0$ and $\xi_1$ are constants.
For the de Sitter space $a$ behaves as $a=a_0\e^{H_0 t}$.
For matter with a constant equation of state $w$,
we find
\be
\label{NN1}
\rho_\mathrm{matter} = \rho_0 \e^{-3(w+1)H_0 t}\, .
\ee
Then, after setting $\xi_0=0$, by substituting (\ref{NLdS1}), 
(\ref{NLdS3}), and (\ref{NN1}) into (\ref{nl7a}), one obtains
\be
\label{NLdS4}
0 = - 3H_0^2 \left(1 + \xi_1\right) + 6H_0^2 f_0
\left( \frac{2}{\beta} - 1 \right)\e^{-\frac{4H_0 t}{\beta}}
+ \Lambda + \kappa^2 \rho_0 \e^{-3(w+1)H_0 t} \, .
\ee
When $\rho_0=0$, with the choice
\be
\label{NLdS5}
\beta = 2\, ,\quad
\xi_1 = - 1 + \frac{\Lambda}{3H_0^2} \, ,
\ee
the de Sitter space can be a solution.
Even if $\rho_\mathrm{matter}\neq 0$, if we choose
\be
\label{NN2}
\beta =\frac{4}{3(1+w)} \, ,\quad f_0
= \frac{\kappa^2 \rho_0}{3H_0^2 \left(1 + 3w \right)}\, ,
\quad \xi_1 = -1 + \frac{\Lambda}{3H_0^2} \, ,
\ee
there is a de Sitter solution.

Eq.~(\ref{NLdS5}) or (\ref{NN2}) shows that
\be
\label{NNNN1}
H_0^2 = \frac{3 \Lambda}{ 1 + \xi_1 }\, .
\ee
This indicates that the cosmological constant $\Lambda$ is effectively
screened by $\xi$.
We should also note that in the case without the cosmological constant
$\Lambda=0$, if we choose $\xi_1 = -1$, $H_0$ can be arbitrary.
Then, $H_0$ can be determined by the initial condition.
Because $H_0$ can be small or large, the theory with 
the function (\ref{NLdS2}) with $\beta =2$ in (\ref{NLdS5})
can describe the early-time inflation or current cosmic acceleration.

In the presence of matter with $w\neq 0$, the de Sitter solution
$H=H_0$ is realized even if $f(\eta)$ given by
\be
\label{nLL1}
f(\eta)=f_0\e^{\eta/2} + f_1\e^{3(w+1)\eta/4}\, .
\ee
The following solution exists:
\be
\label{nLL2}
\eta= - 4H_0 t\, ,\quad \xi=1 + 3f_0\e^{-2H_0t}+ 
\frac{f_1}{w}\e^{-3(w+1)H_0
t}\, , \quad
\rho_\mathrm{matter}= - \frac{3(3w+1)H_0^2f_1}{\kappa^2}
\e^{-3(1+w)H_0t}\, .
\ee
The theory with multiple de Sitter solutions may describe the unification 
of inflation with dark energy \cite{Nojiri:2007uq}. Other
cosmological aspects of such non-local gravity were studied in
refs.~\cite{nonlocal}.

One can construct more general non-local gravity along the above line.
Let us start from a general non-local action:
\be
\label{nl9}
S = \int d^4 x
\sqrt{-g} \left[ F\left(R, \Box R, \Box^2 R , \cdots, \Box^m R,
\Box^{-1} R, \Box^{-2} R,
\cdots , \Box^{-n} R \right) + \mathcal{L}_\mathrm{matter} \right]\, ,
\ee
with $m$ and $n$ being the positive integers. By introducing scalar
fields $A$, $B$, $\chi_k$, $\eta_k$, $\left(k=1,2,\cdots,m \right)$,
and $\xi_l$, $\phi_l$, $\left(l=1,2,\cdots,n\right)$, the action 
(\ref{nl9}) can be rewritten in the following form:
\bea
\label{nsigma1}
S &=& \int d^4 x \sqrt{-g} \left[ BR - BA + F\left(A, \eta_1, \eta_2,
\cdots, \eta_m, \phi_1, \phi_2, \cdots , \phi_n \right) + \partial_\mu 
\chi \partial^\mu A + \sum_{k=2}^m
\partial_\mu \chi_k \partial^\mu \eta_{k-1} \right. \nn
&& \left. + \sum_{l=1}^n
\partial_\mu \xi_l \partial^\mu \phi_l + \sum_{k=1}^m \chi_1 \eta_1
+ A\xi_1 + \sum_{l=2}^n \xi_l \phi_{l-1} 
+ \mathcal{L}_\mathrm{matter} \right] \, .
\eea
The variations over $A$, $B$, $\chi_k$, and $\xi_l$ lead to 
the following equations
\be
\label{nsigma2}
0=R-A=\eta_1 - \Box A = \eta_k - \Box \eta_{k-1}
= \Box \phi_1 - A = \Box \phi_l - \phi_{l-1} \quad \left(k=2,
\cdots, m\, , \ l=2, \cdots, n\right)\, ,
\ee
which give $\eta_k = \Box^k R$ and $\Box^l \phi_l = R$. Thus,
the actions (\ref{nsigma1}) and (\ref{nl9}) are equivalent.

Let us now consider a scale transformation given by
\be
\label{nsigma3}
g_{\mu\nu}\to \frac{1}{2 B} g_{\mu\nu}\, .
\ee
Then, the action (\ref{nsigma1}) is transformed into the action in the
Einstein frame:
\bea
\label{nsigma4}
S &=& \int d^4 x \sqrt{-g}
\left[ \frac{R}{2} - \frac{3}{2B^2}\partial_\mu B \partial^\mu B
+ \frac{1}{2 B} \left(\partial_\mu \chi \partial^\mu A
+ \sum_{k=2}^m \partial_\mu \chi_k \partial^\mu \eta_{k-1}
+ \sum_{l=1}^n \partial_\mu \xi_l \partial^\mu \phi_l\right) \right. \nn
&& \left. + \frac{1}{4 B^2} \left( - BA + F\left(A, \eta_1, \cdots, 
\eta_m, \phi_1, \cdots , \phi_n \right) + \sum_{k=1}^m \chi_1 \eta_1 
+ A\xi_1 + \sum_{l=2}^n \xi_l \phi_{l-1}
\right) + \mathcal{L}^A_\mathrm{matter} \right] \, .
\eea
Here, $\mathcal{L}^A_\mathrm{matter}$ can be obtained by 
scale transformation of the metric tensor in 
$\mathcal{L}_\mathrm{matter}$ by (\ref{nsigma3}).
The above action can be regarded as a non-linear $\sigma$ model coupled to 
gravity. The action (\ref{nl1}) can be considered as a
particular case of the general non-local action. In this case,
however, the continuity equation (\ref{CEm}) is no longer
valid; additional terms on the right-hand side of this equation are
generated and are responsible for inducing the interaction of matter
within the background. This aspect is crucial for the discussion
of local gravity constraints for such models based on the modified
theories of gravity.

%%%%
%%%%%
%%%%%
%%%%%
%%%%%%

\subsubsection{Non-local Gauss-Bonnet gravity}

It may be interesting to discuss the extension of the model presented 
in the above section by including the Gauss-Bonnet
invariant \cite{Capozziello:2008gu,Cognola:2009jx}.

The corresponding non-local action generalizing 
the modified Gauss-Bonnet gravity is
\be
\label{nlGB1}
S=\int d^4 x \sqrt{-g} \left(\frac{R}{2\kappa^2}
 - \frac{\kappa^2}{2\alpha} \mathcal{G}\Box^{-1}\mathcal{G} +
\mathcal{L}_\mathrm{matter} \right)\, .
\ee
Here, $\mathcal{L}_\mathrm{matter}$ is the matter Lagrangian, and 
$\mathcal{G}$ is the Gauss-Bonnet invariant.
By introducing the scalar field $\phi$, one may rewrite the action
(\ref{nlGB1}) in a local form:
\be
\label{nlGB3}
S=\int d^4 x \left(\frac{R}{2\kappa^2} -
\frac{\alpha}{2\kappa^2}\partial_\mu\phi \partial^\mu \phi
+ \phi \mathcal{G} + \mathcal{L}_\mathrm{matter} \right)\, .
\ee
In fact, from the $\phi$ equation, it follows that 
$\phi = - \frac{\kappa^2}{\alpha} \Box^{-1} \mathcal{G}$.
By substituting this expression into (\ref{nlGB3}), we re-obtain 
(\ref{nlGB1}).
It is remarkable that the above action belongs to a specific class of
string-inspired, scalar-Gauss-Bonnet gravity that is known to be
consistent with the local tests and may serve as a realistic candidate for 
dark energy \cite{Nojiri:2005vv} (see section C of this chapter). 
Unlike the situation in standard stringy compactifications,
the scalar-Gauss-Bonnet coupling is not an exponential scalar function. 
The scalar potential is zero, as it should be in the low-energy, 
perturbative approach to the string effective action.

Let us now study the de Sitter solutions that may describe late-time
universe acceleration in such non-local gravity.
Considering the FRW metric with a flat spatial part and a $\phi$ 
that only depends on the cosmological time, the equations of motion are
\be
\label{nlGB5}
0= - \frac{3H^2}{\kappa^2} 
+ \frac{\alpha}{2\kappa^2}{\dot\phi}^2 - 24\dot \phi
H^3 + \rho_\mathrm{matter} \, ,\quad
0= - \frac{\alpha}{\kappa^2}\left(\ddot\phi + 3H \dot\phi\right)
+ 24 \left(\dot H H^2 + H^4 \right)\, .
\ee
When the matter contribution is neglected, assuming $\dot \phi$ and $H$ 
are constant $\dot\phi=c_\phi$, $H=H_0$, one can solve 
Eqs.~(\ref{nlGB5}) as follows: $c_\phi = \frac{8\kappa^2}{\alpha}H_0^3$, 
$H_0^4 = - \frac{3\alpha}{160\kappa^4}$.
Therefore, if $\alpha$ is allowed to be negative, there is a de Sitter 
universe solution, namely
\be
\label{nlGB8}
H=H_0 = \left( - \frac{3\alpha}{160\kappa^4} \right)^{1/4}\, .
\ee
On the other hand, a negative $\alpha$ means that the scalar field has a
`wrong' kinetic term, like a ghost or phantom.
As we will see later in (\ref{nlGBr14}), the de Sitter space solution
corresponds to the case $a\to \infty$ (late-time acceleration) when 
matter is included.
Even in the early universe, before matter was created, this solution 
can be used for the description of the inflationary epoch.

Note that the second equation (\ref{nlGB5}) yields
\be
\label{nlGBr1}
\left( -\frac{\alpha}{\kappa^2} \dot\phi + 8H^2 \right)a^3 = C\, .
\ee
Here, $C$ is a constant. The action (\ref{nlGB3}) is invariant
under a constant shift of the scalar field 
$\phi$: $\phi\to \phi + c$ ($c$ is a constant)
because $\mathcal{G}$ is a total derivative. The quantity $C$ 
is the Noether charge corresponding to the invariance. 
By using (\ref{nlGBr1}), we may delete $\dot\phi$ in 
the first equation in (\ref{nlGB5}) and obtain
\be
\label{nlGBr2}
0= - \frac{3H^2}{\kappa^2} - \frac{160\kappa^2 H^6}{\alpha}
+ \frac{16\kappa^2 C H^3}{\alpha a^3} + \frac{\kappa^2 C^2}{\alpha a^6} 
+ \rho_\mathrm{matter}\, .
\ee
When $\rho_\mathrm{matter}$ is given by the sum of the contributions from 
all forms of matter witha constant EoS parameter $w_i$, one gets 
$\rho_\mathrm{matter}(a) = \sum_i \rho_{0i} a^{-3\left(w_i + 1\right)}$.
Then, we can solve algebraically Eq.~(\ref{nlGBr2}) with respect to $H$ as 
a function of $a$: $H=\mathcal{H}(a)$. Using the definition of $H$, 
the equation has the form $\dot a = a \mathcal{H}(a)$. Thus,
$t=\int \frac{da}{a\mathcal{H}(a)}$, which gives, at least formally, the
solution of (\ref{nlGBr2}) and, therefore, the FRW universe.

As a simple case, one may consider $C=0$. Then, Eq.~(\ref{nlGBr2})
reduces to the following form
\be
\label{nlGBr5}
0 = \mathcal{F}(x) \equiv x^3
+ \frac{3\alpha}{160\kappa^4} 
x - \frac{\alpha \rho_\mathrm{matter} (a)}{160 \kappa^2}\, .
\ee
Here, $x\equiv H^2$ and, therefore, $x\geq 0$. Because
$\mathcal{F}'(x)=3 \left(x^2 + \frac{\alpha}{160\kappa^4}\right)$,
when $\alpha>0$, $\mathcal{F}(x)$ is a monotonically increasing 
function of $x$.
On the other hand, when $\alpha<0$, we find
\be
\label{nlGBr7}
\mathcal{F}'\left(x_\pm\right) = 0\, ,\quad
\mathcal{F}\left(x_\pm\right)
= \mp \left( - \frac{\alpha}{160\kappa^4}
\right)^{3/2} - \frac{\alpha \rho_\mathrm{matter}}
{160\kappa^2}\, ,\quad x_\pm \equiv
\pm \left( - \frac{\alpha}{160\kappa^4}\right)^{1/2}\, .
\ee
Because
\be
\label{nlGBr8}
\mathcal{F}(0)= - \frac{\alpha \rho_\mathrm{matter}}{160 \kappa^2}\, ,
\ee
when $\alpha>0$, because $\mathcal{F}(0)<0$ and $\mathcal{F}(x)$ is a
monotonically increasing function,
there is only one positive solution in (\ref{nlGBr5}):
\be
\label{nlGBr9}
H^2 = x = \alpha_+ + \alpha_-\, ,\quad \alpha_\pm^3 
= \frac{\alpha \rho_\mathrm{matter}}{320\kappa^2}
\left( 1 \pm \sqrt{1 + \frac{\alpha}{40\kappa^8
\rho_\mathrm{matter}^2}}\right)\, .
\ee
We now consider the case that $\rho_\mathrm{matter}$ comes from a unique 
kind of matter with a constant EoS parameter $w$.
Then, when $a\to 0$, one obtains
$\alpha_+^3 \to \frac{\alpha \rho_0}{160 \kappa^2} a^{-3(1+w)}$ and 
$\quad \alpha_-\to 0$.
Therefore, $H\propto a^{-(1+w)/2}$ and $a \propto t^{2/(1+w)}$,
which could be compared with the usual FRW universe in Einstein 
gravity, where $a\propto t^{2/3(1+w)}$.
On the other hand, when $a\to \infty$, Eq.~(\ref{nlGBr9}) implies that
$H\propto a^{-3(1+w)/2}$ and, therefore, $a\propto t^{2/3(1+w)}$, 
which reproduces the result in general relativity.
This indicates that such a non-local term becomes less dominant as compared
with the Einstein-Hilbert term in the late universe.

We may consider the case when $\alpha<0$. 
Because $\mathcal{F}(0)>0$ in this case, 
if $\mathcal{F}\left(x_+\right)>0$, there is no positive solution for $x$. 
On the other hand, if $\mathcal{F}\left(x_+\right)<0$, that is, when
\be
\label{nlGBr10}
 - \alpha > 40 \kappa^6 \rho_\mathrm{matter}^2 \, ,
\ee
there are two positive solutions and one negative solution in 
(\ref{nlGBr5}), namely
\be
\label{nlGBr11}
x = \tilde\alpha_+ + \tilde\alpha_-\, ,\
\tilde\alpha_+ \xi + \tilde\alpha_- \xi^2\, ,
\ \tilde\alpha_+ \xi^2 + \tilde\alpha_- \xi\, ,
\quad \tilde\alpha_\pm^3 = \frac{\alpha \rho_\mathrm{matter}}{320\kappa^2}
\left( 1 \pm i \sqrt{ - 1 - \frac{\alpha}{40\kappa^8
\rho_\mathrm{matter}^2}}\right)\, .
\ee
Eq.~(\ref{nlGBr10}) indicates that there is a lower bound in $a$ as
$a \geq a_\mathrm{min} \equiv \left( - \frac{\alpha}{40\kappa^4
\rho_0^2}\right)^{1/6(1+w)}$.
Here, it is assumed that $w>-1$.
Note that even if $a=a_\mathrm{min}$, $x$ and, therefore, $H$ do not
vanish, which means that the existence of $a_\mathrm{min}$ does not 
imply the bounce solution.
It is likely that a universe with $a=a_\mathrm{min}$ could appear, 
for example, as a quantum correction.

When $a$ is large, we find that $\mathcal{F}(0)$ in (\ref{nlGBr8}) 
vanishes and $\alpha_\pm$ tends to a constant:
$\alpha_\pm^3 \to \pm i \frac{1}{640 \kappa^6} \sqrt{ - \frac{ 
\alpha^3}{10}}$,
which gives
\be
\label{nlGBr14}
H^2 = x = 0\, ,\ \pm\sqrt{ - \frac{3\alpha}{160\kappa^4}}\, .
\ee
The positive solution of $x$ in (\ref{nlGBr14}) corresponds to the de 
Sitter solution (\ref{nlGB8}).

When the Einstein-Hilbert term in (\ref{nlGB1}) or (\ref{nlGB3}) can be
neglected, by assuming
\be
\label{nlGB9}
\phi=\frac{a_\phi}{t^2}\, ,\quad H = \frac{h_0}{t}\ 
\left(a=a_0 t^{h_0}\right)\, ,
\ee
Eqs.~(\ref{nlGB5}) reduce to the following algebraic equations
\be
\label{nlGB10}
0=a_\phi\left(\frac{\alpha}{\kappa^2}a_\phi + 24 h_0^3\right)\, ,\quad
0= \left(h_0 - 1\right)\left(\frac{\alpha}{\kappa^2}a_\phi + 4 
h_0^3\right)\, .
\ee
The equations in (\ref{nlGB10}) have a non-trivial solution ($a_\phi\neq 0$
and $h_0\neq 0$): $h_0=1$ and $a_\phi = - \frac{24\kappa^2}{\alpha}$.

Matter can also be accounted in the model; in that case, 
the first equation in (\ref{nlGB10}) is modified as
\be
\label{nlGB12}
0= - \frac{3H^2}{\kappa^2} + \frac{\alpha}{2\kappa^2}
{\dot\phi}^2 - 24\dot \phi H^3 + \rho_\mathrm{matter}\, .
\ee
When one may neglect the Einstein-Hilbert term in (\ref{nlGB1}) or
(\ref{nlGB3}), and if we assume (\ref{nlGB9}), there is 
a non-trivial solution when $h_0 = \frac{4}{3(w+1)}$.
In fact, Eqs.~(\ref{nlGB5}) reduce to the following algebraic expressions
\be
\label{nlGB14}
0 = \frac{2\alpha}{\kappa^2}a_\phi^2 + \rho_0 a_0^{-3(w+1)}
+ 48 a_\phi \left(\frac{4}{3(1+w)}\right)^3\, ,
\quad
0= \left(\frac{4}{3(1+w)} - 1\right)\left(\frac{\alpha}{\kappa^2}a_\phi
+ 4\left(\frac{4}{3(1+w)}\right)^3 \right)\, .
\ee
If the matter is radiation with $w=1/3$, the second equation is satisfied
and the first equation acquires the following form:
$0 = \frac{2\alpha}{\kappa^2}a_\phi^2 + \rho_0 a_0^{-4} + 48 a_\phi$.
Then, if $\tilde D \equiv \left(24\right)^2 - \frac{2\alpha 
\rho_0}{\kappa^2 a_0^4} > 0$,
there are non-trivial solutions:
$a_\phi = \frac{\kappa^2}{2\alpha}\left(-24 \pm \sqrt{\tilde D}\right)$.
On the other hand, when $w\neq 1/3$, we find
$a_\phi = - \frac{4\kappa^2}{\alpha} \left(\frac{4}{3(1+w)}\right)^3$,
$\rho_0 a_0^{-4} = \frac{160 \kappa^2}{\alpha}
\left(\frac{4}{3(1+w)}\right)^6$.
The second equation requires $\alpha$ to be positive.
Thus, a phantom like universe solution from purely non-local
gravity, without the $R$ term, has been found.
In the same way, in the case where several de Sitter solutions exist 
one can expect the unified description of inflation with dark energy.

\subsubsection{Other classical non-local Gauss-Bonnet models and their de
Sitter solutions}

Another model giving rise to an interesting non-local action is the 
following
\be
\label{nlGBrrr1}
S=\int d^4 x \sqrt{-g}\left[
\frac{R}{2\kappa^2} - \frac{\kappa^2}{2 a} F(\mathcal{G})
\Box^{-1}F (\mathcal{G}) \right]\, ,
\ee
where $a$ is a dimensional constant and $F(\mathcal{G})$ is an adequate
function of $\mathcal{G}$.
By introducing three scalar fields, $\phi$, $\xi$,
and $\eta$, one can rewrite the action (\ref{nlGBrrr1}) in the 
following form:
\be
\label{nlGBrrr2}
S=\int d^4 x \sqrt{-g} \left(\frac{R}{2\kappa^2}
+ \frac{ a}{2\kappa^2}\partial_\mu \phi \partial^\mu \phi
+ \phi F(\eta)
+ \xi \left(\eta - \mathcal{G}\right)\right)\, .
\ee
We may further add a potential $V(\phi)$ to the action
\be
\label{nlGBrrr2b}
S=\int d^4 x \sqrt{-g} \left(\frac{R}{2\kappa^2}
+ \frac{ a}{2\kappa^2}\partial_\mu \phi \partial^\mu \phi
+ \phi F(\eta)
+ \xi \left(\eta - \mathcal{G}\right) - V(\phi)\right)\, .
\ee
In the FRW universe, this action leads to the following equations
\bea
\label{nlGBrrr3}
&& 0 = - \frac{3}{\kappa^2}H^2 - \frac{a}{2\kappa^2}{\dot\phi}^2 
+ 24 \dot\xi H^3 + V(\phi) \, ,\quad
0 = \frac{a}{\kappa^2} \left(\ddot\phi + 3H\dot\phi\right) 
+ F(\eta) - V'(\phi)\, ,\nonumber \\
&& 0 = \phi F(\eta) + \xi\, ,\quad
0 = \eta - 24 \left(H^4 + \dot H H^2\right)\, .
\eea
However, after adding the potential $V(\phi)$, it is difficult to obtain 
the corresponding non-local action explicitly.

When $V(\phi)=0$, assuming that $\phi = c_\phi t$, $\eta = \eta_0$,
$\xi = c_\xi t$, and $H=H_0$, with constant $c_\phi$, $\eta_0$, 
$c_\xi$, and $H_0$, the equations in (\ref{nlGBrrr3}) reduce
to the algebraic equations
\be
\label{nlGBrrr5}
0 = - \frac{3}{\kappa^2}H_0^2 - \frac{a}{2\kappa^2}c_\phi^2
+ 24 c_\xi H_0^3\, ,\quad
0 = \frac{3a}{\kappa^2} H_0 c_\phi + F(\eta_0)\, ,\quad
0 = c_\phi F'(\eta_0) + c_\xi\, ,\quad
0 = \eta_0 - 24 H_0^4 \, .
\ee
One can solve Eqs.~(\ref{nlGBrrr5}) with respect to $c_\phi$, $c_\xi$, 
and $\eta_0$ as follows: 
$c_\phi = - \frac{\kappa^2 F\left(24 H_0^4 \right)}{3a H_0}$,
$c_\xi = - \frac{\kappa^2}{3 a H_0}F\left(24 H_0^4 \right)
F'\left(24 H_0^4\right) $, $\eta_0 = 24 H_0^4$, and we find
\be
\label{H0}
0 = - \frac{G_0}{8\kappa^2} - \frac{\kappa^2}{18a} F\left(G_0 \right)^2
+ \frac{\kappa^2}{9a} G_0 F\left(G_0 \right) F'\left(G_0 \right)\, .
\ee
Here, $G_0 = 24 H_0^4$.

For example, if
\be
\label{H1}
F\left(\mathcal{G} \right) = f_0 \mathcal{G}^2\, ,
\ee
Eq.~(\ref{H0}) gives
\be
\label{H2}
0 = G_0 \left( - \frac{1}{8\kappa^2} 
+ \frac{11 \kappa^2 f_0^2}{18 a} G_0^3 \right) \, ,
\ee
which has a trivial solution $G_0 = 0$ corresponding to the flat 
background and
\be
\label{H3}
G_0^3 = \frac{9a}{44 \kappa^4 f_0^2}\, ,
\ee
which corresponds to the de Sitter universe.

As another example, one can choose
\be
\label{H4}
F\left(\mathcal{G} \right)^2 = - g_0 \left(\mathcal{G} + g_1\right) 
\left( \mathcal{G} - g_2\right) \, .
\ee
Here, $g_0$, $g_1$, and $g_2$ are positive constants.
For $F\left(\mathcal{G} \right)$ to be real, the value
of $\mathcal{G}$ is restricted as $-g_1 < \mathcal{G} < g_2$. 
Then, Eq.~(\ref{H0}) yields
\be
\label{H5}
0 = G_0^2 + \frac{9a}{4 g_0 \kappa^2} G_0 + g_1 g_2\, ,
\ee
which can be solved as
\be
\label{H6}
G_0 = G_{0\,\pm} \equiv \frac{1}{2}\left( - \frac{9a}{4 g_0 \kappa^2} 
\pm \sqrt{\left(\frac{9a}{4 g_0 \kappa^2}\right)^2 + 4g_1 g_2} 
\right)\, .
\ee
The above solution is positive, provided that $a<0$.
For $-g_1 < G_{0\,\pm} < g_2$, we require 
\be
\label{H7}
g_1> - \frac{9a}{4 \kappa^4 g_0}\, ,\quad g_2 
> - \frac{9a}{8 \kappa^4 g_0}\, .
\ee
Then, as long as Eq.~(\ref{H7}) is satisfied, there are two solutions
describing the de Sitter universe.
By choosing a more general $F(\mathcal{G})$, we may find
multiple de Sitter universe solutions,
among which the solution with the largest value of $H_0$ could perfectly
correspond to the inflation epoch and the one with the smallest 
value of $H_0$ to the dark energy era.
Note that, in the model (\ref{H4}), as long as Eq.~(\ref{H7}) is satisfied,
there is no singularity, for example at $\mathcal{G}=0$. 
Thus, a smooth transition could occur and there exists a true 
possibility to realize matter dominance before late-time acceleration.
The important lesson from the above investigation is again
the fact that the inflation and dark energy epochs may be natural 
solutions of the field equations.

%%%%%%%%%%%
%%%%%%%%%%%%%
%%%%%%%%%%%%

\subsection{Gravity non-minimally coupled with 
the matter Lagrangian \label{IIE}}

\subsubsection{Non-minimally coupled scalar theory}

It could be interesting to investigate the role of the non-standard
non-minimal coupling of modified gravity with the matter Lagrangian 
in the cosmic acceleration epoch.
The corresponding, quite general theory has been suggested in
refs.~\cite{Nojiri:2004bi,Allemandi:2005qs} 
(see also \cite{Capozziello:1999xt}).
As an example of such a theory, the following action is considered:
\be
\label{LR1}
S=\int d^4 x \sqrt{-g}\left\{ \frac{R}{2\kappa^2} 
+ f \left(R\right) L_d \right\}\, .
\ee
Here, $L_d$ is a standard matter like action that produces dark energy due 
to non-trivial coupling with the curvature. 
In principle, the first term may be more general, for instance, of 
the $f(R)$ form.
However, for illustrative purposes it is enough to consider this simpler
variant of the non-minimal theory.
The second term in the above action describes the non-linear coupling of
matter with gravity.
Very often, such terms may be induced by quantum effects such 
as the non-linear effective action.
Thus, it is natural to consider that the second term 
belongs to matter sector.

By the variation over $g_{\mu\nu}$, the equation of motion follows as:
\be
\label{LR2}
0= \frac{1}{\sqrt{-g}} \frac{\delta S}{\delta g_{\mu\nu}}
= \frac{1}{2\kappa^2}\left\{ \frac{1}{2}g^{\mu\nu}R - R^{\mu\nu}\right\}
+ \tilde T^{\mu\nu}\, .
\ee
Here, the effective energy-momentum tensor $\tilde T_{\mu\nu}$ 
is defined by
\bea
\label{w5}
\tilde T^{\mu\nu}&\equiv& - f'(R) R^{\mu\nu} L_d
+ \left(\nabla^\mu \nabla^\nu - g^{\mu\nu}\nabla^2 \right)
\left(f'(R) L_d\right) + f(R) T^{\mu\nu}\, ,\nn
T^{\mu\nu}&\equiv& \frac{1}{\sqrt{-g}}\frac{\delta}{\delta g_{\mu\nu}}
\left(\int d^4x\sqrt{-g} L_d\right)\, .
\eea
As a matter we consider the massless scalar 
\be
\label{LR4}
L_d = - \frac{1}{2}g^{\mu\nu}\partial_\mu \phi \partial_\nu \phi\, .
\ee
Then, the equation given by the variation over $\phi$ has 
the following form:
\be
\label{LR5}
0 = \frac{1}{\sqrt{-g}}\frac{\delta S}{\delta \phi}= \frac{1}{\sqrt{-g}}
\partial_\mu \left( f(R) \sqrt{-g} g^{\mu\nu}\partial_\nu \phi \right)\, .
\ee
It is enough to consider the case that
\be
\label{fLd}
f(R) = \left( \frac{R}{\mu^2} \right)^\alpha\, .
\ee
We again assume the spatially flat FRW universe (\ref{JGRG14}) and 
that the scalar $\phi$ only depends on $t$ $\left(\phi=\phi(t)\right)$. 
The solution of the scalar field equation (\ref{LR5}) is given by
\be
\label{LR9}
\dot \phi = q a^{-3} R^{-\alpha}\, .
\ee
Here, $q$ is a constant of the integration.
Thus, $R^\alpha L_d = \frac{q^2}{2 a^6 R^\alpha}$,
which becomes dominant when $R$ is small (large) compared with the
Einstein term $\frac{R}{2\kappa^2}$ if 
$\alpha>-1$ $\left(\alpha <-1\right)$.
Thus, one arrives at the remarkable possibility 
\cite{Nojiri:2004bi,Allemandi:2005qs}
that dark energy grows to asymptotic dominance over the usual matter 
with a decrease of the curvature.
In the current universe, this solves the coincidence problem 
(the equality of the energy density for dark energy and for matter) 
simply by the fact of universe expansion.

Substituting (\ref{LR9}) into (\ref{LR2}), the $(\mu,\nu)=(t,t)$
component of the equation of motion has the following form:
\bea
\label{LR11}
0&=& - \frac{3}{2\kappa^2}H^2
+ \frac{36q^2}{\mu^{2\alpha} a^6\left(6\dot H + 12 H^2\right)^{\alpha 
+ 2}} \left\{ \frac{\alpha (\alpha + 1)}{4}\ddot H H 
+ \frac{\alpha + 1}{4}{\dot H}^2
\right. \nn
&& \left. + \left(1 + \frac{13}{4}\alpha + \alpha^2\right)\dot H H^2
+ \left(1 + \frac{7}{2}\alpha\right) H^4 \right\}\, .
\eea
The accelerated FRW solution of (\ref{LR11}) exists as  
\cite{Nojiri:2004bi,Allemandi:2005qs}
\be
\label{LR13}
a = a_0 t^{\frac{\alpha + 1}{3}} \quad \left(H= \frac{\alpha + 1}{3t}
\right)\, ,\quad
a_0^6 \equiv \frac{2\kappa^2 q^2 \left(2\alpha - 1\right)
\left(\alpha - 1\right) }
{\mu^{2\alpha} 3\left(\alpha + 1\right)^{\alpha + 1}
\left( \frac{2}{3}\left(2\alpha - 1\right)\right)^{\alpha + 2}}\, .
\ee
Eq.~(\ref{LR13}) shows that the universe accelerates, that is, $\ddot a>0$
if $\alpha>2$. If $\alpha<-1$, the solution (\ref{LR13}) describes 
a shrinking universe if $t>0$.
If the time is shifted as $t\to t - t_s$ ($t_s$ is a constant), the
accelerating and expanding universe occurs when $t<t_s$.
In the solution with $\alpha<-1$ a Big Rip singularity appears at
$t=t_s$. 
For $a\propto t^{h_0}$, the effective EoS parameter $w_\mathrm{eff}$ in
(\ref{JGRG18}) is given by $w_\mathrm{eff} = -1 + \frac{2}{3h_0}$,
and the accelerating universe expansion ($h_0>1$) occurs if
$-1< w_\mathrm{eff} <- \frac{1}{3}$.
For the case of (\ref{LR13}), one finds
\be
\label{LRa2}
w_\mathrm{eff} = \frac{1 - \alpha}{1 + \alpha}\, .
\ee
Then, if $\alpha<-1$, we have $w_\mathrm{eff}<-1$, which is an effective
phantom regime.
For general matter with the constant EoS parameter $w$,
the energy $E$ and the energy density $\rho_\mathrm{matter}$ behave as
$E\sim a^{-3w}$ and $\rho_\mathrm{matter}\sim a^{-3\left(w + 1\right)}$.
Thus, for the standard phantom with $w<-1$, the density becomes large with
time and might generate the Big Rip.
Thus, it is demonstrated that the non-linear coupling of gravity with 
matter may produce an accelerated cosmological expansion with 
the effective EoS parameter close (above, equal or below) to $-1$. 
More complicated couplings \cite{Nojiri:2004bi,Allemandi:2005qs,nonlinear} 
may be considered in the same fashion. 
Note that such theories show some very interesting properties, 
like the induction of extra gravitational force \cite{extra}.
Other cosmological aspects of such non-minimally coupled non-linear 
theories are discussed in refs.~\cite{nonlinear1}.

%%%%%%%%%%%
%%%%%%%%%%%
%%%%%%%%%%%%

\subsubsection{Non-minimally coupled vector model}

The non-minimal coupling of gravity with the matter Lagrangian of the 
previous section may be considered for more complicated models.
Let us investigate the abelian vector field Lagrangian non-minimally 
coupled with gravity \cite{Bamba:2008ja}:
\be
\label{V1}
S = \int d^4 x \sqrt{-g} \left\{ \frac{R}{2\kappa^2} - \frac{I(R)}{4}
F_{\mu\nu} F^{\mu\nu} \right\}\, .
\ee
Here, $I(R)$ is an appropriate function of the scalar curvature $R$, and
$F_{\mu\nu}$ is the curvature of the $U(1)$ gauge field. 
Making the variation of the action Eq.~(\ref{V1}) with respect to the
metric $g_{\mu\nu}$ and the $U(1)$ gauge field $A_{\mu}$, we obtain the
following equations:
\bea
\label{eq:2.5}
&& R_{\mu \nu} - \frac{1}{2}g_{\mu \nu}R 
= \kappa^2 T^{(\mathrm{EM})}_{\mu\nu}\, , \\
&& T^{(\mathrm{EM})}_{\mu \nu}
= I(R) \left( g^{\alpha\beta} F_{\mu\beta}
F_{\nu\alpha} - \frac{1}{4} g_{\mu\nu} F_{\alpha\beta}
F^{\alpha\beta} \right) \nn
&& \quad +\frac{1}{2} \left\{ f'(R) F_{\alpha\beta} F^{\alpha\beta} 
R_{\mu\nu} + g_{\mu \nu} \Box \left[ f' (R) F_{\alpha\beta}
F^{\alpha\beta} \right] - {\nabla}_{\mu} {\nabla}_{\nu} \left[ f' (R)
F_{\alpha\beta}F^{\alpha\beta} \right] \right\} \, ,
\label{eq:2.6}
\eea
and
\be
 -\frac{1}{\sqrt{-g}}{\partial}_{\mu}
\left( \sqrt{-g} I(R) F^{\mu\nu} \right) = 0\, .
\ee
In the FRW universe (\ref{JGRG14}), the $(\mu,\nu)=(0,0)$ component 
and $(\mu,\nu)=(i,j)$ component of Eq.~(\ref{eq:2.5}) gives 
\bea
H^2 &=& \frac{\kappa^2}{3} \left\{
I(R) \left( g^{\alpha\beta} F_{0\beta} F_{0\alpha}
+ \frac{1}{4} F_{\alpha\beta}F^{\alpha\beta} \right) \right. \nn
&& + \frac{3}{2} \left[ -f'(R) \left( \dot{H} + H^2 \right)
+ 6 f''(R) H \left( \ddot{H} + 4H\dot{H} \right)
\right] F_{\alpha\beta}F^{\alpha\beta} \nn
&& \left. + \frac{3}{2} f'(R) H \frac{d}{dt}\left(
F_{\alpha\beta}F^{\alpha\beta}
\right) -\frac{1}{2} f'(R) \frac{1}{a^2}
{\mathop{\Delta}\limits^{(3)}} \left( F_{\alpha\beta} F^{\alpha\beta} 
\right) \right\}\, , \\
2\dot{H} + 3H^2 &=& \frac{\kappa^2}{2}
\left\{ \frac{1}{6} I(R) F_{\alpha\beta}F^{\alpha\beta}
+ \left[ -f' (R) \left( \dot{H} + 3H^2 \right) \right. \right. \nn
&& \left. + 6f'' (R) \left(\dddot{H}+7H\ddot{H}+4\dot{H}^2+12H^2\dot{H} 
\right) + 36f'''(R) \left( \ddot{H} + 4H \dot{H} \right)^2 \right]
F_{\alpha\beta}F^{\alpha\beta} \nn
&& +3\left[f'(R)H + 4f''(R) \left( \ddot{H} + 4H \dot{H} \right) \right]
\frac{d}{dt} \left( F_{\alpha\beta}F^{\alpha\beta} \right)
+ f'(R) \frac{d^2}{dt^2} \left( F_{\alpha\beta}F^{\alpha\beta} \right) \nn
&& \left. -\frac{2}{3}f^{\prime}(R) \frac{1}{a^2}
{\mathop{\Delta}\limits^{(3)}} \left( F_{\alpha\beta}F^{\alpha\beta} 
\right) \right\}\, .
\label{eq:2.34}
\eea
By properly choosing $I(R)$, a model unifying the late
acceleration and the inflation emerges. However, 
the properties of the model may be investigated only qualitatively 
\cite{Bamba:2008ja} or numerically because 
the equations of motion are very complicated. Alternatively,
as will be shown in the next chapter the cosmological reconstruction 
of the theory under investigation may be conducted 
so that unified cosmic history is realized. 
Different properties of the non-minimally coupled vector theory have 
been studied in refs.~\cite{bal}.

\subsubsection{Non-minimal Yang-Mills theory}

Here, we consider the cosmology in the non-abelian non-minimal
vector-$F(R)$ gravity based on ref.~\cite{Bamba:2008xa}. Different 
cosmological proposals regarding the non-minimal Yang-Mills theory 
may be found in refs.~\cite{bal1}.

The proposed action follows:
\bea
&& \bar{S}_{\mathrm{MG}} = \int d^{4}x \sqrt{-g}
\left[ {\mathcal{L}}_{\mathrm{MG}} + {\mathcal{L}}_{\mathrm{V}} 
\right]\, , \nn
\label{eq:4.1}
&& {\mathcal{L}}_{\mathrm{MG}} = \frac{1}{2\kappa^2} \left[ R+F(R) 
\right]\, , \quad
{\mathcal{L}}_{\mathrm{V}} = I(R) 
\left\{ -\frac{1}{4} F_{\mu\nu}^{a} F^{a\mu\nu} - V[A^2]
\right\}\, ,\quad A^2 \equiv {A^a}^2
\eea
where
$F_{\mu\nu}^{a}$ is given by
\be
F_{\mu\nu}^{a} = {\partial}_{\mu}A_{\nu}^{a} - {\partial}_{\nu}A_{\mu}^{a}
+ f^{abc} A_{\mu}^{b} A_{\nu}^{c}\,,
\label{eq:2.6B}
\ee
and $A^2= g^{\mu \nu} A_{\mu}^{a} A_{\nu}^{a}$.
We should note that the last term $V[A^2]$ in the action (\ref{eq:4.1}) 
is not gauge invariant, but it can be rewritten in a gauge-invariant way.
For example, if the gauge group is a unitary group, one may introduce a
$\sigma$ model like field $U$, which satisfies $U^\dagger U=1$. 
Then, the last term could be rewritten in the gauge-invariant form:
\be
\label{sn1}
V[A^2] \to V\left[ \bar{c} \tr \left(U^\dagger A_{\mu}^{a} U\right)
\left(U^\dagger A^{a\mu} U\right) \right]\, .
\ee
Here, $\bar{c}$ is the normalization constant. 
Choosing the unitary gauge $U=1$, the term in (\ref{sn1}) reduces to
the original one: $V[A^2]$. This may indicate that the action 
(\ref{eq:4.1}) describes the theory in which the gauge group 
is spontaneously broken.

The field equations can be derived by taking variations of the
action (\ref{eq:4.1}) with respect to the
metric $g_{\mu\nu}$ and the vector field $A_{\mu}^{a}$ as follows:
\be
\left[ 1+F' (R) \right] R_{\mu \nu} - \frac{1}{2}g_{\mu \nu}
\left[ R+F(R) \right] + g_{\mu \nu}
\Box F'(R) - {\nabla}_{\mu} {\nabla}_{\nu} F'(R)
= \kappa^2 T^{(\mathrm{V})}_{\mu \nu}\, ,
\label{eq:4.3}
\ee
with
\begin{eqnarray}
T^{(\mathrm{V})}_{\mu \nu} &=&
I(R) \left\{ g^{\alpha\beta} F_{\mu\beta}^{a} F_{\nu\alpha}^{a}
+ 2 A_{\mu}^{a} A_{\nu}^{a}
\frac{d V[A^2]}{d A^2} - \frac{1}{4} g_{\mu\nu}
\bar{\mathcal{F}} \right\} \nonumber \\
&& +\frac{1}{2} \left\{ f'(R) \bar{\mathcal{F}} R_{\mu \nu}
+ g_{\mu \nu} \Box \left[ f'(R) \bar{\mathcal{F}}
\right] - {\nabla}_{\mu} {\nabla}_{\nu}
\left[ f'(R) \bar{\mathcal{F}} \right] \right\}\,,
\label{eq:4.4} \\
\bar{\mathcal{F}} &=& F_{\mu\nu}^{a} F^{a\mu\nu} + 4V[A^2]\,,
\label{eq:4.5}
\end{eqnarray}
and
\be
\frac{1}{\sqrt{-g}}{\partial}_{\mu}
\left[ \sqrt{-g} I(R) F^{a\mu\nu} \right] - I(R) \left\{
f^{abc} A_{\mu}^{b} F^{c\mu\nu}
+ 2 \frac{d V[A^2]}{d A^2} A^{a \nu} \right\} = 0\,,
\label{eq:4.6}
\ee
where $T^{(\mathrm{V})}_{\mu \nu}$ is the contribution to
the energy-momentum tensor from $A_{\mu}^{a}$.
As will be shown in the next chapter, the cosmological reconstruction of
the above theory may result in the accelerating expanding universe.
The direct solution of the equations of motion is extremely complicated.

%%%%%%%%%%%
%%%%%%%%%%%
%%%%%%%%%%%

\subsection{Modified $F(R)$ Ho\v{r}ava-Lifshitz gravity \label{IIF}}

\subsubsection{General formulation}

This section is devoted to a review of the $F(R)$
Ho\v{r}ava-Lifshitz gravity \cite{Carloni:2010nx}. 
The FRW equations for this theory are also formulated. 
The Ho\v{r}ava-Lifshitz version of general relativity
was originally introduced as a candidate for renormalizable 
quantum gravity at the price of the violation of Lorentz 
invariance. The below model represents natural generalization 
of the standard $F(R)$ gravity (\ref{JGRG6}) discussed 
at the beginning of this chapter.

Let us present the general formulation of the theory.
By using the ADM decomposition \cite{Arnowitt:1962hi,Gao:2009er}
(for reviews, see \cite{Wald:1984,Gourgoulhon:2007}),
one can write the metric in the following form:
\be
\label{HLF2}
ds^2 = - N^2 d t^2 +
g^{(3)}_{ij}\left(d x^i + N^i d t \right)\left(d x^j + N^j d t \right),
\quad i=1,2,3\, .
\ee
Here, $N$ is called the lapse variable and $N^i$'s are the shift variables.
The scalar curvature $R$ becomes:
\be
\label{HLF3} R= K^{ij}
K_{ij} - K^2 + R^{(3)} + 2 \nabla_\mu
\left( n^\mu \nabla_\nu n^\nu - n^\nu \nabla_\nu n^\mu \right) \, ,
\ee
and $\sqrt{-g} = \sqrt{g^{(3)}} N$.
Here, $R^{(3)}$ is the three-dimensional scalar
curvature defined by the metric $g^{(3)}_{ij}$, and $K_{ij}$ is the
extrinsic curvature defined by
\be
\label{HLF4}
K_{ij}=\frac{1}{2N}\left(\dot
g^{(3)}_{ij}-\nabla^{(3)}_iN_j-\nabla^{(3)}_jN_i\right) \, ,\quad K
=K^i_{\ i}\, .
\ee
$n^\mu$ is a unit vector perpendicular to the
three-dimensional hypersurface $\Sigma_t$ defined by $t=\text{constant}$
and $\nabla^{(3)}_i$ expresses the covariant derivative on
the hypersurface $\Sigma_t$.

The first attempt of the extension of $F(R)$ gravity to 
a Ho\v{r}ava-Lifshitz-type theory \cite{Horava:2009uw} which 
is known to be a power-counting renormalizable model was proposed 
in ref.~\cite{Kluson:2009xx}. The starting action is
\be
\label{HLF5}
S_{F_\mathrm{HL}(R)} = \int d^4 x \sqrt{g^{(3)}} N
F(R_\mathrm{HL})\, ,\quad R_\mathrm{HL} 
\equiv K^{ij} K_{ij} - \lambda K^2 - E^{ij}\mathcal{G}_{ijkl} E^{kl} \, .
\ee
Here, $\lambda$ is a real constant in the ``generalized 
De~Witt metric'' or ``super-metric'',
\be
\label{HLF6}
\mathcal{G}^{ijkl} = \frac{1}{2}\left( g^{(3) ik} g^{(3) jl} + g^{(3) il}
g^{(3) jk} \right) - \lambda g^{(3) ij} g^{(3) kl}\, ,
\ee
defined on the three-dimensional hypersurface $\Sigma_t$, and 
$E^{ij}$ can be defined by the so-called ``detailed balance condition''
by using an action $W[g^{(3)}_{kl}]$ on the hypersurface $\Sigma_t$
\be
\label{HLF7}
\sqrt{g^{(3)}}E^{ij} = \frac{\delta W[g^{(3)}_{kl}]}{\delta g_{ij}}\, ,
\ee
and the inverse of $\mathcal{G}^{ijkl}$ is written as
\be
\mathcal{G}_{ijkl} = \frac{1}{2}\left( g^{(3)}_{ik} g^{(3)}_{jl}
+ g^{(3)}_{il} g^{(3)}_{jk} \right)
- \tilde{\lambda} g^{(3)}_{ij} g^{(3)}_{kl}\, ,\quad
\tilde{\lambda} = \frac{\lambda}{3\lambda - 1}\, .
\ee
The action $W[g^{(3)}_{kl}]$ is assumed to be defined by the metric and
the covariant derivatives on the hypersurface $\Sigma_t$.
The original motivation for the detailed balance condition is its ability 
to simplify the quantum behavior and renormalization properties
of theories that satisfy the detailed balance condition. 
Otherwise, there is no a priori physical
reason to restrict $E^{ij}$ to be defined by (\ref{HLF7}).

There exists an anisotropy between space and time in the 
Ho\v{r}ava-Lifshitz approach. 
In the ultraviolet (high-energy) region, the time
coordinate and the spatial coordinates are assumed to behave as
\be
\label{HLF7b}
\bm{x}\to b\bm{x}\, ,\quad t\to b^z t\, ,\quad
z=2,3,\cdots\, ,
\ee
under the scale transformation.
In ref.~\cite{Horava:2009uw}, $W[g^{(3)}_{kl}]$ is explicitly given for 
the case $z=2$,
\be
\label{HLF7c} W=\frac{1}{\kappa_W^2}\int
d^3\bm{x}\,\sqrt{g^{(3)}}(R-2\Lambda_W)\, ,
\ee
and for the case $z=3$,
\be
\label{HLF7d}
W=\frac{1}{w^2}\int_{\Sigma_t}\omega_3(\Gamma)\, .
\ee
Here, $\kappa_W$ is a coupling constant of dimension $-1/2$ and
$w^2$ is the dimensionless coupling constant.
$\omega_3(\Gamma)$ is given by
\be
\label{HLF7e}
\omega_3(\Gamma) = \mathrm{Tr}\left(\Gamma\wedge
d\Gamma+\frac{2}{3}\Gamma\wedge\Gamma \wedge\Gamma\right) \equiv
\varepsilon^{ijk}\left(\Gamma^{m}_{il}\partial_j
\Gamma^{l}_{km}+\frac{2}{3}\Gamma^{n}_{il}\Gamma^{l}_{jm}
\Gamma^{m}_{kn}\right)d^3\bm{x}\, .
\ee
A general $E^{ij}$ consists of all contributions to $W$ up 
to the chosen value $z$.

%%%%%%%%%%%%
%%%%%%%%%%%%
%%%%%%%%%%%%

In the Ho\v{r}ava-Lifshitz like $F(R)$ gravity, it is assumed that $N$
can only depend on the time coordinate $t$, which is called the
``projectability condition''. The reason for this is 
that Ho\v{r}ava-Lifshitz gravity does not have the full diffeomorphism
invariance, but rather is invariant only under ``foliation-preserving''
diffeomorphisms, i.e., under the transformations
\be
\label{fpd1}
\delta x^i=\zeta^i(t,\bm{x})\,, \, \quad \delta t=f(t)\, .
\ee
If $N$ depends on the spatial coordinates, it could not be fixed to be
unity ($N=1$) by using the foliation-preserving diffeomorphisms.
There exists a version of Ho\v{r}ava-Lifshitz gravity without
the projectability condition, but it is suspected to possess a few
additional consistency problems \cite{Henneaux:2009zb,Li:2009}.
Therefore, we prefer to assume that $N$ depends only on the time
coordinate $t$.

Let us consider the FRW universe with a flat spatial part
and the lapse function $N$:
\be
\label{HLF8}
ds^2 = - N^2 d t^2 + a(t)^2 \sum_{i=1,2,3} \left( d x^i \right)^2\, .
\ee
It is clear from the explicit
expressions (\ref{HLF7c}) and (\ref{HLF7d}) that $W[g^{(3)}_{kl}]$
vanishes identically if $\Lambda_W =0$, which is assumed because
a non-vanishing $\Lambda_W$ gives a cosmological constant. 
Then, one can obtain
\be
\label{HLF9}
R= \frac{ 12 H^2}{N^2} +
\frac{6}{N}\frac{d}{d t}\left(\frac{H}{N}\right) 
= - \frac{6H^2}{N} + \frac{6}{a^3 N} \frac{d}{d t}\left(
\frac{Ha^3}{N} \right)\, ,\quad R_\mathrm{HL} = \frac{ \left(3 - 9
\lambda \right) H^2}{N^2} \, .
\ee
In the case of general relativity, the second term in the 
last expression for $R$ becomes a total derivative:
\be
\label{HLF10}
\int d^4 x \sqrt{-g} R = \int d^4 x\ a^3 N
\left\{ - \frac{6H^2}{N} + \frac{6}{a^3 N} \frac{d}{d t}
\left( \frac{Ha^3}{N} \right) \right\}
= \int d^4 x \left\{ - 6H^2 a^3 + 6 \frac{d}{d t}
\left( \frac{Ha^3}{N} \right) \right\} \, .
\ee
Therefore, this term can be dropped in Einstein gravity. 
The total derivative term comes from the last term
$2\nabla_\mu \left( n^\mu \nabla_\nu n^\nu - n^\nu \nabla_\nu n^\mu
\right)$ in (\ref{HLF3}), which is dropped in the usual
Ho\v{r}ava-Lifshitz gravity. In $F(R)$ gravity, however, this
term cannot be dropped due to its non-linearity. Then, if we consider
FRW cosmology with the flat spatial part, there is almost no
qualitative difference between Einstein gravity and 
Ho\v{r}ava-Lifshitz gravity, except that an effective dark matter 
could appear as a kind of integration constant in 
Ho\v{r}ava-Lifshitz gravity \cite{Mukohyama:2009mz}. The effective
dark matter appears because the constraint given by the variation over
$N$ becomes global in the projectable Ho\v{r}ava-Lifshitz gravity.

Now, we introduce very general Ho\v{r}ava-Lifshitz like
$F(R)$ gravity \cite{Carloni:2010nx} by
\be
\label{HLF11}
S_{F(\tilde R)} = \int d^4 x \sqrt{g^{(3)}} N F(\tilde R)\, , \quad
\tilde R \equiv K^{ij} K_{ij} - \lambda K^2
+ 2 \mu \nabla_\mu \left( n^\mu \nabla_\nu n^\nu - n^\nu \nabla_\nu
n^\mu \right) - E^{ij}\mathcal{G}_{ijkl} E^{kl} \, .
\ee
In the FRW universe with a flat spatial part, $\tilde R$ 
has the following form:
\be
\label{HLF12}
\tilde R= \frac{ \left(3 - 9 \lambda \right)
H^2}{N^2} + \frac{6\mu }{a^3 N} \frac{d}{d t}\left(
\frac{Ha^3}{N} \right) = \frac{ \left(3 - 9 \lambda + 18 \mu \right)
H^2}{N^2} + \frac{6\mu }{N} \frac{d}{d t}\left( \frac{H}{N}
\right)\, .
\ee
The case one obtains with the choice of parameters
$\lambda=\mu=1$ corresponds to the usual $F(R)$ gravity as long as
we consider spatially flat FRW cosmology, because $\tilde R$ reduces
to $R$ in (\ref{HLF9}). On the other hand, in the case of $\mu=0$,
$\tilde R$ reduces to $R_\mathrm{HL}$ (\ref{HLF9}) and, therefore, 
the action (\ref{HLF11}) becomes identical with the action
(\ref{HLF5}) of the Ho\v{r}ava-Lifshitz like $F(R)$ gravity in
ref.~\cite{Kluson:2009xx}. Thus, the $\mu=0$ version corresponds to some
degenerate limit of the above general $F(R)$ Ho\v{r}ava-Lifshitz
gravity. We call this limit degenerate because it is very
difficult to obtain the FRW equations when
$\mu=0$ is set from the very beginning. In our theory the FRW equations
can be obtained quite easily, and then $\mu=0$ is a simple limit.

For the action (\ref{HLF11}), the FRW equation given by the variation
over $g^{(3)}_{ij}$ has the following form after assuming
the FRW space-time (\ref{HLF8}) and setting $N=1$:
\be
\label{HLF13}
0 = F\left(\tilde R\right) - 2 \left(1 - 3\lambda + 3\mu \right)
\left(\dot H + 3 H^2\right)
F'\left(\tilde R\right) - 2\left(1 - 3\lambda \right) H \frac{d
F'\left(\tilde R\right)}{d t}
+ 2\mu \frac{d^2 F'\left(\tilde R\right)}{d t^2} + p_\mathrm{matter}\, .
\ee

On the other hand, the variation over $N$ gives the global constraint:
\be
\label{HLF14}
0 = \int d^3 \bm{x} \left[ F\left(\tilde R\right) - 6 \left\{ 
\left(1 - 3\lambda + 3\mu\right) H^2 + \mu \dot H
\right\} F'\left(\tilde R\right) + 6 \mu H \frac{d F'\left(
\tilde R\right)}{d t} - \rho_\mathrm{matter} \right]\, ,
\ee
after setting $N=1$.
Because $N$ only depends on $t$, and thus does not depend on the spatial 
coordinates, we only obtain the global constraint given 
by the integration. If the standard conservation law (\ref{CEm}) 
is used, Eq.~(\ref{HLF13}) can be integrated to give
\be
\label{HLF16}
0 = F\left(\tilde R\right) - 6 \left\{ \left(1 - 3\lambda
+ 3\mu\right) H^2 + \mu \dot H
\right\} F'\left(\tilde R\right) + 6 \mu H \frac{d F'\left(\tilde
R\right)}{d t} - \rho_\mathrm{matter} - \frac{C}{a^3}\, .
\ee
Here, $C$ is the
integration constant. Using (\ref{HLF14}), one finds $C=0$. In 
\cite{Mukohyama:2009mz}, however, it was claimed that $C$ need
not always vanish in a local region, because (\ref{HLF14}) needs to be
satisfied throughout the entire universe. In the region $C>0$, 
the $Ca^{-3}$ term in (\ref{HLF16}) may be regarded as dark matter.

Note that Eq.~(\ref{HLF16}) corresponds to the first FRW equation
and (\ref{HLF13}) to the second one. Specifically, if we choose
$\lambda=\mu=1$ and $C=0$, Eq.~(\ref{HLF16}) reduces to
\bea
\label{HLF17}
0 &=& F\left(\tilde R\right) - 6 \left(H^2
+ \dot H \right) F'\left(\tilde R\right)
+ 6 H \frac{d F'\left(\tilde R\right)}{d t} - \rho_\mathrm{matter} \nn
&=& F\left(\tilde R\right) - 6 \left(H^2 + \dot H \right)
F'\left(\tilde R\right) + 36 \left(4H^2 \dot H + \ddot H\right)
F''\left(\tilde R\right) - \rho_\mathrm{matter} \, .
\eea
This coincides with the corresponding FRW equation for traditional $F(R)$
gravity introduced in the first section of this chapter.
Note that in the degenerate $\mu=0$ case 
\cite{Kluson:2009xx}, the action (\ref{HLF11}) or (\ref{HLF5}) does
not contain any term with second derivatives with respect to the
coordinates. Such a term appears in the standard $F(R)$ gravity. The
existence of the second derivatives in the usual $F(R)$ gravity
induces the third and fourth derivatives in the FRW equation as in
(\ref{HLF13}). Due to such higher derivatives, an extra scalar mode 
appears, which is often called the scalaron in the usual
$F(R)$ gravity. This scalar mode often affects the correction to 
Newton's law as well as other solar tests. Therefore, such a scalar
mode does not appear in the $F(R)$ Ho\v{r}ava-Lifshitz gravity
with $\mu=0$. Thus, we have formulated a general
Ho\v{r}ava-Lifshitz $F(R)$ gravity that describes the standard
$F(R)$ gravity or its non-degenerate Ho\v{r}ava-Lifshitz extension
in a consistent way.

\subsubsection{FRW cosmology}

This section is devoted to the study of the FRW Eqs.~(\ref{HLF13})
and (\ref{HLF14}), which admit a de Sitter universe solution. We follow
ref.~\cite{Carloni:2010nx}. It is convenient to neglect the matter
contribution: $p_\mathrm{matter}=\rho_\mathrm{matter}=0$.
Assuming $H=H_0$, both of Eqs.~(\ref{HLF13}) and (\ref{HLF14}) lead
to the same equation
\be
\label{HLF18}
0 = F\left( 3 \left(1 - 3 \lambda + 6 \mu \right) H_0^2
\right) - 6 \left(1 - 3\lambda + 3\mu\right) H_0^2
F'\left( 3 \left(1 - 3 \lambda + 6 \mu \right) H_0^2 \right) \, ,
\ee
as long as the integration constant vanishes ($C=0$) in Eq.~(\ref{HLF16}).

First, we consider the popular case that
\be \label{HLF19}
F\left(\tilde R\right) \propto \tilde R + \beta \tilde R^2 \, .
\ee
Then, Eq.~(\ref{HLF18}) gives
\be
\label{HLF20}
0 = H_0^2 \left\{ 1 -
3\lambda + 9\beta \left(1 - 3\lambda + 6 \mu \right)
\left( 1 - 3\lambda + 2\mu \right) H_0^2 \right\}\, .
\ee
In the case of usual $F(R)$ gravity, where $\lambda=\mu=1$ and, 
therefore, $1 - 3\lambda + 2\mu =0$, there is formally only 
a trivial solution $H_0^2 = 0$. For our general
case, however, there exists the non-trivial solution
\be
\label{HLF21}
H_0^2 = - \frac{ 1 - 3\lambda }{\beta \left(1 -
3\lambda + 6 \mu \right) \left( 1 - 3\lambda + 2\mu \right)}\, ,
\ee
as long as the right-hand side of (\ref{HLF21}) is positive. 
If the magnitude of this non-trivial solution is small enough, 
this solution might correspond to the late-time accelerating expansion.
Thus, the $R^2$ term may generate the late-time acceleration. On
the other hand, the above solution may serve as an inflationary
solution for the early universe (with the corresponding choice of
parameters).

Instead of (\ref{HLF19}) one may consider the following model:
\be
\label{HLF22}
F\left(\tilde R\right) \propto \tilde R + \beta \tilde
R^2 + \gamma \tilde R^3\, .
\ee
Eq.~(\ref{HLF18}) becomes
\be
\label{HLF23}
0 = H_0^2 \left\{ 1 - 3\lambda 
+ 9\beta \left(1 - 3\lambda + 6 \mu \right) 
\left( 1 - 3\lambda + 2\mu \right) H_0^2 
+ 9\gamma \left(1 - 3\lambda + 6 \mu \right)^2 
\left( 5 - 15\lambda + 12 \mu \right) H_0^4 \right\}\, ,
\ee
which has the following two non-trivial solutions,
\be
\label{HLF24} H_0^2 = - \frac{ 
\left( 1 - 3\lambda + 2\mu \right) \beta }{2 
\left(1 - 3\lambda + 6 \mu \right)
\left( 5 - 15\lambda + 12 \mu \right) \gamma} 
\left( 1 \pm \sqrt{ 1 - \frac{4 \left(1 - 3\lambda \right)
\left( 5 - 15\lambda + 12 \mu
\right) \gamma} { 9 \left( 1 - 3\lambda + 2\mu \right)^2 \beta^2} }
\right)\, ,
\ee
as long as the r.h.s. is real and positive. If
\be
\label{HLF25}
\left| \frac{4 \left(1 - 3\lambda \right)\left( 5 -
15\lambda + 12 \mu \right) \gamma} { 9 \left( 1 - 3\lambda + 2\mu
\right)^2 \beta^2} \right| \ll 1\, ,
\ee
one of the two solutions is much smaller than the other solution. 
Then, one may regard that the larger solution corresponds 
to the inflation and the smaller one to the
late-time acceleration, similarly to the
modified gravity model \cite{Nojiri:2003ft}, where such unification
had  been first proposed. Note that some of the above models may 
possess the future singularity in the same way as the usual 
$F(R)$ gravity. However, it would be
possible to demonstrate that adding terms with even higher derivatives
might cure this singularity, as the addition of the
$R^2$ term did in the standard $F(R)$ gravity.
Thus, we demonstrated the qualitative possibility to unify
the early-time inflation with the late-time acceleration in the
modified Ho\v{r}ava-Lifshitz $F(R)$ gravity.

\subsubsection{Unified inflation and dark energy in modified
Ho\v{r}ava-Lifshitz gravity}

Let us consider here some viable $F(\tilde{R})$ gravities that admit
the unification \cite{Nojiri:2003ft} of inflation with late-time 
acceleration.
In the traditional $F(R)$ theory, a number of viable models
(see \cite{Nojiri:2007as,Nojiri:2007cq,Cognola:2007zu}), 
which pass all local tests and are able to unify the inflationary
and the current cosmic accelerated epochs,  have been proposed. Here, 
following ref.~\cite{Elizalde:2010ep}, it is shown that this class of
models may be extended to the Ho\v{r}ava-Lifshitz gravity. The starting 
action is
\be
F(\tilde{R})=\tilde{R}+f(\tilde{R})\, ,
\label{3.1}
\ee
where it is assumed that the term $f(\tilde{R})$ becomes important at
cosmological scales, whereas for scales comparable to that of 
the solar system, the theory becomes linear on $\tilde{R}$. As an example,
the following function \cite{Nojiri:2007cq} is considered
\be
f(\tilde{R})=\frac{\tilde{R}^n(\alpha\tilde{R}^n-\beta)}
{1+\gamma \tilde{R}^n}\, ,
\label{3.2}
\ee
where $\alpha$, $\beta$, and $\gamma$ are constants and $n>1$. This theory
reproduces the inflationary and cosmic acceleration epochs in convenient
$F(R)$ gravity (see ref.~\cite{Nojiri:2007cq}).
During inflation, it is assumed that the curvature scalar tends 
to infinity. In this case, the model
(\ref{3.1}), with (\ref{3.2}), behaves as
\be
\lim_{\tilde{R}\rightarrow\infty}F(\tilde{R})=\alpha\tilde{R}^n\, .
\label{3.3}
\ee
Then, by solving the FRW equation (\ref{HLF16}) with $C=0$,
this kind of function yields a power-law solution of the type
\be
H(t)=\frac{h_1}{t}\, , \quad \mbox{where} \quad
h_1=\frac{2\mu(n-1)(2n-1)}{1-3\lambda+6\mu-2n(1-3\lambda+3\mu)}\, .
\label{3.4}
\ee
This solution produces acceleration during the inflationary epoch if
the parameters of the theory are properly defined. The acceleration
parameter is given by $\frac{\ddot{a}}{a}=h_1(h_1-1)/t^2$; thus, for
$h_1>1$, the inflationary epoch is well reproduced by the model
(\ref{3.2}). Alternatively, the function (\ref{3.2}) has a
minimum at $\tilde{R}_0$, given by
\be
\tilde{R}_0 \sim \left( \frac{\beta}{\alpha\gamma}\right)^{1/4}\, ,
\qquad f'(\tilde{R})=0\, , \qquad f(\tilde{R})
= -2\Lambda\sim -\frac{\beta}{\gamma}\, ,
\label{3.5}
\ee
where the condition $\beta\gamma/\alpha\gg 1$ is imposed. Then, at
the current epoch, the scalar curvature acquires a small value that 
can be fixed to coincide with the minimum (\ref{3.5}), such that the
FRW equations (\ref{HLF13}) and (\ref{HLF16}) with $C=0$ yield
\be
H^2=\frac{\kappa^2}{3(3\lambda-1)}\rho_\mathrm{matter}
+\frac{2\Lambda}{3(3\lambda-1)}\, \quad
\dot{H}=-\kappa^2\frac{\rho_\mathrm{matter}
+p_\mathrm{matter}}{3\lambda-1}\, ,
\label{3.6}
\ee
which look very similar to the standard FRW equations in general
relativity, except for the parameter $\lambda$.
(For a first consideration of the FRW dynamics in Ho\v{r}ava-Lifshitz
gravity based on general relativity, see
refs.~\cite{Takahashi:2009wc,Kiritsis:2009sh,Brandenberger:2009yt,
Mukohyama:2009zs,Sotiriou:2009bx,Saridakis:2009bv,
Minamitsuji:2009ii,Calcagni:2009qw,Wang:2009rw,Park:2009zra,
Nojiri:2009th,Jamil:2009sq,Bogdanos:2009uj,Boehmer:2009yz,
Bakas:2009ku,Calcagni:2009ar,Carloni:2009jc,Gao:2009wn,Myung:2009if,
Son:2010qh,Wang:2010mw,Ali:2010sv}.)
As has been pointed out, at the current
epoch, the scalar $\tilde{R}$ is small, so the theory is
in the IR limit where the parameter $\lambda\sim1$, and the equations
approach the usual equations for $F(R)$ gravity. Thus, the FRW
equations (\ref{3.6}) reproduce the behavior of the well-known
$\Lambda$CDM model with no need to introduce a dark energy fluid
to explain the current universe acceleration.

As another example of the models described by (\ref{3.1}), one
can consider the function \cite{Nojiri:2007as,Cognola:2007zu,Hu:2007nk},
\be
f(\tilde{R})=-\frac{(\tilde{R}-\tilde{R}_0)^{2n+1}
+\tilde{R}_0^{2n+1}}{f_0+f_1\left[(\tilde{R}-\tilde{R}_0)^{2n+1}
+\tilde{R}_0^{2n+1}\right]}=-\frac{1}{f_1}
+\frac{f_0/f_1}{f_0 +f_1\left[(\tilde{R}-\tilde{R}_0)^{2n+1}
+\tilde{R}_0^{2n+1}\right]}\, .
\label{3.7}
\ee
This function could also serve as the unification of inflation and
cosmic acceleration; however, in this case, when one takes the limit
$\tilde{R}\rightarrow\infty$, one gets
\be
\lim_{\tilde{R}\rightarrow\infty}F(\tilde{R})=\tilde{R}-2\Lambda_i\, ,
\quad \text{where} \quad \Lambda_i=1/2f_1\, ,
\label{3.8}
\ee
where the subscript $i$ denotes the inflationary. 
By inserting this into Eqs.~(\ref{HLF13}) and  (\ref{HLF16}) with $C=0$,
the FRW equations take the same form as in (\ref{3.6}). 
Then, for the function (\ref{3.7}), the inflationary epoch 
is produced by an effective cosmological constant, 
which implies that the parameter
$\lambda>1/3$, or the equations themselves will present
inconsistencies, as was discussed in the above section. For the
current epoch, it is easy to see that the function (\ref{3.7})
exhibits a minimum for $\tilde{R}=\tilde{R}_0$, which implies, as in
the model above, an effective cosmological constant for late time
that can produce the cosmic acceleration. The emergence of matter 
dominance before the dark energy epoch can be exhibited,
in analogy with the case of the convenient theory. Thus, we have
shown that the model (\ref{3.7}) also unifies the cosmic expansion 
history, although with different properties during the inflationary 
epoch as compared with the model (\ref{3.2}).

It is also interesting to explore the de Sitter solutions allowed by
the theory (\ref{HLF5}). By taking $H(t)=H_0$, the FRW equation
(\ref{HLF16}), in the absence of any kind of matter and with $C=0$,
reduces to
\be
0=F(\tilde{R}_0)-6H^2_0(1-3\lambda+3\mu)F'(\tilde{R}_0)\, ,
\label{3.9}
\ee
which represents an algebraic equation that can be solved to yield 
the possible de Sitter points allowed by the
theory. As an example, let us consider the model (\ref{3.2}), where
Eq.~(\ref{3.9}) takes the form
\be
\tilde{R}_0+\frac{\tilde{R}_0^n(\alpha\tilde{R}_0^n-\beta)}
{1 +\gamma\tilde{R}_0^n}
+\frac{6H_0^2(-1+3\lambda-3\mu)\left[1+n\alpha\gamma\tilde{R}_0^{3n-1}
+\tilde{R}_0^{n-1}(2\gamma\tilde{R}_0-n\beta)
+\tilde{R}_0^{2n-1}(\gamma^2\tilde{R}_0
+2n\alpha)\right]}{(1+\gamma\tilde{R}_0^n)^2}=0\, .
\label{3.10}
\ee
Here, $\tilde{R}_0=3(1-3\lambda+6\mu)H^2_0$. By specifying the free
parameters of the theory, one can solve Eq.~(\ref{3.10}), which
yields several de Sitter points, as the one studied above. 
The de Sitter points can be used to explain the coincidence problem, 
with the argument that the present will not be the only 
late-time accelerated epoch experienced by our universe. 
In standard $F(R)$ gravity, it was found that this model contains 
at least two de Sitter points along the cosmic history 
(see \cite{Elizalde:2009gx}). In the same way, the second
model studied here (\ref{3.7}) provides several de Sitter points in
the course of cosmic history. Note that when $\mu=\lambda-\frac{1}{3}$,
Eq.~(\ref{3.9}) turns out to be much more simple; it reduces
to $F(\tilde{R}_0)=0$, where the de Sitter points are the roots.
For example, for (\ref{3.10}) we have
$\tilde{R}_0(1+\gamma\tilde{R}_0^n)+\tilde{R}_0^n(\alpha\tilde{R}_0^n-\beta)
=0$, where the number of positive roots (de Sitter points) depends
on the free parameters of the theory.

In summary, it has been demonstrated here that, as in $F(\tilde{R})$
Ho\v{r}ava-Lifshitz gravity, the so-called viable models, as in 
(\ref{3.2}) or (\ref{3.7}), can in fact reproduce the whole
cosmological history of the universe, in the same way as convenient 
modified gravity.

\subsection{Covariant power-counting renormalizable gravity \label{IIG}}

Motivated by the idea of ref.~\cite{Horava:2009uw},
a model of covariant, power-counting renormalizable gravity was 
proposed \cite{Nojiri:2010tv,Nojiri:2010kx}. Its prototype using the 
perfect fluid is given in ref.~\cite{Nojiri:2009th}. 
Such a theory seems to be more natural than 
models \cite{Horava:2009uw,Carloni:2010nx} because Lorentz
invariance is not broken explicitly. Moreover, the theory looks very 
similar to $R^2$ gravity with a scalar. The present section is devoted 
to review of such covariant model in the FRW universe.

\subsubsection{Perfect fluid formulation}

Let us briefly review the covariant, power-counting renormalizable gravity 
of ref.~\cite{Nojiri:2009th}.
The starting action is
\be
\label{Hrv1}
S = \int d^4 x \sqrt{-g} \left\{ \frac{R}{2\kappa^2} - \alpha \left( 
T^{\mu\nu} R_{\mu\nu} + \beta T R \right)^2 \right\}\, .
\ee
Here, $T_{\mu\nu}$ is the energy-momentum tensor of some string-inspired 
fluid.
The action (\ref{Hrv1}) is fully diffeomorphism invariant.
We consider the perturbation from the flat background
$g_{\mu\nu} = \eta_{\mu\nu} + h_{\mu\nu}$.
The following gauge conditions are chosen:
$h_{tt} = h_{ti} = h_{it} = 0$.
The fluid energy-momentum tensor in the flat background has
the following form:
\be
\label{Hrv5}
T_{tt} = \rho \, ,\quad T_{ij} = p \delta_{ij} = w \rho \delta_{ij}\, .
\ee
Here, $w$ is the EoS parameter.
Then, one finds
\bea
\label{Hrv6}
&& T^{\mu\nu} R_{\mu\nu} + \beta T R \nn
&& = \rho \left[ \left\{ - \frac{1}{2} + \frac{w}{2} + \left( - 1 + 3w 
\right) \beta \right\}
\partial_t^2 \left(\delta^{ij} h_{ij} \right)
+ \left( w - \beta + 3w \beta \right) \partial^i \partial^j h_{ij}
\right. \nn
&& \qquad + \left( - w + \beta - 3w \beta \right)
\partial_k \partial^k \left(\delta^{ij} h_{ij} \right) \Bigr]\, .
\eea
If
\be
\label{Hrv7}
\beta = - \frac{w-1}{2\left(3w - 1\right)}\, ,
\ee
the second term in the action (\ref{Hrv1}) becomes
\be
\label{Hrv8}
\alpha \left( T^{\mu\nu} R_{\mu\nu} + \beta T R \right)^2
= \alpha \rho^2 \left(\frac{w}{2} + \frac{1}{2} \right)^2 \left\{
\partial^i \partial^j h_{ij} - \partial_k \partial^k
\left(\delta^{ij} h_{ij} \right) \right\}^2\, ,
\ee
which does not contain the derivative with respect to $t$ and breaks the
Lorentz invariance.
Imagine $\rho$ is almost constant.
Then, in the ultraviolet region, in which $\bm{k}$ is large, the second term 
in the action (\ref{Hrv1}) gives the propagator behaving as 
$1/\left| \bm{k} \right|^4$.
It renders the ultraviolet behavior (compared with Eq.~(1.4) in
\cite{Horava:2009uw}).
Note that the form (\ref{Hrv7}) indicates that the longitudinal mode does
not propagate; only the transverse mode propagates.

The action (\ref{Hrv1}) gives $z=2$ theory.
For the theory to be ultraviolet, power-counting 
renormalizable in $3+1$ dimensions, $z=3$ theory is necessary. 
To obtain such a theory, we note that, for any scalar quantity $\Phi$ 
with the choice
\be
\label{Hrv13}
\gamma = \frac{1}{3 w - 1}\, ,
\ee
one obtains
\be
\label{Hrv14}
T^{\mu\nu}\nabla_\mu \nabla_\nu \Phi 
+ \gamma T \nabla^\rho \nabla_\rho \Phi
= \rho \left( w + 1 \right) \partial_k \partial^k \Phi \, ,
\ee
which does not contain the derivative with respect to the time coordinate $t$.
This is true even if the coordinate frame is not the local Lorentz frame.
The derivative with respect to the time coordinate $t$ is not contained in any
coordinate frame where the perfect fluid does not flow.
Thus, the proposed action looks as
\be
\label{Hrv15}
S = \int d^4 x \sqrt{-g} \left\{ \frac{R}{2\kappa^2} - \alpha \left( 
T^{\mu\nu} R_{\mu\nu} + \beta T R \right)
\left(T^{\mu\nu}\nabla_\mu \nabla_\nu + \gamma T \nabla^\rho 
\nabla_\rho\right)
\left( T^{\mu\nu} R_{\mu\nu} + \beta T R \right)\right\}\, ,
\ee
with $\beta = - \frac{w-1}{2\left(3w - 1\right)}$ and
$\gamma = \frac{1}{3 w - 1}$. It corresponds to
$z=3$ theory, which seems to be (power-counting) renormalizable.
In general, for the case
\be
\label{Hrv18}
S = \int d^4 x \sqrt{-g} \left[ \frac{R}{2\kappa^2} - \alpha \left\{
\left(T^{\mu\nu}\nabla_\mu \nabla_\nu + \gamma T \nabla^\rho
\nabla_\rho\right)^n
\left( T^{\mu\nu} R_{\mu\nu} + \beta T R \right) \right\}^2 \right]\, ,
\ee
with a constant $n$, $z = 2 n + 2$ theory, which is
super-renormalizable for $n\geq 1$ emerges.

The second terms in the actions (\ref{Hrv1}), (\ref{Hrv15}), 
and (\ref{Hrv18}), which effectively break the Lorentz symmetry, are relevant 
only in the high-energy/UV region because they contain higher-derivative terms. 
In the IR region, these terms do not dominate, and the usual Einstein gravity 
follows as a limit.

\subsubsection{Covariant field theory of gravity with a Lagrange 
multiplier}

We now show that the perfect fluid, which appeared in (\ref{Hrv1}),
(\ref{Hrv15}), and (\ref{Hrv18}), can be realized by a scalar field and 
the Lagrange multiplier field.
As a result, we obtain a covariant, power-counting renormalizable field 
theory of gravity.

The following constrained action for the scalar field $\phi$ is proposed: 
\be
\label{LagHL1}
S_\phi = \int d^4 x \sqrt{-g} \left\{ - \lambda \left( \frac{1}{2}
\partial_\mu \phi \partial^\mu \phi + U(\phi) \right) \right\} \, .
\ee
Here, $\lambda$ is the Lagrange multiplier field, which gives a constraint of 
\be
\label{LagHL2}
\frac{1}{2} \partial_\mu \phi \partial^\mu \phi + U(\phi) = 0\, ,
\ee
that is, the vector $(\partial_\mu \phi)$ is time like.
At least locally, one can choose the direction of time to be parallel to
$(\partial_\mu \phi)$.
Then, Eq.~(\ref{LagHL2}) has the following form:
\be
\label{LagHL3}
\frac{1}{2} \left(\frac{d\phi}{dt}\right)^2 = U(\phi)\, .
\ee
The equation given by the variation over $\phi$ will be discussed later.

The scalar field energy-momentum tensor $T^\phi_{\mu\nu}$ with an arbitrary 
scalar potential is defined as follows:
\be
\label{LagHL4}
T^\phi_{\mu\nu} = \partial_\mu \phi \partial_\nu \phi - g_{\mu\nu}
\left( \frac{1}{2} \partial_\rho \phi \partial^\rho \phi
+ V(\phi) \right) \, .
\ee
The ``energy density'' $\rho_\phi$ and ``pressure'' $p_\phi$ become:
\be
\label{LagHL5}
p_\phi = \frac{1}{2} \left(\frac{d\phi}{dt}\right)^2 - V(\phi) 
= U(\phi) - V(\phi) \, ,\quad
\rho_\phi = \frac{1}{2} \left(\frac{d\phi}{dt}\right)^2 + V(\phi) 
= U(\phi) + V(\phi) \, .
\ee
Here, the constraint (\ref{LagHL3}) is used.
Note that (unknown, for the moment) $V(\phi)$ is not identical with 
$U(\phi)$: $V(\phi)\neq U(\phi)$.
In the case that $V(\phi) = U(\phi)$, Eq.~(\ref{LagHL5}) shows that $p_\phi =0$, 
which corresponds to dust with $w_\phi \equiv p_\phi/\rho_\phi = 0$.
Of course, the classical and quantum dynamics of such constrained theories
are quite non-trivial \cite{masud}.

For simplicity, we choose $V(\phi)$ and $U(\phi)$ to be constants:
\be
\label{LagHL6}
U(\phi) = U_0\, ,\quad V(\phi) = V_0\, .
\ee
Then, if $U_0 = V_0$, the EoS parameter $w_\phi$ vanishes. 
In the general case, one has $w_\phi = \frac{U_0 - V_0}{U_0 +V_0}$.
Let us now use $T^\phi_{\mu\nu}$ as an energy-momentum tensor in the 
previous section.
Eq.(\ref{Hrv7}) shows
$\beta = - \frac{w-1}{2\left(3w - 1\right)} = \frac{V_0}{2U_0 - 4V_0}$.
One can simplify
\be
\label{HrvHL9}
T^{\phi\, \mu\nu} R_{\mu\nu} + \beta T^\phi R
= \partial^\mu \phi \partial^\nu \phi R_{\mu\nu} + U_0 R\, .
\ee
Here, Eqs.~(\ref{LagHL2}), (\ref{LagHL4}), and (\ref{LagHL6}) are used.
Similarly, $\gamma$ in (\ref{Hrv13}) has the following form:
$\gamma = \frac{U_0 - V_0}{2U_0 - 4V_0}$,
which gives, by using (\ref{Hrv14}),
\be
\label{HrvHL10}
T^{\phi\,\mu\nu}\nabla_\mu \nabla_\nu \Phi + \gamma T^\phi \nabla^\rho
\nabla_\rho \Phi
= \partial^\mu \phi \partial^\nu \phi \nabla_\mu \phi \nabla_\nu \Phi
+ 2 U_0 \nabla^\rho \nabla_\rho \Phi\, .
\ee
Eq.~(\ref{HrvHL9}) enables us to write down the $z=2$ total 
action corresponding to (\ref{Hrv1}) as
\be
\label{HrvHL11}
S = \int d^4 x \sqrt{-g} \left\{ \frac{R}{2\kappa^2} - \alpha
\left( \partial^\mu \phi \partial^\nu \phi R_{\mu\nu} + U_0 R
\right)^2 - \lambda \left( \frac{1}{2}
\partial_\mu \phi \partial^\mu \phi + U_0 \right) \right\} \, .
\ee
On the other hand, the $z=3$ total action corresponding to 
(\ref{Hrv15}) is:
\bea
\label{HrvHL12-0}
S &=& \int d^4 x \sqrt{-g} \left\{ \frac{R}{2\kappa^2} - \alpha \left(
\partial^\mu \phi \partial^\nu \phi R_{\mu\nu} + U_0 R \right)
\left(\partial^\mu \phi \partial^\nu \phi \nabla_\mu \nabla_\nu
+ 2 U_0 \nabla^\rho \nabla_\rho \right)
\left( \partial^\mu \phi \partial^\nu \phi R_{\mu\nu} + U_0 R \right) 
\right. \nn
&& \left. - \lambda \left( \frac{1}{2} \partial_\mu \phi \partial^\mu \phi
+ U_0 \right) \right\}\, ,
\eea
and
\bea
\label{HrvHL12}
S_{2n+2} &=& \int d^4 x \sqrt{-g} \left\{ \frac{R}{2\kappa^2} - \alpha 
\left\{ \left(\partial^\mu \phi \partial^\nu \phi \nabla_\mu \nabla_\nu
+ 2 U_0 \nabla^\rho \nabla_\rho \right)^n
\left( \partial^\mu \phi \partial^\nu \phi R_{\mu\nu} + U_0 R 
\right)\right\}^2 \right. \nn
&& \left. - \lambda \left( \frac{1}{2} \partial_\mu \phi \partial^\mu \phi
+ U_0 \right) \right\}\, .
\eea
for the $z=2n + 2$ model $\left( n=0,1,2,\cdots\right)$, and
\bea
\label{HrvHL13}
S_{2n+3} &=& \int d^4 x \sqrt{-g} \left\{ \frac{R}{2\kappa^2} - \alpha 
\left\{ \left(\partial^\mu \phi \partial^\nu \phi \nabla_\mu \nabla_\nu
+ 2 U_0 \nabla^\rho \nabla_\rho \right)^n
\left( \partial^\mu \phi \partial^\nu \phi R_{\mu\nu} + U_0 R 
\right)\right\} \right. \nn
&& \left. \times \left\{
\left(\partial^\mu \phi \partial^\nu \phi \nabla_\mu \nabla_\nu
+ 2 U_0 \nabla^\rho \nabla_\rho \right)^{n+1}
\left( \partial^\mu \phi \partial^\nu \phi R_{\mu\nu}
+ U_0 R \right)\right\} - \lambda \left( \frac{1}{2}
\partial_\mu \phi \partial^\mu \phi
+ U_0 \right) \right\}\, ,
\eea
for the $z=2n + 3$ model $\left( n=0,1,2,\cdots\right)$ 
(compare with \cite{Carloni:2010nx})\footnote{
Note that the model (\ref{HrvHL12}) or (\ref{HrvHL13}) has better 
renormalization properties and no scalar graviton when the action is 
modified a little bit and the scalar projector is inserted in second 
term \cite{Kluson:2011rs}.}.
Note that the actions (\ref{HrvHL11}), (\ref{HrvHL12-0}), (\ref{HrvHL12}),
and (\ref{HrvHL13}) are totally diffeomorphism invariant and are only given 
in terms of the local fields. The actions (\ref{HrvHL11}), (\ref{HrvHL12-0}),
(\ref{HrvHL12}), and (\ref{HrvHL13}) do not depend on $V_0$.

By the variation over $\phi$, for example for the $z=2$ $\left( n=0 \right)$ 
case in (\ref{HrvHL12}), one finds
\be
\label{CRG_phi1}
0 = 4 \alpha \partial^\mu \left\{ \partial^\nu \phi R_{\mu\nu}
\left( \partial^\rho \phi \partial^\sigma \phi R_{\rho\sigma} + U_0 R 
\right) \right\} + \partial^\mu \left( \lambda \partial_\mu \phi \right) \, .
\ee
For the $z\geq 3$ ($n \geq 1$ case in (\ref{HrvHL12}) or $n\geq 0$ case in
(\ref{HrvHL13})), one obtains a rather complicated equation.

The dispersion relation of the graviton is now given by
\be
\label{sym29}
\omega = \alpha c_0 k^z\, ,
\ee
in the high-energy region.
Here, $c_0$ is a constant, $\omega$ is the angular frequency corresponding 
to the energy and $k$ is the wave number corresponding to momentum.
If $\alpha<0$, the dispersion relation becomes inconsistent and, therefore, 
$\alpha$ should be positive.

\subsubsection{FRW cosmology}

The gravitational terms different from general relativity in 
(\ref{HrvHL12}) and (\ref{HrvHL13}) are relevant
in the high-energy region. Such terms might affect the inflationary era.
In this section, we briefly study FRW cosmology within the theory under
discussion.
To obtain the FRW equations, the following form of the metric is assumed:
\be
\label{bFRW}
ds^2 = - \e^{2b(t)}dt^2 + a(t)^2 \sum_{i=1,2,3} \left(dx^i\right)^2\, ,
\ee
and it is assumed that the scalar field $\phi$ only depends on time. 
Then, Eq.~(\ref{LagHL2}) looks as:
\be
\label{BBFRW}
\frac{1}{2} \left(\frac{d\phi}{dt}\right)^2 = \e^{2b(t)} U_0 \, .
\ee
Thus, one gets 
\be
\label{HrvCos1}
\partial^\mu \phi \partial^\nu \phi R_{\mu\nu} + U_0 R
= 6 U_0 \e^{-2b} H^2 \, ,\quad
\partial^\mu \phi \partial^\nu \phi \nabla_\mu \nabla_\nu
+ 2 U_0 \nabla^\rho \nabla_\rho = - 6 U_0 \e^{-2b} H \partial_t\, ,
\ee
and the actions (\ref{HrvHL12}) and (\ref{HrvHL13}) have the following 
form:
\bea
\label{sym34}
S_{2n+2} &=& \int d^4 x a^3 \left[ \frac{\e^{-b}}{2\kappa^2}
\left(6\dot H + 12 H^2 - 6\dot b H\right) - \left( 6 U_0 \right)^{2n+2} 
\e^b \left\{ \left( \e^{-2b} H \partial_t \right)^n
\left( H^2 \e^{-2b} \right) \right\}^2 \right. \nn
&& \left. - \lambda \left( - \frac{\e^{-b}}{2}\left( \frac{d\phi}{dt} 
\right)^2 + \e^b U_0 \right) \right]\, ,\\
\label{sym35}
S_{2n+3} &=& \int d^4 x a^3 \left[ \frac{\e^{-b}}{2\kappa^2} \left(6\dot H 
+ 12 H^2 - 6\dot b H\right) \right. \nn
&& - 2^{2n+3}\cdot 3^2 \alpha\ U_0^{2n+2} \e^b \left\{ \left( \e^{-2b} H
\partial_t \right)^n \left( H^2 \e^{-2b} \right) \right\}
\left\{ \left( \e^{-2b} H \partial_t \right)^{n+1} \left( H^2 \e^{-2b} 
\right) \right\} \nn
&& \left. - \lambda \left( - \frac{\e^{-b}}{2}\left( \frac{d\phi}{dt} 
\right)^2 + \e^b U_0 \right) \right]\, .
\eea
The first FRW equation looks as follows: 
\bea
\label{sym36}
0 &=& \frac{3}{\kappa^2}H^2 - \left( 6 U_0 \right)^{2n+2} \alpha a^{-3}
\left[ a^3 \left( D^n \left(H^2\right) \right)^2 - 4 (-1)^n H^2 {\bar D}^n
\left( a^3 D^n \left(H^2\right) \right) \right. \nn
&& \left. -4 \sum_{k=0}^{n-1} \left( -1 \right)^k \left( {\bar D}^{n-k}
\left(H^2\right) \right)
\left( {\bar D}^k \left( a^3 D^n \left(H^2\right) \right) \right)
\right] - 2\lambda U_0 - \rho_\mathrm{matter}\, ,
\eea
for (\ref{sym34}) and as follows:
\bea
\label{sym37}
0 &=& \frac{3}{\kappa^2}H^2 - \left( 6 U_0 \right)^{2n+3} \alpha a^{-3}
\left[ a^3 \left( D^n \left(H^2\right) \right) \left( D^{n+1} 
\left(H^2\right) \right) - 2 (-1)^n H^2 {\bar D}^n 
\left( a^3 D^{n+1} \left(H^2\right) \right) \right. \nn
&& - 2 (-1)^{n+1} H^2 {\bar D}^{n+1} 
\left( a^3 D^n \left(H^2\right) \right) -2 \sum_{k=0}^{n-1}
\left( -1 \right)^k \left( {\bar D}^{n-k} \left(H^2\right) \right)
\left( {\bar D}^k \left( a^3 D^{n+1} \left(H^2\right) \right) \right) \nn
&& \left. -2 \sum_{k=0}^{n} \left( -1 \right)^k \left( {\bar D}^{n+1-k}
\left(H^2\right) \right)
\left( {\bar D}^k \left( a^3 D^n \left(H^2\right) \right) \right)
\right] - 2\lambda U_0 - \rho_\mathrm{matter}\, ,
\eea
for (\ref{sym35}).
We have also put $b=0$ after the variation over $b$, where the metric
(\ref{bFRW}) reduces to the standard FRW metric and the operations of $D$ 
and $\bar D$ for a scalar $\varphi$ are defined by
\be
\label{sym38}
D\varphi \equiv H \frac{d\varphi}{dt}\, ,\quad
\bar D\varphi \equiv \frac{d\left( H\varphi\right)}{dt}\, .
\ee
On the other hand, by the variation over $a$, we get
\bea
\label{sym39}
0 &=& \frac{1}{\kappa^2} \left( 2\dot H + 3 H^2 \right)
+ 2^{2n+2} 3^{2n+1} \alpha U_0^{2n+2} \left\{ - 3 \left(D^n  \left(H^2\right)
\right)^2 + 4 \left(-1\right)^n a^{-3}\frac{d}{dt}\left( H {\bar D}^n
\left( a^3 {\bar D}^n \left(H^2\right) \right) \right) \right. \nn
&& \left. + 2\sum_{k=1}^n a^{-3}\frac{d}{dt}\left(\left(\frac{d}{dt}
\left( D^{n-k} \left(H^2\right)\right)\right)
\left( {\bar D}^{k-1}\left( a^3 D^n\left(H^2\right)\right) \right)\right)
\right\} + p_\mathrm{matter}\, ,
\eea
for (\ref{sym34}) and
\bea
\label{sym40}
0 &=& \frac{1}{\kappa^2} \left( 2\dot H + 3 H^2 \right)
+ 2^{2n+3} 3^{2n+2} \alpha U_0^{2n+2} \left\{ - 3 \left(D^n \left(H^2\right)
\right)\left(D^{n+1} \left(H^2\right) \right) \right. \nn
&& + 2 \left(-1\right)^n a^{-3}\frac{d}{dt}\left( H {\bar D}^n
\left( a^3 {\bar D}^{n+1} \left(H^2\right) \right) \right)
+ 2 \left(-1\right)^{n+1} a^{-3}\frac{d}{dt}\left( H {\bar D}^{n+1}
\left( a^3 {\bar D}^n \left(H^2\right) \right) \right) \nn
&& + \sum_{k=1}^n a^{-3}\frac{d}{dt}\left(\left(\frac{d}{dt}
\left( D^{n-k} \left(H^2\right)\right)\right)
\left( {\bar D}^{k-1}\left( a^3 D^{n+1}\left(H^2\right)\right) 
\right)\right) \nn
&& \left. + \sum_{k=1}^{n+1} a^{-3}\frac{d}{dt}\left(\left(\frac{d}{dt}
\left( D^{n-k+1} \left(H^2\right)\right)\right)
\left( {\bar D}^{k-1}\left( a^3 D^n\left(H^2\right)\right) \right)\right)
\right\} + p_\mathrm{matter}\, ,
\eea
for (\ref{sym35}).
The simplest case is $n=0$ in (\ref{sym34}) when the FRW equations are
\bea
\label{HrvHL16}
\frac{3}{\kappa^2} H^2 &=& - 108 \alpha U_0^2 H^4 + 2 \lambda U_0
+ \rho_\mathrm{matter} \, , \\
\label{HrvHL17}
 - \frac{1}{\kappa^2} \left( 2\dot H + 3 H^2 \right)
&=& 36 \alpha U_0^2 \left( 3H^4 + 4H^2 \dot H \right)
+ p_\mathrm{matter} \, .
\eea
In the early universe, in which the curvature was large, the contribution from
the Einstein term, which corresponds to the right-hand side in (\ref{HrvHL16})
and (\ref{HrvHL17}), and the contributions from the 
matter $\rho_\mathrm{matter}$ and $p_\mathrm{matter}$, can be neglected.
Then, a solution of (\ref{HrvHL17}) is given by
\be
\label{sym41}
H=\frac{4}{3t}\, ,
\ee
which expresses the (power-law) accelerated universe expansion 
corresponding to that with a perfect fluid with $w=-1/2$.
Eq.~(\ref{HrvHL16}) gives
\be
\label{sym42}
\lambda = \frac{32 \alpha U_0}{3t^4}\, .
\ee
For more complicated versions of the above theory, one can expect to find 
a more rich class of background cosmological solutions. The cosmological 
reconstruction may also be used for such models.

\subsection{Dark fluid with an inhomogeneous equation of state \label{IIH}}

Usually, a perfect fluid with a constant EoS parameter $w$ is discussed in
the cosmological-related literature.
One may, however, consider a general equation of state fluid:
\be
\label{GEoS1}
p = w\left(\rho,a,H,\dot H, \ddot H, \cdots\right) \rho
+ \left(\rho,a,H,\dot H, \ddot H, \cdots\right)\, .
\ee
It is clear that any modified gravity may be presented as such an effective
gravitational fluid coupled with general relativity as was already 
mentioned in section A of the present chapter. Such an approach to modified 
gravity can make things simpler when background evolution is investigated.
However, it may lead to a number of problems in the study of cosmological
perturbation theory or structure formation. Indeed, when developing such 
an approach, one often forgets that the initial modified gravity is 
a higher-derivative theory and that perturbation theory should also 
be a higher-derivative theory. Non-linearities are also effectively dropped. 
That is why we prefer to work with the explicit modified gravity formulation 
but not with general relativity coupled to gravitational fluid.

The above fluid formulation is very general. Moreover, even
if we restrict it to the case that the pressure $p$ depends only on the
energy density $\rho$, the EoS fluid, which arbitrarily reproduced in FRW 
cosmology \cite{Nojiri:2005sr}, may be constructed. 
Needless to say, the unification of inflation with the dark energy epoch 
may already be achieved for such a fluid. 
The FRW equations (\ref{JGRG11}) show that, 
for the perfect fluid with the generalized equation of state
\be
\label{ma3}
p =-\rho - \frac{2}{\kappa^2}f'\left(f^{-1}\left(
\kappa\sqrt{\frac{\rho}{3}}\right)\right) \, ,
\ee
the solution of the FRW equations is given by $H=f(t)$.
Here, $f(t)$ is an arbitrary function of the cosmological time $t$.
Therefore, the arbitrary universe evolution given by $H=f(t)$ can
be realized via the perfect fluid whose equation of state is given 
by (\ref{ma3}).

It is often convenient to use e-folding $N =\ln\frac{a}{a_0}$ instead of
the cosmological time $t$ because e-folding is related with the redshift 
$z$, which is directly given by observations (for more details, 
see the next chapter).
One may consider the reconstruction using $N$ instead of the cosmological
time $t$.
Then, the second equation (\ref{JGRG11}) can be rewritten as
\be
\label{ma2b}
p_\mathrm{total} = - \frac{1}{\kappa^2} \left(2H H' + 3H^2\right)\, .
\ee
Here, $'$ expresses the derivative with respect to $N$: $'\equiv d/dN$.
Thus, the fluid with the generalized equation of state
\be
\label{ma3b}
p=-\rho - \frac{2}{\kappa}\tilde f'\left(\tilde
f^{-1}\left(\kappa\sqrt{\frac{\rho}{3}}\right)\right)
\sqrt{\frac{\rho}{3}}\, ,
\ee
gives the solution of the FRW equations by $H=\tilde f(N)$.

It was already mentioned in section A that a phantom fluid whose EoS 
parameter $w$ is less than $-1$ leads to the future Big Rip
singularity \cite{Caldwell:2003vq}. Several types of softer future 
singularities may occur for an effective quintessence fluid. 
The classification of four future finite-time singularities is given 
in section A following ref.~\cite{{Nojiri:2005sx}}.

The Type I singularity corresponds to the Big Rip 
singularity \cite{Caldwell:2003vq,ref5}, which emerges when $w<-1$.
The Type II singularity corresponds to the sudden future
singularity \cite{barrow} at which $a$ and $\rho$ are finite but $p$ diverges.
The Type III singularity appears, for instance, for the model with 
$p=-\rho-A \rho^{\alpha}$ \cite{Nojiri:2004pf}, which is different from 
the sudden future singularity in the sense that $\rho$ diverges.
This type of singularity was discovered in the model
of ref.~\cite{Nojiri:2004pf}, where the corresponding Lagrangian
model of a scalar field with a potential was constructed.

Let us show that all above four types of future singularities may naturally 
occur for fluid dark energy.
As an example, one may start from the following fluid dark energy
\be
\label{EOS1}
p=-\rho - f(\rho)\, ,
\ee
where $f(\rho)$ is generally an arbitrary function.
The case of $f(\rho)\propto \rho^\alpha$ with a constant $\alpha$
was proposed and investigated in detail in ref.~\cite{Nojiri:2004pf}.
Using the conservation law (\ref{CEm}) for such a choice,
the scale factor $a$ is given by
\be
\label{EOS4}
a=a_0\exp\left(\frac{1}{3} \int \frac{d\rho}{f(\rho )} \right)\, .
\ee
Using the first equation (\ref{JGRG11}),
the cosmological time $t$ is
\be
\label{tint}
t=\int \frac{d \rho}{\kappa \sqrt{3\rho} f(\rho)}\, .
\ee
In the case
\be
\label{ppH6}
f(\rho)=A\rho^\alpha\, ,
\ee
by using Eq.~(\ref{EOS4}), it follows that 
\be
\label{EOS26}
a = a_0 \exp \left[\frac{\rho^{1-\alpha}}{3(1-\alpha)A} \right]\, .
\ee
When $\alpha> 1$, the scale factor remains finite even if $\rho$ goes to
infinity. 
When $\alpha<1$, $a\to \infty$ $(a\to 0)$ as $\rho\to \infty$ for $A>0$
$(A<0)$.
Because the pressure is now given by
\be
\label{EOS27}
p= -\rho - A\rho^\alpha\, ,
\ee
$p$ always diverges when $\rho$ becomes infinite.
If $\alpha>1$, the EoS parameter $w=p/\rho$ also goes to infinity, 
that is, $w \to +\infty$ ($-\infty$) for $A<0$ $(A>0$).
When $\alpha<1$, we have $w\to -1+0$ ($-1-0$)
for $A<0$ $(A>0$) as $\rho \to \infty$.

Using Eq.~(\ref{tint}) for (\ref{ppH6}), one finds \cite{{Nojiri:2005sx}}
\be
\label{EOS31}
t= t_s + \frac{2}{\sqrt{3}\kappa A}
\frac{\rho^{-\alpha+1/2}}{1-2\alpha}\, ,\quad
\mbox{for} \quad \alpha \neq \frac12\, ,
\ee
and
\be
\label{EOS32}
t= t_s + \frac{\mathrm{ln}\left(\frac{\rho}{\rho_0}\right)}
{\sqrt{3}\kappa A}\, ,
\quad \mbox{for}\quad \alpha=\frac12\,.
\ee
Therefore if $\alpha\leq 1/2$, $\rho$ diverges into an infinite future or 
past.
On the other hand, if $\alpha>1/2$, the divergence of $\rho$
corresponds to a finite future or past.
In the case of a finite future, the singularity could be regarded as a Big 
Rip or a Type I singularity (compare with the case in section A).

For the equation of state (\ref{ppH6}), the following cases were 
discussed \cite{{Nojiri:2005sx}}:
\begin{itemize}
\item In the case $\alpha=1/2$ or $\alpha=0$, there does not appear any
singularity.
\item In the case $\alpha>1$, Eq.~(\ref{EOS31}) indicates that when $t\to 
t_s$, the energy density behaves as $\rho\to\infty$ and, therefore, 
$|p|\to \infty$ due to (\ref{EOS27}).
Eq.~(\ref{EOS26}) shows
that the scale factor $a$ is finite even if $\rho\to \infty$. Therefore, 
the $\alpha>1$ case corresponds to a Type III singularity.
\item The $\alpha=1$ case corresponds to $p=w\rho$ with a constant $w$
if we replace $-1 - A$ with $w$.
Therefore, if $A>0$, either the Big Rip or a Type I singularity occurs, 
but if $A\leq 0$, no future singularity appears.
\item In the case that $1/2<\alpha<1$, when $t\to t_s$, all of $\rho$, 
$|p|$, and $a$ diverge if $A>0$; this corresponds to a Type I singularity.
\item In the case that $0<\alpha<1/2$, when $t\to t_s$, we find $\rho$, 
$|p|\to 0$ and $a\to a_0$, but by combining (\ref{EOS26}) 
and (\ref{EOS31}), we find
\be
\label{typeIV}
\ln a \sim \left|t-t_s\right|^{\frac{\alpha-1}{\alpha - 1/2}}\, .
\ee
Because the exponent $(\alpha -1)/(\alpha - 1/2)$ is not always an integer, 
even if $a$ is finite, the higher derivatives of $H$ diverge in general.
Therefore, this case corresponds to a Type IV singularity.
\item In the case that $\alpha<0$, when $t\to t_s$, we find $\rho\to 0$, 
$a\to a_0$, but $|p|\to \infty$.
Therefore, this case corresponds to a Type II singularity.
\end{itemize}

Thus, it is explicitly demonstrated that simple fluid dark energy may
bring the universe evolution to one of four possible future singularities
as well as to an asymptotically non-singular universe. The particular future
behavior is defined by the choice of the equation of state and of the 
parameters of this equation. Precisely the same phenomenon occurs for 
any field model of dark energy or modified gravity, so there is no 
qualitative difference between different dark energy models in this respect. 
The emergence of a finite-time future singularity in different modified 
gravities will be described in the fourth chapter.

In summary, this chapter provided a review of several popular models of 
modified gravity is given. Of course, the explicit choice of the models is related 
to our scientific interests. In particular, the following models were discussed:
$F(R)$ gravity, modified Gauss-Bonnet theory, scalar-Einstein-Gauss-Bonnet
gravity, non-local gravity and non-minimally coupled gravity.
More exotic theories with explicit Lorentz symmetry breaking like
Ho\v{r}ava-Lifshitz $F(R)$ gravity or with apparent Lorentz symmetry 
breaking like covariant power-counting renormalizable field gravity 
were also presented.
In all cases, the spatially flat FRW cosmology was investigated. The 
qualitative possibility of a unified description of early-time inflation 
with late-time acceleration was also demonstrated. It turns out that the easiest 
and very realistic unification of these two eras is possible in $F(R)$ gravity.

\section{Cosmological reconstruction of modified gravity \label{SecIII}}

To explain the unified cosmic history of the universe, especially 
the inflation and/or the late-time accelerating expansion of the present 
universe, several models of extended or modified gravity
were proposed in the previous chapter.
Usually, we start from a theory, which is defined by the action, and solve
equations of motion to define the background dynamics.
In this chapter, however, we consider the inverse problem, i.e., the
cosmological reconstruction of gravitational theories. In other words,
using the fact that modified gravity is defined in terms of some arbitrary
function (or with the help of some potential), we show how the 
complicated background cosmology, which complies with
observational data, may be reconstructed. Eventually, this leads to
a particular choice of the above arbitrary function (or potential).
This provides the qualitative difference between general relativity and 
modified gravity; the latter theory (more exactly, some class of it) may have as
a solution any given gravitational space-time. 
The problem is, of course, how to define the specific version of the theory
that has this particular space-time as an explicit (exact or approximate)
solution.
The reconstruction scheme is explicitly constructed here for several 
modified gravities introduced in the previous chapter.
The general approach to reconstruction in modified gravity and dark energy 
models was developed in
refs.~\cite{Nojiri:2006gh,Capozziello:2006dj,Nojiri:2006be}.
This kind of reconstruction may also be used for a spherically symmetric 
solution like black holes (see, for example, \cite{Nojiri:2009uu}).

\subsection{Scalar-tensor gravity \label{IIIA}}

As the first example, let us consider the reconstruction of 
scalar-Einstein gravity (or scalar-tensor theory),
whose action can be written as
\be
\label{ma7}
S=\int d^4 x \sqrt{-g}\left\{
\frac{1}{2\kappa^2}R - \frac{1}{2}\omega(\phi)\partial_\mu \phi
\partial^\mu\phi - V(\phi) + L_\mathrm{matter} \right\}\, .
\ee
Here, $\omega(\phi)$ and $V(\phi)$ are functions of the scalar $\phi$.
The function $\omega(\phi)$ is not relevant and can be absorbed into the
redefinition of the scalar field $\phi$. In fact, if one redefines 
the scalar field $\phi$ by
\be
\label{ma13}
\varphi \equiv \int^\phi d\phi \sqrt{\left|\omega(\phi)\right|} \, ,
\ee
the kinetic term of the scalar field in the action (\ref{ma7}) has the
following form:
\be
\label{ma13b}
 - \omega(\phi) \partial_\mu \phi \partial^\mu\phi
= \left\{ \begin{array}{ll}
 - \partial_\mu \varphi \partial^\mu\varphi & 
\mbox{when $\omega(\phi) > 0$} \\
\partial_\mu \varphi \partial^\mu\varphi & \mbox{when $\omega(\phi) < 0$}
\end{array} \right. \, .
\ee
The case of $\omega(\phi) > 0$ corresponds to the quintessence or 
non-phantom scalar field, but the case of $\omega(\phi) < 0$ corresponds 
to the phantom scalar.
Although $\omega(\phi)$ can be absorbed into the redefinition of the 
scalar field, we keep $\omega(\phi)$ for a later convenience.

\subsubsection{The reconstruction using the cosmological time}

This section is based on the studies 
\cite{Nojiri:2005pu,Capozziello:2005tf,Vikman:2004dc}.

For the action (\ref{ma7}), in the FRW equations (\ref{JGRG11}), the
energy density and the pressure are as follows:
\be
\label{ma8}
\rho = \frac{1}{2}\omega(\phi){\dot \phi}^2 + V(\phi)\, ,\quad
p = \frac{1}{2}\omega(\phi){\dot \phi}^2 - V(\phi)\, .
\ee
Then
\be
\label{ma9}
\omega(\phi) {\dot \phi}^2 = - \frac{2}{\kappa^2}\dot H\, ,\quad
V(\phi)=\frac{1}{\kappa^2}\left(3H^2 + \dot H\right)\, .
\ee
Assuming $\omega(\phi)$ and $V(\phi)$ are given by a single function
$f(\phi)$, as follows,
\be
\label{ma10}
\omega(\phi)=- \frac{2}{\kappa^2}f'(\phi)\, ,\quad
V(\phi)=\frac{1}{\kappa^2}\left(3f(\phi)^2 + f'(\phi)\right)\, ,
\ee
the exact solution of the FRW equations (\ref{JGRG11}) with
(\ref{ma8}) (when we neglect the contribution from the matter) 
has the following form:
\be
\label{ma11}
\phi=t\, ,\quad H=f(t)\, .
\ee
It can be confirmed that the equation given by the variation over $\phi$
\be
\label{ma12}
0=\omega(\phi)\ddot \phi + \frac{1}{2}\omega'(\phi){\dot\phi}^2 
+ 3H\omega(\phi)\dot\phi + V'(\phi)\, ,
\ee
is also satisfied by the solution (\ref{ma11}).
Then, the arbitrary universe evolution expressed by $H=f(t)$ can be
realized by an appropriate choice of $\omega(\phi)$ and $V(\phi)$.
In other words, by defining the particular type of universe evolution,
the corresponding scalar-Einstein gravity may be found.

As mentioned in (\ref{ma13}) and (\ref{ma13b}), $\omega(\phi)$ can be 
absorbed into the redefinition of the scalar field $\phi$.
By keeping $\omega(\phi)$, however, we can construct a model that exhibits  
a smooth transition from the non-phantom phase to the phantom phase (and vice versa).

Let the matter energy density $\rho_\mathrm{matter}$ and the matter 
pressure $p_\mathrm{matter}$ be given by the sum of the contributions 
from the matter whose EoS parameters are
$w^i_\mathrm{matter}=p^i_\mathrm{matter}/\rho^i_\mathrm{matter}$is a 
constants.
Here, the suffix ``$i$'' specifies the type of matter.
Using the conservation law of the energy density (\ref{CEm}),
we find $\rho_\mathrm{matter}=\sum \rho^i_0 
a^{-3(1+w^i_\mathrm{matter})}$.
Here, $\rho^i_0$ is a constant. If $\omega(\phi)$
and $V(\phi)$ are given in terms of a single function $g(\phi)$ as
\bea
\label{any5}
\omega(\phi) &=& - \frac{2}{\kappa^2}g''(\phi) - \sum_i 
\frac{w^i_\mathrm{matter} + 1}{2}\rho^i_0
a_0^{-3(1+w^i_\mathrm{matter})}
\e^{-3(1+w^i_\mathrm{matter})g(\phi)}\, ,\nn
V(\phi) &=& \frac{1}{\kappa^2}\left(3g'(\phi)^2 + g''(\phi)\right)
+ \sum_i \frac{w^i_\mathrm{matter} -1}{2}\rho^i_0
a_0^{-3(1+w^i_\mathrm{matter})} \e^{-3(1+w^i_\mathrm{matter})g(\phi)} \, ,
\eea
the solution of the FRW equations (\ref{JGRG11}) is \be
\label{any6}
\phi=t\, ,\quad H=g'(t)\, ,\quad
\left(a=a_0 \e^{g(t)}\right)\, .
\ee
Thus, even in the presence of matter, any given cosmology
defined by $H=g'(t)$ can be realized by potentials(\ref{any5}).

As an example, the model with only one kind of matter with the
EoS parameter $w_\mathrm{matter}$ (\ref{any5}) can be considered.
It is also assumed that $w_\mathrm{matter}>-1$. Choosing $g(\phi)$ as
\be
\label{any11}
g(\phi)=\frac{2}{3\left(w_\mathrm{matter} + 1\right)}
\ln \left(\frac{\phi}{t_s - \phi}\right)\, ,
\ee
one obtains
\bea
\label{any12}
\omega(\phi)&=&- \frac{4}{3\left(w_\mathrm{matter} + 1\right)\kappa^2}
\left(- \frac{1}{\phi^2} + \frac{1}{\left(t_s - \phi\right)^2}
\right) - \frac{w_\mathrm{matter}+ 1}{2}\rho_0
a_0^{-3(1+w_\mathrm{matter})} \frac{\left(t_s - \phi\right)^2}{\phi^2}
\, ,\nn
V(\phi) &=& \frac{1}{\kappa^2}\left\{\frac{4}{3\left(w_\mathrm{matter} 
+ 1\right)^2}\left( \frac{1}{\phi} + \frac{1}{t_s - \phi}\right)^2
+ \frac{2}{3\left(w_\mathrm{matter} + 1\right)}\left( - \frac{1}{\phi^2}
+ \frac{1}{\left(t_s - \phi\right)^2}\right)\right\} \nn
&& + \frac{w_\mathrm{matter} - 1}{2}\rho_0 a_0^{-3(1+w_\mathrm{matter})}
\frac{\left(t_s - \phi\right)^2}{\phi^2}\, .
\eea
The Hubble rate $H$ follows as
\be
\label{any13}
H=\frac{2}{3\left(w_\mathrm{matter} + 1\right)}\left(\frac{1}{t}
+ \frac{1}{t_s - t}\right) \, .
\ee
Because
\be
\label{any14}
\dot H=\frac{2}{3\left(w_\mathrm{matter} + 1\right)}\left(-\frac{1}{t^2}
+ \frac{1}{\left(t_s - t\right)^2}\right)\, ,
\ee
the EoS parameter $w_\mathrm{eff}$ (\ref{JGRG12})
goes to $w_\mathrm{matter}>-1$ when $t\to 0$ and goes to 
$-2 - w_\mathrm{matter}<-1$ at late times. The
crossing $w_\mathrm{eff}=-1$ occurs when $\dot H=0$, that is, $t=t_s/2$.
Note that
\be
\label{any15}
\frac{\ddot a}{a}
= \frac{16t_s}{27 \left(w_\mathrm{matter} + 1\right)^3
\left(t_s - t\right)^2 t^2}
\left\{ t - \frac{\left(3w_\mathrm{matter} + 1\right)t_s}{4}\right\}\, .
\ee
Thus, if $w_\mathrm{matter}>-1/3$, the deceleration of the universe turns 
to the acceleration at 
$t= t_a \equiv \left(3w_\mathrm{matter} + 1\right)t_s/4$.

The second example in (\ref{any5}) is given by
\be
\label{any19}
g(\phi) = - \alpha \ln \left(1 - \beta \ln \frac{\phi}{\kappa}\right)\, .
\ee
Here, $\alpha$ and $\beta$ are dimensionless positive constants.
Let us choose $\beta \sim {\cal O}\left(10^{-2}\right)$ for later 
convenience.
Note that the parameter order of $10^{-2}$ is not so unnatural.
When only one kind of matter with the
EoS parameter $w_\mathrm{matter}$ is included, for simplicity,
Eq.~(\ref{any19}) gives the following expressions for $\omega(\phi)$ and
$V(\phi)$:
\bea
\label{any20}
\omega(\phi)&=&-\frac{2\alpha\beta\left(\beta - 1 + \beta\ln
\frac{\phi}{\kappa}\right)}
{\kappa^2\left(1 - \beta\ln \frac{\phi}{\kappa}\right)^2
\phi^2} - \frac{w_\mathrm{matter}
+ 1}{2}\rho_0 a_0^{-3(1+w_\mathrm{matter})}
\left(1 - \beta\ln \frac{\phi}{\kappa}
\right)^{3(w_\mathrm{matter} + 1)\alpha}\, ,\nn
V(\phi)&=&\frac{\alpha\beta\left(3\alpha\beta + \beta - 1 + \beta\ln
\frac{\phi}{\kappa}\right)}
{\kappa^2\left(1 - \beta\ln \frac{\phi}{\kappa}\right)^2\phi^2}
+ \frac{w_\mathrm{matter} - 1}{2} \rho_0 a_0^{-3(1+w_\mathrm{matter})}
\left(1 - \beta\ln \frac{\phi}{\kappa}
\right)^{3(w_\mathrm{matter} + 1)\alpha}\, .
\eea
Supposing $\rho_0 a_0^{-3(1+w_\mathrm{matter})}\sim {\cal
O}\left(\kappa^{-2}\right)$,
any small parameter, compared with the Planck scale does not appear.
Now, the Hubble rate is given by
$H=\alpha\beta/\left(1 - \beta\ln \left(t/\kappa\right)\right)t$,
which is positive if
\be
\label{any22}
0<t<t_s\equiv \kappa \e^{\frac{1}{\beta}}\, ,
\ee
and has a Big Rip-type singularity at $t=t_s$.
Because $10^{61}\sim \e^{140}$, with the choice $\beta \sim 1/140$, 
we obtain 
$t_s\sim \kappa\times 10^{61}\sim \left(10^{-33}\,\mathrm{eV}\right)^{-1}$,
whose order is that of the age of the present universe.
Thus, due to the property of the exponential function (or logarithmic
function), the small scale like $t_s$ appears rather naturally.
We should also note that if $\alpha\beta\sim {\cal O}(10^{0-2})$ and
$t$ is a present age of the universe 
$t\sim \left(10^{-33}\,\mathrm{eV}\right)^{-1}$, the
observed value of the Hubble rate $H\sim 10^{-33}\,\mathrm{eV}$ can also be
reproduced. 
Because
\be
\label{any23}
\frac{\ddot a}{a}=\frac{\alpha \beta^2\left(\ln\frac{t}{\kappa} + \alpha 
+  1 - \frac{1}{\beta}\right)}
{\left(1 - \beta \ln \frac{t}{\kappa}\right)^2 t^2}\, ,
\ee
the universe changes to an accelerating from a decelerating 
universe when
\be
\label{any24}
t=t_a \equiv \kappa\e^{\frac{1}{\beta} - \alpha -1}<t_s\, .
\ee
Because the energy density $\rho$ of the scalar field $\phi$ and
that of the matter $\rho_\mathrm{matter}$ are given by
\bea
\label{any25}
\rho&=& \frac{3 \alpha^2 \beta^2}{\kappa^2 \left(1 - \beta \ln
\frac{t}{\kappa}\right)^2 t^2} - \rho_0 a_0^{-3(1+w_\mathrm{matter})}
\left(1 - \beta\ln \frac{t}{\kappa}
\right)^{3(w_\mathrm{matter} + 1)\alpha}\, , \nn
\rho_\mathrm{matter} &=& \rho_0 a_0^{-3(1+w_\mathrm{matter})}
\left(1 - \beta\ln \frac{t}{\kappa}\right)^{3(w_\mathrm{matter} 
+ 1)\alpha}\, ,
\eea
the coincidence time $t_c$ could be given by solving the following 
equation:
\be
\label{any26}
\left(1 - \beta\ln \frac{t_c}{\kappa}\right)^{3(w_\mathrm{matter} 
+ 1)\alpha+ 2}t_c^2
=\frac{3\alpha^2\beta^2}{\kappa^2 \rho_0 
a_0^{-3(1+w_\mathrm{matter})}}\, .
\ee
One may regard $\rho_\mathrm{matter}$ as the sum of the energy density of 
usual matter, like baryons, and that of (cold) dark matter. 
If $\rho$ corresponds to the energy density of the dark energy,
the current data indicate that $\rho:\rho_\mathrm{matter}\sim 7:3$. 
Then, in the present universe, it follows that 
\be
\label{any27}
\left(1 - \beta\ln \frac{t}{\kappa}\right)^{3(w_\mathrm{matter} 
+ 1)\alpha+ 2}t^2 \sim \frac{9\alpha^2\beta^2}{10\kappa^2 g_0}\, .
\ee
Thus, in the model (\ref{any19}), the acceleration of the
present universe and the coincidence problem seem to be explained rather
naturally. This is achieved via explicit cosmological reconstruction of 
the scalar potentials.

\subsubsection{The reconstruction of the scalar-tensor models using e-folding}

In this subsection, the formulation using the e-folding $N$ instead of
cosmological time is considered as in Eqs.~(\ref{ma2b}) and (\ref{ma3b}).
The FRW equations (\ref{JGRG11}) with (\ref{ma8}) (when the matter 
contribution is neglected) are given in terms of $N$, as follows: 
\be
\label{ma15}
\frac{3}{\kappa^2} H^2 = \frac{H^2 \omega\left(\phi\right){\phi'}^2 }{2}
+ V\left(\phi\right)\, ,\quad - \frac{1}{\kappa^2}
\left(2H H' + 3 H^2 \right)
= \frac{H^2 \omega\left(\phi\right){\phi'}^2 }{2} - V\left(\phi\right)\, .
\ee
We now identify $\phi=N$; then, 
\be
\label{ma16}
\omega(\phi) = - \frac{2H'}{\kappa^2 H}\, ,\quad
V\left(\phi\right) = \frac{1}{\kappa^2}\left( H H' + 3 H^2 \right)\, .
\ee
The above equations (\ref{ma16}) show that if one studies the model
\be
\label{ma17}
\omega(\phi) = - \frac{2{\tilde f}'\left(\phi\right)}{\kappa^2 
\tilde f\left(\phi\right)}\, ,\quad
V\left(\phi\right) = \frac{1}{\kappa^2}\left( \tilde f\left(\phi\right) 
\tilde f'\left(\phi\right) + 3 \tilde f\left(\phi\right)^2 \right)\, ,
\ee
which is given in terms of a single function $\tilde f$, then the exact 
solution is given by $H\left(N\right) = \tilde f\left(N \right)$.

Let us consider an example by using the following function $\tilde f$: 
\be
\label{ma18}
\tilde f(\phi) = H_1 \phi^{-\gamma}\, .
\ee
Here, $H_1$ and $\gamma$ are positive constants. In this case
\be
\label{ma18b}
\omega(\phi) = \frac{2\gamma}{\kappa^2 \phi}\, ,\quad
V\left(\phi\right) = \frac{H_1^2}{\kappa^2}
\left( - \gamma \phi^{-2\gamma  -1} + 3 \phi^{-2\gamma} \right)\, ,
\ee
and the exact solution is given by $H = H_1 N^{-\gamma}$.
With a redefinition of $\phi$ by 
$\phi\to \varphi = \frac{2 \sqrt{2\gamma\phi}}{\kappa}$,
the kinetic term becomes canonical, and the potential is given by
\be
\label{ma20}
V\left(\varphi\right) = \frac{H_1^2}{\kappa^2}\left\{ - \gamma
\left(\frac{\kappa^2}{8\gamma}\right)^{- 2\gamma - 1} 
\varphi^{-4\gamma - 2}
+ 3 \left(\frac{\kappa^2}{8\gamma}\right)^{- 2\gamma} \varphi^{-4\gamma}
\right\}\, .
\ee
At late times (large $N$), only the last term is relevant.
Conversely, at late times, if the potential is given by
$V \propto \varphi^{-4\gamma}$, the solution has the form of 
$H \propto N^{-\gamma}$.
The effective EoS parameter (\ref{JGRG12}) is equal to 
\be
\label{ma19}
w_\mathrm{eff} = - 1 + \frac{2\gamma}{3N} \, ,
\ee
which goes to $-1$ when $N$ goes to infinity.

Because we naively find 
$C \equiv \frac{\mbox{Plank scale}}{H_0} \sim 10^{61} \sim \e^{140}$,
if $\frac{H_1}{H_0} \sim C$, it follows that $N \sim \ln C$ and, therefore, 
$\gamma \sim 28$.
Thus, the fine-tuning problem might be relaxed, but in this case one 
obtains $w_\mathrm{eff}\sim - 0.87$, which is a little bigger than 
the observational value $-0.14 < 1+w < 0.12$ \cite{Komatsu:2008hk}.

In the case that $\tilde C \equiv \frac{\mbox{Weak scale}}{H_0}
\sim 10^{44} \sim \e^{101}$, if $\frac{H_1}{H_0} \sim \tilde C$, 
it follows that $N \sim \ln \tilde C$ and $\gamma \sim 21$. 
This finishes the cosmological reconstruction of 
scalar-tensor theory using e-folding.

\subsubsection{Two-scalar model}

Note that when studying the transition from the non-phantom phase to 
the phantom phase, the instability of
the single scalar-tensor gravity becomes infinite at the transition 
point \cite{Nojiri:2005pu,Capozziello:2005tf}.
Thus, we only investigate two scalar-tensor theory, which 
has an instability that can always finite. 
Moreover, with two scalars it is easier to realize unified universe
history expansion.

The action of two scalar-tensor gravity is
\be
\label{A1}
S=\int d^4 x \sqrt{-g}\left\{\frac{1}{2\kappa^2}R
 - \frac{1}{2}\omega(\phi)\partial_\mu \phi \partial^\mu \phi
 - \frac{1}{2}\eta(\chi)\partial_\mu \chi\partial^\mu \chi
 - V(\phi,\chi)\right\}\, .
\ee
Here, $\omega(\phi)$ is a function of the scalar field $\phi$, and 
$\eta(\chi)$ is a function of another scalar field $\chi$.
Note that the above theory parameterizes a large number of popular dark energy 
models. 
Indeed, if both scalars could be canonical scalars, then a 
kind of the quintessence model emerges. 
One of scalars may be phantom (wrong sign of the kinetic term) 
while another scalar remains the canonical one.
This is dark energy sometimes called quintom theory.
Finally, both scalars could be phantom. Moreover, two scalars may be 
organized so that a charged scalar with real and imaginary parts appears.

It is also assumed that the FRW metric is spatiallyflat (\ref{JGRG14}) and
that $\phi$ and $\chi$ only depend on the time coordinate $t$.
The FRW equations lead to
\be
\label{A2}
\omega(\phi) {\dot \phi}^2 + \eta(\chi) {\dot \chi}^2
= - \frac{2}{\kappa^2}\dot H\, ,\quad
V(\phi,\chi)=\frac{1}{\kappa^2}\left(3H^2 + \dot H\right)\, .
\ee
If
\be
\label{A3}
\omega(t) + \eta(t)=- \frac{2}{\kappa^2}f'(t)\, ,\quad
V(t,t)=\frac{1}{\kappa^2}\left(3f(t)^2 + f'(t)\right)\, ,
\ee
the explicit solution follows as:
\be
\label{A4}
\phi=\chi=t\, ,\quad H=f(t)\, .
\ee
One may choose $\omega$ to always be positive and $\eta$ to always
be negative, for example
\be
\label{A5}
\omega(\phi) =-\frac{2}{\kappa^2}\left\{f'(\phi)
- \sqrt{\alpha^2 + f'(\phi)^2} \right\}>0\, ,\quad
\eta(\chi) = -\frac{2}{\kappa^2}\sqrt{\alpha^2 + f'(\chi)^2}<0\, .
\ee
Here, $\alpha$ is a constant. Define a new function $\tilde f(\phi,\chi)$ 
by
\be
\label{A6}
\tilde f(\phi,\chi)\equiv - \frac{\kappa^2}{2}\left(\int d\phi 
\omega(\phi) + \int d\chi \eta(\chi)\right)\, .
\ee
The constant of the integration could be fixed to require
\be
\label{A7}
\tilde f(t,t)=f(t)\, .
\ee
If $V(\phi,\chi)$ is given using $\tilde f(\phi,\chi)$ as
\be
\label{A8}
V(\phi,\chi)=\frac{1}{\kappa^2}\left(3{\tilde f(\phi,\chi)}^2
+ \frac{\partial \tilde f(\phi,\chi)}{\partial \phi}
+ \frac{\partial \tilde f(\phi,\chi)}{\partial \chi} \right)\, ,
\ee
the FRW and the scalar field equations are also satisfied:
\be
\label{A9}
0 = \omega(\phi)\ddot\phi + \frac{1}{2}\omega'(\phi) {\dot \phi}^2
+ 3H\omega(\phi)\dot\phi + \frac{\partial \tilde V(\phi,\chi)}{\partial 
\phi} \, , \quad
0 = \eta(\chi)\ddot\chi + \frac{1}{2}\eta'(\chi) {\dot \chi}^2
+ 3H\eta(\chi)\dot\chi + \frac{\partial \tilde V(\phi,\chi)}{\partial 
\chi}\, .
\ee
Thus, for given cosmology the cosmological reconstruction (the 
calculation of scalar potentials) is made.

It is now interesting to investigate the (in)stability of the model under
discussion.
By introducing the new quantities, $X_\phi$, $X_\chi$, and $Y$ as
\be
\label{A10}
X_\phi \equiv \dot \phi\, ,\quad X_\chi \equiv \dot \chi\, ,\quad
Y\equiv \frac{\tilde f(\phi,\chi)}{H} \, ,
\ee
the FRW equations and the scalar field equations (\ref{A9}) are:
\bea
\label{A11}
&& \frac{dX_\phi}{dN} = - \frac{\omega'(\phi)}{2H 
\omega(\phi)}\left(X_\phi^2 - 1 \right) - 3(X_\phi-Y)\, ,\quad
\frac{dX_\chi}{dN} = - \frac{\eta'(\chi)}{2H \eta(\chi)}
\left(X_\chi^2 - 1\right) - 3(X_\chi-Y)\, ,\nn
&& \frac{dY}{dN} = \frac{1}{2\kappa^2H^2}
\left\{X_\phi \left(X_\phi Y -1\right)
+ X_\chi\left(X_\chi Y -1\right)\right\}\, .
\eea
Here, $d/dN\equiv H^{-1}d/dt$. In the solution (\ref{A4}), 
$X_\phi=X_\chi=Y=1$.
The following perturbation may be considered
\be
\label{A12}
X_\phi=1+\delta X_\phi\, ,\quad X_\chi=1 + \delta X_\chi\, ,
\quad Y=1 + \delta Y\, .
\ee
Thus,
\be
\label{A13}
\frac{d}{dN}\left(\begin{array}{c}
\delta X_\phi \\
\delta X_\chi \\
\delta Y
\end{array}\right)
= M \left(\begin{array}{c}
\delta X_\phi \\
\delta X_\chi \\
\delta Y
\end{array}\right)\, ,\quad
M \equiv \left(\begin{array}{ccc}
 - \frac{\omega'(\phi)}{H\omega(\phi)} - 3 & 0 & 3 \\
0 & - \frac{\eta'(\chi)}{H\eta(\chi)} - 3 & 3 \\
\frac{1}{2\kappa^2H^2} & \frac{1}{2 \kappa^2 H^2} & \frac{1}{\kappa^2 H^2}
\end{array}\right)\, .
\ee
The eigenvalues of the matrix $M$ are given by solving the following 
eigenvalue equation
\bea
\label{A14}
0 &=& \left(\lambda + \frac{\omega'(\phi)}{H\omega(\phi)} + 3\right)
\left(\lambda + \frac{\eta'(\chi)}{H\eta(\chi)} + 3\right)
\left(\lambda - \frac{1}{\kappa^2 H^2}\right)
+ \frac{3}{2\kappa^2 H^2}\left(\lambda 
+ \frac{\omega'(\phi)}{H\omega(\phi)}
+ 3\right) \nn
&& + \frac{3}{2\kappa^2 H^2}\left(\lambda 
+  \frac{\eta'(\chi)}{H\eta(\chi)}
+  3\right)\, .
\eea
The eigenvalues (\ref{A14}) for the two-scalar model are clearly finite.
Thus, the instability could be finite.
In fact, right at the transition point where $\dot H=f'(t)=0$ and,  
therefore, $f'(\phi)=f'(\chi)=0$, for the choice in (\ref{A5}), one gets
\be
\label{AA1}
\omega(\phi)=-\eta(\chi)=\frac{2\alpha}{\kappa^2}\, ,\quad
\omega'(\phi)=-\frac{2\ddot H}{\kappa^2}\, ,\quad \eta'(\chi)=0\, .
\ee
Then, the eigenvalue equation (\ref{A14}) reduces to
\be
\label{AA2}
0=\lambda^3 + \left(-A-B + 6\right) \lambda^2
+ \left(AB - 3A - 3B + 9\right) \lambda - \frac{3}{2}AB + 9B\, ,\quad
A\equiv \frac{\ddot H}{\alpha}\, ,\quad B\equiv \frac{1}{\kappa^2 H^2}\, .
\ee
Here, $\alpha>0$ is chosen. The eigenvalues are surely finite, which
shows that even if the solution (\ref{A4}) is not stable, the solution
has a non-vanishing measure and, therefore, the transition from 
the non-phantom phase to the phantom phase can occur. 
We should also note that the solution (\ref{A4})
can be stable. For example, for the case
$A,B\to 0$. Eq.~(\ref{AA2}) further reduces to
\be
\label{AA3}
0=\lambda\left(\lambda + 3\right)^2\, .
\ee
Then, the eigenvalues are given by $0$ and $-3$. Because there is no positive
eigenvalue, the solution (\ref{A4}) is stable in this case.

It is not difficult to extend the above formulation to the multi-scalar  
model, whose action is given by
\be
\label{mA1}
S=\int d^4 x \sqrt{-g}\left\{\frac{1}{2\kappa^2}R - \frac{1}{2}
\sum_i\omega_i(\phi_i)\partial_\mu \phi_i \partial^\mu \phi_i - V(\phi_i)
\right\}\, .
\ee
Here, $\omega_i(\phi_i)$ is a function of the scalar field $\phi_i$.
We now choose $\omega_i$ to satisfy
\be
\label{mA2}
\sum_i \omega_i(t)=-\frac{2}{\kappa^2}f'(t) \, ,
\ee
by a proper function. The potential $V(\phi_i)$ is chosen as
\be
\label{mA3}
V(\phi_i)=\frac{1}{\kappa^2}\left(3\tilde f(\phi_i)
+ \sum_i\frac{\partial\tilde f}{\partial \phi_i}\right)\, .
\ee
Here
\be
\label{mA4}
\tilde f(\phi_i)\equiv - \frac{\kappa^2}{2}\sum_i \int d\phi_i
\omega_i(\phi_i)\, .
\ee
The constant of the integration in (\ref{mA4}) is determined to satisfy
\be
\label{mA5}
\left. \tilde f(\phi_i) \right|_{\mathrm{all}\ \phi_i=t}=f(t)\, .
\ee
Then, a solution of the FRW equations and the scalar field equations
is given by
\be
\label{mA6}
\phi_i=t\, ,\quad H=f(t)\, .
\ee
The remaining arguments coincide with those given for the two-scalar model.

It is interesting to demonstrate that the transition from the 
matter-dominant period to the acceleration period can be realized in the present
formulation.
In the following, the matter contribution can be neglected because the 
ratio of the matter with the (effective) dark matter may be small.

The first example is as follows:
\be
\label{STm12}
H=f(t)=g'(t)=g_0 + \frac{g_1}{t}\, .
\ee
When $t$ is large, the first term in (\ref{STm12}) dominates and
the Hubble rate $H$ becomes a constant.
Therefore, the universe is asymptotically the de Sitter space, which 
corresponds to an accelerating universe. On the
other hand, when $t$ is small, the second term in (\ref{STm12}) dominates
and the scale factor behaves as $a\sim t^{g_1}$. Therefore, if $g_1=2/3$,
the matter-dominated period emerges.
Here, one of the scalars may be considered as usual matter.

Because
\be
\label{2s1}
f'(t)=-\frac{g_1}{t}<0\, ,
\ee
instead of (\ref{A5}), by using a positive constant $\alpha$, one may 
choose
\be
\label{2s2}
\omega(\phi)=-\frac{2\left(1+\alpha\right)}{\kappa^2}f'(\phi)>0\, ,\quad
\eta(\chi) = \frac{2\alpha}{\kappa^2}f'(\phi)<0\, ,
\ee
that is, 
\be
\label{2s3}
\omega(\phi)=\frac{2(1+\alpha)}{\kappa^2}\frac{g_1}{\phi^2}\, ,\quad
\eta(\chi)=\frac{2\alpha}{\kappa^2}\frac{g_1}{\phi^2}\, .
\ee
One should note that in the limit $\alpha\to 0$, the scalar field $\chi$
decouples and the single-scalar model follows.
Thus, one gets
\be
\label{2s4}
\tilde f(\phi,\chi) = g_0
+ \frac{(1+\alpha)g_1}{\phi} - \frac{\alpha g_1}{\chi}\, ,\quad
V(\phi,\chi) = \frac{1}{\kappa^2}\left\{3\left(g_0
+ \frac{(1+\alpha)g_1}{\phi} - \frac{\alpha g_1}{\chi}
\right)^2 - \frac{(1+\alpha)g_1}{\phi^2}
+ \frac{\alpha g_1}{\chi^2}\right\}\, .
\ee
Thus, for the scalar-tensor theory with such potentials the matter-dominant 
era occurs before the acceleration epoch. This fact is easily established 
within the reconstruction scheme.

Before going to the second example,
we consider the Einstein gravity with cosmological constant and with
matter characterized by the EOS parameter $w$.
The FRW equation is as follows:
\be
\label{LCDM1}
\frac{3}{\kappa^2}H^2 = \rho_0 a^{-3(1+w)} + \frac{3}{\kappa^2 l^2}\, .
\ee
Here, $l^2$ is the inverse cosmological constant.
The solution of (\ref{LCDM1}) is given by
\be
\label{LCDM2}
a=a_0\e^{g(t)}\, ,\quad
g(t)=\frac{2}{3(1+w)}\ln \left(\alpha \sinh
\left(\frac{3(1+w)}{2l}\left(t -
t_s \right)\right)\right)\, .
\ee
Here, $t_s$ is a constant of the integration and
\be
\label{LCDM3}
\alpha^2\equiv \frac{1}{3}\kappa^2 l^2 \rho_0 a_0^{-3(1+w)}\, .
\ee
Thus, this is the $\Lambda$CDM era as it follows in the commonly accepted 
approach.

As the second example, the reconstruction of the $\Lambda$CDM metric 
(\ref{LCDM2}) for scalar-tensor theory is done below.
Because
\be
\label{2s5}
f(t) \equiv g'(t)=\frac{1}{l}\coth\left(\frac{3(1+w)}{2l}\left(t - t_s
\right)\right)\, ,\quad
f'(t)=g''(t)=- \frac{3(1+w)}{2l^2}\sinh^{-2}\left(\frac{3(1+w)}{2 l}
\left(t - t_s \right)\right)<0\, ,
\ee
it is convenient to use (\ref{2s2}) instead of (\ref{A5}).
Then, one arrives at
\bea
\label{2s6}
\omega(\phi) &=& \frac{3(1+w)(1+\alpha)}{\kappa^2 l^2}
\sinh^{-2}\left(\frac{3(1+w)}{2l}\left(\phi - t_s \right)\right) > 0
\, ,\nn
\eta(\chi) &=& - \frac{3(1+w)\alpha}{\kappa^2 l^2}
\sinh^{-2}\left(\frac{3(1+w)}{2l}\left(\chi - t_s \right)\right) < 0
\, ,\nn
\tilde f(\phi,\chi) &=& \frac{1+\alpha}{l}\coth\left(\frac{3(1+w)}{2l}
\left(\phi - t_s \right)\right) - \frac{\alpha}{l}
\coth\left(\frac{3(1+w)}{2l}\left(\chi - t_s \right)\right) \, ,\nn
V(\phi,\chi) &=& \frac{1}{\kappa^2}\left[ \frac{3}{l^2}\left\{
(1+\alpha)\coth\left(\frac{3(1+w)}{2l}\left(\phi - t_s \right)
\right) - \alpha \coth\left(\frac{3(1+w)}{2l}
\left(\chi - t_s \right)\right) \right\}^2 \right. \\
&& \left. - \frac{3(1+w)(1+\alpha)}{l^2}
\sinh^{-2}\left(\frac{3(1+w)}{2l}\left(\phi - t_s \right)\right)
+ \frac{3(1+w)\alpha}{l^2}\sinh^{-2}
\left(\frac{3(1+w)}{2l}\left(\chi - t_s \right)\right) \right]\, .
\nonumber
\eea
This completes the reconstruction of scalar potentials to reproduce 
the $\Lambda$CDM epoch.

Thus, in both examples, (\ref{STm12}) and (\ref{LCDM2}), 
the matter dominant stage, the transition from the matter dominant phase to 
the acceleration phase and acceleration epoch occur.
In the acceleration phase of the above examples, the universe asymptotically
approaches the de Sitter space. This does not conflict with 
the Wilkinson Microwave Anisotropy Probe (WMAP) data.
Indeed, three years WMAP data have been analyzed in
ref.~\cite{Spergel:2003cb,Peiris:2003ff,Spergel:2006hy}.
The combined analysis of WMAP data with the Supernova Legacy
Survey (SNLS) constrains the dark energy equation of state 
$w_\mathrm{DE}$, pushing it toward the cosmological constant. 
The marginalized best fit values of the
equation of state parameter at the 68$\%$ confidence level
are given by $-1.14\leq w_\mathrm{DE} \leq -0.93$. 
In the prior case of a flat universe, the combined data give 
$-1.06 \leq w_\mathrm{DE} \leq -0.90$.
In the examples (\ref{STm12}) and (\ref{LCDM2}),
the universe goes asymptotically to the de Sitter space, which gives
$w_\mathrm{DE}\to -1$, which does not conflict
with the above constraints.
Note, however, that one needs to fine-tune $g_0$ in (\ref{STm12})
and $1/l$ in (\ref{LCDM2}) to be $g_0\sim 1/l \sim 10^{-33}$ eV, 
to reproduce the observed Hubble rate
$H_0\sim 70$ km$\,$s$^{-1}$Mpc$^{-1}\sim 10^{-33}$ eV.

As a final remark, there is no problem regarding the inclusion of the usual 
matter (for example, ideal fluid and dust) into the above reconstruction scenario 
and/or finding the scalar potentials corresponding to realistic
cosmology in the multi-scalar case. Moreover, in the same way, one may include
the radiation-dominated epoch where quantum effects may still be neglected
in the above scenario. In the next section it is shown how similar
reconstruction scheme may be developed for modified gravity.

\subsubsection{Brans-Dicke gravity}

The action of the original Brans-Dicke model \cite{Brans:1961sx} has the
following form:
\be
\label{BD1}
S_\mathrm{BD}=\frac{1}{2\kappa^2}\int d^4 x \sqrt{-g}
\left( \phi R - \omega
\frac{\partial_\mu \phi \partial^\mu \phi}{\phi}\right)\, ,
\ee
which can be generalized as
\be
\label{BD2}
S=\int d^4x\sqrt{-g}\left[\frac{
\e^{\varphi(\phi)}R}{2\kappa^2} -\frac{1}{2}\omega(\phi)\partial_{\mu}
\phi\partial^{\mu}\phi-V(\phi)+\mathcal{L}_\mathrm{m}\right]\, .
\ee
This kind of model was applied to the dark energy problem in 
\cite{Elizalde:2004mq}, and it was found that even if $\omega(\phi)>0$, 
that is, the scalar field is canonical, the phantom universe
can be realized.

The FRW equations are as follows:
\be
\label{BD3}
H^2 = 
\frac{\e^{-\varphi}\kappa^2}{3}\left[\frac{1}{2}\omega(\phi)\dot\phi^2
+V(\phi) \right] \, , \quad
\dot H = -\frac{\e^{-\varphi}\kappa^2}{2}\omega(\phi)
\dot\phi^2 -\frac{1}{2}\left(\ddot\varphi+\dot\varphi^2\right) \, ,
\ee
This can be rewritten as
\be
\label{BD4}
\omega(\phi)\dot\phi^2 = -\frac{2\e^{\varphi}}{\kappa^2}\left[ \dot H
+ \frac{1}{2}\left(\ddot\varphi+\dot\varphi^2\right) \right]\, , \quad
V(\phi) = \frac{2\e^{\varphi}}{\kappa^2}\left[ \dot H + 3 H^2
+ \frac{1}{2}\left(\ddot\varphi+\dot\varphi^2\right) \right] \, .
\ee
Then, if the following choice of scalar potentials using a function
$f(\phi)$ \cite{yoshioka} is taken:
\be
\label{BD5}
\omega(\phi) = -\frac{2\e^{\varphi(\phi)}}{\kappa^2}\left[ f'(\phi)
+ \frac{1}{2}\left(\varphi''(\phi) + \varphi'(\phi)^2\right) \right]\, , 
\quad
V(\phi) = \frac{2\e^{\varphi(\phi)}}{\kappa^2}\left[ f'(\phi) + 3 
f(\phi)^2 
+ \frac{1}{2}\left(\varphi''(\phi) + \varphi'(\phi)^2\right) \right]\, ,
\ee
the explicit solution looks like
\be
\label{BD6}
H=f(t)\, ,\quad \phi = t\, .
\ee
Note that $\varphi(\phi)$ can be an arbitrary function of $\phi$.
By this reconstruction formalism, the phantom crossing ($w=-1$ crossing) 
can be realized for $\omega(\phi)>0$, that is, when the scalar field 
is canonical (non-ghost).
A slightly different version of the reconstruction formalism for above 
theory is given in \cite{Elizalde:2008yf}.

\subsection{The $k$-essence model \label{IIIB}}

Here, based on \cite{Matsumoto:2010uv}, we consider the reconstruction of 
the $k$-essence model, which represents the evident generalization of 
quintessence theory.
The $k$-essence model is a rather general model that includes only one scalar 
field and the action is given by
\be
\label{KK1}
S= \int d^4 x \sqrt{-g} \left( \frac{R}{2\kappa^2} - K \left(
\phi, X \right) + L_\mathrm{matter}\right)\, ,\quad X \equiv \partial^\mu 
\phi
\partial_\mu \phi \, .
\ee
Here, $\phi$ is a scalar field.

Now, the Einstein equation has the following form:
\be
\label{Sch2}
\frac{1}{\kappa^2}\left( R_{\mu\nu} - \frac{1}{2}g_{\mu\nu} R \right)
= - K \left( \phi, X \right) g_{\mu\nu} + 2 K_X \left( \phi, X \right)
\partial_\mu \phi \partial_\nu \phi
+ T_{\mu\nu}\, .
\ee
Here, $K_X \left( \phi, X \right) \equiv \partial K \left( \phi, X \right) 
/\partial X$.
On the other hand, the variation over $\phi$ gives
\be
\label{Sch3}
0= - K_\phi \left( \phi, X \right) + 2 \nabla^\mu \left( K \left( \phi, X
\right) \partial_\mu \phi \right)\, .
\ee
Here, $K_\phi \left( \phi, X \right) \equiv \partial K \left( \phi, X 
\right) / \partial \phi$ and it is
assumed that the scalar field $\phi$ does not directly couple with the
matter.

\subsubsection{The reconstruction using the cosmological time}

Let us start from the FRW universe (\ref{JGRG14}) and assume that the 
scalar field $\phi$ only depends on time.
The FRW equations are given by
\be
\label{KK2}
\frac{3}{\kappa^2} H^2 = 2 X \frac{\partial K\left( \phi, X \right)}
{\partial X} - K\left( \phi, X \right)
+ \rho_\mathrm{matter}(t)\, ,\quad - \frac{1}{\kappa^2}
\left(2 \dot H + 3 H^2 \right)
= K\left( \phi, X \right) + p_\mathrm{matter}(t)\, .
\ee
The FRW equations (\ref{KK2}) show that for the following model
\be
\label{k4}
K(\phi,X) = \sum_{n=0}^\infty \left(X+1\right)^n K^{(n)} (\phi)\, ,\quad
K^{(0)} (\phi) = - \frac{1}{\kappa^2}\left(2 f'(\phi) + 3 f(\phi)^2 
\right)\, ,\quad K^{(1)} (\phi) = \frac{1}{\kappa^2} f'(\phi) \, ,
\ee
there exists a solution given by
\be
\label{k5}
H = f(t)\, ,\quad \phi = t\, .
\ee
Note that in (\ref{k4}), $K^{(n)} (\phi)$ $n\geq 2$ can be an arbitrary
function.
To investigate the role of $K^{(n)} (\phi)$ ($n\geq 2$),
the following perturbation may be discussed: 
\be
\label{k6}
H = f(t) + \delta H\, ,\quad \phi = t + \delta \phi\, ,
\ee
The stability of the solution (\ref{k5}) should now be investigated.
In (\ref{k6}), it is assumed that $\delta H$ and $\delta\phi$ do not depend on 
the spatial coordinates.
Then, $\delta H$ is solved as a function of $\delta \phi$: 
$\delta H = \delta H(\delta\phi)$, and one obtains
\be
\label{k7}
\frac{d \delta \dot \phi}{dN} = \left[ - 3 - \frac{g^{\prime \prime}}
{g^{\prime 2}} - \frac{d}{dN}
\left\{ \frac{ \kappa ^2}{6g^{\prime 2}}
\left(8K^{(2)} - \frac{2}{\kappa^2}
g^{\prime \prime}\right) \right\} \right]
\delta \dot \phi \, .
\ee
If the quantity inside $[\ ]$ is negative, the solution (\ref{k5}) becomes
stable.
Because Eq.~(\ref{k7}) contains $K^{(2)}$, if $K^{(2)}$ is chosen properly, 
the solution corresponding to an arbitrary evolution law becomes stable.

Let us consider the condition that there is a Schwarzschild or
Schwarzschild-(A)dS solution.
Let $\phi$ be a constant:
\be
\label{Sch4}
\phi = \phi_0\, .
\ee
Eqs.~(\ref{Sch2}) and (\ref{Sch3}) reduce to
\bea
\label{Sch5}
&& \frac{1}{\kappa^2}\left( R_{\mu\nu} - \frac{1}{2}g_{\mu\nu} R \right)
= - K \left( \phi_0, 0 \right) g_{\mu\nu} + T_{\mu\nu}\, , \\
\label{Sch6}
&& 0= K_\phi \left( \phi_0, 0 \right) \, .
\eea
When $T_{\mu\nu}=0$, if $K \left( \phi_0, 0 \right)$ does not vanish, a
solution of (\ref{Sch5}) is given by Schwarzschild-(A)dS space-time.
On the other hand, if $K \left( \phi, 0 \right)$ vanishes, the 
Schwarzschild solution, which is asymptotically
Minkowski space-time, is a solution.
The equation (\ref{Sch6}) requires that, in general, 
$K_\phi \left( \phi,0\right)$ has an extremum or
$K \left( \phi, 0 \right)$ is a constant independent of $\phi$.
Especially if $K \left( \phi, 0 \right)$ vanishes identically, 
Eq.~(\ref{Sch5}) gives the Schwarzschild solution.

$K \left( \phi, X \right)$ can be written as
\be
\label{Sch7}
K \left( \phi, X \right) = \sum_{n=0} \left( 1 + X \right)^n 
K_n (\phi)\, .
\ee
Then, if $K \left( \phi, 0 \right)=0$, it follows that 
\be
\label{Sch8}
\sum_{n=0} K_n (\phi) = 0\, .
\ee
In particular, one may choose
\be
\label{Sch9}
K_3 (\phi) = - K_0 (\phi) - K_1 (\phi) - K_2 (\phi) \, .
\ee

Thus, if the condition (\ref{Sch8}) or (\ref{Sch9}) is satisfied, the
Schwarzschild space-time is always a solution
independent of the value of $\phi$ as long as $\phi$ is a constant.
Here, any point source of matter creates the Schwarzschild space-time which
generates the Newton potential.
The correction to Newton's law does not appear.
Note that the value of the scalar field $\phi$ changes by the evolution of 
the universe, but as long as
the condition (\ref{Sch8}) or (\ref{Sch9}) is satisfied, in a local region
where $\phi$ is almost constant,
the correction to Newton's law is negligible.

\subsubsection{The reconstruction using e-folding}

Because it is often convenient to use the redshift $z$ or e-folding $N$,
we rewrite the FRW equations (\ref{KK2}) as
\be
\label{KK2N}
\frac{3}{\kappa^2} H^2 = 2 X \frac{\partial K\left( \phi, X \right)}
{\partial X} - K\left( \phi, X \right)
+ \rho_\mathrm{matter}(N)\, ,\quad - \frac{1}{\kappa^2}
\left(2 H H' + 3 H^2 \right)
= K\left( \phi, X \right) + p_\mathrm{matter}(N)\, .
\ee
Here, $H' \equiv dH/dN$.
If the matter energy density is given by a sum of the contributions
from the matter with constant EoS parameters $w_i$, one finds
\bea
\label{KGC1}
&& \rho_\mathrm{matter}(N) = \sum _i \rho_{0i}a^{-3(1+w_i)}
= \sum _i \rho_{0i}\e^{-3(1+w_i)\left(N-N_0\right)} \, ,\nn
&& p_\mathrm{matter}(N) = \sum _i w_i \rho_{0i}a^{-3(1+w_i)}
= \sum _i w_i \rho_{0i}\e^{-3(1+w_i)\left(N-N_0\right)} \, .
\eea
Here, $\rho_{0i}$'s are constants.
For general energy density $\rho_\mathrm{matter}(N)$,
because the conservation law (\ref{CEm})
can be rewritten in terms of $N$ as
\be
\label{KGC3}
\rho_\mathrm{matter}'(N) + 3 \left( \rho_\mathrm{matter}(N) 
+ p_\mathrm{matter} (N) \right) = 0\, ,
\ee
it follows that 
\be
\label{KGC4}
p_\mathrm{matter} (N) = - \rho_\mathrm{matter}(N) - \frac{1}{3}
\rho_\mathrm{matter}'(N)\, .
\ee
With the help of the FRW equations (\ref{KK2N}), we find
\bea
\label{KGC5}
&& K\left( \phi, X \right) = - \frac{1}{\kappa^2}\left(2 H \frac{d H}{dN} 
+ 3 H^2 \right)
+ \rho_\mathrm{matter}(N) + \frac{1}{3} \rho_\mathrm{matter}'(N) \, ,\nn
&& H^2 \frac{\partial K\left( \phi, X \right)}{\partial X}
= \frac{1}{\kappa^2} H \frac{d H}{dN}
- \frac{1}{6} \rho_\mathrm{matter}'(N)\, .
\eea
If a new variable $G(N) = H(N)^2$ is defined, the equations (\ref{KGC5}) can
be rewritten as
\bea
\label{KGC6}
&& K\left( \phi, X \right) = - \frac{1}{\kappa^2}\left( G'(N) 
+ 3 G(N) \right)
+ \rho_\mathrm{matter}(N) + \frac{1}{3} \rho_\mathrm{matter}'(N) \, ,\nn
&& H^2 \frac{\partial K\left( \phi, X \right)}{\partial X}
= \frac{1}{2 \kappa^2} G'(N) - \frac{1}{6} \rho_\mathrm{matter}'(N)\, .
\eea
Using the appropriate function $g_\phi (\phi)$, if we choose
\bea
\label{KGC7}
&& K(\phi,X) = \sum_{n=0}^\infty \left(\frac{X}{g_\phi (\phi) 
+ \frac{\kappa^2}{3} \rho_\mathrm{matter}'(\phi)}
+ 1\right)^n \tilde K^{(n)} (\phi) \, , \nn
&& \tilde K^{(0)} (\phi) \equiv - \frac{1}{\kappa^2}
\left( g_\phi'(\phi) + 3 g_\phi(\phi) \right)\, \quad
\tilde K^{(1)} (\phi) = \frac{1}{2 \kappa^2} g_\phi'(\phi) \, ,
\eea
the explicit solution for the FRW equations (\ref{KK2}) is found as 
\be
\label{KGC8}
G(N) = H(N)^2 = g_\phi (N) 
+ \frac{\kappa^2}{3} \rho_\mathrm{matter}(N)\, ,
\quad \phi = N \quad \left(X=-H^2\right)\, .
\ee
Here, $\tilde K^{(n)} (\phi)$ with $n\geq 2$ can be arbitrary.
The reconstruction is fulfilled.

Let us now investigate the (in)stability of the solution (\ref{KGC8}).
For this purpose, the perturbation from the solution (\ref{KGC8}) is 
taken as follows:
\be
\label{KGC9}
G(N) = G_0(N) + \delta G(N) \quad
\left( G_0(N) \equiv g_\phi (N) + \frac{\kappa^2}{3} 
\rho_\mathrm{matter}'(N) \right) \, ,
\quad \phi = N + \delta \phi \, .
\ee
Note that $N$-dependence in the energy density
$\rho_\mathrm{matter}$ is usually given by a fixed function as in 
(\ref{KGC1}).
Therefore, $\delta \rho_\mathrm{matter}=0$.
Thus, the equations (\ref{KGC6}) give
\bea
\label{KGC10}
&& - \frac{1}{\kappa^2} \left( g_\phi''(N)
+ 3 g_\phi'(N) \right) \delta\phi(N) - \frac{g_\phi'(N)}{2\kappa^2}
\left(\frac{\delta G(N)}{G_0(N)} + 2\delta
\phi'(N) - \frac{G_0'(N)}{G_0(N)} \delta\phi(N) \right)
= - \frac{1}{\kappa^2} \left( \delta G'(N) + 3 \delta G(N) \right) \, , 
\nn
&& \frac{g_\phi'(N)}{2\kappa^2} \frac{\delta G(N)}{G_0(N)} 
+ \frac{g_\phi''(N)}{2\kappa^2} \delta\phi(N) - \frac{g_\phi'(N)}{2\kappa^2}
\frac{G_0'(N)}{G_0(N)}\delta\phi(N) - 2 \tilde K^{(2)} (N)
\left(\frac{\delta G(N)}{G_0(N)}
+ 2\delta \phi'(N) - \frac{G_0'(N)}{G_0(N)} \delta\phi(N) \right) \nn
&& \qquad = \frac{1}{2\kappa^2} \delta G'(N)\, .
\eea
Then, we find
\bea
\label{KGC11}
&& \left( \begin{array}{c}
\delta \phi'(N) \\ \delta G'(N)
\end{array} \right)
= \frac{1}{L(N)} \left( \begin{array}{cc} A & B \\
C & D \end{array} \right)
\left( \begin{array}{c}
\delta \phi(N) \\ \delta G(N)
\end{array} \right) \, , \nn
&& A \equiv - \left( 3 + \frac{G_0'(N)}{G_0(N)} \right) g_\phi'(N) 
+ \frac{G_0'(N)}{2G_0(N)} L(N) \, , \quad B \equiv 
\left(3 + \frac{g_\phi'(N)}{G_0(N)}\right) - \frac{L(N)}{2G_0(N)} \, , \\
&& C \equiv \left(g_\phi''(N) + 3g_\phi'(N) \right) L(N) - g_\phi'(N)^2
\left( 3 + \frac{G_0'(N)}{G_0(N)} \right) \, , \quad
D \equiv -3L(N) + g_\phi'(N) \left(3 + \frac{g_\phi'(N)}{G_0(N)}\right) 
\, . \nonumber
\eea
Here, 
\be
\label{KGC12}
L(N) \equiv g_\phi'(N) + 8\kappa^2 \tilde K^{(2)} (N)\, .
\ee
The stability of the solution requires $A+D < 0$ and $AD - BC > 0$, which 
gives
\bea
\label{KGC13}
&& \frac{1}{L(N)} \frac{g_\phi'(N)}{G_0(N)} \left( g_\phi'(N) - G_0' (N)
\right) < 3 - \frac{G_0'(N)}{2G_0(N)} \, , \\
&& \frac{1}{L(N)} \left( -2 g_\phi'(N) 
g_\phi''(N) - 9 g_\phi'(N)^2 - 6 g_\phi''(N) G_0(N)
+ 3 G_0'(N) g_\phi''(N) + 6 G_0'(N) g_\phi'(N) \right) > g_\phi''(N) \, .
\eea
It might be possible to make the system stable by choosing $L(N)$ and, 
therefore, $\tilde K^{(2)} (N)$ properly.
Neglecting the matter, one finds $g_\phi(N) = G_0(N)$ and
obtains 
\be
\label{KGC14}
3 > \frac{G_0'(N)}{2G_0(N)} \, , \quad
\frac{1}{L(N)} \left( G_0'(N) G_0''(N) - 3 G_0'(N)^2 - 6 G_0''(N) G_0(N)
\right) > G_0''(N) \, .
\ee
Thus, as long as the first condition is satisfied, the second condition 
can be always satisfied  by choosing $L(N)$, and therefore 
$\tilde K^{(2)} (N)$, as an appropriate function. 
Using the above-developed general reconstruction scheme, any 
specific accelerating universe may be formulated as a solution of such a theory.

%%%%%%%%
%%%%%%%%
%%%%%%%%

\subsection{Modified $F(R)$ gravity \label{IIIC}}

\subsubsection{The reconstruction using the cosmological time \label{recFR}}

It is easy now to generalize the cosmological reconstruction scheme
for modified gravity with $F(R)$ action (\ref{JGRG7}) 
\cite{Nojiri:2006gh,Capozziello:2006dj,Nojiri:2006be}.
The action (\ref{JGRG7}) is equivalently rewritten as
\be
\label{PQR1}
S=\int d^4 x \sqrt{-g} \left\{\frac{1}{2\kappa^2} \left( P(\phi) R 
+  Q(\phi) \right) + \mathcal{L}_\mathrm{matter}\right\}\, .
\ee
Here, $P$ and $Q$ are proper functions of the auxiliary scalar $\phi$.
The presence of an auxiliary scalar may simplify the formulation below, 
making it more similar to the one for scalar-tensor gravity.
In fact, by the variation over $\phi$, it follows that 
$0=P'(\phi)R + Q'(\phi)$,
which may be solved with respect to $\phi$ as $\phi=\phi(R)$.
By substituting the obtained expression of $\phi(R)$ into (\ref{PQR1}),
one arrives again at the $F(R)$-gravity:
\be
\label{PQR4}
S=\int d^4 x \sqrt{-g} \left\{\frac{F(R)}{2\kappa^2} +
\mathcal{L}_\mathrm{matter}\right\}\, , \quad
F(R)\equiv P(\phi(R)) R + Q(\phi(R))\, .
\ee
By the variation of the action (\ref{PQR1}) with respect
to the metric $g_{\mu\nu}$, the standard spatially flat FRW equations are
obtained as: 
\bea
\label{PQR6}
0&=&-6 H^2 P(\phi) - Q(\phi) - 6H\frac{dP(\phi(t))}{dt} + 2\kappa^2
\rho_\mathrm{matter} \, ,\\
\label{PQR7}
0&=&\left(4\dot H + 6H^2\right)P(\phi) + Q(\phi) 
+ 2\frac{d^2 P(\phi(t))}{dt}
+ 4H\frac{d P(\phi(t))}{dt} + 2\kappa^2 p_\mathrm{matter}\, .
\eea
Simple algebra leads to the following equation
\be
\label{PQR7b}
0=2\frac{d^2 P(\phi(t))}{dt^2} - 2 H \frac{dP(\phi(t))}{dt} 
+ 4\dot H P(\phi)
+ 2\kappa^2 \left( p_\mathrm{matter} + \rho_\mathrm{matter} \right) \, .
\ee
As one can redefine the scalar field $\phi$ freely, a very natural choice is
$\phi=t$.
It is assumed that $\rho_\mathrm{matter}$ and $p_\mathrm{matter}$ are the 
sum from the contributions of the matters with constant equation of state 
parameters $w_i$.
In particular, when a combination of the radiation and dust is assumed, 
one gets the standard expression
\be
\label{PQR9}
\rho_\mathrm{matter}=\rho_{r0} a^{-4} + \rho_{d0} a^{-3}\, ,\quad
p_\mathrm{matter}=\frac{\rho_{r0}}{3}a^{-4}\, ,
\ee
with constant $\rho_{r0}$ and $\rho_{d0}$. If the scale factor $a$ is 
given by a proper function $g(t)$ as $a=a_0\e^{g(t)}$ with a constant $a_0$, 
Eq.~(\ref{PQR7}) reduces to the second-rank differential equation:
\be
\label{PQR11}
0 = 2 \frac{d^2 P(\phi)}{d\phi^2} - 2 g'(\phi) \frac{dP(\phi))}{d\phi} 
+ 4g''(\phi) P(\phi)
+ 2\kappa^2 \sum_i \left(1 + w_i\right) \rho_{i0} a_0^{-3(1+w_i)} 
\e^{-3(1+w_i)g(\phi)} \, .
\ee
In principle, by solving (\ref{PQR11}) the form of $P(\phi)$ may be found.
Using (\ref{PQR6}) (or equivalently (\ref{PQR7})), the form of $Q(\phi)$
follows:
\be
\label{eq:2.12}
Q(\phi)=-6 \left(g'(\phi)\right)^2 P(\phi) - 6g'(\phi) 
\frac{dP(\phi)}{d\phi} 
+ 2\kappa^2 \sum_i \rho_{i0} a_0^{-3(1+w_i)} \e^{-3(1+w_i)g(\phi)} \, .
\ee
Thus, in principle, any given cosmology expressed as $a=a_0\e^{g(t)}$
can be realized as the solution of some specific (reconstructed)
$F(R)$ gravity. Some explicit examples of such a reconstruction 
are presented below. 
Of course, the reconstruction is applied to cosmological eras that 
are fundamentally important in modern cosmology.

\

\noindent
{\it Model reproducing $\Lambda$CDM-type cosmology}

Let us investigate whether $\Lambda$CDM-type cosmology can be reconstructed
exactly by $F(R)$ gravity in the present formulation when we include dust,
which can be a sum of the baryon and dark matter, and radiation.

In Einstein gravity, when there is matter with the EOS parameter $w$
and cosmological constant, the FRW equation has the form (\ref{LCDM1}).
The solution of (\ref{LCDM1}) is given by (\ref{LCDM2}).
Let us show how it is possible to reconstruct $F(R)$ gravity by reproducing
such an epoch (\ref{LCDM2}).
When matter is included, Eq.~(\ref{PQR11}) has the following form:
\bea
\label{LCDM4}
0 &=& 2\frac{d^2 P(\phi)}{d\phi^2}
 - \frac{2}{l}\coth \left(\frac{3(1+w)}{2l}\left(\phi- t_s \right)\right)
\frac{dP(\phi)}{d\phi}
 - \frac{6(1+w)}{l^2} \sinh^{-2} \left(\frac{3(1+w)}{2l}\left(\phi - t_s
\right)\right) P(\phi) \\
&& + \frac{4}{3}\rho_{r0} a_0^{-4} \left(\alpha \sinh
\left(\frac{3(1+w)}{2l}\left(\phi
 - t_s \right)\right)\right)^{-8/3(1+w)}
+ \rho_{d0}a_0^{-3}\left(\alpha \sinh \left(\frac{3(1+w)}{2l}
\left(\phi - t_s \right)\right)\right)^{-2/(1+w)}\, . \nonumber
\eea
Because this equation is a linear inhomogeneous equation,
its general solution is given by the sum of the
special solution and of the general solution that corresponds to the
homogeneous equation.
For the case without matter, by changing the variable from $\phi$ to $z$ as
follows,
\be
\label{LCDM5}
z\equiv - \sinh^{-2} \left(\frac{3(1+w)}{2l}\left(t - t_s \right)\right) 
\, ,
\ee
Eq.~(\ref{LCDM4}) without matter can be rewritten in the form of the Gauss 
hypergeometric differential equation:
\bea
\label{LCDM6}
&& 0=z(1-z)\frac{d^2 P}{dz^2} + \left[\tilde\gamma - \left(\tilde\alpha
+ \tilde \beta + 1\right)z\right] \frac{dP}{dz} - \tilde\alpha
\tilde\beta P\, , \nn
&& \tilde\gamma \equiv 4 + \frac{1}{3(1+w)},\
\tilde\alpha + \tilde\beta + 1 \equiv 6 + \frac{1}{3(1+w)},\
\tilde\alpha \tilde\beta \equiv - \frac{1}{3(1+w)}\, ,
\eea
whose solution is given by Gauss' hypergeometric function:
\be
\label{LCDM7}
P= P_0 F(\tilde\alpha,\tilde\beta,\tilde\gamma;z)
\equiv P_0 \frac{\Gamma(\tilde\gamma)}{\Gamma(\tilde\alpha)
\Gamma(\tilde\beta)} 
\sum_{n=0}^\infty \frac{\Gamma(\tilde\alpha + n) \Gamma(\beta 
+ n)}{\Gamma(\tilde\gamma + n)} \frac{z^n}{n!}\, .
\ee
Here, $\Gamma$ is the $\Gamma$ function. There is one more 
linearly independent solution like
$(1-z)^{\tilde\gamma - \tilde\alpha - \tilde\beta}
F(\tilde\gamma - \tilde\alpha, \tilde\gamma - \tilde\beta, 
\tilde\gamma;z)$, 
but we drop it here, for simplicity.
Using (\ref{eq:2.12}), one finds the form of $Q(\phi)$:
\be
\label{LCDM8}
Q = - \frac{6(1-z)P_0}{l^2}
F( \tilde\alpha,\tilde\beta,\tilde\gamma;z) - \frac{3(1+w) z(1-z)P_0}
{l^2(13 + 12w)} F(\tilde\alpha+1,\tilde\beta+1,\tilde\gamma+1;z)\, .
\ee
From (\ref{LCDM5}), it follows that $z\to 0$ when $t=\phi\to + \infty$. 
Then in the limit of $t=\phi\to + \infty$, one arrives at 
$P(\phi)R + Q(\phi) \to P_0 R - 6P_0/l^2$.
Identifying $P_0=1/2\kappa^2$ and $\Lambda = 6/l^2$,
the Einstein theory with cosmological constant $\Lambda$ can be 
reproduced.
The action is not singular even in the limit of $t\to \infty$.
Therefore, even without the cosmological constant nor cold dark matter,
the cosmology of the $\Lambda$CDM model
can be reproduced by $F(R)$-gravity.

We now investigate the special solution of (\ref{LCDM4}).
By changing the variable as in (\ref{LCDM5}), the
inhomogeneous differential equation looks as follows:
\bea
\label{LCDMa1}
&& 0=z(1-z)\frac{d^2 P}{dz^2} + \left[\tilde\gamma - \left(\tilde\alpha
+ \tilde \beta + 1\right)z\right] \frac{dP}{dz} - \tilde\alpha \tilde\beta 
P + \eta\left(-z\right)^{-2(1+3w)/3(1+w)}
+ \xi\left(-z\right)^{-\frac{1+2w}{1+w}}\, , \nn
&& \eta \equiv \frac{4l^2}{27(1+w)}\rho_{r0}a_0^{-4}
\alpha^{-8/3(1+w)}\, ,\quad
\zeta \equiv \frac{l^2}{3(1+w)}\rho_{d0}a_0^{-3}\alpha^{-2/(1+w)}\, .
\eea
It is not trivial to find the solution of (\ref{LCDMa1}).
Let us consider the case that $w=0$ and $z\to -\infty$, that is,
$t\to t_s$. In the limit of $t\to t_s$,
Eq.~(\ref{LCDMa1}) reduces to
\be
\label{LCDMa2}
0= - z^2\frac{d^2 P}{dz^2} + \tilde\gamma z \frac{dP}{dz} - \tilde\alpha
\tilde\beta P + \eta\left(-z\right)^{-2/3} \, ,
\ee
whose special solution is given by
\be
\label{LCDMa3}
P=P_0(-z)^{-2/3}\, ,\quad P_0 
= \frac{\eta}{\frac{10}{9} - \frac{2\left(\tilde\alpha + \tilde\beta 
+ 1\right)}{3} + \tilde\alpha\tilde\beta} = - \frac{9\eta}{25}\, .
\ee
In principle, other special solutions of Eq.~(\ref{LCDMa1}) could be found.
This proves that, in the presence (or in the absence) of traditional matter, 
the standard $\Lambda$CDM cosmology
can be reproduced exactly by $F(R)$ gravity.

\subsubsection{Reconstructed model realizing the phantom divide crossing}

Using the above-developed method, we reconstruct an explicit model in
which a crossing of the phantom divide can be realized \cite{Bamba:2008hq}.
In other words, the modified $F(R)$ theory that describes
the transition from acceleration to superacceleration (phantom era).
will be presented. 

A solution of Eq.~(\ref{PQR11}) without matter can be given by
\bea
P(\phi) &=& \e^{\tilde{g}(\phi)/2} \tilde{p}(\phi)\, ,
\label{PDF2} \\
\tilde{g}(\phi) &=& - 10 \ln \left[ \left(\frac{\phi}{t_0}
\right)^{-\gamma} - C \left(\frac{\phi}{t_0}\right)^{\gamma+1} \right]\, ,
\label{PDF4} \\
\tilde{p}(\phi) &=& \tilde{p}_+ \phi^{\beta_+} 
+ \tilde{p}_- \phi^{\beta_-}\,,
\label{PDF6} \\
\beta_\pm &=& \frac{1 \pm \sqrt{1 + 100 \gamma (\gamma + 1)}}{2}\,,
\label{PDF7}
\end{eqnarray}
where $\gamma$ and $C$ are positive constants,
$t_0$ is the present time, and $\tilde{p}_\pm$ are arbitrary constants.

 From Eq.~(\ref{PDF4}), it follows that $\tilde{g}(\phi)$ diverges at
finite $\phi$ when
\be
\label{PDF8}
\phi = t_s \equiv t_0 C^{-1/(2\gamma + 1)}\, ,
\ee
which shows that the Big Rip singularity could occur at
$t=t_s$ \cite{Caldwell:2003vq}.
One only needs to consider the period $0<t<t_s$ because
$\tilde{g}(\phi)$ should be a real number.
Eq.~(\ref{PDF4}) also gives the following Hubble rate $H(t)$:
\be
\label{PDF9}
H(t)= \frac{d \tilde{g}(\phi)}{d \phi}
= \left(\frac{10}{t_0}\right) \left[ \frac{ \gamma
\left(\frac{\phi}{t_0}\right)^{-\gamma-1 }
+ (\gamma+1) C \left(\frac{\phi}{t_0}\right)^{\gamma}
}{\left(\frac{\phi}{t_0}\right)^{-\gamma} - C
\left(\frac{\phi}{t_0}\right)^{\gamma+1}}\right]\, ,
\ee
where it is taken that $\phi=t$.

In the FRW background (\ref{JGRG14}), as in (\ref{JGRG11}), the effective
energy density and pressure of the universe are given by 
$\rho_\mathrm{eff} = 3H^2/\kappa^2$ and
$p_\mathrm{eff} = -\left(2\dot{H} + 3H^2 \right)/\kappa^2$, respectively.
Then, the effective EoS parameter $w_\mathrm{eff}$ is defined as in 
(\ref{JGRG12}).
For the case of $H(t)$ of Eq.~(\ref{PDF9}), from Eq.~(\ref{JGRG12}), we
find that $w_\mathrm{eff}$ is expressed as
\be
w_\mathrm{eff} = -1 + U(t)\, ,
\label{eq:I0-1-1}
\ee
where
\be
U(t) \equiv -\frac{2\dot{H}}{3H^2} 
= - \frac{-\gamma + 4\gamma \left( \gamma+1 \right)
\left( \frac{t}{t_s} \right)^{2\gamma+1} 
+ \left( \gamma+1 \right) \left( \frac{t}{t_s}
\right)^{2\left( 2\gamma+1 \right)} }
{15 \left[ \gamma + \left( \gamma+1 \right)
\left( \frac{t}{t_s} \right)^{2\gamma+1} \right]^2}\, .
\label{eq:I0-1-2}
\ee
For the case of Eq.~(\ref{PDF9}), the scalar curvature
$R=6\left( \dot{H} + 2H^2 \right)$ is given by
\be
R = \frac{60 \left[ \gamma \left( 20\gamma -1 \right) 
+ 44\gamma \left( \gamma+1 \right)
\left( \frac{t}{t_s} \right)^{2\gamma+1} 
+ \left( \gamma+1 \right) \left( 20\gamma+21 \right)
\left( \frac{t}{t_s} \right)^{2\left( 2\gamma+1 \right)}
\right] } {t^2 \left[ 1- \left( \frac{t}{t_s} \right)^{2\gamma+1} 
\right]^2} \, .
\label{eq:I0-2}
\ee
In deriving Eqs.~(\ref{eq:I0-1-2}) and (\ref{eq:I0-2}), Eq.~(\ref{PDF8}) 
was used.

When $t\to 0$, i.e., $t \ll t_s$, $H(t)$ behaves as
\be
\label{PDF10}
H(t) \sim \frac{10\gamma}{t}\, .
\ee
In this limit, it follows from Eq.~(\ref{JGRG12}) that
the effective EoS parameter is given by
\be
\label{PDF11}
w_\mathrm{eff} = -1 + \frac{1}{15\gamma}\, .
\ee
This behavior is identical to that in Einstein gravity with matter
with an EoS parameter greater than $-1$.

On the other hand, when $t\to t_s$, one finds that 
\be
\label{PDF12}
H(t) \sim \frac{10}{t_s - t}\, .
\ee
In this case, the scale factor is given by
$a(t) \sim a_0 \left( t_s - t \right)^{-10}$.
When $t\to t_s$, therefore, $a \to \infty$, namely, the Big Rip 
singularity appears.
In this limit, the effective EoS parameter is given by
\be
\label{PDF13}
w_\mathrm{eff} = - 1 - \frac{1}{15} = -\frac{16}{15}\, .
\ee
This behavior is identical with that of the case in which there is phantom 
matter that has an EoS parameter smaller than $-1$.
Thus, we have obtained an explicit model showing a crossing of
the phantom divide.

\begin{figure}[tbp]
\begin{center}
%\resizebox{!}{8cm}{
\includegraphics{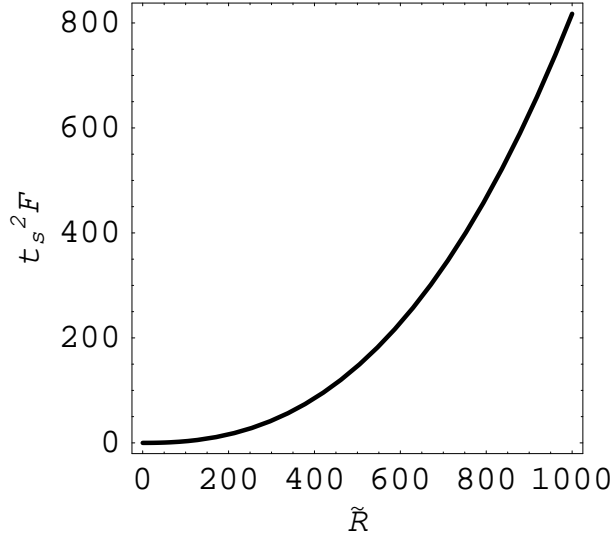}
% }
\caption{Behavior of $t_s^2F(\tilde{R})$ as a function of $\tilde{R}$ for
$\gamma =1/2$, $\tilde{p}_+ =-1/t_s^{\beta_+}$,
$\tilde{p}_- =0$, $\beta_+ = \left(1+2\sqrt{19}\right)/2$ and $t_s =2t_0$.
}
\end{center}
\label{fg:1}
\end{figure}

It follows from Eq.~(\ref{JGRG12}) that the effective EoS parameter
$w_\mathrm{eff}$ is equal to $-1$ when $\dot{H}=0$.
Solving $w_\mathrm{eff} = -1$ with respect to
$t$ by using Eq.~(\ref{eq:I0-1-1}), namely, $U(t)=0$,
we find that the effective EoS parameter crosses the phantom divide at
$t=t_\mathrm{c}$ given by
\be
t_\mathrm{c} = t_s \left( -2\gamma +
\sqrt{4\gamma^2 + \frac{\gamma}{\gamma+1}}
\right)^{1/\left( 2\gamma + 1 \right)}\, .
\label{eq:I1}
\ee
From Eq.~(\ref{eq:I0-1-2}), one sees that when $t<t_\mathrm{c}$, $U(t)>0$
because $\gamma >0$.
Moreover, the time derivative of $U(t)$ is given by
\be
\frac{d U(t)}{dt} = -\frac{2\gamma \left( \gamma+1 \right) 
\left( 2\gamma+1 \right)^2}{15\left[ \gamma + \left( \gamma+1 \right)
\left( \frac{t}{t_s} \right)^{2\gamma+1} \right]^3}
\left( \frac{1}{t_s} \right)
\left( \frac{t}{t_s} \right)^{2\gamma}
\left[ 1 - \left( \frac{t}{t_s} \right)^{2\gamma+1} \right]\,.
\label{eq:I1-2}
\ee
Eq.~(\ref{eq:I1-2}) shows that the relation $d U(t)/\left(dt\right) <0$ is
always satisfied because only the period $0<t<t_s$ is considered as 
mentioned above.
This means that $U(t)$ decreases monotonically. Thus, the value of $U(t)$
evolves from positive to negative. From Eq.~(\ref{eq:I0-1-1}), we see that
the value of $w_\mathrm{eff}$ crosses $-1$.
Once the universe enters the phantom phase, it stays in this phase,
namely, the value of $w_\mathrm{eff}$ remains less than $-1$, and
finally the Big Rip singularity occurs because $U(t)$ decreases
monotonically.
Note that other types of finite-time future singularities in modified
gravity are possible as demonstrated after Eq.~(\ref{XIV}) in the 
second chapter (see also ref.~\cite{Bamba:2008ut}).
It follows from Eqs.~(\ref{PDF2}), (\ref{PDF4}), (\ref{PDF6}) and 
(\ref{PDF8}) that $P(t)$ is given by
\be
P(t) = \left[ \frac{\left( \frac{t}{t_0} \right)^\gamma}
{1-\left( \frac{t}{t_s} \right)^{2\gamma+1}} \right]^5
\sum_{j=\pm} \tilde{p}_j t^{\beta_j}\, .
\label{eq:I2}
\ee
Using Eq.~(\ref{eq:I2}), one gets
\be
Q(t) = -6H \left[ \frac{\left( \frac{t}{t_0} \right)^\gamma}
{1-\left( \frac{t}{t_s} \right)^{2\gamma+1}} \right]^5
\sum_{j=\pm} \left( \frac{3}{2}H + \frac{\beta_j}{t} \right)
\tilde{p}_j t^{\beta_j}\,.
\label{eq:I3}
\ee
If one can solve Eq.~(\ref{eq:I0-2}) with respect to $t$ as $t=t(R)$,
in principle we can obtain the form of $F(R)$ by using this solution and
Eqs.~(\ref{PQR4}), (\ref{eq:I2}) and (\ref{eq:I3}).
In fact, however, for the general case it is difficult to solve
Eq.~(\ref{eq:I0-2}) as $t=t(R)$. Thus, as a solvable example, we show 
the behavior of $t_s^2 F(\tilde{R})$ as a function of 
$\tilde{R} \equiv t_s^2 R$ in Fig.~1 for $\gamma =1/2$,
$\tilde{p}_+ = -1/t_s^{\beta_+}$, $\tilde{p}_- =0$,
$\beta_+ = \left(1+2\sqrt{19}\right)/2$ and $t_s =2t_0$.
The quantities in Fig.~\ref{fg:1} are shown as dimensionless value.
The horizontal and vertical axes show $\tilde{R}$ and $t_s^2 F$, 
respectively.
(Here, $\tilde{R} = t_s^2R = 4R/R_0$, where $R_0$ is the current 
curvature.
In deriving this relation, we have used $t_s =2t_0$, 
$t_0 \approx H_0^{-1}$,
where $H_0$ is the present Hubble parameter.) From Fig.~1, we see that
the value of $F(R)$ increases as that of $R$ becomes larger.

%%%%%% Fig. 1 %%%%%%%%%
%%%%%%%%%%%%%%%%%%%%%

To examine the analytic form of $F(R)$ for the general case, we 
will now investigate the behavior of $F(R)$ in the limits $t \to 0$ 
and $t \to t_s$.
When $t \to 0$, from Eq.~(\ref{PDF10}) we find
\be
t \sim \sqrt{ \frac{60\gamma \left( 20\gamma -1 \right)}{R} }\, .
\label{eq:I4}
\ee
In this limit, it follows from Eqs.~(\ref{PQR4}), (\ref{PDF10}),
(\ref{eq:I2}), (\ref{eq:I3}) and (\ref{eq:I4}) that
the form of $F(R)$ is given by
\bea
F(R) &\sim & \left\{
\frac{\left[\frac{1}{t_0} \sqrt{60\gamma \left( 20\gamma -1 \right)} 
R^{-1/2} \right]^\gamma}{1 - \left[\frac{1}{t_s} 
\sqrt{60\gamma \left( 20\gamma -1 \right)} R^{-1/2}
\right]^{2\gamma+1}} \right\}^5 R \nonumber \\
&& \times \sum_{j=\pm}
\biggl\{ \left( \frac{5\gamma -1 -\beta_j}{20\gamma -1} \right) 
\tilde{p}_j \left[60\gamma \left( 20\gamma -1 \right) 
\right]^{\beta_j /2} R^{-\beta_j /2} \biggr\}\, .
\label{eq:I5}
\eea
On the other hand, when $t\to t_s$, from Eq.~(\ref{PDF12}) we obtain
\be
t \sim t_s - 3\sqrt{ \frac{140}{R} }\, .
\label{eq:I6}
\ee
In this limit, it follows from Eqs.~(\ref{PQR4}), (\ref{PDF12}),
(\ref{eq:I2}), (\ref{eq:I3}) and (\ref{eq:I6}) that
the form of $F(R)$ is given by
\bea
F(R) &\sim& \left( \frac{ \left\{ \frac{1}{t_0}
\left[ t_s - 3\sqrt{140} R^{-1/2} \right] \right\}^\gamma}
{1 - \left[ 1 - \frac{3\sqrt{140}}{t_s} R^{-1/2}
\right]^{2\gamma+1} } \right)^5 R
\sum_{j=\pm} \tilde{p}_j
\left[ t_s - 3\sqrt{140} R^{-1/2} \right]^{\beta_j} \nn
&& \times \left\{ 1- \sqrt{\frac{20}{7}}
\left[ \sqrt{\frac{15}{84}} t_s + \left( \beta_j - 15 \right) R^{-1/2} 
\right] \frac{1}{t_s - 3\sqrt{140} R^{-1/2}} \right\}\,.
\label{eq:I7}
\eea
The above modified gravity may be considered as an approximated form of
a more realistic and viable theory.
For large $R$, namely, $t_s^2R \gg 1$, the expression of $F(R)$ in
(\ref{eq:I7}) can be written approximately as
\be
F(R) \approx \frac{2}{7}
\left[ \frac{1}{3\sqrt{140} \left( 2\gamma +1 \right)}
\left( \frac{t_s}{t_0} \right)^\gamma \right]^5
\left( \sum_{j=\pm} \tilde{p}_j t_s^{\beta_j} \right) t_s^5 R^{7/2}\, .
\label{eq:I8}
\ee

Note also that viable models that lead to a Big Rip singularity as is 
discussed in the second chapter necessarily predict the phantom 
divide crossing. This gives exact (non-asymptotic) and realistic 
modified gravities that show such phenomena.
Moreover, some of the viable models with accelerating solutions that are
asymptotically non-singular or tend to softer Type II, Type III and Type 
IV singularities may also show phantom divide crossing. 
However, in this case the phantom era turns out to be transient.

\subsubsection{The reconstruction using e-folding}

Let us demonstrate the reconstruction of $F(R)$ gravity in terms of e-folding. 
Moreover, unlike the previous sub-section, the slightly different approach is 
developed without the introduction of an auxiliary scalar.
We introduce the e-folding instead of the cosmological time $t$ as:
$N=\ln \frac{a}{a_0}$.
The variable $N$ is related to the redshift $z$ by
\be
\label{Nz}
\e^{-N}=\frac{a_0}{a} = 1 + z\, .
\ee
Because $\frac{d}{dt} = H \frac{d}{dN}$ and, therefore, 
$\frac{d^2}{dt^2} = H^2 \frac{d^2}{dN^2} + H \frac{dH}{dN} \frac{d}{dN}$,
one can rewrite (\ref{JGRG15}) by
\be
\label{RZ4}
0 = - \frac{F(R)}{2} + 3 \left( H^2 + H H'\right) F'(R)
 - 18 \left(4 H^3 H' + H^2 \left(H'\right)^2 + H^3 H''\right) F''(R)
+ \kappa^2 \rho_\mathrm{matter}\, .
\ee
Here, $H'\equiv dH/dN$ and $H''\equiv d^2 H/dN^2$.
As usual, the matter energy density $\rho_\mathrm{matter}$ is given by the 
sum of the fluid densities with the constant EoS parameter $w_i$
\be
\label{RZ6}
\rho_\mathrm{matter} = \sum_i \rho_{i0} a^{-3(1+w_i)}
= \sum_i \rho_{i0} a_0^{-3(1+w_i)} \e^{-3(1+w_i)N}\, .
\ee
Let the Hubble rate be given in terms of $N$ via the function $g(N)$ as
\be
\label{RZ7}
H=g(N) = g \left(- \ln\left(1+z\right)\right)\, .
\ee
Then, the scalar curvature takes the form: $R = 6 g'(N) g(N) + 12 g(N)^2$,
which can be solved with respect to $N$ as $N=N(R)$. 
Using (\ref{RZ6}) and (\ref{RZ7}), one can rewrite (\ref{RZ4}) as
\bea
\label{RZ9}
0 &=& -18 \left(4g\left(N\left(R\right)\right)^3 
g'\left(N\left(R\right)\right)
+ g\left(N\left(R\right)\right)^2 g'\left(N\left(R\right)\right)^2
+ g\left(N\left(R\right)\right)^3g''\left(N\left(R\right)\right)\right)
\frac{d^2 F(R)}{dR^2} \nn
&& + 3 \left( g\left(N\left(R\right)\right)^2
+ g'\left(N\left(R\right)\right) g\left(N\left(R\right)\right)\right)
\frac{dF(R)}{dR} - \frac{F(R)}{2}
+ \kappa^2 \sum_i \rho_{i0} a_0^{-3(1+w_i)} \e^{-3(1+w_i)N(R)}\, ,
\eea
which constitutes a differential equation for $F(R)$, where the variable
is the scalar curvature $R$.
Instead of $g$, if $G(N) \equiv g\left(N\right)^2 = H^2$ is used,
the expression (\ref{RZ9}) can be simplified as:
\bea
\label{RZ11}
0 &=& -9 G\left(N\left(R\right)\right)
\left(4 G'\left(N\left(R\right)\right)
+ G''\left(N\left(R\right)\right)\right) \frac{d^2 F(R)}{dR^2}
+ \left( 3 G\left(N\left(R\right)\right)
+ \frac{3}{2} G'\left(N\left(R\right)\right) \right) \frac{dF(R)}{dR} \nn
&& - \frac{F(R)}{2}
+ \kappa^2 \sum_i \rho_{i0} a_0^{-3(1+w_i)} \e^{-3(1+w_i)N(R)}\, .
\eea
Note that $R= 3 G'(N) + 12 G(N)$.
Thus, when we find an $F(R)$ that satisfies the differential equation
(\ref{RZ9}) or (\ref{RZ11}), and such $F(R)$ theory 
yields a solution (\ref{RZ7}). Thus, such reconstructed $F(R)$ gravity
realizes the above cosmological solution. It is interesting that the above scheme 
may be applied not only to general relativity 
but also to modified gravity (partial reconstruction). 
In other words, one can start from modified gravity which
consistently describes the entire sequence of the universe evolution eras
and satisfies local tests; however, some of its predictions seem 
to contradict the observational data. 
In this case, one can apply the partial reconstruction
of such a model (for instance, slightly modifying it at the very early 
universe) so that the above contradiction is resolved 
but all good properties of the theory are preserved. 
This provides a way to avoid any future exclusion of a non-realistic 
modified gravity to arrive at more narrow and more realistic sub-class 
of such theories.

\subsubsection{Stability of the cosmological solution \label{stability}}

In this section, we investigate the condition of the stability of 
the given cosmological solution.
The FRW equation (\ref{JGRG15}) can be rewritten by using $G=H^2$
and the e-folding $N$ as follows (\ref{RZ11}):
\bea
\label{FR_2}
0 &=& -9 G(N) \left(4 G'(N) + G''(N) \right) \left. \frac{d^2 F(R)}{dR^2}
\right|_{R=3 G'(N) + 12 G(N)}
+ \left( 3 G(N) + \frac{3}{2} G'(N) \right) \left. \frac{dF(R)}{dR}
\right|_{R=3 G'(N) + 12 G(N)} \nn
&& - \left. \frac{F(R)}{2} \right|_{R=3 G'(N) + 12 G(N)}
+ \kappa^2 \rho_\mathrm{matter} \left(N\right)\, .
\eea
Assume that the $N$-dependence of $\rho_\mathrm{matter}$ is known as in
(\ref{RZ6}).
Let a solution of (\ref{FR_2}) be $G=G_0(N)$, and consider the perturbation
from the following solution:
\be
\label{FR_3}
G(N) = G_0(N) + \delta G(N)\, .
\ee
Then, Eq.~(\ref{FR_2}) gives
\bea
\label{FR_4}
0 &=& G_0(N) \left. \frac{d^2 F(R)}{dR^2} \right|_{R=3 G_0'(N) 
+ 12 G_0(N)} \delta G'' (N)
+ \left\{ 3 G_0(N) \left( 4G_0'(N) + G_0''(N) \right)
\left. \frac{d^3 F(R)}{dR^3} \right|_{R=3 G_0'(N) + 12 G_0(N)} \right. \nn
&& \left. + \left( 3 G_0(N) - \frac{1}{2} G_0'(N) \right)
\left. \frac{d^2 F(R)}{dR^2} \right|_{R=3 G_0'(N) + 12 G_0(N)}
\right\} \delta G'(N) \nn
&& + \left\{ 12 G_0 (N) \left( 4G_0'(N) + G_0''(N) \right)
\left. \frac{d^3 F(R)}{dR^3} \right|_{R=3 G_0'(N) + 12 G_0(N)}
\right. \nn
&& \left. + \left( - 4G_0(N) + 2 G_0'(N) + G_0''(N) \right)
\left. \frac{d^2 F(R)}{dR^2} \right|_{R=3 G_0'(N) + 12 G_0(N)}
+ \frac{1}{3} \left. \frac{d F(R)}{dR} \right|_{R=3 G_0'(N) + 12 G_0(N)}
\right\} \delta G \, .
\eea
Note that in this formulation, as it is assumed that 
the $N$-dependence of $\rho_\mathrm{matter}$ is known,
we do not need to consider the fluctuation of $\rho_\mathrm{matter}$. 
Using the cosmological time $t$ instead of $N$, we need to include 
the fluctuation of $\rho_\mathrm{matter}$.
Then, the conditions of the stability are given by
\bea
\label{FR_5}
&& 6 \left( 4 G_0'(N) + G_0''(N)\right) \left. \frac{d^3 F(R)}{dR^3}
\right|_{R=3 G_0'(N) + 12 G_0(N)}
\left( \left. \frac{d^2 F(R)}{dR^2} \right|_{R=3 G_0'(N) + 12 G_0(N)}
\right)^{-1} \nn
&& + 6 - \frac{G_0'(N)}{G_0(N)} >0 \, , \\
\label{FR_6}
&& 36 \left( 4G_0'(N) + G_0''(N) \right) \left. \frac{d^3 F(R)}{dR^3}
\right|_{R=3 G_0'(N) + 12 G_0(N)}
\left( \left. \frac{d^2 F(R)}{dR^2} \right|_{R=3 G_0'(N) + 12 G_0(N)}
\right)^{-1} - 12
+ \frac{6 G_0'(N)}{G_0(N)} \nn
&& + \frac{3 G_0''(N)}{G_0(N)} + \frac{1}{G_0(N)} \left. \frac{d F(R)}{dR}
\right|_{R=3 G_0'(N) + 12 G_0(N)}
\left( \left. \frac{d^2 F(R)}{dR^2} \right|_{R=3 G_0'(N) + 12 G_0(N)}
\right)^{-1} >0 \, .
\eea
In the case of de Sitter space, where $H$ and, therefore, $G_0$ and
$R=R_0 \equiv 12 G_0$ are constants, Eq.~(\ref{FR_5}) gives $6>0$ 
and is trivially satisfied. Eq.~(\ref{FR_6}) becomes
\be
\label{FR7}
 - 12 G_0 + \left. \frac{d F(R)}{dR} \right|_{R=12 G_0}
\left( \left. \frac{d^2 F(R)}{dR^2} \right|_{R=12 G_0} \right)^{-1}
= - R_0 + \left. \frac{d F(R)}{dR} \right|_{R=R_0}
\left( \left. \frac{d^2 F(R)}{dR^2} \right|_{R=R_0} \right)^{-1} >0 \, ,
\ee
which is the standard result.

One may consider the case that $F(R)$ gravity admits two de Sitter 
solutions. If one is stable but another
is unstable, there could be a solution in which the unstable de Sitter 
solution could transit to the stable solution.
If the Hubble rate $H$ in the unstable solution is much larger than the 
Hubble rate in the stable one,
the unstable solution may correspond to the inflation and the stable one 
to the late-time acceleration.
Note that we may directly construct an $F(R)$ theory that admits the
transition from an asymptotic de Sitter
universe with a large Hubble rate to another asymptotic de Sitter
universe with a small Hubble rate. Then, if the solution and $F(R)$
satisfy the conditions (\ref{FR_5})
and (\ref{FR_6}), such a transition occurs.

Consider the stability of a de Sitter universe satisfying the condition
(\ref{JGRG16}), or equivalently (\ref{FRV2}). Eq.~(\ref{JGRG16}) can 
be rewritten as
\be
\label{dS1}
0 = \frac{d}{dR}\left( \frac{F(R)}{R^2} \right) \, .
\ee
Let $R=R_0$ be a solution of (\ref{dS1}). Then, $F(R)$ has the
following form:
\be
\label{dS2}
\frac{F(R)}{R^2} = f_0 + f (R) \left( R - R_0 \right)^n \, .
\ee
Here, $f_0$ is a constant, which should be positive if we require $F(R)>0$
and $n$ to be integers greater than or equal to two: $n\geq 2$.
Assume that the function $f(R)$ does not vanish at $R=R_0$, $f(R_0)\neq 0$.
When $n=2$,
\be
\label{dS3}
 - R_0 + \frac{F'(R_0)}{F''(R_0)} = - \frac{f(R_0)A_0}{f_0 + f(A_0)}\, .
\ee
Then, Eq.~(\ref{FR7}) shows that the de Sitter solution is stable if
\be
\label{dS4}
 - f_0 < f(A_0) < 0\, .
\ee
When $n\geq 3$, one finds
\be
\label{dS5}
 - R_0 + \frac{F'(R_0)}{F''(R_0)} = 0\, .
\ee
It is necessary to conduct a more detailed investigation to check the stability.
Using the expression of $m_\sigma^2$ in (\ref{JGRG24}), one can investigate 
the sign of $m_\sigma^2$ in the region $R\sim R_0$.
Note that the expression (\ref{FRV3}) is not used because
the expression is only valid at the point $R=R_0$.
Thus, one gets
\be
\label{dS6}
m_\sigma^2 \sim - \frac{3n(n-1) R_0^2f(R_0)}{2f_0^2}
\left( R - R _0\right)^{n-2} \, .
\ee
Eq.~(\ref{dS6}) indicates that when $n$ is an even integer, the de Sitter
solution is stable if $f(R_0)<0$ but unstable if $f(R_0)>0$.
On the other hand, when $n$ is an odd integer, the de Sitter solution is
always unstable. Note, however, that when $f(R_0)<0$ $\left( f(R_0)>0 
\right)$, we find $m_\sigma^2 >0$ $\left(m_\sigma^2 < 0\right)$ if $R>R_0$ but
$m_\sigma^2 <0$ $\left(m_\sigma^2 > 0\right)$ if $R<R_0$.
Therefore, when $f(R_0)<0$, $R$ becomes smaller, but when
$f(R_0)>0$, $R$ becomes larger. The stability condition may be used to obtain 
realistic (unstable) de Sitter inflation for specific $F(R)$ gravity.

\subsubsection{FRW cosmology in $F(R)$ gravity with a Lagrange multiplier}

Here, we consider $F(R)$ gravity with a Lagrange multiplier field, following
ref.~\cite{Capozziello:2010uv}.
In usual $F(R)$ gravity, the scalar mode called scalaron appears,
which often affects Newton's law. In this section, we try to 
suppress the propagation of the scalaron by imposing the constraint under 
the Lagrange multiplier field. 
As a result, however, a propagating mode in the Lagrange multiplier field 
seems to appear, which may break Newton's law but in an easier way. 
The solution of this question may require additional modifications 
of the constraints.
Another purpose of this section is the reconstruction. In usual 
$F(R)$ gravity,
one needs to solve the complicated differential equation developed in the
previous sections to realize the reconstruction.
It is shown that the reconstruction can be realized more easily in the
model with the Lagrange multiplier field.

The starting action is given by
\be
\label{Lag1} S = \int d^4 x \sqrt{-g} \left\{ F_1(R) - \lambda
\left( \frac{1}{2} \partial_\mu R \partial^\mu R
+ F_2 (R) \right) \right\}\, .
\ee
Here, $\lambda$ is the Lagrange multiplier field, which
gives a constraint of 
\be
\label{Lag2}
\frac{1}{2} \partial_\mu R \partial^\mu R + F_2 (R) = 0\, .
\ee
On the other hand, by the variation over the metric
$g_{\mu\nu}$, the equation of motion follows:
\be
\label{Lag3}
0 = \frac{1}{2} g_{\mu\nu} F_1(R)
+ \frac{\lambda}{2} \partial_\mu R \partial_\nu R
+ \left( - R_{\mu\nu} + \nabla_\mu \nabla_\nu - g_{\mu\nu} \nabla^2 
\right) \left( F_1'(R) - \lambda F_2' (R) - \nabla^\mu \left(\lambda
\nabla_\mu R \right) \right)\, .
\ee
If the Ricci tensor is covariantly constant and the scalar curvature 
is a constant, 
\be
\label{Lag4}
R_{\mu\nu} = \frac{R_0}{4}g_{\mu\nu}\, , \quad
R=R_0 \, ,
\ee
Eqs.~(\ref{Lag2}) and (\ref{Lag3}) reduce to
\bea
\label{Lag5}
0 &=& F_2 (R_0)\, , \\
\label{Lag6} 0 &=& F_1(R_0) - \frac{1}{2} R_0 \left( F_1' (R_0) -
\lambda F_2' (R_0) \right)\, .
\eea
If Eq.~(\ref{Lag5}) has a solution, Eq.~(\ref{Lag6}) can be solved 
with respect to the Lagrange multiplier field:
\be
\label{Lag6b}
\lambda = \frac{ - F_1 (R_0) + R_0
F_1'(R_0)}{F_2'(R_0)}\, .
\ee
Then, if $R_0$ is positive, the above solution describes de Sitter space-time, 
which may correspond to dark energy or the inflationary epoch. 
The spatially flat FRW metric Eqs.~(\ref{Lag2}) and 
$(\mu,\nu) = (0,0)$-component of (\ref{Lag3}) have the following form:
\bea
\label{Lag8}
0 &=& - \frac{1}{2}{\dot R}^2 + F_2(R) \, ,\\
\label{Lag9}
0 &=& - \frac{1}{2} F_1(R) + 18 \lambda \left(\ddot H 
+ 4 H \dot H \right)^2 \nn
&& + \left\{3\left( \dot H 
+ H^2 \right) - 3H\frac{d}{dt} \right\} \left\{ F_1'(R) - \lambda
F_2' (R) + \left(\frac{d}{dt} + 3H \right) \left( \lambda
\frac{dR}{dt} \right) \right\} \, .
\eea
When $F_2(R)>0$, Eq.~(\ref{Lag8}) results in 
\be
\label{Lag10}
t = \int^R
\frac{dR}{\sqrt{2F_2(R)}}\, ,
\ee
which can be solved with respect
to $R$ as a function of $t$ $R=F_R(t)$. Because
\be
\label{Lag11}
R = 6 \frac{dH}{dt} + 12 H^2\, ,
\ee
one can find the behavior of $H=\frac{\dot a}{a}$ by solving the
differential equation
\be
\label{Lag12}
6 \frac{dH}{dt} + 12 H^2 =F_R(t) \, .
\ee
Using the obtained solution for $H=H(t)$ (and $R=F_R(t)$),
Eq.~(\ref{Lag9}) becomes a differential equation of the multiplier
field $\lambda$, and the behavior of $\lambda$:
$\lambda=\lambda(t)$ follows. (Note that FRW cosmology in scalar theory 
with a Lagrange multiplier constraint was studied in refs.~\cite{vikman}).

Conversely, when the behavior of $H(t)$ is known from the
observational data, one may reconstruct $F_2(R)$ to reproduce
$H(t)$ by using (\ref{Lag8}). $H(t)$ gives the
behavior of $R$ as $R=R(t)$, which can be solved with respect to
$t$ as $t=t(R)$. Using (\ref{Lag8}), the explicit form of $F_2(R)$
is found to be
\be
\label{Lag14}
F_2(R) = \frac{1}{2}\left.
\left(\frac{d R}{dt}\right)^2 \right|_{t=t(R)}\, .
\ee
Note that $F_1(R)$ can be an arbitrary function.
Thus, the cosmological reconstruction of the model can be performed more
easily than that in the usual $F(R)$ gravity.

As an explicit example, one may consider
\be
\label{Lag15} H(t) = \frac{h_0}{t}\, ,
\ee
where $h_0$ is a constant. Here, 
\be
\label{LAg16} R = \frac{- 6h_0 + 12
h_0^2}{t^2}\,\quad \mbox{or} \quad t = \sqrt{\frac{- 6h_0 + 12
h_0^2}{R}}\, .
\ee
Therefore, we find
\be
\label{Lag17}
\frac{dR}{dt} = - \frac{12\left( - h_0 + 2 h_0^2\right)}{t^3}
= - \frac{2 R^{\frac{3}{2}}}{\sqrt{6 \left( - h_0 + 2 h_0^2 \right) }}\, ,
\ee
which gives
\be
\label{Lag18}
F_2(R) = \frac{R^3}{12 \left( - h_0 + 2 h_0^2 \right) }\, .
\ee

Another example is given by
\be
\label{dS1Lag}
R = \frac{R_-}{2} \left( 1 - \tanh \omega t \right)
+ \frac{R_+}{2} \left( 1 + \tanh \omega t \right)\, .
\ee
Here, $R_\pm$ and $\omega$ are constants.
When $t\to \pm \infty$, $R\to \pm R_\pm$ and, therefore, the space-time
becomes asymptotically de Sitter. One may identify the epoch of 
$t\to -\infty$ as inflation and $t\to +\infty$ as late acceleration.
Because
\be
\label{dS2Lag}
\dot R = \frac{\left(R_- - R_+\right) \omega}{2\cosh^2 \omega t}
= \frac{\left(R_- - R_+\right) \omega}{2}
\left( 1 - \frac{\left( R_- + R_+ - 2R \right)^2}
{\left( R_- - R_+ \right)^2} \right)\, ,
\ee
from Eq.~(\ref{Lag14}), one gets
\be
\label{dS3Lag}
F_2(R) = \frac{\left(R_- - R_+\right)^2 \omega^2}{8}
\left( 1 - \frac{\left( R_- + R_+ - 2R \right)^2}
{\left( R_- - R_+ \right)^2}
\right)^2\, ,
\ee
Thus, the unification of early-time inflation with dark energy epoch is
possible also in Lagrange multiplier constrained modified gravity.

Thus, the universe evolution only depends on the constraint
equation (\ref{Lag8}) and not on $F_1(R)$. $F_1(R)$
can affect the correction to Newton's law. In convenient
$F_1(R)$ cosmology, all of the dynamics are defined by the form of
this function. With the constraint (\ref{Lag8}), $F_1(R)$ becomes
irrelevant. The cosmological dynamics are defined by the form of
$F_2(R)$. In convenient $F(R)$ gravity, the scalar propagating
mode appears, which often violates Newton's law. 
One can show that such a scalar does not propagate in the theory 
under investigation. Indeed, Eq.~(\ref{Lag3}) contains the second 
derivative of the multiplier field $\lambda$. We now need to solve 
the second-order differential
equation to find $\lambda$, which can indicate that $\lambda$ could
propagate and that a correction to Newton's law can be found. 
The magnitude of the correction depends on the choice of
$F_1(R)$ and/or $F_2(R)$.

To investigate the regime of Newton's law, $F_1(R)$ is chosen as
\be
\label{Lag19}
F_1(R) = \frac{R}{2\kappa^2}\, .
\ee
The matter presence is assumed. Eq.~(\ref{Lag3}) has the
following form:
\be
\label{Lag20} 0 = \frac{1}{2\kappa^2}
\left(\frac{1}{2} g_{\mu\nu} R - R_{\mu\nu} \right)
+ \frac{1}{2} T_{\mu\nu} - \left( - R_{\mu\nu}
+ \nabla_\mu \nabla_\nu - g_{\mu\nu} \nabla^2 \right)
\left( \lambda F_2' (R) - \nabla^\mu \left(\lambda \nabla_\mu R
\right) \right)\, .
\ee
For the solution with $\lambda=0$,
Eq.~(\ref{Lag3}) reduces to the Einstein equation,
\be
\label{Lag21} 0 = \frac{1}{\kappa^2} \left(\frac{1}{2} g_{\mu\nu}
R - R_{\mu\nu} \right) + T_{\mu\nu} \, .
\ee
Here, $T_{\mu\nu}$ is the matter energy-momentum tensor. Without matter
$T_{\mu\nu}=0$, the Schwarzschild space-time with $R=R_{\mu\nu}=0$ 
is a solution, which also satisfies the
constraint equation (\ref{Lag2}) if $F_2(0)=0$. In the case with
matter $T_{\mu\nu}\neq 0$, however, the Einstein equation
(\ref{Lag21}) gives
\be
\label{Lag22} R = - \kappa^2 T\, .
\ee
Here, $T$ is the trace of the energy-momentum tensor. The constraint
equation (\ref{Lag2}) is rewritten to
\be
\label{Lag23}
0 = \frac{\kappa^4}{2} \partial_\mu
T \partial^\mu T + F_2 \left( - \kappa^2 T \right)\, ,
\ee
which is not always satisfied. Thus, in the presence of matter, the
constraint equation (\ref{Lag2}) should be modified to be
\be
\label{Lag24}
0 = \frac{1}{2} \partial_\mu R \partial^\mu R + F_2
(R) - \frac{\kappa^4}{2} \partial_\mu T \partial^\mu T - F_2
\left( - \kappa^2 T \right)\, .
\ee
This indicates that the total constrained action with matter
should be, instead of (\ref{Lag1}),
\be
\label{Lag25}
S = \int d^4
x \sqrt{-g} \left[ \frac{R}{2\kappa^2} - \lambda \left\{
\frac{1}{2} \partial_\mu R \partial^\mu R 
+ F_2 (R) - \frac{\kappa^4}{2} \partial_\mu T \partial^\mu T - F_2
\left( - \kappa^2 T \right) \right\} + \mathcal{L}_\mathrm{matter}
\right]\, .
\ee
In this case, Newton's law can be easily reproduced.
Of course, some other form of the constraint
may also solve this problem.

We also note that the form of the constraint could be correct when
$F_1(R)$ is given by (\ref{Lag19}).
For general form of $F_1(R)$, the constraint should be changed.

Because $T$ vanishes in the vacuum, the constraint (\ref{Lag24}) 
reduces to
\be
\label{Lag24b}
0 = \frac{1}{2} \partial_\mu R \partial^\mu R
+ F_2 (R) - F_2 \left( 0 \right)\, .
\ee
If $F_2 \left( 0 \right)=0$, the constraint (\ref{Lag24}) gives
(\ref{Lag2}) and (\ref{Lag8}) and the cosmological evolution can 
be generated.
Note that $F_2 \left( R \right)$ (\ref{Lag18}) satisfies the condition
$F_2 \left( 0 \right)=0$ but $F_2 \left( R \right)$ in (\ref{dS3}) does 
not.
When the condition $F_2 \left( 0 \right)=0$ is satisfied, there are
two classes of solutions for the constraint (\ref{Lag8}). One is a trivial
solution $R=0$, and another corresponds to the non-trivial cosmological
evolution given by (\ref{Lag10}).
Near solar systems and galaxies, the solution could correspond
to the trivial one with $R=0$ to reproduce Newton's law, but at
large scales, the solution should correspond to (\ref{Lag10}), so that the
evolution of the universe can be generated.
It is not so trivial to prove or to deny that the two solutions are
connected in the intermediate region between the galaxy scales
and large-scale universe.
More careful (possibly numerical) analysis should be done in order 
to solve this problem.

\subsection{$F(\mathcal{G})$ gravity \label{IIID}}

The cosmological reconstruction scheme developed above can be applied to
an arbitrary modified gravity. As one more example of its application, we 
discuss the reconstruction in modified Gauss-Bonnet gravity.
The action (\ref{GB1b}) can be rewritten by using the auxiliary scalar 
field $\phi$:
\be
\label{ma32}
S=\int d^4 x \sqrt{-g}\left[
\frac{R}{2\kappa^2} - V(\phi) - \xi(\phi) \mathcal{G}\right]\, .
\ee

By the variation over the scalar field $\phi$, an algebraic equation is
obtained:
\be
\label{ma33}
0=V'(\phi) + \xi'(\phi) \mathcal{G} \, .
\ee
It can be solved, at least locally, with respect to $\phi$ as $\phi=
\phi(\mathcal{G})$.
Inserting the obtained expression of $\phi(\mathcal{G})$ into the
action (\ref{ma32}), we find the action (\ref{GB1b}):
\be
\label{33b}
S=\int d^4 x \sqrt{-g}\left[ \frac{R}{2\kappa^2} + f(\mathcal{G}) 
\right]\, .
\ee
Here, 
\be
\label{33c}
f(\mathcal{G}) \equiv - V\left(\phi(\mathcal{G})\right) +
\xi\left(\phi(\mathcal{G})\right)\mathcal{G}\, .
\ee

In the following, we use the results obtained in
refs.~\cite{Nojiri:2005jg,Nojiri:2005am,Cognola:2006eg} and consider
the reconstruction of scalar-$F(\mathcal{G})$ gravity.
By the variation over the metric in the FRW universe (\ref{JGRG14}),
one obtains the following equations:
\bea
\label{ma34}
0&=& - \frac{3}{\kappa^2}H^2 + V(\phi) 
+ 24 H^3 \frac{d \xi (\phi(t))}{dt} \, ,\nn
0&=& \frac{1}{\kappa^2}\left(2\dot H + 3 H^2 \right) - V(\phi) - 8H^2
\frac{d^2 \xi (\phi(t))}{dt^2} - 16H \dot H \frac{d\xi
(\phi(t))}{dt} - 16 H^3 \frac{d \xi (\phi(t))}{dt}\, .
\eea
Using (\ref{ma34}), it follows that 
\bea
\label{ma35}
&& \xi (\phi(t)) = \frac{1}{8}\int^t dt_1 \frac{a(t_1)}{H(t_1)^2} W(t_1)
\, ,\nn
&& V(\phi(t)) = \frac{3}{\kappa^2}H(t)^2 - 3a(t) H(t) W(t) \, ,\quad
W(t) \equiv \frac{2}{\kappa^2} \int^{t} \frac{dt_1}{a(t_1)} \dot H (t_1) 
\, .
\eea
Because there is no kinetic term of $\phi$, one can redefine $\phi$ properly 
and even identify $\phi$ with the cosmological time: $\phi=t$. 
In the same way as in the previous section, if we consider the following 
$V(\phi)$ and $\xi$ given in term of a single function $g$,
\bea
\label{ma36}
&& V(\phi) = \frac{3}{\kappa^2}g'\left(\phi
\right)^2 -  3g'\left(\phi\right)
\e^{g\left(\phi\right)} U(\phi) \, , \nn
&& \xi (\phi) = \frac{1}{8}\int^\phi d\phi_1
\frac{\e^{g\left(\phi_1\right)}}{g'(\phi_1)^2} U(\phi_1)\, ,\quad
U(\phi) \equiv \frac{2}{\kappa^2}\int^\phi d\phi_1 
\e^{-g\left(\phi_1\right)} g''\left(\phi_1\right) \, ,
\eea
the explicit solution follows:
\be
\label{ma37}
a=a_0\e^{g(t)}\ \left(H= g'(t)\right)\, .
\ee

In terms of e-folding $N$ 
\bea
\label{ma38}
&& \xi (\phi(N)) = \frac{1}{8}\int^N dN_1 \frac{\e^{N_1}}{H(N_1)^3} 
\tilde W(N_1) \, ,\nn
&& V(\phi(N)) = \frac{3}{\kappa^2}H(N)^2 - 3\e^N H(N) \tilde W(N) \, , \
\tilde W(N) \equiv \frac{2}{\kappa^2} \int^{N} \frac{dN_1}{\e^{N_1}} 
\dot H (N_1) \, .
\eea
Now we identify $\phi$ with the e-folding $N$: $\phi=N$ instead of 
$\phi=t$. Then, if $V(\phi)$ and $\xi$ are given by
\bea
\label{ma39}
&& V(\phi) = \frac{3}{\kappa^2}h\left(\phi\right)^2 - 3h\left(\phi\right)
\e^{\phi} \tilde U(\phi) \, , \nn
&& \xi (\phi) = \frac{1}{8}\int^\phi d\phi_1 
\frac{\e^{\phi_1} }{h(\phi_1)^3} \tilde U(\phi_1)\, ,\quad
U(\phi) \equiv \frac{2}{\kappa^2}\int^\phi d\phi_1 \e^{-\phi_1}
h'\left(\phi_1\right) \, ,
\eea
we obtain the solution $H=h(N)$. This general scheme may now be applied to
explicitly reconstruct the modified Gauss-Bonnet gravity by realizing any 
given FRW metric with a particular scale factor.

\subsection{Scalar-Einstein-Gauss-Bonnet gravity \label{IIIE}}

In the same way as above, let us
reconstruct string-inspired scalar-Einstein-Gauss-Bonnet gravity in \ref{IIC}. 
The reconstructed theory realizes the solution with the arbitrary
Hubble rate $H$.
Combining (\ref{ma24}) and (\ref{GBany5}) and deleting $V(\phi)$, we 
obtain
\bea
\label{ma25}
0 &=& \frac{2}{\kappa^2}\dot H 
+ {\dot\phi}^2 - 8H^2 \frac{d^2 \xi(\phi(t))}{dt^2}
 - 16 H\dot H \frac{d\xi(\phi(t))}{dt} + 8H^3 \frac{d\xi(\phi(t))}{dt} \nn
&=& \frac{2}{\kappa^2}\dot H + {\dot\phi}^2
 - 8a\frac{d}{dt}\left(\frac{H^2}{a}\frac{d\xi(\phi(t))}{dt}\right)\, .
\eea
The following expressions of $\xi\left(\phi(t)\right)$
and $V\left(\phi(t)\right)$ may be obtained:
\bea
\label{ma26}
\xi(\phi(t)) &=& \frac{1}{8}\int^t dt_1 \frac{a(t_1)}{H(t_1)^2} \int^{t_1}
\frac{dt_2}{a(t_2)}
\left(\frac{2}{\kappa^2}\dot H (t_2) + {\dot\phi(t_2)}^2 \right)\, , \nn
V(\phi(t)) &=& \frac{3}{\kappa^2}H(t)^2 - \frac{1}{2}{\dot\phi (t)}^2
 - 3a(t) H(t) \int^t \frac{dt_1}{a(t_1)}
\left(\frac{2}{\kappa^2}\dot H (t_1) + {\dot\phi(t_1)}^2 \right)\, .
\eea
Therefore, for the model where $V(\phi)$ and $\xi(\phi)$ are given
by adequate functions $g(t)$ and $f(\phi)$ as follows,
\bea
\label{ma27}
V(\phi) &=& \frac{3}{\kappa^2}g'\left(f(\phi)\right)^2 - 
\frac{1}{2f'(\phi)^2}
 - 3g'\left(f(\phi)\right) \e^{g\left(f(\phi)\right)} \int^\phi 
d\phi_1 f'(\phi_1 ) \e^{-g\left(f(\phi_1)\right)} 
\left(\frac{2}{\kappa^2}g''\left(f(\phi_1)\right)
+ \frac{1}{f'(\phi_1 )^2} \right)\, ,\nn
\xi(\phi) &=& \frac{1}{8}\int^\phi d\phi_1 \frac{f'(\phi_1)
\e^{g\left(f(\phi_1)\right)} }{g'(\phi_1)^2}
\int^{\phi_1} d\phi_2 f'(\phi_2) \e^{-g\left(f(\phi_2)\right)}
\left(\frac{2}{\kappa^2}g''\left(f(\phi_2)\right) + \frac{1}{f'(\phi_2)^2}
\right)\, ,
\eea
the solution of (\ref{ma24}) and (\ref{GBany5}) is given by
\be
\label{ma28}
\phi=f^{-1}(t)\quad \left(t=f(\phi)\right)\, ,\quad
a=a_0\e^{g(t)}\ \left(H= g'(t)\right)\, .
\ee

Similarly, the reconstruction of the model in terms of $N$, instead of the
cosmological time $t$ may be presented.
Using the same steps that led to Eq.(\ref{ma26}), we find
\bea
\label{ma29}
\xi(\phi(N)) &=& \frac{1}{8}\int^N dN_1 \frac{\e^{N_1}}{H(N_1)^3} 
\int^{N_1} \frac{dN_2}{\e^{N_2}}
\left(\frac{2}{\kappa^2}H' (N_2) + H(N_2) {\phi'(N_2)}^2 \right)\, , \nn
V(\phi(N)) &=& \frac{3}{\kappa^2}H(N)^2 - \frac{1}{2}H(N)^2 
\phi' (N)^2 - 3\e^N H(N) \int^N \frac{dN_1}{\e^{N_1}}
\left(\frac{2}{\kappa^2} H' (N_1) + H(N_1) \phi'(N_1)^2 \right)\, .
\eea
In terms of functions $h(\phi)$ and $\tilde f(\phi)$, if $V(\phi)$ and
$\xi(\phi)$ are given by
\bea
\label{ma30}
V(\phi) &=& \frac{3}{\kappa^2}h \left(\tilde f(\phi)\right)^2
 - \frac{h\left(\tilde f\left(\phi\right)\right)^2}{2\tilde f'(\phi)^2}
 - 3h\left(\tilde f(\phi)\right) \e^{\tilde f(\phi)} \int^\phi d\phi_1
\tilde f'( \phi_1 )
\e^{-\tilde f(\phi_1)} \left(\frac{2}{\kappa^2}h'\left(\tilde 
f(\phi_1)\right)
+ \frac{h'\left(\tilde f\left(\phi_1\right)\right)}{f'(\phi_1 )^2} 
\right)\, ,\nn
\xi(\phi) &=& \frac{1}{8}\int^\phi d\phi_1
\frac{\tilde f'(\phi_1) \e^{\tilde f(\phi_1)} }{h\left(\tilde
f\left(\phi_1\right)\right)^3}
\int^{\phi_1} d\phi_2 f'(\phi_2) \e^{-\tilde f(\phi_2)}
\left(\frac{2}{\kappa^2}h'\left(\tilde f(\phi_2)\right)
+ \frac{h\left(\tilde f(\phi_2)\right)}{\tilde f'(\phi_2)^2} \right)\, ,
\eea
one obtain the following solution:
\be
\label{ma31}
\phi=\tilde f^{-1}(N)\quad \left(N=\tilde f(\phi)\right)\, ,\quad H = h(N) 
\, .
\ee
This may now be applied to obtain an explicit reconstruction of scalar-Gauss-Bonnet
gravity, which admits any given FRW cosmology as a solution.

\subsection{$F(R)$ Ho\v{r}ava-Lifshitz gravity \label{IIIF}}

In this section, we consider the reconstruction of $F(R)$ 
Ho\v{r}ava-Lifshitz gravity. 
It is based on the results of ref.~\cite{Carloni:2010nx}.

To start, let us analyze the simple model $F(\tilde{R})=\tilde{R}$,
of which the cosmology was studied in 
\cite{Takahashi:2009wc,Kiritsis:2009sh,Brandenberger:2009yt,
Mukohyama:2009zs,Sotiriou:2009bx,Saridakis:2009bv,Minamitsuji:2009ii,
Calcagni:2009qw,Wang:2009rw,Park:2009zra,Nojiri:2009th,Jamil:2009sq,
Bogdanos:2009uj,Boehmer:2009yz,Bakas:2009ku,Calcagni:2009ar,Carloni:2009jc,
Gao:2009wn,Myung:2009if,Son:2010qh,Wang:2010mw,Ali:2010sv,Gong:2010xp}. 
In such a case, the FRW equations are similar to 
those of general relativity,
\be
H^2=\frac{\kappa^2}{3(3\lambda-1)}\rho_\mathrm{matter}\, , \quad
\dot{H}=-\frac{\kappa^2}{2(3\lambda-1)}
(\rho_\mathrm{matter}+p_\mathrm{matter})\, ,
\label{2.1}
\ee
where, for $\lambda\rightarrow 1$, the standard FRW equations are
recovered. Note that the constant $\mu$ is now irrelevant because, as
pointed out above, the term in front of $\mu$ in (\ref{HLF11}) becomes a
total derivative. For such a theory, one has to introduce a dark energy 
source as well as an inflaton field, to reproduce the cosmic and inflationary
accelerated epochs, respectively. It is also important to note that,
for this case, the coupling constant is restricted to be
$\lambda>1/3$; otherwise, Eqs.~(\ref{2.1}) become
inconsistent. It seems reasonable to think that, for the current epoch,
where $\tilde{R}$ has a small value, the IR limit of the theory is
satisfied by $\lambda \sim 1$, but for the inflationary epoch, when the
scalar curvature $\tilde{R}$ goes to infinity, $\lambda$ will take a
different value. It has been realized that for $\lambda=1/3$, the
theory develops an anisotropic Weyl invariance (see \cite{Horava:2009uw}),
and thus it takes a special role, although for the present model this
value is not allowed.

We now discuss some cosmological solutions of
$F(\tilde{R})$ Ho\v{r}ava-Lifshitz gravity. The first FRW equation,
given by (\ref{HLF16}) with $C=0$, can be rewritten as a function of
the number of e-foldings $N =\ln\frac{a}{a_0}$ instead of the usual
time $t$. Here, we extend the reconstruction formalism to 
the Ho\v{r}ava-Lifshitz $F(R)$ gravity.
Because $\frac{d}{dt}=H\frac{d}{dN }$ and 
$\frac{d^2}{dN ^2}=H^2\frac{d^2}{dN ^2}+H\frac{dH}{dN }\frac{d}{dN }$,
the first FRW equation (\ref{HLF16}) is rewritten as
\be
0=F(\tilde{R}) - 6\left[(1-3\lambda+3\mu) H^2
+ \mu HH'\right]\frac{dF(\tilde{R})}{d\tilde{R}}
+ 36\mu H^2\left[(1-3\lambda+6\mu)HH'+\mu H'^2+\mu H''H \right]
\frac{d^2F(\tilde{R})}{d^2\tilde{R}}-\kappa^2\rho_\mathrm{matter}\, ,
\label{D1}
\ee
where the primes denote derivatives with
respect to $N$. Thus, in this case, there is no restriction on the
values of $\lambda$ or $\mu$. By using the energy conservation
equation and assuming a perfect fluid with an EoS of 
$p_\mathrm{matter}=w_\mathrm{matter}\rho_\mathrm{matter}$, the energy 
density yields
\be
\rho_\mathrm{matter}=\rho_0 a^{-3(1+w_\mathrm{matter})}
=\rho_0 a_0^{-3(1+w_\mathrm{matter})}\e^{-3(1+w_\mathrm{matter})N }\, .
\label{D2}
\ee
As the Hubble parameter can be written as a function of the number of
e-foldings, $H=H(N)$, the scalar curvature in (\ref{HLF11}) takes the
form
\be
\tilde{R}=3(1-3\lambda+6\mu)H^2+6\mu HH'\, ,
\label{D3}
\ee
which can be solved with respect to $N $ as $N =N (\tilde{R})$,
and one obtains an expression (\ref{D1}) that gives an equation of
$F(\tilde{R})$ with the variable $\tilde{R}$. This can be
simplified by writing $G(N)=H^2$ instead of the Hubble parameter. In
such a case, the differential equation (\ref{D1}) yields
\be
0=F(\tilde{R})-6\left[(1-3\lambda+3\mu)G+\frac{\mu}{2}G'\right]
\frac{dF(\tilde{R})}{d\tilde{R}}
+18\mu\left[(1-3\lambda+6\mu)GG'+\mu GG''\right]
\frac{d^2F(\tilde{R})}{d^2\tilde{R}}-\kappa^2\rho_0a_0^{-3(1+w)}
\e^{-3(1+w)N}\, ,
\label{D4}
\ee
and the scalar curvature is now written as
$\tilde{R}=3(1-3\lambda+6\mu)G+3\mu G'$.
Thus, for a given cosmological solution $H^2=G(N)$, one can resolve
Eq.~(\ref{D4}), and the $F(\tilde{R})$ that reproduces such solution
is obtained.

As an example, we consider the Hubble parameter that reproduces the
$\Lambda$CDM epoch. It is expressed as
\be
H^2 =G(N)= H_0^2 + \frac{\kappa^2}{3}\rho_0 a^{-3} 
= H_0^2 + \frac{\kappa^2}{3}\rho_0 a_0^{-3} \e^{-3N } \, ,
\label{D5}
\ee
where $H_0$ and $\rho_0$ are constant. In general
relativity, the terms on the r.h.s. of Eq.~(\ref{D5}) correspond to an
effective cosmological constant $\Lambda=3H_0^2$ and to cold dark
matter with an EoS parameter of $w=0$. The corresponding
$F(\tilde{R})$ can be reconstructed by following the same steps as
described above. Using the expression for the scalar curvature
$\tilde{R}=3(1-3\lambda+6\mu)G+3\mu G'$,
the relation between $\tilde{R}$ and $N$ is obtained,
\be
\e^{-3N }=\frac{\tilde{R}
 -3(1-3\lambda+6\mu)H_0^2}{3k(1+3(\mu-\lambda))}\, ,
\label{D7}
\ee
where $k=\frac{\kappa^2}{3}\rho_0 a_0^{-3}$. Then, substituting
(\ref{D5}) and (\ref{D7}) into Eq.~(\ref{D4}), one obtains the
differential expression
\bea
0 &=& (1-3\lambda+3\mu)F(\tilde{R}) - 2\left(1 - 3\lambda 
+ \frac{3}{2}\mu\right)
\tilde{R} +9\mu(1-3\lambda)H^2_0\frac{dF(\tilde{R})}{d\tilde{R}} \nn
&& -6\mu(\tilde{R}-9\mu H^2_0)(\tilde{R}-3H^2_0(1-3\lambda+6\mu))
\frac{d^2F(\tilde{R})}{d^2\tilde{R}}-R-3(1-3\lambda+6\mu)H^2_0\, ,
\label{D8}
\eea
where, for simplicity, we have considered a pressureless fluid $w=0$
in Eq.~(\ref{D4}). Performing the change of
variable $x=\frac{\tilde{R}-9\mu H^2_0}{3H^2_0(1+3(\mu-\lambda))}$,
the homogeneous part of Eq.~(\ref{D8}) can be easily identified 
as a hypergeometric differential equation
\be
0=x(1-x)\frac{d^2 F}{dx^2} + \left(\gamma - \left(\alpha + \beta 
+ 1\right)x\right)\frac{dF}{dx} - \alpha \beta F\, ,
\label{D9}
\ee
with the set of parameters $(\alpha,\beta,\gamma)$ being given by
\be
\gamma=-\frac{1}{2}\, , \quad \alpha+\beta
= \frac{1-3\lambda-\frac{3}{2}\mu}{3\mu}\, , \quad
\alpha\beta=-\frac{1+3(\mu-\lambda)}{6\mu}\, .
\label{D10}
\ee
The complete solution of Eq.~(\ref{D9}) is a Gauss hypergeometric 
function plus a linear term and a cosmological constant from the particular
solution of Eq.~(\ref{D8}), namely
\be
F(\tilde{R}) = C_1 F(\alpha,\beta,\gamma;x) + C_2 x^{1-\gamma} 
F(\alpha - \gamma + 1, \beta - \gamma + 1,2-\gamma;x)
+\frac{1}{\kappa_1}\tilde{R}-2\Lambda\, ,
\label{D11}
\ee
where $C_1$ and $C_2$ are constants, $\kappa_1=3\lambda-1$ and
$\Lambda=-\frac{3H_0^2(1-3\lambda+9\mu)}{2(1-3\lambda+3\mu)}$.
Note that for the metric (\ref{D5}), the convenient $F(R)$
gravity was reconstructed and studied
in refs.~\cite{Nojiri:2009kx,Capozziello:2006dj,
Nojiri:2006gh,Nojiri:2006be,Dunsby:2010wg}.
In this case, the solution (\ref{D11}) behaves similarly
to the convenient $F(R)$ theory, except that now the parameters of
the theory depend on $(\lambda,\mu)$, which are
allowed to vary, as was noted above.
Thus, the cosmic evolution
described by the Hubble parameter (\ref{D5}) is reproduced by this class 
of theories.

One can also explore the solution (\ref{D5}) for a particular choice
of the parameters $\mu=\lambda-\frac{1}{3}$, which plays a special role 
in the cosmological expansion as is shown below. 
In this case, the scalar $\tilde{R}$ turns out to be a
constant, and Eq.~(\ref{D8}), in the presence of a pressureless fluid, 
has the solution
\be
F(\tilde{R})=\frac{1}{\kappa_1}\tilde{R}-2\Lambda\, , \quad \text{with}
\quad \Lambda=\frac{3}{2}(3\lambda-1)H_0^2\, .
\label{DD11}
\ee
Thus, for this constraint on the parameters, the only consistent solution
reduces to the Ho\v{r}ava linear theory with a cosmological constant.

As a further example, we consider the phantom (super)accelerating
expansion.
Such a system can be easily expressed in general relativity,
where the FRW equation reads as $H^2=\frac{\kappa^2}{3}\rho_\mathrm{ph}$. 
Here, the subscript ``$\mathrm{ph}$'' denotes the phantom nature of the fluid, 
which has an EoS given by $p_\mathrm{ph}=w_\mathrm{ph}\rho_\mathrm{ph}$
with $w_\mathrm{ph}<-1$. By using the energy conservation equation, 
the solution for the Hubble parameter turns out to be
\be
H(t)=\frac{H_0}{t_s-t}\, ,
\label{D12}
\ee
where $H_0=-1/3(1+w_\mathrm{ph})$, and $t_s$ is the Rip time.
As in the above example, one can rewrite the Hubble parameter 
as a function of the number of e-foldings; this yields
\be
G(N )=H^2(N )=H^2_0\e^{2N /H_0}\, .
\label{D13}
\ee
Then, by using the expression of the scalar curvature, the relation
between $\tilde{R}$ and $N $ is given by
\be
\e^{2N /H_0}=\frac{R}{H_0(AH_0+6\mu)}\, .
\label{D14}
\ee
By inserting (\ref{D13}) and (\ref{D14}) into the differential
equation (\ref{D4}), we get (without matter)
\be
\tilde{R}^2\frac{d^2F(\tilde{R})}{d\tilde{R}^2}+k_1\tilde{R}
\frac{dF(\tilde{R})}{d\tilde{R}}+k_0F(\tilde{R})=0\, ,
\label{D15}
\ee
where
\be
k_1=-\frac{(AH_0+6\mu)((AH_0+3\mu))}{6\mu(AH_0+12\mu)}\, , \quad
k_0=\frac{(AH_0+6\mu)^2}{12\mu(AH_0+12\mu)} \, .
\label{D16}
\ee
Eq.~(\ref{D15}) is an Euler equation, whose solution is well-known
\be
F(R)=C_1R^{m_+}+C_2R^{m_-}\, , \quad \text{where} \quad
m_{\pm}=\frac{1-k_1\pm\sqrt{(k_1-1)^2-4k_0}}{2}\, .
\label{D17}
\ee
Such a theory belongs to the class of models with positive and negative
powers of the curvature introduced in ref.~\cite{Nojiri:2003ft} as is 
discussed in the second chapter.
Thus, an $F(\tilde{R})$ Ho\v{r}ava-Lifshitz gravity has been reconstructed 
that reproduces the phantom dark epoch with no need of any exotic fluid.
In the same way, any given cosmology may be reconstructed, including the
unified inflation within dark energy era.

\subsection{Non-minimal Yang-Mills theory \label{IIIG}}

We now develop a reconstruction scheme of the Yang-Mills theory by 
starting with the following action
\be
\label{1}
S = \int dx^4 \sqrt{-g} \left[ \frac{R}{2 \kappa^2}
+ \mathcal{F}\left( F_{\mu \nu}^a F^{a \, \mu \nu} \right) \right]\, ,
\ee
where $F_{\mu \nu}^a = \partial_\mu A^a_\nu - \partial_\nu A^a_\mu 
+ f^{abc} A^b_\mu A^c_\nu$, and $\mathcal F$ may be assumed to be 
a continuously differentiable function.

Here, we review the reconstruction of the model(\ref{1}) based on
ref.~\cite{Elizalde:2010xq}.

We concentrate here on the $SU(2)$ case where $f^{abc}=\epsilon^{abc}$.
Because
\bea
&& \frac{\delta \left( F_{\mu \nu}^a F^{a \, \mu \nu} \right)}
{\delta A_\beta^h}
= - 4 \epsilon^{hbc} A^b_\gamma F^{c \, \gamma \beta}\, , \\
\label{3}
&& \frac{\delta \left( F_{\mu \nu}^a F^{a \, \mu \nu} \right)}
{\delta \left( \partial_\alpha A_\beta^h \right)}
= 4 F^{h \, \alpha \beta}\, ,
\eea
the equation of motion for the field potential $A_\mu^a$ turns into
\be
\label{4}
\partial_\nu \left[ \frac{\delta S}{\delta 
\left( \partial_\nu A^a_\mu \right)} 
\right] - \frac{\delta S}{\delta A^a_\mu} = 0 \, ,
\ee
and, from here, 
\be
\label{4B}
\partial_\nu \left[ \sqrt{-g} \, \mathcal F\,' \left( F^a_{\alpha \beta}
F^{a \, \alpha \beta} \right) \, F^{a \, \nu \mu} \right] + \sqrt{-g} \,
\mathcal F\,' \left( F^a_{\alpha \beta} F^{a \, \alpha \beta} \right) \,
\epsilon^{abc} A^b_\nu F^{c \, \nu \mu} = 0\, .
\ee
The variation of (\ref{1}) with respect to $g^{\mu \nu}$ yields 
the following equation of motion
\be
\label{5}
\frac{1}{2 \kappa^2} \left( R_{\mu \nu} - \frac{1}{2} g_{\mu \nu} R 
\right) - \frac{1}{2} g_{\mu \nu} \mathcal F \left( F^a_{\alpha \beta} 
F^{a \, \alpha \beta} \right)
+ 2 \mathcal F\,' \left( F^a_{\alpha \beta} F^{a \, \alpha \beta} \right) 
\, F^a_{\mu \rho} F^{a \, \rho}_\nu = 0 \, ,
\ee
where the following relation is used
$\frac{\delta \left( F^a_{\rho \sigma} F^{a \, \rho \sigma} 
\right)}{\delta g^{\mu \nu}}
= 2 \mathcal F\,' \left( F^a_{\alpha \beta} F^{a \, \alpha \beta} \right) \,
F^a_{\mu \gamma} F^{a \, \gamma}_\nu$.
Considering now a FRW universe (\ref{JGRG14}), and the following Ansatz 
for the gauge field,
\be
\label{gauge}
A_\mu^a = \left\{
\begin{array}{ll}
\bar \alpha e^{\lambda (t)} \delta^a_\mu, & \mu = i, \\
0, & \mu = 0,
\end{array}
\right.
\ee
the $\mu = 0$ component of (\ref{4}) becomes an identity, the $\mu = i$
component yields
\be
\label{6}
\partial_t \left[ a(t) \, \mathcal F\,' \left( F^a_{\alpha \beta} 
F^{a \, \alpha \beta} \right)\, \dot \lambda (t) \, \e^{\lambda (t)} \right]
+ \frac{2 \bar \alpha^2}{a(t)} \,
\mathcal F\,' \left( F^a_{\alpha \beta} F^{a \, \alpha \beta} \right) \, 
\e^{3 \lambda(t)} = 0\, ,
\ee
and the $(t,t)$ component of (\ref{5}) is
\be
\label{7}
\frac{3 H^2(t)}{2 \kappa^2} + \frac{1}{2} \mathcal F 
\left( F^a_{\alpha \beta} F^{a \, \alpha \beta} \right)
+ 2 \bar \alpha^2 \, \mathcal F\,' \left( F^a_{\alpha \beta} 
F^{a \, \alpha \beta} \right) \,
\frac{\dot \lambda^2 (t) \e^{2 \lambda(t)}}{a^2(t)} = 0\, ,
\ee
and the $(i,i)$ component of (\ref{5}) reduces to
\be
\label{8}
\frac{1}{2 \kappa^2} \left[ 2 \dot H(t) + 3 H^2(t) \right] - \frac{1}{2}
\mathcal F \left( F^a_{\alpha \beta} F^{a \, \alpha \beta}
\right) - 2 \bar \alpha^2 \frac{\e^{2 \lambda (t)}}{a^2 (t)}
\mathcal F\,' \left( F^a_{\alpha \beta} F^{a \, \alpha \beta} \right)
\left[ \dot \lambda^2 (t) - 2 \bar \alpha^2 
\frac{e^{2 \lambda (t)}}{a^2 (t)} \right] = 0\, .
\ee
Adding (\ref{7}) to (\ref{8}), one arrives at
\be
a^4 (t) \dot H (t) - 4 \kappa^2 \bar \alpha^4 \,
\mathcal F\,' \left( F^a_{\alpha \beta} F^{a \, \alpha \beta} \right) \, 
\e^{4 \lambda (t)} = 0\, ,
\ee
and then
\be
\label{10}
\mathcal F\,' \left( F^a_{\alpha \beta} F^{a \, \alpha \beta} \right)
= \frac{a^4 (t) \, \dot H (t)}{4 \kappa^2 \bar \alpha^4}
\e^{-4 \lambda (t)}\, .
\ee
Using (\ref{10}), Eq.~(\ref{6}) reduces to:
\be
\label{11}
2 \bar \alpha^2 \dot H (t) \, \e^{2 \lambda (t)} \, + \, \left[ 5 a(t) \, 
\dot a(t) \, \dot H(t) \, 
+ \, a^2(t) \, \ddot H(t) \right] \dot \lambda (t) \, - \, 
3 a^2(t) \dot H(t) \dot \lambda^2 (t) \, + \, a^2(t) \dot H(t) 
\ddot \lambda (t) = 0\, ,
\ee
which constitutes a differential equation for $\lambda (t)$. Thus, by 
using Eq.~(\ref{gauge}), once we have the function $\lambda (t)$, given 
by (\ref{11}), we can obtain the corresponding Yang-Mills theory that 
reproduces the selected cosmology. The Ansatz considered above actually
leads to a mathematical solution of the problem.

As an example, we consider the case of a power-law expansion: 
$a(t) = \left(\frac{t}{t_1} \right)^{h_1}$,
where $t_1$ and $h_1$ are constant, and assuming
$\lambda (t) = (h_1 - 1) \ln{\left( \frac{t}{t_1} \right)} + \lambda_1$,
where $\lambda_1$ is again a constant, Eq.~(\ref{11}) reduces to the 
following algebraic equation
\be
\label{12}
h_1 (h_1 - 1) + \bar \alpha^2 t_1^2 \, e^{2 \lambda_1} = 0\, ,
\ee
thus, 
\be
\label{13}
\lambda_1 = \frac{1}{2} \ln{\left( \frac{h_1 (1 - h_1)}
{\bar \alpha^2 t_1^2} \right)}\, ,
\ee
and
\be
\label{13b}
\lambda (t) = (h_1 - 1) \ln{\left( \frac{t}{t_1} \right)}
+ \frac{1}{2} \ln{\left( \frac{h_1 (1 - h_1)}{\bar \alpha^2 t_1^2} 
\right)}\, .
\ee
With the help of this reconstruction scheme, the function
$\lambda (t)$ given by (\ref{13b}) is obtained. 
Using (\ref{gauge}), one is able to  reproduce the cosmology
given by the power-law expansion: 
$a(t) = \left( \frac{t}{t_1} \right)^{h_1}$.

In summary, a cosmological reconstruction scheme was developed 
in the current chapter. It was presented in a very general form 
and was applied to a number of modified gravities introduced 
in the second chapter. Several examples of important cosmological 
epochs are considered. 
The cosmological reconstruction of modified gravities that realize 
such epochs as solutions was successfully made. 
Despite the fact that we limited ourselves to only a spatially flat FRW universe, 
this scheme may be easily used for the generation of other background 
solutions, i.e., an anisotropic universe, black holes, wormholes, etc.

\section{Finite-time future singularities in modified gravity \label{SecIV}}

This chapter is devoted to the study of finite-time future singularities,  
which often occur in the current dark energy models. 
It is demonstrated that this is a quite general phenomenon; 
the effective phantom and some effective quintessence dark energy models 
including fluid, scalar, and modified models may show such behavior 
in the finite future. 
We start from a simple dark fluid universe and demonstrate that coupling 
with dark matter cannot cure such future singularities. 
The introduction of a specific form of modified gravity may
naturally solve the problem of future singularities. The occurrence of
singularities in modified gravity is considered in detail, as is the 
possible avoidance of the singularities via the addition 
of extra gravitational terms.

\subsection{Future singularities in a dark, fluid universe: dark matter 
versus modified gravity \label{IVA}}

Let us reiterate several simple facts about coupled phantom dark energy.
First, we assume that the energy density $\rho_\mathrm{DE}$
and the pressure $p_\mathrm{DE}$ of the dark
matter satisfy the standard conservation law (\ref{CEm}).
Let the EoS parameter $w_\mathrm{DE}$:
$w_\mathrm{DE}=p_\mathrm{DE}/\rho_\mathrm{DE}<-1$. 
The phantom dark energy solution of the FRW equation
(\ref{JGRG11}) (after replacing $p_\mathrm{matter}$ and 
$\rho_\mathrm{matter}$ with $p_\mathrm{DE}$ and $\rho_\mathrm{DE}$, 
respectively) is given by
\be
\label{OIII}
H = \frac{\frac{2}{3\left( 1 + w_\mathrm{DE} \right)}}{t_s - t}\, ,
\ee
with the famous Big Rip singularity, at $t=t_s$.

Following ref.~\cite{Nojiri:2009pf}, the model
where phantom dark energy couples with dark matter is considered.
Then, the conservation law is modified as
\be
\label{I}
\dot \rho_\mathrm{DE} + 3H \left(1 + w_\mathrm{DE} \right)\rho_\mathrm{DE}
= - Q \rho_\mathrm{DE}\, ,
\quad \dot \rho_\mathrm{DM} + 3H \rho_\mathrm{DM} = Q \rho_\mathrm{DE}\, .
\ee
Here, $Q$ is assumed to be a constant. The first equation can be solved as
\be
\label{II}
\rho_\mathrm{DE} = \rho_\mathrm{DE(0)}
a^{-3\left(1 + w_\mathrm{DE} \right)} \e^{-Qt}\, .
\ee
Here, $\rho_\mathrm{DE(0)}$ is a constant of the integration. The
second equation (\ref{I}) gives
\be
\label{III}
\rho_\mathrm{DM} = Q a(t)^{-3} \int^t dt'
\rho_\mathrm{DE(0)} a^{- 3 w_\mathrm{DE}} \e^{-Qt}\, .
\ee
Thus, the second FRW equation is 
\be
\label{IVBe}
 - \frac{1}{\kappa^2}\left(2\dot H + 3H^2\right)
= p_\mathrm{DE} = w_\mathrm{DE} \rho_\mathrm{DE}
= w \rho_\mathrm{DE(0)} a^{-3\left(1 + w_\mathrm{DE} \right)} \e^{-Qt}\, .
\ee
An exact solution of (\ref{IVBe}) is the de Sitter space
\be
\label{V}
a(t) = a_0 \e^{- \frac{Q}{3\left(1 + w_\mathrm{DE} \right)} t}\quad
\left(H= - \frac{Q}{3\left(1 + w_\mathrm{DE} \right)} \right)\, ,
\ee
where $a_0$ is given by
\be
\label{VIB}
 - \frac{3}{\kappa^2} \left(\frac{Q}{3
\left(1 + w_\mathrm{DE}\right)}\right)^2
= w \rho_\mathrm{DE(0)} a_0^{-3\left(1 + w_\mathrm{DE} \right)}\, .
\ee
Note that because we are considering the phantom dark energy 
with $w_\mathrm{DE}<-1$, $H$ (\ref{V}) is positive and 
Eq.~(\ref{VIB}) has a real solution.

Thus, the coupling of the dark matter with the phantom dark energy may 
give the de Sitter solution instead of the Big Rip solution (\ref{OIII}).
This does not always mean that the Big Rip singularity can be avoided, 
but it opens a possibility that the universe could evolve into a de Sitter 
universe instead of the Big Rip.

Let us show that the above construction may help to solve 
the coincidence problem.
One may identify the Hubble rate $H$ with the present value 
of the Hubble rate
$ H = - \frac{Q}{3\left(1 + w_\mathrm{DE} \right)}
= H_0 \sim 10^{-33}\, \mathrm{eV}$.
Eq.~(\ref{II}) shows that the dark energy density is a constant
\be
\label{VII}
\rho_\mathrm{DE} = \rho_\mathrm{DE(0)}
a_0^{-3\left(1 + w_\mathrm{DE} \right)} \, .
\ee
Eq.~(\ref{III}) can be integrated as
\be
\label{VIIIB}
\rho_\mathrm{DM} = \rho_\mathrm{DM\, 0} a^{-3} - \left( 1
+ w_\mathrm{DE} \right) \rho_\mathrm{DE(0)}
a_0^{-3\left(1 + w_\mathrm{DE} \right)}\, .
\ee
Here, $\rho_\mathrm{DM\, 0}$ is a constant of integration, 
but the first FRW equation
\be
\label{IX}
\frac{3}{\kappa^2}H^2 = \rho_\mathrm{DM} + \rho_\mathrm{DE}\, ,
\ee
shows $\rho_\mathrm{DM\, 0}=0$ and, therefore, the dark matter density
$\rho_\mathrm{DM}$ is also constant:
\be
\label{XB}
\rho_\mathrm{DM} = - \left( 1+ w_\mathrm{DE} \right)
\rho_\mathrm{DE(0)} a_0^{-3\left(1 + w_\mathrm{DE} \right)}
= - \left( 1+ w_\mathrm{DE} \right) \rho_\mathrm{DE}\, .
\ee
Then, if the de Sitter solution (\ref{V}) is an attractor, by choosing
\be
\label{XIB}
 - \left( 1+ w_\mathrm{DE} \right) \sim \frac{1}{3}\, ,\quad \mbox{i.e.}
\quad w_\mathrm{DE} \sim - \frac{4}{3}\, ,
\ee
the coincidence problem could be solved. That is, even starting with a 
wide range of initial conditions, the solution approaches 
the de Sitter solution, in which the ratio of the dark energy and 
the dark matter is approximately $1/3$ and is almost
independent of the initial condition.

To check whether or not the de Sitter solution (\ref{V}) is an attractor, 
the following perturbation is considered
\be
\label{XIIB}
a = a_0 \e^{- \frac{Q}{3\left(1 + w_\mathrm{DE} \right)} t 
+ \Delta(t)}\, .
\ee
Here, $\Delta(t)$ is assumed to be small.
The second FRW equation (\ref{IVBe}) gives
\be
\label{XIIIB}
 - \frac{1}{\kappa^2}\left(2\ddot \Delta - \frac{2Q}
{\left(1 + w_\mathrm{DE} \right)}\dot\Delta\right)
= -3 \left(1 + w_\mathrm{DE} \right)w \rho_\mathrm{DE(0)}
a_0^{-3\left(1 + w_\mathrm{DE} \right)} \Delta
= \frac{3\left(1 + w_\mathrm{DE} \right)}{\kappa^2}
\left(\frac{Q}{3\left(1 + w_\mathrm{DE} \right)}\right)^2 \Delta \, ,
\ee
which is a very simple linear differential equation with a constant 
coefficient. Here, Eq.~(\ref{VIB}) is used.
Assuming $\Delta = \e^{\lambda t}$, it follows that 
\be
\label{XIIIBB}
0= \lambda^2 - \frac{Q}{1 + w_\mathrm{DE}} \lambda
+ 3\left(1 + w_\mathrm{DE} \right)
\left(\frac{Q}{3\left(1 + w_\mathrm{DE} \right)}\right)^2 \, ,
\ee
that is, 
\be
\label{XIVB}
\lambda = \lambda_\pm \equiv \frac{Q}{2\left(1 + w_\mathrm{DE} \right)} 
\pm \frac{1}{2} \left\{ \left( \frac{Q}{1 + w_\mathrm{DE}} 
\right)^2 - \frac{4 \left(1 + w_\mathrm{DE} \right)}{3}
\left(\frac{Q}{\left(1 + w_\mathrm{DE} \right)}
\right)^2 \right\}^\frac{1}{2}\, .
\ee
Then, $\lambda_- < 0$ but $\lambda_+>0$ and, therefore, de Sitter solution
(\ref{V}) is not stable. Because $\lambda_- < 0$, however, the solution is 
saddle point and, therefore, with the appropriate initial condition, there 
is a solution that approaches the saddle point de Sitter solution (\ref{V}).
Although we need to study the global structure of the space of the 
solutions, such an initial condition
could correspond to the direction toward $\lambda_-$.
Thus, generally speaking, the coupling of the dark energy
with dark matter does not prohibit the existence
of the (Big Rip) singular solution.
If the singularity corresponds to the stable solution, because 
the de Sitter space is not completely stable, the solution
will finally evolve into the singular solution.
Thus, the appropriate choice of initial conditions may help 
to realize the non-singular de Sitter cosmology, 
which also solves the coincidence problem.
At least, one may expect that the future singularity would occur 
at more later times compared with the case in which 
the dark matter does not couple with dark energy.

Let us consider the simple example of a perfect fluid 
with the following EoS \cite{Barrow:1990vx}:
\be
\label{EOSB1}
p = - \rho + A \rho^\alpha\, ,
\ee
with a constant $A$ and $\alpha$.
The EoS (\ref{EOSB1}) is a special case of (\ref{EOS1}).
In the spatially flat FRW space-time (\ref{JGRG14}),
the Hubble rate is found to be
\be
\label{EOSB6}
H = \left\{ \begin{array}{ll}
\frac{\frac{3}{2}A}{t}\, ,\quad & \mbox{when}\ \alpha = 1\, ,\quad A>0 \\
\frac{-\frac{3}{2}A}{t_s - t}\, ,\quad & \mbox{when}\ \alpha = 1\, ,
\quad A<0 \\
B \e^{- \frac{\sqrt{3}\kappa A t}{2}}\, ,\quad &
\mbox{when}\ \alpha = \frac{1}{2}\, ,\quad A<0 \\
\begin{array}{l}
C t^{1/(1-2\alpha)} \\
\mbox{or}\ \tilde C \left(t_s - t\right)^{1/(1-2\alpha)}
\end{array} & \mbox{when}\ \alpha \neq 1,\ \frac{1}{2}
\end{array} \right.
\ee
Here, $B$, $C$, and $\tilde C$ are positive constants.
Now, one can describe the future, finite-time singularities of the universe
filled with the above dark fluid for different choices of theory
parameters (see ref.~\cite{Nojiri:2009uu}).
When $\alpha<0$, a Type II or sudden future singularity occurs.
When $0<\alpha<1/2$ and $1/(1-2\alpha)$ is not an integer, 
a Type IV singularity occurs.
When $\alpha=0$, there is no singularity.
When $1/2<\alpha<1$ or $\alpha=1$ and $A<0$, a Type I or Big Rip-type 
singularity occurs (see also the last section of the second chapter).
When $\alpha>1$, a Type III singularity occurs 
(for the classification of all four types of future singularities, 
see \cite{Nojiri:2005sx}).

In the case of a Type II singularity, where $\alpha<0$, $H$ vanishes as
$H \sim \left(t_s - t\right)^{1/(1-2\alpha)}$ when $t\to t_s$
and, therefore, $\rho$ vanishes as follows from the FRW equation. 
Then, near the singularity, the EoS (\ref{EOS1}) is reduced to
\be
\label{EoSt1}
p \sim A\rho^{\alpha}\, .
\ee
On the other hand, in the case of a Type I singularity, where 
$1/2<\alpha<1$ or $\alpha=1$ and $A<0$, $H$ and, therefore, $\rho$ 
diverge when $t\to t_s$.
Then, the EoS (\ref{EOS1}) reduces to
\be
\label{EoSt2}
p \sim - \rho\ \mbox{or}\ p\sim -(1-A)\rho\, .
\ee
In the case of a Type III singularity, where $\alpha>1$,
$H$ and $\rho$ diverge when $t\to t_s$ and, therefore, the EoS
(\ref{EOS1}) reduces to
\be
\label{EoSt3}
p \sim A \rho^\alpha\, .
\ee

Let us now consider the dark energy (\ref{EOSB1}) coupled
with the dark matter as in (\ref{I}).
The conservation law is given by
\be
\label{AI}
\dot \rho_\mathrm{DE} + 3H A \rho_\mathrm{DE}^\alpha 
= - Q \rho_\mathrm{DE}\, ,
\quad \dot \rho_\mathrm{DM} + 3H \rho_\mathrm{DM} 
= Q \rho_\mathrm{DE}\, .
\ee
The solution of (\ref{AI}) is
\be
\label{AII}
\rho_\mathrm{DE}(t) = \e^{-Qt} \left( - 3A\left(1-\alpha\right)
\int^t dt' H(t') \e^{\left(1-\alpha\right) Qt'}
\right)^{\frac{1}{1-\alpha}}\, , \quad
\rho_\mathrm{DM} = Q a(t)^{-3} \int^t dt' a(t')^3 \rho_\mathrm{DE}(t')\, .
\ee
If $A<0$ (and $Q>0$), there is a de Sitter solution $H=H_0$ with
a constant $H_0$
\be
\label{AIII}
\frac{3}{\kappa^2} H_0^2 = \left( 1 + \frac{Q}{3H_0}\right)
\left( - \frac{3A H_0}{Q}\right)^{\frac{1}{1-\alpha}}\, ,
\ee
and $\rho_\mathrm{DE}$ and $\rho_\mathrm{DM}$ are constants
\be
\label{AIV}
\rho_\mathrm{DE}
= \left( - \frac{3A H_0}{Q}\right)^{\frac{1}{1-\alpha}}\, ,\quad
\rho_\mathrm{DM}
= \frac{Q}{3H_0}\left( - \frac{3A H_0}{Q}\right)^{\frac{1}{1-\alpha}}
\, .
\ee
This demonstrates that if $H_0\sim Q$, we find
$\rho_\mathrm{DM}/\rho_\mathrm{DE} \sim 1/3$, and the
coincidence problem may be solved. If $H_0= Q$, Eq.~(\ref{AIII}) 
determines the value of $A$:
\be
\label{AV}
A = - \frac{1}{3}\left( \frac{9}{4\kappa^2} H_0^2 \right)^{1-\alpha}\, .
\ee

One can now investigate the (in)stability of the de Sitter solution $H=H_0$
by putting $H=H_0 + \delta H$.
The perturbation of the energy density is
\be
\label{AVI}
\delta \rho_\mathrm{DE} = -3A \left( - \frac{3AH_0}{Q}\right)^{
\frac{\alpha}{1-\alpha}}
\e^{-\left(1 - \alpha\right) Qt}\int^t dt' \delta H (t')
\e^{\left(1 - \alpha\right) Qt'}\, .
\ee
The second FRW equation becomes
\be
\label{AVII}
 - \frac{1}{\kappa^2}\left(\delta \dot H + 6 H_0 \delta H\right)
\e^{\left(1 - \alpha\right) Qt'}
= 3A \left( 1 + \frac{\alpha Q}{3H_0} \right)
\left( - \frac{3AH_0}{Q}\right)^{ \frac{\alpha}{1-\alpha}}
\int^t dt' \delta H (t') \e^{\left(1 - \alpha\right) Qt'}\, .
\ee
By differentiating both sides of (\ref{AVII}), one gets
\bea
\label{AVIII}
0 &=& \delta \ddot H + \left\{ 6H_0 + \left(1 - \alpha\right) Q \right\} 
\delta \dot H + \left\{ 6H_0 \left(1 - \alpha\right) Q 
+ 3A \kappa^2 \left( 1 + \frac{\alpha Q}{3H_0} \right)
\left( - \frac{3AH_0}{Q}\right)^{ \frac{\alpha}{1-\alpha}} \right\} 
\delta H \nn
&=& \delta \ddot H + \left\{ 6H_0 + \left(1 - \alpha\right) Q \right\} 
\delta \dot H + 3H_0 \left( 1 - 2\alpha \right) Q \delta H \, .
\eea
In the second equality, Eq.~(\ref{AIII}) is used.
Assuming $\delta H \propto \e^{\lambda t}$, one gets
\be
\label{AVIIIA}
0 = \lambda^2 + \left\{ 6H_0 + \left(1 - \alpha\right) Q \right\} \lambda
+ 3H_0 \left( 1 - 2\alpha \right) Q \, ,
\ee
whose solution is given by
\be
\label{AIX}
\lambda = \lambda_\pm \equiv - \frac{6H_0 + \left(1 - \alpha\right) Q}{2}
\pm \frac{1}{2} \left\{ \left\{ 6H_0 
+ \left(1 - \alpha\right) Q \right\}^2 - 12 H_0
\left( 1 - 2\alpha \right) Q \right\}^{\frac{1}{2}}\, .
\ee
Because $H_0$, $Q>0$, if
\be
\label{AX}
\alpha < \frac{1}{2}\, ,
\ee
both $\lambda_\pm$ are real and negative if
\be
\label{AXI}
D= \left(6H_0 + \left(1 - \alpha\right) Q\right)^2 - 12H_0 
\left( 1 - 2\alpha \right) Q > 0\, ,
\ee
or complex, but the real part is negative if
\be
\label{AXII}
D= \left(6H_0 + \left(1 - \alpha\right) Q\right)^2 - 12H_0 \left( 1 - 
2\alpha
\right) Q < 0\, .
\ee
Thus, as long as $\alpha < \frac{1}{2}$ (\ref{AX}), the de Sitter 
solution is stable and, therefore, the singularity can be resolved.
On the other hand, if
\be
\label{AXIII}
6H_0 + \left(1 - \alpha\right) Q > 0\, ,\quad
3H_0 \left( 1 - 2\alpha \right) Q < 0\, ,
\ee
that is, 
\be
\label{AIXX}
1 + \frac{6H_0}{Q} > \alpha > \frac{1}{2}\, ,
\ee
we find $\lambda_+ >0$ and $\lambda_- <0$, as at the beginning of this 
section.
If
\be
\label{AXX}
6H_0 + \left(1 - \alpha\right) Q < 0\, ,\quad
3H_0 \left( 1 - 2\alpha \right) Q < 0\, ,
\ee
that is, 
\be
\label{AXXI}
\alpha > 1 + \frac{6H_0}{Q} \, ,
\ee
we find $\lambda_\pm >0$ and the de Sitter solution is completely unstable.

Thus, the Type II singularity, where $\alpha<0$, and the Type IV singularity, 
where $0<\alpha<1/2$ and $1/(1-2\alpha)$ is not an integer,
can be resolved by the coupling of the dark energy with the dark matter.
As mentioned above, even with a coupling of the dark matter with the 
dark energy, there could be a singular solution. The de Sitter solution 
with $\alpha < \frac{1}{2}$ is, however, at least a local minimum. 
Then, if the universe started with an appropriate initial condition, 
the universe would evolve into the de Sitter universe (asymptotically 
evolving into a de Sitter universe).
A Type I (Big Rip) singularity, where $1/2<\alpha<1$, and a Type III 
singularity, where $\alpha>1$, could not be removed by the coupling 
of the dark matter with the dark energy. Even if a solution goes
near the de Sitter point, the solution could evolve into the singular 
solution if the singular solution is stable.
Nevertheless, the ability to avoid some future singularities due to
the coupling of the dark energy with the dark matter looks quite promising.

Because the coupling of the dark energy with the dark matter does not always
remove the singularity, we now consider
what kind of fluid could cure the future singularity. In the case of the 
Big Rip singularity, for example, the energy density of the dark energy 
diverges like $\rho_\mathrm{DE} \sim 1/\left( t_s - t \right)^2 \sim R$ 
when $t\to t_s$.
Here, $R$ is the scalar curvature. Then, one becomes interested in a fluid 
with a positive pressure (and a positive energy density) that grows 
more rapidly than the dark energy pressure.
There is no such fluid with a constant EoS parameter. However, one can 
consider a pressure that is proportional to a power of the curvature; 
for example,
\be
\label{B1}
p_\mathrm{fluid} \propto R^{1 + \epsilon}\, ,
\ee
with $\epsilon>0$.
Then, the total EoS parameter becomes greater than $-1$ for a large curvature, 
and the Big Rip does not occur.

This kind of inhomogeneous effective fluid \cite{Nojiri:2005sr} could be
realized by quantum effects (for instance, taking account of conformal anomaly)
or by modified gravity as was explained in the second chapter.
As an example, we consider $F(R)$ gravity, where $F(R)=R + f(R)$ behaves 
as $f(R) \propto R^m$. 
In the case of $m=2$, the power-law solution, like the Big Rip 
singularity is prohibited.
It can happen that the (asymptotic) de Sitter solution occurs, instead of 
the power-law solution and, therefore, the singularity would be resolved.
Therefore, the introduction of the special form of $f(R)$ (precisely, 
the $R^2$ term) prevents the future singularity in the universal way. 
Note that as with other dark energies, the modified gravity itself 
(see next section) may lead to all four possible future singularities, 
which may again be resolved by the addition of the $R^2$ term
\cite{Abdalla:2004sw}. Thus, curing all future singularities of fluid dark
energy may be done via modifying the gravity. This seems to be quite a 
fundamental reason to start from modified gravity as a gravitational 
alternative for dark energy.

\subsection{Finite-time singularities in $F(R)$ gravity \label{IVB}}

Let us show that $F(R)$ gravity generates all four known types 
of finite-time singularities. Moreover, the resolution of
future singularities may be done again by adding the
$R^2$ term \cite{Abdalla:2004sw,Capozziello:2009hc}. 
This section is based on
results of refs.~\cite{Nojiri:2008fk,Bamba:2008ut}.

As the first example, we consider the case of the Big Rip singularity
(\ref{OIII}).
To find the $F(R)$ gravity that generates the Big Rip-type
singularity, it is convenient to use the reconstruction scheme 
developed in subsection \ref{recFR}.
Let $h_0 \equiv \frac{2}{3\left( 1 + w_\mathrm{DE} \right)}$.
In the case of (\ref{OIII}), the general solution of (\ref{PQR11}) 
when the matter contribution is neglected, $\rho_{i0}=0$, is given by
\be
\label{frlv11}
P(\phi) = P_+ \left(t_s - \phi\right)^{\alpha_+}
+ P_- \left(t_s - \phi\right)^{\alpha_-}\, ,\quad
\alpha_\pm \equiv \frac{- h_0 + 1 \pm \sqrt{h_0^2 - 10h_0 +1}}{2}\, ,
\ee
when $h_0 > 5 + 2\sqrt{6}$ or $h_0 < 5 - 2\sqrt{6}$ and
\be
\label{rlv12}
P(\phi) = \left(t_s - \phi \right)^{-(h_0 + 1)/2}
\left( \hat A \cos \left( \left(t_s - \phi \right)
\ln \frac{ - h_0^2 + 10 h_0 -1}{2}\right)
+ \hat B \sin \left( \left(t_s - \phi \right)
\ln \frac{ - h_0^2 + 10 h_0 -1}{2}\right) \right)\, ,
\ee
when $5 + 2\sqrt{6}> h_0 > 5 - 2\sqrt{6}$.
One finds the form of $F(R)$ when $R$ is large as
\be
\label{rlv13}
F(R) \propto R^{1 - \alpha_-/2}\ ,
\ee
for the $h_0 > 5 + 2\sqrt{6}$ or $h_0 < 5 - 2\sqrt{6}$ case and
\be
\label{rlv14}
F(R) \propto R^{\left(h_0 + 1\right)/4} \times \left(\mbox{oscillating
parts}\right)\ ,
\ee
for the $5 + 2\sqrt{6}> h_0 > 5 - 2\sqrt{6}$ case.

Let us investigate a more general singularity
\be
\label{frlv9}
H \sim h_0 \left(t_s - t\right)^{-\beta}\, .
\ee
Here, $h_0$ and $\beta$ are constants, $h_0$ is assumed to be positive, and
$t<t_s$, as it should be for an expanding universe.
Even for a non-integer $\beta<0$, some derivative of
$H$ and, therefore, the curvature become singular. Assume $\beta\neq 1$.
Furthermore, because $\beta=0$ corresponds to the de Sitter space, it is 
proposed that $\beta\neq 0$.
When $\beta>1$, the scalar curvature $R$ behaves as
\be
\label{rlv16B}
R \sim 12 H^2 \sim 12h_0^2 \left( t_s - t \right)^{-2\beta}\, .
\ee
On the other hand, when $\beta<1$, the scalar curvature $R$ behaves as
\be
\label{rlv16C}
R \sim 6\dot H \sim 6h_0 \beta \left( t_s - t \right)^{-\beta-1}\, .
\ee

If we write
\be
\label{rlv17}
P(\phi) = \e^{-h_0 \left( t_s - \phi \right)^{-\beta + 1}/2
\left(1 - \beta\right)} S(\phi)\, ,
\ee
Eq.~(\ref{PQR11}), without the matter contribution, has the following 
Schr\"{o}dinger equation like form:
\be
\label{rlv18}
0 = \frac{d^2 S(\phi)}{d\phi^2} + \left( \frac{5\beta h_0}{2}
\left( t_s - \phi \right)^{-\beta - 1} - \frac{h_0^2}{4}
\left( t_s - \phi \right)^{-2\beta}\right) S\, .
\ee
When $\phi=t \to t_s$, in the case that $\beta>1$, one finds
\be
\label{rlv19}
\left|\frac{5\beta h_0}{2}
\left( t_s - \phi \right)^{-\beta - 1}\right| \ll
\left|\frac{h_0^2}{4}\left( t_s - \phi \right)^{-2\beta}\right|\, .
\ee
On the other hand, in the case that $\beta<1$, it follows that 
\be
\label{rlv20}
\left|\frac{5\beta h_0}{2}
\left( t_s - \phi \right)^{-\beta - 1}\right| \gg
\left|\frac{h_0^2}{4}\left( t_s - \phi \right)^{-2\beta}\right|\, .
\ee
In either case, Eq.~(\ref{rlv18}) reduces to the following form:
\be
\label{rlv21}
0 = \frac{d^2 S(\phi)}{d\phi^2} - V_0 
\left( t_s - \phi \right)^{-\alpha} S\, ,
\ee
when $\phi=t \to t_s$. Here
\be
\label{rlv22}
\begin{array}{llll}
V_0 = - \frac{5\beta h_0}{2}\, , & \alpha = \beta + 1 & \mbox{when} &
\beta<1 \\
V_0 = \frac{h_0^2}{4}\, , & \alpha = 2\beta & \mbox{when} & \beta>1
\end{array}\, .
\ee
With a further redefinition
\be
\label{rlv23}
y \equiv \left( t_s - \phi \right)^{1 - \alpha/2}\, ,\quad
S = y^{\left(\alpha/4\right) \left(1 - \alpha/2\right)^{-1}} \varphi\, ,
\ee
Eq.~(\ref{rlv21}) has the following form:
\be
\label{rlv24}
0 = \frac{d^2\varphi}{dy^2} - \left\{ \left(\frac{\alpha^2}
{16} - \frac{\alpha}{4} \right) \frac{1}{y^2} 
+ \frac{4V_0}{\left(2 - \alpha\right)^2} \right\} \varphi\, .
\ee
Note that $y\to 0$ when $\phi\to t_s$ if $1 - \alpha/2 > 0$, but
$y\to \infty$ when $\phi\to t_s$ if $1 - \alpha/2 < 0$.
Then, if $1 - \alpha/2 > 0$, Eq.~(\ref{rlv24}) reduces to the following 
form
when $\phi\to t_s$:
\be
\label{rlv25}
0 = \frac{d^2\varphi}{dy^2} - \left(\frac{\alpha^2}{16} - \frac{\alpha}{4}
\right) \frac{1}{y^2} \varphi\, ,
\ee
whose general solution is given by
\be
\label{rlv26}
\varphi = A y^{\alpha/4 - 1} + B y^{-\alpha/4}\, .
\ee
Here, $A$ and $B$ are constants of integration.
On the other hand, if $1 - \alpha/2 < 0$, Eq.~(\ref{rlv24}) reduces to the
following form when $\phi\to t_s$:
\be
\label{rlv27}
0 = \frac{d^2\varphi}{dy^2} + \frac{4V_0}{\left(\alpha - 2\right)^2}
\varphi\, .
\ee
When $V_0 > 0$, the general solution of (\ref{rlv27}) is given by
\be
\label{rlv28}
y = \tilde A \cos \left(\omega y\right) 
+ \tilde B \sin \left(\omega y\right) \, ,\quad
\omega \equiv \frac{2\sqrt{V_0}}{\alpha - 2}\, .
\ee
Here, $\tilde A$ and $\tilde B$ are constants of integration.
On the other hand, if $V_0 < 0$, the general solution has the following 
form
\be
\label{rlv28b}
y = \hat A \e^{\hat\omega y} + \hat B \e^{-\hat\omega y} \, ,\quad
\hat\omega \equiv \frac{2\sqrt{-V_0}}{\alpha - 2}\, .
\ee

 From the above analysis, one may obtain the asymptotic solution for $P$
when $\phi\to t_s$.
\begin{itemize}
\item {\it $\beta>1$ case:} From (\ref{rlv22}), $\alpha=2\beta>2$ and,  
therefore, $1 - \alpha/2 = 1 - \beta < 0$, which corresponds 
to (\ref{rlv27}).
Because we also find $V_0>0$, the solution is given by (\ref{rlv28}).
Combining (\ref{frlv9}), (\ref{rlv17}), (\ref{rlv23}), and (\ref{rlv28}), 
we find the following asymptotic expression of $P(\phi)$:
\bea
\label{rlv29}
P(\phi) &\sim& \e^{\left(h_0/2\left(\beta - 1\right)\right)
\left(t_s - \phi\right)^{-\beta + 1}} \left(t_s - \phi\right)^{\beta/2}
\left(\tilde A \cos \left(\omega \left(t_s - \phi\right)^{-\beta 
+ 1}\right)
+ \tilde B \sin \left(\omega \left(t_s - \phi\right)^{-\beta + 1}\right)
\right)\, ,\nn
\omega &\equiv& \frac{h_0}{2\left(\beta - 1\right)}\, .
\eea
When $\phi\to t_s$, $P(\phi)$ tends to vanish very rapidly.
For large $R$, $F(R)$ looks like (at large $R$)
\be
\label{rlv29B}
F(R) \propto \e^{\left(h_0/2\left(\beta - 1\right)\right)
\left(\frac{R}{12h_0}\right)^{(\beta - 1)/2\beta}}
R^{-1/4}\times\left( \mbox{oscillating part} \right)\, .
\ee
\item {\it $1 > \beta > 0$ case:} From (\ref{rlv22}), it follows that 
$\alpha=\beta +1$ and, therefore, $1 - \alpha/2 = 1/2 - \beta/2 > 0$, which
corresponds to (\ref{rlv25}). Because
\be
\label{rlv30}
\frac{\alpha}{4} - 1 - \left( - \frac{\alpha}{4} \right)
= \beta>0\, ,
\ee
the second term in (\ref{rlv26}) dominates when $\phi\to t_s$ 
if $B\neq 0$.
Then, by combining (\ref{frlv9}), (\ref{rlv17}),
(\ref{rlv23}), and (\ref{rlv26}), the following asymptotic expression
of $P(\phi)$ is obtained:
\be
\label{rlv31}
P(\phi) \sim B \e^{-\left(h_0/2\left(1 - \beta\right)\right)
\left(t_s - \phi\right)^{1-\beta}}\left(t_s - \phi\right)^{\left(\beta 
+ 1\right)/8}\, .
\ee
Therefore,
\be
\label{rlv31C}
F(R) \sim \e^{-\left(h_0/2\left(1-\beta\right)\right)
\left( - 6\beta h_0 R \right)^{(\beta - 1)/(\beta + 1)} } R^{7/8}\, .
\ee
Eq.~(\ref{rlv16C}) shows that when $\phi=t \to t_s$, $R\to \infty$ 
in the case that $\beta > -1$, but $R\to 0$ in the case that $\beta < -1$.
\item {\it $\beta<0$ case:} As in {\it $1 > \beta > 0$}
case, one gets $\alpha=\beta +1$ and, therefore, $1 - \alpha/2 > 0$; 
however, because
\be
\label{rlv32}
\frac{\alpha}{4} - 1 - \left( - \frac{\alpha}{4} \right)
= \beta <0\, ,
\ee
the first term in (\ref{rlv26}) dominates when $\phi\to t_s$ 
if $A\neq 0$.
Thus, the asymptotic expression of $P(\phi)$ is as follows:
\be
\label{rlv33}
P(\phi) \sim A \e^{-\left(h_0/2\left(1 - \beta\right)\right)
\left(t_s - \phi\right)^{1-\beta}}
\left(t_s - \phi\right)^{- \left(\beta^2 - 6\beta + 1\right)/8}\, .
\ee
Then, $F(R)$ is given by
\be
\label{rlv35}
F(R) \sim
\left( -6h_0 \beta R \right)^{\left(\beta^2 + 2\beta 
+ 9\right)/8\left(\beta + 1\right)}
\e^{-\left(h_0/2\left(1 - \beta\right)\right)
\left( -6h_0 \beta R \right)^{\left(\beta-1\right)/\left(\beta 
+ 1\right)}} \, .
\ee
Note that $-6h_0 \beta R >0$ when $h_0, R>0$.
\end{itemize}

When $\beta>1$ in (\ref{frlv9}), $R$ behaves as in (\ref{rlv16B}), and
when $\beta<1$, the scalar curvature $R$ behaves as in (\ref{rlv16C}).
Conversely, when $R$ behaves as
\be
\label{R1}
R \sim 6\dot H \sim R_0 \left(\beta + 1\right) \left( t_s - t
\right)^{-\gamma}\, ,
\ee
if $\gamma>2$, which corresponds to $\beta = \gamma/2 >1$,
$H$ behaves as
\be
\label{R2}
H \sim \sqrt{\frac{R_0}{12}} \left( t_s - t \right)^{-\gamma/2}\, .
\ee
On the other hand, if $2>\gamma>1$, which corresponds to 
$1> \beta = \gamma -1 >0$, $H$ is given by
\be
\label{R3}
H \sim \frac{R_0}{6\left( \gamma - 1\right)} \left( t_s - t 
\right)^{-\gamma + 1}\, ,
\ee
and if $\gamma<1$, which corresponds to $\beta = \gamma -1 <0$, one
obtains
\be
\label{R4}
H \sim H_0 + \frac{R_0}{6\left( \gamma - 1\right)} \left( t_s - t
\right)^{-\gamma + 1}\, .
\ee
Here, $H_0$ is an arbitrary constant, which is chosen to vanish in
(\ref{frlv9}).
Then, because $H=\dot a(t)/a(t)$, if $\gamma>2$, we find
\be
\label{R5}
a(t) \propto \exp \left( \left(\frac{2}{\gamma} -1 \right)
\sqrt{\frac{R_0}{12}} \left( t_s - t \right)^{-\gamma/2 + 1}\right)\, ,
\ee
when $2>\gamma>1$, $a(t)$ behaves as
\be
\label{R6}
a(t) \propto \exp \left( \frac{R_0}{6\gamma\left( \gamma - 1\right)}
\left( t_s - t \right)^{-\gamma}\right)\, ,
\ee
and if $\gamma<1$,
\be
\label{R7}
a(t) \propto \exp \left( H_0 t + \frac{R_0}{6\gamma\left( \gamma - 
1\right)}
\left( t_s - t \right)^{-\gamma}\right)\, .
\ee
In any case, there appears a sudden future singularity at
$t=t_s$.

Because the second term in (\ref{R4}) is smaller than the first term, the
asymptotic solution of Eq.(\ref{PQR11}) looks as follows:
\be
\label{RR1}
P\sim P_0 \left( 1 + \frac{2h_0}{1-\beta}
\left(t_s - \phi\right)^{1-\beta}\right)\, ,
\ee
with a constant $P_0$, which gives
\be
\label{RR2}
F(R) \sim F_0 R + F_1 R^{2\beta/\left(\beta + 1\right)}\, .
\ee
Here, $F_0$ and $F_1$ are constants.
When $0>\beta>-1$, we find that $2\beta/\left(\beta + 1\right)<0$.
On the other hand, when $\beta<-1$, it follows 
$2\beta/\left(\beta + 1\right)>2$.
As is seen from (\ref{rlv13}), $F(R)$ generates the Big Rip singularity 
when $R$ is large.
Thus, even if $R$ is small, the $F(R)$ generates a singularity where 
higher derivatives of $H$ diverge.

Let us also investigate how the effective EoS parameter
$w_\mathrm{eff}$ for (\ref{R4}) behaves when $t\sim t_s$.
In Einstein gravity, the FRW equations are given by
\be
\label{R8}
\frac{3}{\kappa^2}H^2 = \rho\, ,\quad - \frac{1}{\kappa^2}
\left(2\dot H + 3H^2\right) = p\, .
\ee
As was discussed in section A of the second chapter, for 
$F(R)$ gravity, we may introduce the EoS parameter $w_\mathrm{eff}$ by
\be
\label{R9}
w_\mathrm{eff} = -1 - \frac{2\dot H}{3H^2}\, .
\ee
Then, if $\beta = \gamma/2 >1$, it follows that 
\be
\label{R10}
w_\mathrm{eff} \sim -1 - \frac{2\beta}{3h_0} 
\left( t_s - t \right)^{-1 + \beta} \to -1 \, ,
\ee
when $t\to t_s$.
If $1> \beta = \gamma -1 >0$, we find
\be
\label{R11}
w_\mathrm{eff} \sim - \frac{2\beta}{3h_0} 
\left( t_s - t \right)^{-1 + \beta} \to - \infty \, .
\ee
Finally, if $\beta = \gamma -1 <0$, one gets the expression (\ref{R11})
when $H_0=0$. When $H_0\neq 0$, on the other hand, we obtain
\be
\label{R12}
w_\mathrm{eff} \sim -1 - \frac{2\beta h_0}{3H_0^2}
\left( t_s - t \right)^{-1 - \beta}\, ,
\ee
when $H_0$ vanishes. Then, if $-1< \beta <0$, $w_\mathrm{eff}\to +\infty$
when $t\to t_s$. On the other hand, if $\beta<-1$, $w_\mathrm{eff}\to -1$.

Eq.~(\ref{R8}) also shows that, even for $F(R)$ gravity, we may define the
effective energy density $\rho_\mathrm{eff}$ and the effective 
pressure $p_\mathrm{eff}$ by
\be
\label{R12B}
\rho_\mathrm{eff}\equiv \frac{3}{\kappa^2}H^2 ,\quad
p_\mathrm{eff} \equiv - \frac{1}{\kappa^2}\left(2\dot H + 3H^2\right)\, .
\ee
We now assume that $H$ behaves as (\ref{frlv9}).
Then, if $\beta> 1$, when $t\to t_s$,
$a\sim \exp( h_0\left( t_s - t \right)^{1-\beta}/\left( \beta -1 \right) )
\to \infty$ and
$\rho_\mathrm{eff} ,\, |p_\mathrm{eff}| \to \infty$.
If $\beta=1$, we find
$a\sim \left(t_s - t\right)^{-h_0} \to \infty$ and
$\rho_\mathrm{eff} ,\, |p_\mathrm{eff}| \to \infty$.
If $0<\beta<1$, $a$ goes to a constant but $\rho ,\, |p| \to \infty$.
If $-1<\beta<0$, we find that $a$ and $\rho$ vanish 
but $|p_\mathrm{eff}| \to \infty$.
When $\beta<0$, instead of (\ref{frlv9}), as in (\ref{R3}), one may assume
\be
\label{R13}
H \sim H_0 + h_0 \left(t_s - t\right)^{-\beta}\, .
\ee
Thus, $\rho_\mathrm{eff}$ has a finite value $3H_0^2/\kappa^2$
in the limit $t\to t_s$ when $-1<\beta<0$.
If $\beta<-1$ but $\beta$ is not any integer, $a$ is finite and
$\rho_\mathrm{eff}$ and $p_\mathrm{eff}$
vanish if $H_0=0$ or both of $\rho_\mathrm{eff}$ and $p_\mathrm{eff}$ 
are finite if $H_0\neq 0$ but higher derivatives of $H$ diverge.

In the classification of the singularities in \cite{Nojiri:2005sx},
Type I corresponds to the $\beta>1$ or $\beta=1$ case, Type II to 
the $-1<\beta<0$ case, Type III to the $0<\beta<1$ case, and Type IV 
to the $\beta<-1$ case (where $\beta$ is not an integer).
Thus, we have constructed $F(R)$ gravity examples that show all four 
types of the above finite-time singularities. 
Viable $F(R)$ gravities (see second chapter) may also show (soft) 
future singular behavior. 
It is interesting to mention that future singularities manifest 
themselves in different presentations of
$F(R)$ gravity. For instance, if the scalar-tensor version shows 
the Big Rip-type singularity, then its $F(R)$ counterpart becomes 
a complex theory precisely at the rip time \cite{Briscese:2006xu}.

Near the future singularity at $t=t_s$, the curvature becomes large in
general. As a result, near the singularity, the quantum fields/quantum
gravity effects become very important again.
All classical considerations are not valid, and all speculations about 
a future cosmic doomsday cannot restrict the classical theory structure
because quantum effects can stop (or shift) the future singularity.
Moreover, the quantum corrections usually contain the curvature powers,
which become important near the singularity. Thus, any claim about the
appearance of the effective phantom/quintessence phase in modified gravity
that subsequently enters the future singularity is not justified without
quantum effects occurring near the singularity.
One may include the massless quantum effects by taking
into account the conformal anomaly contribution as 
a back-reaction near the singularity.
The conformal anomaly $T_A$ has the following well-known form:
\be
\label{OVII}
T_A=b\left(\mathcal{F} + \frac{2}{3}\Box R\right) + b' \mathcal{G} 
+ b''\Box R\, ,
\ee
where $\mathcal{F}$ is the square of the 4D Weyl tensor, and 
$\mathcal{G}$ is the Gauss-Bonnet invariant, which are given by
\be
\label{GF}
\mathcal{F} = \frac{1}{3}R^2 -2 R_{\mu\nu}R^{\mu\nu}+
R_{\mu\nu\rho\sigma}R^{\mu\nu\rho\sigma}\, , \quad
\mathcal{G}=R^2 -4 R_{\mu\nu}R^{\mu\nu}+
R_{\mu\nu\rho\sigma}R^{\mu\nu\rho\sigma}\, .
\ee
In general, with $N$ scalars, $N_{1/2}$ spinors, $N_1$ vector fields, $N_2$ 
($=0$ or $1$) gravitons and $N_\mathrm{HD}$ higher-derivative conformal 
scalars, $b$ and $b'$ are given by
\be
\label{bs}
b= \frac{N +6N_{1/2}+12N_1 + 611 N_2 - 8N_\mathrm{HD}}{120(4\pi)^2}
\, ,\quad
b'=- \frac{N+11N_{1/2}+62N_1 + 1411 N_2 -28 N_\mathrm{HD}}{360(4\pi)^2}\ .
\ee
As is seen, $b>0$ and $b'<0$ for the usual matter, except the 
higher-derivative conformal scalars.
Notice that $b''$ can be shifted by the finite renormalization of the
local counterterm $R^2$, so $b''$ can be an arbitrary coefficient.

By including the trace anomaly, the trace of (\ref{JGRG13}) is modified as
\be
\label{CA1}
R + 2f(R) - Rf'(R) - 3\Box f'(R) = - \frac{\kappa^2}{2} 
\left(T_\mathrm{matter}
+ T_A \right)\, .
\ee
For the FRW universe, we find
\be
\label{CA2}
\mathcal{F}=0\ ,\quad \mathcal{G}=24\left(\dot H H^2 + H^4\right)\ .
\ee
We now assume that $H$ behaves as (\ref{frlv9}) and neglect 
the contribution from matter by putting $T_\mathrm{matter}=0$.
Then, in the case that $f(R)$ behaves as $f(R)\sim -\frac{\alpha}{R^n}$,
Eq.~(\ref{CA1}) becomes
\be
\label{CA3}
R + 3\Box \left(\frac{\alpha n}{R^{n+1}}\right) 
= - \frac{\kappa^2}{2} T_A\, .
\ee
First, we consider the case that $2b/3 + b''=0$ and, therefore, 
$T_A=\mathcal{G}$.
The solution (\ref{R13}) is assumed.
If $-1<\beta<0$, $R$ behaves as $\left(t_s - t \right)^{-\beta - 1}$ and
$\mathcal{G}$ behaves as 
$\mathcal{G}\sim 24 \dot H H^2 \sim \left(t_s - t \right)^{-3\beta -1}$.
Because $-3\beta -1 > -\beta - 1$ when $\beta<0$, $T_A$ is negligible 
compared with $R$. Here, $\Box \left(\alpha n/R^{n+1}\right) \sim
\left(t_s - t \right)^{\left(\beta + 1\right)\left(n+1\right) -2}$.
Then, the curvature singularity appears in a finite time, and the
quantum correction does not prevent the singularity when $2b/3 + b''=0$.
One may, however, consider the case that $2b/3 + b''\neq 0$. 
In this case, $T_A$ behaves as
\be
\label{CA4}
T_A \sim \left(\frac{2}{3}b + b'' \right) \Box R \sim \left(t_s - t
\right)^{-\beta - 3}\, .
\ee
Because $R \sim \left(t_s - t \right)^{-\beta - 1}$ and
$\Box \left(\alpha n/R^{n+1}\right) \sim
\left(t_s - t \right)^{\left(\beta + 1\right)\left(n+1\right) -2}$,
the terms in the l.h.s of (\ref{CA3}) are always less singular than $T_A$
because
\be
\label{CA5}
-\beta - 1,\ \left(\beta + 1\right)\left(n+1\right) -2 > -\beta - 3\, .
\ee
This indicates that Eq.~(\ref{CA3}) does not allow the singular solution
and the curvature singularity does not appear.
Therefore, in the case that $2b/3 + b''\neq 0$, the quantum effects 
prevent the singularity appearance.

In the above analysis, the $\Box R$ term acts against the singularity.
The $\Box R$ term is generated by a local term $R^2$, which shows that if
one modifies $F(R)$ by adding the $R^2$ term as
\be
\label{R2A}
F(R) \to F(R) + \gamma R^2\, ,
\ee
with a constant $\gamma$, the curvature singularity cannot be 
generated. 
Thus, the addition of $R^2$ term which is relevant at the very early universe
gives the universal tool for resolving the finite-time future singularities.
Moreover, as was mentioned in section A of the second chapter, this term is 
the source of early-time inflation. Thus, removal of the singularity gives 
the natural prescription for the unification of the inflation with 
the dark energy epoch in frames of modified gravity.

\subsection{Future singularities in $F(\mathcal{G})$ gravity \label{IVC}}

Let us show that the occurrence of future singularities is a general 
phenomenon in modified gravity. In this section, which is based on 
ref.~\cite{Bamba:2009uf}, it is demonstrated that modified Gauss-Bonnet 
gravity can also drive the universe evolution to the singularity.

We start with the $F(\mathcal{G})$ models, in which the
finite-time future singularities can occur when the form of $H$
is taken as Eq.~(\ref{R13}).
To find such $F(\mathcal{G})$ gravities, the reconstruction method is used 
in a form which is different from that in subsection \ref{IIID}.

Using proper functions $P(t)$ and $Q(t)$ of a scalar field $t$
which is identified with the cosmic time,
the action in Eq.~(\ref{GB1b}) can be rewritten as
\be
S=\int d^{4}x \sqrt{-g} \left[
\frac{1}{2\kappa^2}\left( R+P(t)\mathcal{G} +Q(t) \right)
+{\mathcal{L}}_{\mathrm{matter}}\right]\,.
\label{onebis}
\ee
The variation with respect to $t$ yields
\be
\frac{d P(t)}{dt}\mathcal{G}+\frac{d Q(t)}{dt}=0\,,
\label{PQ}
\ee
from which one can find $t=t(\mathcal{G})$.
Substituting $t=t(\mathcal{G})$ into Eq.~(\ref{onebis}),
the action in terms of $F(\mathcal{G})$ follows
\be
F(\mathcal{G})=P(t)\mathcal{G}+Q(t)\,.
\label{F}
\ee
The scale factor is presented as
\be
a(t)= a_0 \exp \left( g(t) \right)\,,
\label{eq:Scale-Factor}
\ee
where $a_0$ is a constant and $g(t)$ is a proper function.
Using Eqs.~(\ref{GB7b}), (\ref{ma24}), (\ref{GBany5}),
(\ref{eq:Scale-Factor}),
the matter conservation law (\ref{CEm}) and then neglecting the
contribution from matter, we get the differential equation
\be
2 \frac{d}{d t} \left( \dot{g}^2(t) \frac{d P(t)}{d t} \right) -2
\dot{g}^{3}(t) \frac{d P(t)}{d t} + \ddot{g}(t)=0\,.
\label{P}
\ee
With the help of Eq.~(\ref{GB7b}), $Q(t)$ is given by
\be
Q(t)= -24 \dot{g}^{3}(t)\frac{d P(t)}{d t}-6\dot{g}^2(t)\,.
\label{Q}
\ee

First, the Big Rip singularity can be analyzed.
If $\beta=1$ in Eq.~(\ref{R13}) with $H_{0}=0$,
$H$ and $\mathcal{G}$ are given by
\be
H = \frac{h_0}{(t_s-t)}\, ,
\quad \label{H} %\\
\mathcal{G} = \frac{24h_0^3}{(t_s-t)^4}(1+h_0)\, .
%\label{G}
\ee
The most general solution of Eq.~(\ref{P}) is 
\be
P(t)=\frac{1}{4h_0(h_0-1)}(2t_s-t)t+c_{1}
\frac{(t_s-t)^{3-h_0}}{3-h_0}+c_{2}\, ,
\ee
where $c_{1}$ and $c_{2}$ are constants. From Eq.~($\ref{Q}$), one gets
\be
Q(t)=-\frac{6h_0^{2}}{(t_s-t)^2}-\dfrac{24 h_0^{3} \left[
\frac{(t_s-t)}{2h_0(h_0-1)}-c_{1}(t_s-t)^{2-h_0} \right]}{(t_s-t)^3}\,.
\ee
Furthermore, from Eq.~(\ref{PQ}), we obtain
\be
t= \left[ \dfrac{24(h_0^{3}+h_0^{4})}{\mathcal{G}} \right]^{1/4}+t_s\,.
\ee
Solving Eq.~(\ref{F}), we find the most general form of 
$F(\mathcal{G})$, which realizes the Big Rip singularity
\be
F(\mathcal{G})=\frac{\sqrt{6h_0^{3}(1+h_0)}}{h_0(1-h_0)}\sqrt{\mathcal{G}}
+c_{1}\mathcal{G}^{\frac{h_0+1}{4}}+c_{2}\mathcal{G}\,.
\label{Garr}
\ee
This is an exact solution in the case of Eq.~(\ref{H}).
In general, if, for large values of $\mathcal{G}$,
$F(\mathcal{G})\sim \alpha \mathcal{G}^{1/2}$, where
$\alpha (\neq 0)$ is a constant,
the Big Rip singularity occurs for any value of $h_0\neq 1$.
In the case of $h_0=1$, the solution of $\mathcal{G}(H, \dot{H}...)$
is zero for $F(\mathcal{G})=\alpha \mathcal{G}^{1/2}$.
Note that $\alpha \mathcal{G}^{(1+h_0)/4}$ is an invariant with
respect to the Big Rip solution.

In the case of $h_0=1$, it is possible to find
another exact solution for $P(t)$
\be
P(t)=\alpha(t_s-t)^{q}\ln \left[ \gamma(t_s-t)^{z} \right]\, ,
\ee
where $\gamma ( > 0 )$ is a positive constant and $q$ and $z$ are
constants.
Eq.~(\ref{P}) is satisfied for the case of Eq.~(\ref{H})
if $q=3-h_0=2$ (and, therefore, $h_0=1$) and 
$z\alpha=-1/4$. From Eq.~(\ref{Q}), we have
\be
Q(t)=-\frac{12}{(t_s-t)^{2}}\ln \left[ \gamma (t_s-t) \right]\,.
\label{schiappa}
\ee
The form of $F(\mathcal{G})$ is given by
\be
F(\mathcal{G})= \frac{\sqrt{3}}{2}\sqrt{\mathcal{G}} 
\ln(\gamma \mathcal{G})\,.
\ee
In general, if, for large values of $\mathcal{G}$,
$F(\mathcal{G})\sim \alpha\sqrt{\mathcal{G}} \ln(\gamma \mathcal{G})$
with $\alpha>0$ and $\gamma>0$, the Big Rip singularity occurs.
The same result is found for
$F(\mathcal{G})\sim \alpha\sqrt{\mathcal{G}} \ln(\gamma 
\mathcal{G}^{z}+G_{0})$
with $\alpha>0$, $\gamma>0$ and $z>0$, where $G_{0}$ is a constant.

Next, we investigate the other types of singularities.
If $\beta\neq 1$, Eq.~(\ref{R13}) with $H_{0}=0$
implies that the scale factor $a(t)$ behaves as
\be
a(t) = \exp \left[ \frac{h_0(t_s-t)^{1-\beta}}{\beta-1} \right]\, .
\ee
We consider the case in which $H$ and $\mathcal{G}$ are given by
\be
H = \frac{h_0}{(t_s-t)^{\beta}}\, ,
\quad \beta>1\,,
\label{HII} 
\quad \mathcal{G} \sim \frac{24h_0^4}{(t_s-t)^{4\beta}}\, .
\ee
A solution of Eq.~(\ref{P}) in the limit $t\rightarrow t_s$ is given by
\be
P(t)\simeq \frac{\alpha}{(t_s-t)^{z}}
\ee
with $z=-2\beta$ and $\alpha=-1/4h_0^{2}$. The form of $F(\mathcal{G})$ is
expressed as
\be
F(\mathcal{G})=-12\sqrt{\frac{\mathcal{G}}{24}}\, .
\label{primo}
\ee
Thus, if, for large values of
$\mathcal{G}$, $F(\mathcal{G})\sim -\alpha\sqrt{\mathcal{G}}$
with $\alpha>0$, a Type I singularity could appear.

When $\beta<1$, the forms of $H$ and $\mathcal{G}$ are given by
\be
H = \frac{h_0}{(t_s-t)^{\beta}}\, , \quad 0<\beta<1\,,
\label{Hsuper} 
\quad
\mathcal{G} \sim \frac{24h_0^3\beta}{(t_s-t)^{3\beta+1}}\,.
\ee
An asymptotic solution of Eq.~(\ref{P}) in the limit $t\rightarrow t_s$ is
given by
\be
P(t)\simeq \frac{\alpha}{(t_s-t)^{z}}
\ee
with $z=-(1+\beta)$ and $\alpha=1/2h_0(1+\beta)$. The form of 
$F(\mathcal{G})$ becomes
\be
F(\mathcal{G})=\frac{6h_0^{2}}{(\beta+1)}(3\beta+1)
\left(\frac{|\mathcal{G}|}{24h_0^{3}|\beta|}
\right)^{2\beta/(3\beta+1)}\,.
\label{secondo}
\ee
Thus, if for large values of $\mathcal{G}$, $F(\mathcal{G})$ has the form
\be
F(\mathcal{G})\sim \alpha |\mathcal{G}|^{\gamma}\,,
\quad \gamma = \frac{2\beta}{3\beta+1}\,,
\label{trentatre}
\ee
with $\alpha>0$ and $0<\gamma <1/2$,
we find that $0<\beta<1$ and a Type III singularity could emerge.

If, for $\mathcal{G}\rightarrow -\infty$,
$F(\mathcal{G})$ has the form in Eq.~(\ref{trentatre})
with $\alpha>0$ and $-\infty<\gamma<0$, we find $-1/3<\beta<0$ 
and a Type II (sudden) singularity could occur.
Moreover, if for $\mathcal{G}\rightarrow 0^{-}$,
$F(\mathcal{G})$ has the form in Eq.~(\ref{trentatre})
with $\alpha<0$ and $1<\gamma<\infty$,
it is obtained $-1<\beta<-1/3$ and a Type II singularity could occur.

If, for $\mathcal{G}\rightarrow 0^{-}$,
$F(\mathcal{G})$ has the form in Eq.~(\ref{trentatre})
with $\alpha>0$ and $2/3<\gamma<1$,
we obtain $-\infty<\beta<-1$ and a Type IV singularity could emerge.
It is also required that $\gamma\neq2n/(3n-1)$,
where $n$ is a natural number.

We can generate all of the possible Type II singularities
as shown above except in the case of $\beta=-1/3$,
i.e., $H=h_0/(t_s-t)^{1/3}$.
In this case, the form of $\mathcal{G}$ is:
\be
\mathcal{G}=24h_0^{3}\beta+24h_0^{4}(t_s-t)^{4/3}<0\, .
\ee
To find $t$ in terms of $\mathcal{G}$, one must consider the whole 
expression of $\mathcal{G}$ by also taking into account the
low term of $(t_s-t)$. It follows that 
\be
F(\mathcal{G})\simeq \frac{1}{4 \sqrt{6}h_0^{3}}
\mathcal{G}\left(\mathcal{G}+8h_0^{3}\right)^{1/2}
+\frac{2}{\sqrt{6}}(\mathcal{G}+8h_0^{3})^{1/2}\, .
\ee
As a consequence, the specific model
$F(\mathcal{G})= \sigma_1 \mathcal{G}(\mathcal{G}+c_{3})^{1/2} 
+ \sigma_2 (\mathcal{G}+c_{3})^{1/2}$,
where $\sigma_1$, $\sigma_2$ and $c_{3}$
are positive constants, can generate a Type II singularity.

Let us end this section by considering the following realistic model, 
again for $n>0$,
\be
f_{4}(\mathcal{G}) = \mathcal{G}^\alpha
\frac{a_{1}\mathcal{G}^{n}+b_{1}}{a_{2}\mathcal{G}^{n}+b_{2}}\, .
\label{QUARTO}
\ee
Because for large $\mathcal{G}$, one has $F_{4}(\mathcal{G}) \simeq
a_1/a_2\mathcal{G}^\alpha $ and for small $\mathcal{G}$, 
one has $f_{4}(\mathcal{G}) \simeq b_1/b_2\mathcal{G}^\alpha $,
we do not find any type of singularities for
\be
\frac{1}{2} < \alpha < \frac{2}{3}\, .
\label{sf}
\ee
Eq.~(\ref{QUARTO}) provides an example of a realistic model
free of all possible future singularities when Eq.~(\ref{sf}) is 
satisfied, independently of the coefficients. Moreover, this model suggests the
universal scenario for resolving future finite-time singularities.
Adding the above model to any singular dark energy results in a combined
non-singular model. Thus, unlike convenient dark energy, which may be
singular or not, (non-singular) modified gravity may suggest the universal 
recipe for resolving the future finite-time singularity. In this respect, 
modified gravity seems to be a more fundamental theory than convenient 
dark energies.

%%%%%%%%%%%%%%%%%%%

\subsection{Late-time dynamics in Ho\v{r}ava-Lifshitz gravity \label{IVD}}

Let us show that future singularities may also occur in 
Ho\v{r}ava-Lifshitz dark energy models \cite{Carloni:2010nx}.
Assume the behavior of the Hubble rate is the same as in (\ref{frlv9}).
Let us now consider some simple singularity examples in the power-law
Ho\v{r}ava-Lifshitz $F(R)$ gravity and compare them with the standard case.

In the case that 
\be
\label{F=RRm}
F(\tilde{R})=\tilde{R}+\chi \tilde{R}^m\,,
\ee
one finds the necessary conditions for the presence of the singularities:
\be
\label{singCond_beta}
\begin{array}{lc}
\mbox{Type I} &\beta>1 , \quad m>\frac{\beta-1}{2\beta} \\
\mbox{Type II} & m<0, \quad -1<\beta<0 \\
\mbox{Type III} & 0<\beta<1 , \quad m\neq 0 \\
\mbox{Type IV} & \beta \in \mathbb{Q}-\mathbb{Z},\quad \beta <-1 \,.
\end{array}
\ee
It is important to stress that the conditions (\ref{singCond_beta}) are 
only necessary for the existence of the singularity.
In the following, only the case that $\lambda>\frac{1}{3}$ is considered 
for simplicity. Then, one sees that a solution
\be
\label{SolVac}
a=a_0 t^{\gamma}\,, \qquad \rho=\rho_0 t^{-2}\,,\qquad 
\gamma=\frac{2 (m-1)
(2 m-1) \mu }{(2 m-1) \left(3 \lambda -1\right)-6 (m-1) \mu }\,,\quad
\rho_0 =0\,,
\ee
presents a singularity of Type I for
\be
\left\{
\begin{array}{cc}
0<m<\frac{1}{2}& \frac{-6 m \lambda +2 m+3 \lambda-1}{4 m^2-12 m+8}<\mu
<\frac{6 m \lambda -2 m-3 \lambda +1}{8 m^3-16 m^2+16 m-8}\,, \\
\frac{1}{2}<m<1& \frac{-6 m \lambda +2 m+3 \lambda-1}{4 m^2-12 m+8}<\mu
<\frac{6 m \lambda -2 m-3 \lambda +1}{6 m-6}\,,\\
1<m<2& \frac{6 m \lambda -2 m-3 \lambda +1}{6 m-6}<\mu <\frac{-6 m\lambda 
+2 m+3 \lambda -1}{4 m^2-12 m+8}\,,\\
m=2 & \mu >\frac{1}{6} (9 \lambda -3)\,,\\
m>2 & \mu <\frac{-6 m \lambda+2 m+3 \lambda -1}{4 m^2-12 m+8}\quad \mu
> \frac{6 m \lambda -2 m-3 \lambda +1}{6 m-6} \,,
\end{array}
\right.
\ee
a singularity of Type II for
\be
\left\{
\begin{array}{cc}
m<-1 & \frac{6 m \lambda -2 m-3 \lambda +1}{4 m^2-4}<\mu<0\,,\\
m=-1& \mu <0\,, \\ -1<m<0& \mu <0 \qquad 
\mu >\frac{6 m \lambda -2 m-3 \lambda +1}{4 m^2-4} 
\,,
\end{array}
\right.
\ee
and a singularity of Type III for
\be
\left\{
\begin{array}{cc}
m<0 & \mu <0\quad \mu >\frac{6 m \lambda -2 m-3 \lambda +1}{6 m-6}\, ,\\
0<m<\frac{1}{2}& 0<\mu <\frac{-6 m \lambda +2 m
+3\lambda -1}{4 m^2-12 m+8}\,, \\
\frac{1}{2}<m<1& \mu <\frac{-6 m \lambda +2 m
+3 \lambda -1}{4 m^2-12 m+8}\qquad \mu>0\,,\\
1<m<2& \mu <0\quad \mu >\frac{-6 m \lambda +2 m
+3 \lambda -1}{4 m^2-12 m+8}\,,\\
m=2 & \mu <0\\
m>2 & \frac{-6 m \lambda +2 m+3 \lambda -1}{4 m^2-12 m+8}<\mu <0 \,.
\end{array}
\right.
\ee
This is very different from the standard case, where we have a singularity
of Type I only for $m>2$, and a Type III for $\frac{1}{2}<m<1$ and $m<0$.

For the solution
\bea
\label{SolMat}
&& a=a_0 t^{\gamma}\,, \quad \rho=\rho_0 t^{-2}\,,\quad
\gamma=\frac{2 m}{3 (1 + w)}\,,\nn
&&\rho_0 =\frac{\chi \left[3\mu(m-1) (2 m (w+2)-w-1)-m (2 m-1) 
\left(3 \lambda -1\right)\right]}{\kappa ^2 \left(m 
\left[3 \lambda -6 \mu -1\right)+3 (w+1) \mu \right]} \nn
&& \qquad \qquad \times \left(\frac{4 m^2 \left(-3 \lambda 
+6 \mu +1\right)-12 m (w+1)\mu }{(w+1)^2}\right)^m\,,
\eea
we have a singularity of Type III when
\be
\label{M2}
\left\{
\begin{array}{ccc}
 -3<m\leq -\frac{3}{2}& \frac{1}{3} (-2 m-3)<w\leq 1\,,\\
 -\frac{3}{2}<m<\frac{1}{2}\left(1-\sqrt{13}\right)& 0\leq w\leq 1\,.
\end{array}
\right.
\ee
For the solution
\be
\label{Sol HL}
a=a_0 t^{\gamma}\,, \qquad \rho=\rho_0 t^{-2}\, ,
\quad \gamma=\frac{2 \mu}{-3 \lambda +6 \mu +1}\,,\qquad \rho_0 =0\, ,
\ee
which exists only in the Ho\v{r}ava-Lifshitz version of
$F(R)$ gravity, we have instead a singularity of Type I for
\be
\label{M3}
\left\{
\begin{array}{ccc}
\lambda <\frac{1}{3}&\frac{1}{6} (3 \lambda -1)<\mu <\frac{1}{8}
(3 \lambda -1)&m>1\,,\\
\lambda >\frac{1}{3}&\frac{1}{8} (3 \lambda -1)<\mu <\frac{1}{6}
(3 \lambda -1)& m>1\,,
\end{array}
\right.
\ee
and of Type III for
\be
\label{M4}
\left\{
\begin{array}{ccc}
\lambda <\frac{1}{3}& \frac{1}{8} (3 \lambda -1)<\mu <0 & m>1\,,\\
\lambda >\frac{1}{3} & 0<\mu <\frac{1}{8} (3 \lambda -1)& m>1\,.
\end{array}
\right.
\ee
The results above clearly show that the presence of singularities is 
deeply altered in the Ho\v{r}ava-Lifshitz version of $F(R)$ gravity. 
In particular, it seems that the additional parameters make it much 
easier to realize the singularities.
The intervals we have presented above for the parameters can then be
interpreted as constraints on this type of Ho\v{r}ava-Lifshitz $F(R)$ 
gravity. Thus, we have demonstrated that modified Ho\v{r}ava-Lifshitz 
gravity can be incorporated into the
phantom like or quintessence like accelerating cosmologies, which might
lead to singularities of Type I, II, or III. Of course, as in the case of
conventional theory one can resolve these future singularities by adding 
a higher-derivative gravitational term, which is relevant at very early 
times. However, the structure of this term is different from the case of 
convenient modified gravity. For example, the addition of the $R^2$ term does 
not necessarily remove all types of singularities in the theory under discussion.

In summary, in this chapter, we demonstrated that an effective phantom as well 
as some quintessence dark energy models may drive the universe evolution 
to future finite-time singularities. This is a universal property, 
regardless of whether dark energy is a fluid, field model or modified gravity. 
To resolve such future singularities, one has to add extra gravitational 
terms to the theory action. In other words, resolving the singularities invites 
the modification of gravity, even if the
initial dark energy was not modified gravity. The additional gravitational 
term, which is usually relevant at very early times, also helps in 
the realization of the inflationary phase. In this way, the
additional support for unification of early-time inflation with late-time
acceleration in modified gravity follows.

\section{Discussion \label{SecV}}

In summary, we reviewed a number of popular modified gravities including
$F(R)$ theory, modified Gauss-Bonnet and scalar-Gauss-Bonnet gravities,
non-minimal models, non-local gravity, modified $F(R)$ Ho\v{r}ava-Lifshitz
theory and (power-counting renormalizable) covariant gravity. General
properties and different representations of such theories were considered.
Their spatially flat FRW equations, which often have higher-derivative terms 
and/or non-canonical and non-standard couplings were derived. The
accelerating cosmological solutions including the de Sitter universe 
and an attempt to obtain the unified description of inflation 
with dark energy were proposed.
It was demonstrated that the qualitative possibility of such a unification 
is a very natural property for this large class of alternative gravities. 
This is also the case for the theories with broken Lorentz symmetry 
like Ho\v{r}ava-Lifshitz gravity.

To describe the background evolution of modified gravity, we developed 
the cosmological reconstruction method (inverse problem) in terms of
cosmological time or e-folding. Using the freedom in the choice of scalar
potentials and of the modified term function, which depends on geometrical
invariants, such as curvature and the Gauss-Bonnet term, we arrived to master
differential equations whose solution solves the problem. It was demonstrated
that eventually, any given cosmology may be reconstructed in this way. We
explicitly considered the reconstruction in scalar-tensor theory, Brans-Dicke
gravity, the k-essence model, $F(R)$ theory and the Lagrange multiplier $F(R)$
theory, modified Gauss-Bonnet and scalar-Gauss-Bonnet gravities,
Ho\v{r}ava-Lifshitz gravity and non-minimal Yang-Mills theory.
As an example of the number of universe evolution eras including $\Lambda$CDM
one, the effective quintessence or phantom acceleration, the unified
early-time inflation with late-time acceleration was shown to be the
background solution of the above theories. It should be stressed that 
even if specific modified gravity cannot capture the entire sequence 
of the universe evolution (namely, the inflation, radiation and 
matter dominance, dark energy era) with good accuracy, it may be 
additionally reconstructed so that it can become a viable theory. 
Thus, such a partial reconstruction is an extremely powerful tool that 
may help the theory to pass the known observationally related tests.
Of course, the price for that is some extra modifications of the action.
Fortunately, such extra modifications may be attributed to the gravitational
sector relevant at the very early universe.

Special attention was paid to late-time dynamics of the effective
quintessence/phantom dark energy of arbitrary nature: fluid, particle model
or modified gravity. It is known that all or some of the energy conditions for
such a (phantom or quintessence) dark energy universe are violated. As a
result, the universe may end up with its evolution in one of four known types of
finite-time singularities. The appearance of all four types
of future singularities in coupled fluid dark energy, $F(R)$ theory, modified
Gauss-Bonnet gravity and modified $F(R)$ Ho\v{r}ava-Lifshitz gravity 
was demonstrated. The universal prescription for resolving such singularities 
that may lead to bad phenomenological consequences was proposed.
It consists of the addition of an extra higher-derivative gravitational term
relevant in the very early universe and helping in the realization of the
inflation. In the case of $F(R)$ theory or standard fluid dark energy, this term
is proportional to $R^2$ at large curvatures.

The advantage of the approach proposed in this review is its very general
character. Even if one of the particular gravity models under discussion is 
outside of mainstream research, its properties and background solutions will
remain mathematically and physically correct. Moreover, the developed
reconstruction scheme proposes the way to change the properties of any
particular theory in the desirable way. Thus, we did not
discuss the cosmological perturbation theory or structure formation in
modified gravity. Indeed, the standard way to study the cosmological
perturbations in modified gravity is based on the close analogy with general
relativity. As a result, higher-order differential equations are
approximated by second-order equations.
Clearly, this is not a consistent approach. To analyze the
cosmological perturbations one has to develop 
a powerful numerical technique that should be applied to
higher-derivative differential equations.
We also did not discuss the local tests of the gravity 
(especially, post-Newtonian regime tests) in detail. 
This is due to existence of a number of
observationally related reviews \cite{review} from a point of view.  
From another point of view, only the small curvature regime is relevant 
for the purposes of the local tests. 
Clearly, all modified gravities under consideration may be parametrized by
some effective theory with small deviations from general relativity.
In other words, a very large class of models with qualitatively different
properties may quite easily pass the local tests.

The next important step in the investigation of modified gravity should be
related to a detailed proposal on the description of the sequence of the 
entire evolution of the universe: inflation, radiation/matter dominance, 
and dark energy. As we saw in
this work, the equations of motion for such theories are extremely
complicated, and including all of the universe matter content makes 
them even more involved. 
Presumably, such a study can be done only numerically.
This is an extremely hard task that should give the final answer to the
question: is the evolution of the observable universe  governed by modified
gravity?

\section*{Acknowledgments \label{Ack}}

We are grateful to all of our friends and collaborators, especially,
those listed here: 
M.~C.~B.~Abdalla, G.~Allemandi, K.~Bamba, A.~Borowiec, K.~A.~Bronnikov,
I.~Brevik, F.~Briscese, S.~Carloni, S.~Capozziello,
V. Cardone, M.~Chaichian, G.~Cognola, A.~Dev, P.~K.~S.~Dunsby,
E.~Elizalde, V.~Faraoni,
M.~Francaviglia, C.~Q.~Geng, O.~G.~Gorbunova, R.~Goswami, Y.~Ito,
D.~Jain, S.~Jhingan,
M.~De Laurentis, J.~E.~Lidsey, J.~Matsumoto, M.~Oksanen,
D.~Saez-Gomez, M.~Sami, M.~Sasaki,
L.~Sebastiani, I.~Thongkool, A.~Toporensky,
A.~Troisi, P.~V.~Tretyakov, A.~Tureanu,
L.~Vanzo, P.~Wang, H.~Yoshioka,
S.~Zerbini, and Y.~l.~Zhang
for stimulating discussions of different aspects of modified gravity.
This research has been supported in part
by the MEC (Spain) project FIS2006-02842 and AGAUR(Catalonia) 2009SGR-994 
(SDO), by the Global COE Program of Nagoya University (G07)
provided by the Ministry of Education, Culture, Sports, Science \& 
Technology
and by the JSPS Grant-in-Aid for Scientific Research (S) \# 22224003 (SN).


\begin{thebibliography}{99}

%\cite{Caldwell:2003vq}
\bibitem{Caldwell:2003vq}
R.~R.~Caldwell, M.~Kamionkowski and N.~N.~Weinberg,
%``Phantom Energy and Cosmic Doomsday,''
Phys.\ Rev.\ Lett.\ {\bf 91}, 071301 (2003)
[arXiv:astro-ph/0302506].
%%CITATION = PRLTA,91,071301;%%

%\cite{Nojiri:2003ft}
\bibitem{Nojiri:2003ft}
S.~Nojiri and S.~D.~Odintsov,
%``Modified gravity with negative and positive powers of the
%curvature:
%Unification of the inflation and of the cosmic acceleration,''
Phys.\ Rev.\ D {\bf 68}, 123512 (2003)
[arXiv:hep-th/0307288].
%%CITATION = PHRVA,D68,123512;%%

%\cite{Nojiri:2006ri}
\bibitem{Nojiri:2006ri}
S.~Nojiri and S.~D.~Odintsov,
%``Introduction to modified gravity and gravitational alternative
%for
%dark
%energy,''
eConf {\bf C0602061}, 06 (2006)
[Int.\ J.\ Geom.\ Meth.\ Mod.\ Phys.\ {\bf 4}, 115 (2007)]
[arXiv:hep-th/0601213]; arXiv:0807.0685.
%%CITATION = 00436,4,115;%%

\bibitem{review}
%\cite{Capozziello:2009nq}
%\bibitem{Capozziello:2009nq}
S.~Capozziello, M.~De Laurentis and V.~Faraoni,
%``A bird's eye view of f(R)-gravity,''
arXiv:0909.4672 [gr-qc]; \\
%%CITATION = ARXIV:0909.4672;%%
%\cite{Capozziello:2007ec}
%\bibitem{Capozziello:2007ec}
S.~Capozziello and M.~Francaviglia,
%``Extended Theories of Gravity and their Cosmological and Astrophysical
%Applications,''
Gen.\ Rel.\ Grav.\ {\bf 40}, 357 (2008)
[arXiv:0706.1146 [astro-ph]]; \\
%%CITATION = GRGVA,40,357;%%
%\cite{Sotiriou:2008rp}
%\bibitem{Sotiriou:2008rp}
T.~P.~Sotiriou and V.~Faraoni,
%``f(R) Theories Of Gravity,''
Rev.\ Mod.\ Phys.\ {\bf 82}, 451 (2010)
[arXiv:0805.1726 [gr-qc]]; \\
%%CITATION = RMPHA,82,451;%%
%\cite{Lobo:2008sg}
%\bibitem{Lobo:2008sg}
F.~S.~N.~Lobo,
%``The dark side of gravity: Modified theories of gravity,''
arXiv:0807.1640 [gr-qc].
%%CITATION = ARXIV:0807.1640;%%


\bibitem{ASPECTS1}
%\cite{Nelson:2010wp}
%\bibitem{Nelson:2010wp}
W.~Nelson,
%``Restricting Fourth Order Gravity via Cosmology,''
Phys.\ Rev.\  D {\bf 82}, 124044 (2010)
[arXiv:1012.3353 [gr-qc]].
%%CITATION = PHRVA,D82,124044;%%
%\cite{Lee:2010tm}
%\bibitem{Lee:2010tm}
H.~W.~Lee, K.~Y.~Kim and Y.~S.~Myung,
%``Equations of State in the Brans-Dicke cosmology,''
arXiv:1010.5556 [hep-th]; \\
%%CITATION = ARXIV:1010.5556;%%
%\cite{Amendola:2010bk}
%\bibitem{Amendola:2010bk}
L.~Amendola, K.~Enqvist and T.~Koivisto,
%``Unifying Einstein and Palatini gravities,''
arXiv:1010.4776 [gr-qc]; \\
%%CITATION = ARXIV:1010.4776;%%
%\cite{Zhong:2010ae}
%\bibitem{Zhong:2010ae}
Y.~Zhong, Y.~X.~Liu and K.~Yang,
%``Thick $f(R)$-Brane Solutions in Maximally Symmetric Spaces,''
arXiv:1010.3478 [hep-th]; \\
%%CITATION = ARXIV:1010.3478;%%
%\cite{Gorbunov:2010bn}
%\bibitem{Gorbunov:2010bn}
D.~S.~Gorbunov and A.~G.~Panin,
%``Scalaron the mighty: is a quick mare in time everywhere?,''
arXiv:1009.2448 [hep-ph]; \\
%%CITATION = ARXIV:1009.2448;%%
%\cite{Corda:2010zza}
%\bibitem{Corda:2010zza}
C.~Corda,
%``Massive relic gravitational waves from f(R) theories of gravity:
%Production
%and potential detection,''
Eur.\ Phys.\ J.\ C {\bf 65}, 257 (2010)
[arXiv:1007.4077 [gr-qc]]; \\
%%CITATION = EPHJA,C65,257;%%
%\cite{Jaime:2010kn}
%\bibitem{Jaime:2010kn}
L.~G.~Jaime, L.~Patino and M.~Salgado,
%``A robust approach to f(R) gravity,''
arXiv:1006.5747 [gr-qc]; \\
%%CITATION = ARXIV:1006.5747;%%
%\cite{Farajollahi:2010tc}
%\bibitem{Farajollahi:2010tc}
H.~Farajollahi, M.~Setare, F.~Milani and F.~Tayebi,
%``Cosmic Dynamics in $F(R)$ Modified Gravity,''
arXiv:1005.2026 [physics.gen-ph]; \\
%%CITATION = ARXIV:1005.2026;%%
%\cite{Balcerzak:2010kr}
%\bibitem{Balcerzak:2010kr}
A.~Balcerzak and M.~P.~Dabrowski,
%``Brane f(R) gravity cosmologies,''
Phys.\ Rev.\ D {\bf 81}, 123527 (2010)
[arXiv:1004.0150 [hep-th]]; \\
%%CITATION = PHRVA,D81,123527;%%
%\cite{Capozziello:2010sc}
%\bibitem{Capozziello:2010sc}
S.~Capozziello, P.~Martin-Moruno and C.~Rubano,
%``Physical non-equivalence of the Jordan and Einstein frames,''
Phys.\ Lett.\ B {\bf 689}, 117 (2010)
[arXiv:1003.5394 [gr-qc]]; \\
%%CITATION = PHLTA,B689,117;%%
%\cite{Schmidt:2010jr}
%\bibitem{Schmidt:2010jr}
F.~Schmidt,
%``Dynamical Masses in Modified Gravity,''
Phys.\ Rev.\ D {\bf 81}, 103002 (2010)
[arXiv:1003.0409 [astro-ph.CO]]; \\
%%CITATION = PHRVA,D81,103002;%%
%\cite{Saidov:2010wx}
%\bibitem{Saidov:2010wx}
T.~Saidov and A.~Zhuk,
%``Bouncing inflation in nonlinear $R^2+R^4$ gravitational model,''
Phys.\ Rev.\ D {\bf 81}, 124002 (2010)
[arXiv:1002.4138 [hep-th]]; \\
%%CITATION = PHRVA,D81,124002;%%
%\cite{Brevik:2010jv}
%\bibitem{Brevik:2010jv}
I.~Brevik, S.~Nojiri, S.~D.~Odintsov and D.~Saez-Gomez,
%``Cardy-Verlinde formula in FRW Universe with inhomogeneous
%generalized
%fluid
%and dynamical entropy bounds near the future singularity,''
arXiv:1002.1942 [hep-th]; \\
%%CITATION = ARXIV:1002.1942;%%
%\cite{GarciaSalcedo:2009fb}
%\bibitem{GarciaSalcedo:2009fb}
R.~Garcia-Salcedo, T.~Gonzalez, C.~Moreno, Y.~Napoles, Y.~Leyva and
I.~Quiros,
%``Asymptotic Properties of a Supposedly Regular (Dirac-Born-Infeld)
%Modification of General Relativity,''
JCAP {\bf 1002}, 027 (2010)
[arXiv:0912.5048 [gr-qc]]; \\
%%CITATION = JCAPA,1002,027;%%
%\cite{Dzhunushaliev:2009dt}
%\bibitem{Dzhunushaliev:2009dt}
V.~Dzhunushaliev, V.~Folomeev, B.~Kleihaus and J.~Kunz,
%``Some thick brane solutions in $f(R)$-gravity,''
JHEP {\bf 1004}, 130 (2010)
[arXiv:0912.2812 [gr-qc]]; \\
%%CITATION = JHEPA,1004,130;%%
%\cite{Aghmohammadi:2009xt}
%\bibitem{Aghmohammadi:2009xt}
A.~Aghmohammadi, K.~Saaidi, M.~R.~Abolhassani and A.~Vajdi,
%``Standard Cosmological Evolution in f(R) Model to Kaluza Klein
%Cosmology,''
Phys.\ Scripta {\bf 80}, 065008 (2009)
[arXiv:0912.2510 [gr-qc]]; \\
%%CITATION = PHSTB,80,065008;%%
%\cite{Multamaki:2009us}
%\bibitem{Multamaki:2009us}
T.~Multamaki, J.~Vainio and I.~Vilja,
%``Hamiltonian perturbation theory in f(R) gravity,''
Phys.\ Rev.\ D {\bf 81}, 064025 (2010)
[arXiv:0910.5659 [gr-qc]]; \\
%%CITATION = PHRVA,D81,064025;%%
%\cite{Cai:2009qf}
%\bibitem{Cai:2009qf}
R.~G.~Cai, L.~M.~Cao, Y.~P.~Hu and N.~Ohta,
%``Generalized Misner-Sharp Energy in f(R) Gravity,''
Phys.\ Rev.\ D {\bf 80}, 104016 (2009)
[arXiv:0910.2387 [hep-th]]; \\
%%CITATION = PHRVA,D80,104016;%%
%\cite{Bamba:2009gq}
%\bibitem{Bamba:2009gq}
K.~Bamba, C.~Q.~Geng, S.~Nojiri and S.~D.~Odintsov,
%``Equivalence of modified gravity equation to the Clausius relation,''
Europhys.\ Lett.\ {\bf 89}, 50003 (2010)
[arXiv:0909.4397 [hep-th]]; \\
%%CITATION = EULEE,89,50003;%%
%\cite{Nozari:2009tv}
%\bibitem{Nozari:2009tv}
K.~Nozari and T.~Azizi,
%``Phantom-Like Behavior in $f(R)$-Gravity,''
Phys.\ Lett.\ B {\bf 680}, 205 (2009)
[arXiv:0909.0351 [gr-qc]]; \\
%%CITATION = PHLTA,B680,205;%%
%\cite{Tsujikawa:2009ku}
%\bibitem{Tsujikawa:2009ku}
S.~Tsujikawa, R.~Gannouji, B.~Moraes and D.~Polarski,
%``The dispersion of growth of matter perturbations in f(R) gravity,''
Phys.\ Rev.\ D {\bf 80}, 084044 (2009)
[arXiv:0908.2669 [astro-ph.CO]]; \\
%%CITATION = PHRVA,D80,084044;%%
%\cite{Chakrabarti:2009ku}
%\bibitem{Chakrabarti:2009ku}
S.~K.~Chakrabarti, E.~N.~Saridakis and A.~A.~Sen,
%``A new approach to modified-gravity models,''
arXiv:0908.0293 [astro-ph.CO]; \\
%%CITATION = ARXIV:0908.0293;%%
%\cite{Quiros:2009xx}
%\bibitem{Quiros:2009xx}
I.~Quiros, Y.~Leyva and Y.~Napoles,
%``A Note on de Sitter Embedding of $f(R)$ Theories,''
Phys.\ Rev.\ D {\bf 80}, 024022 (2009)
[arXiv:0906.1190 [gr-qc]]; \\
%%CITATION = PHRVA,D80,024022;%%
%\cite{Hui:2009kc}
%\bibitem{Hui:2009kc}
L.~Hui, A.~Nicolis and C.~Stubbs,
%``Equivalence Principle Implications of Modified Gravity Models,''
Phys.\ Rev.\ D {\bf 80}, 104002 (2009)
[arXiv:0905.2966 [astro-ph.CO]]; \\
%%CITATION = PHRVA,D80,104002;%%
%\cite{Gao:2009er}
%\bibitem{Gao:2009er}
C.~Gao,
%``Modified gravity in Arnowitt-Deser-Misner formalism,''
Phys.\ Lett.\ B {\bf 684}, 85 (2010)
[arXiv:0905.0310 [astro-ph.CO]]; \\
%%CITATION = PHLTA,B684,85;%%
%\cite{Paul:2009nb}
%\bibitem{Paul:2009nb}
B.~C.~Paul, P.~S.~Debnath and S.~Ghose,
%``Accelerating Universe in Modified Theories of Gravity,''
Phys.\ Rev.\ D {\bf 79}, 083534 (2009)
[arXiv:0904.0345 [astro-ph.CO]]; \\
%%CITATION = PHRVA,D79,083534;%%
%\cite{Ketov:2009wc}
%\bibitem{Ketov:2009wc}
S.~V.~Ketov,
%``Scalar potential in F(R) supergravity,''
Class.\ Quant.\ Grav.\ {\bf 26}, 135006 (2009)
[arXiv:0903.0251 [hep-th]]; \\
%%CITATION = CQGRD,26,135006;%%
%\cite{Bauer:2009ke}
%\bibitem{Bauer:2009ke}
F.~Bauer, J.~Sola and H.~Stefancic,
%``Relaxing a large cosmological constant,''
Phys.\ Lett.\ B {\bf 678}, 427 (2009)
[arXiv:0902.2215 [hep-th]]; \\
%%CITATION = PHLTA,B678,427;%%
%\cite{Setare:2009ym}
%\bibitem{Setare:2009ym}
M.~R.~Setare,
%``Holographic modified gravity,''
Int.\ J.\ Mod.\ Phys.\ D {\bf 17}, 2219 (2008)
[arXiv:0901.3252 [hep-th]].
%%CITATION = IMPAE,D17,2219;%%

\bibitem{ASPECTS2}
%\cite{Granda:2008tx}
%\bibitem{Granda:2008tx}
L.~N.~Granda,
%``Reconstructing the f(R) gravity from the holographic principle,''
arXiv:0812.1596 [hep-th]; \\
%%CITATION = ARXIV:0812.1596;%%
%\cite{Durrer:2008in}
%\bibitem{Durrer:2008in}
R.~Durrer and R.~Maartens,
%``Dark Energy and Modified Gravity,''
arXiv:0811.4132 [astro-ph]; \\
%%CITATION = ARXIV:0811.4132;%%
%\cite{Koivisto:2008ak}
%\bibitem{Koivisto:2008ak}
T.~S.~Koivisto,
%``Disformal quintessence,''
arXiv:0811.1957 [astro-ph]; \\
%%CITATION = ARXIV:0811.1957;%%
%\cite{Faraoni:2008ze}
%\bibitem{Faraoni:2008ze}
V.~Faraoni,
%``Extension of the EGS theorem to metric and Palatini f(R) gravity,''
arXiv:0811.1870 [gr-qc]; \\
%%CITATION = ARXIV:0811.1870;%%
%\cite{Dyer:2008hb}
%\bibitem{Dyer:2008hb}
E.~Dyer and K.~Hinterbichler,
%``Boundary Terms, Variational Principles and Higher Derivative
%Modified
%Gravity,''
Phys.\ Rev.\ D {\bf 79}, 024028 (2009)
[arXiv:0809.4033 [gr-qc]]; \\
%%CITATION = PHRVA,D79,024028;%%
%\cite{SaezGomez:2008uj}
%\bibitem{SaezGomez:2008uj}
D.~Saez-Gomez,
%``Modified f(R) gravity from scalar-tensor theory and inhomogeneous
%EoS
%dark
%energy,''
Gen.\ Rel.\ Grav.\ {\bf 41}, 1527 (2009)
[arXiv:0809.1311 [hep-th]]; \\
%%CITATION = GRGVA,41,1527;%%
%\cite{Ishak:2008td}
%\bibitem{Ishak:2008td}
M.~Ishak and J.~Moldenhauer,
%``A minimal set of invariants as a systematic approach to higher order
%gravity models,''
JCAP {\bf 0901}, 024 (2009)
[arXiv:0808.0951 [astro-ph]]; \\
%%CITATION = JCAPA,0901,024;%%
%\cite{Lecian:2008vc}
%\bibitem{Lecian:2008vc}
O.~M.~Lecian and G.~Montani,
%``Implications of non-analytical f(R) gravity at Solar-System
%scales,''
Class.\ Quant.\ Grav.\ {\bf 26}, 045014 (2009)
[arXiv:0807.4428 [gr-qc]]; \\
%%CITATION = CQGRD,26,045014;%%
%\cite{Stefancic:2008zz}
%\bibitem{Stefancic:2008zz}
H.~Stefancic,
%``The solution of the cosmological constant problem from the
%inhomogeneous
%equation of state - a hint from modified gravity?,''
Phys.\ Lett.\ B {\bf 670}, 246 (2009)
[arXiv:0807.3692 [gr-qc]]; \\
%%CITATION = PHLTA,B670,246;%%
%\cite{Nojiri:2008ku}
%\bibitem{Nojiri:2008ku}
S.~Nojiri and S.~D.~Odintsov,
%``Modified gravity as realistic candidate for dark energy, inflation
%and
%dark
%matter,''
AIP Conf.\ Proc.\ {\bf 1115}, 212 (2009)
[arXiv:0810.1557 [hep-th]]; \\
%%CITATION = APCPC,1115,212;%%
%\cite{Cortes:2008fy}
%\bibitem{Cortes:2008fy}
J.~L.~Cortes and J.~Indurain,
%``An extension of the cosmological standard model with a bounded
%Hubble
%expansion rate,''
Astropart.\ Phys.\ {\bf 31}, 177 (2009)
[arXiv:0805.3481 [astro-ph]]; \\
%%CITATION = APHYE,31,177;%%
%\cite{Multamaki:2008tk}
%\bibitem{Multamaki:2008tk}
T.~Multamaki, J.~Vainio and I.~Vilja,
%``Maximal symmetry and metric-affine f(R) gravity,''
Class.\ Quant.\ Grav.\ {\bf 26}, 075005 (2009)
[arXiv:0805.1834 [gr-qc]]; \\
%%CITATION = CQGRD,26,075005;%%
%\cite{Vakili:2008ea}
%\bibitem{Vakili:2008ea}
B.~Vakili,
%``Noether symmetry in $f(R)$ cosmology,''
Phys.\ Lett.\ B {\bf 664}, 16 (2008)
[arXiv:0804.3449 [gr-qc]]; \\
%%CITATION = PHLTA,B664,16;%%
%\cite{Deruelle:2008fs}
%\bibitem{Deruelle:2008fs}
N.~Deruelle, M.~Sasaki and Y.~Sendouda,
%``'Detuned' f(R) gravity and dark energy,''
Phys.\ Rev.\  D {\bf 77}, 124024 (2008)
[arXiv:0803.2742 [gr-qc]].
%%CITATION = PHRVA,D77,124024;%%
%\cite{Appleby:2008tv}
%\bibitem{Appleby:2008tv}
S.~A.~Appleby and R.~A.~Battye,
%``Aspects of cosmological expansion in F(R) gravity models,''
JCAP {\bf 0805}, 019 (2008)
[arXiv:0803.1081 [astro-ph]]; \\
%%CITATION = JCAPA,0805,019;%%
%\cite{delaCruzDombriz:2008cp}
%\bibitem{delaCruzDombriz:2008cp}
A.~de la Cruz-Dombriz, A.~Dobado and A.~L.~Maroto,
%``On the evolution of density perturbations in f(R) theories of
%gravity,''
Phys.\ Rev.\ D {\bf 77}, 123515 (2008)
[arXiv:0802.2999 [astro-ph]]; \\
%%CITATION = PHRVA,D77,123515;%%
%\cite{DeDeo:2007yn}
%\bibitem{DeDeo:2007yn}
S.~DeDeo and D.~Psaltis,
%``Stable, accelerating universes in modified-gravity theories,''
Phys.\ Rev.\ D {\bf 78}, 064013 (2008)
[arXiv:0712.3939 [astro-ph]]; \\
%%CITATION = PHRVA,D78,064013;%%
%\cite{Wu:2007tn}
%\bibitem{Wu:2007tn}
X.~Wu and Z.~H.~Zhu,
%``Reconstructing f(R) theory according to holographic dark energy,''
Phys.\ Lett.\ B {\bf 660}, 293 (2008)
[arXiv:0712.3603 [astro-ph]]; \\
%%CITATION = PHLTA,B660,293;%%
%\cite{Capozziello:2007eu}
%\bibitem{Capozziello:2007eu}
% S.~Capozziello and S.~Tsujikawa,
%``Solar system and equivalence principle constraints on $f(R)$ gravity
%by
%chameleon approach,''
% Phys.\ Rev.\ D {\bf 77}, 107501 (2008)
% [arXiv:0712.2268 [gr-qc]]; \\
%%CITATION = PHRVA,D77,107501;%%
%\cite{Machado:2007ea}
%\bibitem{Machado:2007ea}
P.~F.~Machado and F.~Saueressig,
%``On the renormalization group flow of f(R)-gravity,''
Phys.\ Rev.\ D {\bf 77}, 124045 (2008)
[arXiv:0712.0445 [hep-th]]; \\
%%CITATION = PHRVA,D77,124045;%%
%\cite{Multamaki:2007wb}
%\bibitem{Multamaki:2007wb}
T.~Multamaki, A.~Putaja, I.~Vilja and E.~C.~Vagenas,
%``Energy-momentum complexes in f(R) theories of gravity,''
Class.\ Quant.\ Grav.\ {\bf 25}, 075017 (2008)
[arXiv:0712.0276 [gr-qc]]; \\
%%CITATION = CQGRD,25,075017;%%
%\cite{Tsujikawa:2007tg}
%\bibitem{Tsujikawa:2007tg}
S.~Tsujikawa, K.~Uddin and R.~Tavakol,
%``Density perturbations in f(R) gravity theories in metric and
%Palatini
%formalisms,''
Phys.\ Rev.\ D {\bf 77}, 043007 (2008)
[arXiv:0712.0082 [astro-ph]]; \\
%%CITATION = PHRVA,D77,043007;%%
%\cite{Evans:2007ch}
%\bibitem{Evans:2007ch}
J.~D.~Evans, L.~M.~H.~Hall and P.~Caillol,
%``Standard Cosmological Evolution in a Wide Range of f(R) Models,''
Phys.\ Rev.\ D {\bf 77}, 083514 (2008)
[arXiv:0711.3695 [astro-ph]]; \\
%%CITATION = PHRVA,D77,083514;%%
%\cite{Barenboim:2007bu}
%\bibitem{Barenboim:2007bu}
G.~Barenboim and J.~Lykken,
%``Self-accelerating solutions of scalar-tensor gravity,''
JCAP {\bf 0803}, 017 (2008)
[arXiv:0711.3653 [astro-ph]]; \\
%%CITATION = JCAPA,0803,017;%%
%\cite{Bohmer:2007fh}
%\bibitem{Bohmer:2007fh}
C.~G.~Boehmer, T.~Harko and F.~S.~N.~Lobo,
%``Generalized virial theorem in f(R) gravity,''
JCAP {\bf 0803}, 024 (2008)
[arXiv:0710.0966 [gr-qc]]; \\
%%CITATION = JCAPA,0803,024;%%
%\cite{Saffari:2007zt}
%\bibitem{Saffari:2007zt}
R.~Saffari and S.~Rahvar,
%``f(R) Gravity: From the Pioneer Anomaly to the Cosmic Acceleration,''
Phys.\ Rev.\ D {\bf 77}, 104028 (2008)
[arXiv:0708.1482 [astro-ph]]; \\
%%CITATION = PHRVA,D77,104028;%%
%\cite{Iglesias:2007nv}
%\bibitem{Iglesias:2007nv}
A.~Iglesias, N.~Kaloper, A.~Padilla and M.~Park,
%``How (Not) to Palatini,''
Phys.\ Rev.\ D {\bf 76}, 104001 (2007)
[arXiv:0708.1163 [astro-ph]]; \\
%%CITATION = PHRVA,D76,104001;%%
%\cite{Santos:2007bs}
%\bibitem{Santos:2007bs}
J.~Santos, J.~S.~Alcaniz, M.~J.~Reboucas and F.~C.~Carvalho,
%``Energy conditions in f(R)-gravity,''
Phys.\ Rev.\ D {\bf 76}, 083513 (2007)
[arXiv:0708.0411 [astro-ph]]; \\
%%CITATION = PHRVA,D76,083513;%%
%\cite{Sotiriou:2007zu}
%\bibitem{Sotiriou:2007zu}
T.~P.~Sotiriou, V.~Faraoni and S.~Liberati,
%``Theory of gravitation theories: a no-progress report,''
Int.\ J.\ Mod.\ Phys.\ D {\bf 17}, 399 (2008)
[arXiv:0707.2748 [gr-qc]]; \\
%%CITATION = IMPAE,D17,399;%%
%\cite{Schmidt:2007vj}
%\bibitem{Schmidt:2007vj}
F.~Schmidt, M.~Liguori and S.~Dodelson,
%``Galaxy-CMB Cross-Correlation as a Probe of Alternative Models of
%Gravity,''
Phys.\ Rev.\ D {\bf 76}, 083518 (2007)
[arXiv:0706.1775 [astro-ph]]; \\
%%CITATION = PHRVA,D76,083518;%%
%\cite{de Souza:2007fq}
%\bibitem{de Souza:2007fq}
J.~C.~C.~de Souza and V.~Faraoni,
%``The phase space view of f(R) gravity,''
Class.\ Quant.\ Grav.\ {\bf 24}, 3637 (2007)
[arXiv:0706.1223 [gr-qc]]; \\
%%CITATION = CQGRD,24,3637;%%
%\cite{Rahvar:2007sq}
%\bibitem{Rahvar:2007sq}
S.~Rahvar and Y.~Sobouti,
%``An Inverse $f(R)$ Gravitation for Cosmic Speed up, and Dark Energy
%Equivalent,''
Mod.\ Phys.\ Lett.\ A {\bf 23}, 1929 (2008)
[arXiv:0704.0680 [astro-ph]]; \\
%%CITATION = MPLAE,A23,1929;%%
%\cite{Faraoni:2007yn}
%\bibitem{Faraoni:2007yn}
V.~Faraoni,
%``de Sitter space and the equivalence between f(R) and scalar-tensor
%gravity,''
Phys.\ Rev.\ D {\bf 75}, 067302 (2007)
[arXiv:gr-qc/0703044]; \\
%%CITATION = PHRVA,D75,067302;%%
%\cite{Sokolowski:2007pk}
%\bibitem{Sokolowski:2007pk}
L.~M.~Sokolowski,
%``Physical interpretation and viability of various metric nonlinear
%gravity
%theories applied to cosmology,''
Class.\ Quant.\ Grav.\ {\bf 24}, 3391 (2007)
[arXiv:gr-qc/0702097]; \\
%%CITATION = CQGRD,24,3391;%%
%\cite{Rador:2007wq}
%\bibitem{Rador:2007wq}
T.~Rador,
%``f(R) Gravities \`a la Brans-Dicke,''
Phys.\ Lett.\ B {\bf 652}, 228 (2007)
[arXiv:hep-th/0702081]; \\
%%CITATION = PHLTA,B652,228;%%
%\cite{Capozziello:2007gm}
%\bibitem{Capozziello:2007gm}
S.~Capozziello and R.~Garattini,
%``The cosmological constant as an eigenvalue of f(R)-gravity
%Hamiltonian
%constraint,''
Class.\ Quant.\ Grav.\ {\bf 24}, 1627 (2007)
[arXiv:gr-qc/0702075]; \\
%%CITATION = CQGRD,24,1627;%%
%\cite{Li:2007xn}
%\bibitem{Li:2007xn}
B.~Li and J.~D.~Barrow,
%``The Cosmology of f(R) Gravity in the Metric Variational Approach,''
Phys.\ Rev.\ D {\bf 75}, 084010 (2007)
[arXiv:gr-qc/0701111].
%%CITATION = PHRVA,D75,084010;%%

\bibitem{ASPECTS3}
%\cite{Sotiriou:2006sf}
%\bibitem{Sotiriou:2006sf}
T.~P.~Sotiriou,
%``Curvature scalar instability in f(R) gravity,''
Phys.\ Lett.\ B {\bf 645}, 389 (2007)
[arXiv:gr-qc/0611107]; \\
%%CITATION = PHLTA,B645,389;%%
%\cite{Poplawski:2006kv}
%\bibitem{Poplawski:2006kv}
N.~J.~Poplawski,
%``Interacting dark energy in $f(R)$ gravity,''
Phys.\ Rev.\ D {\bf 74}, 084032 (2006)
[arXiv:gr-qc/0607124]; \\
%%CITATION = PHRVA,D74,084032;%%
%\cite{Borowiec:2006qr}
%\bibitem{Borowiec:2006qr}
A.~Borowiec, W.~Godlowski and M.~Szydlowski,
%``Dark matter and dark energy as a effects of Modified Gravity,''
eConf {\bf C0602061}, 09 (2006)
[Int.\ J.\ Geom.\ Meth.\ Mod.\ Phys.\ {\bf 4}, 183 (2007)]
[arXiv:astro-ph/0607639]; \\
%%CITATION = 00436,4,183;%%
%\cite{delaCruzDombriz:2006fj}
%\bibitem{delaCruzDombriz:2006fj}
A.~de la Cruz-Dombriz and A.~Dobado,
%``A $f(R)$ gravity without cosmological constant,''
Phys.\ Rev.\ D {\bf 74}, 087501 (2006)
[arXiv:gr-qc/0607118]; \\
%%CITATION = PHRVA,D74,087501;%%
%\cite{Bludman:2006cg}
%\bibitem{Bludman:2006cg}
S.~A.~Bludman,
%``Cosmological acceleration: Dark energy or modified gravity?,''
arXiv:astro-ph/0605198; \\
%%CITATION = ASTRO-PH/0605198;%%
%\cite{Sotiriou:2006qn}
%\bibitem{Sotiriou:2006qn}
T.~P.~Sotiriou and S.~Liberati,
%``Metric-affine $f(R)$ theories of gravity,''
Annals Phys.\ {\bf 322}, 935 (2007)
[arXiv:gr-qc/0604006]; \\
%%CITATION = APNYA,322,935;%%
%\cite{Rizzo:2006wf}
%\bibitem{Rizzo:2006wf}
T.~G.~Rizzo,
%``Higher curvature gravity in TeV-scale extra dimensions,''
arXiv:hep-ph/0603242; \\
%%CITATION = HEP-PH/0603242;%%
%\cite{Srivastava:2006ky}
%\bibitem{Srivastava:2006ky}
S.~K.~Srivastava,
%``Cosmic evolution with early and late acceleration inspired by dual
%nature
%of the Ricci scalar curvature,''
Int.\ J.\ Mod.\ Phys.\ D {\bf 17}, 755 (2008)
[arXiv:astro-ph/0602116]; \\
%%CITATION = IMPAE,D17,755;%%
%\cite{Koivisto:2006ie}
%\bibitem{Koivisto:2006ie}
T.~Koivisto,
%``The matter power spectrum in $f(R)$ gravity,''
Phys.\ Rev.\ D {\bf 73}, 083517 (2006)
[arXiv:astro-ph/0602031]; \\
%%CITATION = PHRVA,D73,083517;%%
%\cite{Atazadeh:2006re}
%\bibitem{Atazadeh:2006re}
K.~Atazadeh and H.~R.~Sepangi,
%``Accelerated expansion in modified gravity with a Yukawa-like term,''
Int.\ J.\ Mod.\ Phys.\ D {\bf 16}, 687 (2007)
[arXiv:gr-qc/0602028]; \\
%%CITATION = IMPAE,D16,687;%%
%\cite{Schmidt:2006jt}
%\bibitem{Schmidt:2006jt}
H.~J.~Schmidt,
%``Fourth order gravity: Equations, history, and applications to
%cosmology,''
eConf {\bf C0602061}, 12 (2006)
[Int.\ J.\ Geom.\ Meth.\ Mod.\ Phys.\ {\bf 4}, 209 (2007)]
[arXiv:gr-qc/0602017]; \\
%%CITATION = 00436,4,209;%%
%\cite{Brevik:2006md}
%\bibitem{Brevik:2006md}
I.~H.~Brevik,
%``Crossing of the w = -1 barrier in viscous modified gravity,''
Int.\ J.\ Mod.\ Phys.\ D {\bf 15}, 767 (2006)
[arXiv:gr-qc/0601100]; \\
%%CITATION = IMPAE,D15,767;%%
%\cite{Sotiriou:2005hu}
%\bibitem{Sotiriou:2005hu}
T.~P.~Sotiriou,
%``Unification of inflation and cosmic acceleration in the Palatini
%formalism,''
Phys.\ Rev.\ D {\bf 73}, 063515 (2006)
[arXiv:gr-qc/0509029]; \\
%%CITATION = PHRVA,D73,063515;%%
%\cite{Capozziello:2005ra}
%\bibitem{Capozziello:2005ra}
S.~Capozziello, V.~F.~Cardone, E.~Piedipalumbo and C.~Rubano,
%``Dark energy exponential potential models as curvature
%quintessence,''
Class.\ Quant.\ Grav.\ {\bf 23}, 1205 (2006)
[arXiv:astro-ph/0507438]; \\
%%CITATION = CQGRD,23,1205;%%
%\cite{Ezawa:2005zr}
%\bibitem{Ezawa:2005zr}
Y.~Ezawa, H.~Iwasaki, Y.~Ohkuwa, S.~Watanabe, N.~Yamada and T.~Yano,
%``A canonical formalism of f(T)-type gravity in terms of Lie
%derivatives,''
Class.\ Quant.\ Grav.\ {\bf 23}, 3205 (2006)
[arXiv:gr-qc/0507060]; \\
%%CITATION = CQGRD,23,3205;%%
%\cite{Koivisto:2005yk}
%\bibitem{Koivisto:2005yk}
T.~Koivisto,
%``Covariant conservation of energy momentum in modified gravities,''
Class.\ Quant.\ Grav.\ {\bf 23}, 4289 (2006)
[arXiv:gr-qc/0505128]; \\
%%CITATION = CQGRD,23,4289;%%
%\cite{Easson:2004fq}
%\bibitem{Easson:2004fq}
D.~A.~Easson,
%``Cosmic Acceleration and Modified Gravitational Models,''
Int.\ J.\ Mod.\ Phys.\ A {\bf 19}, 5343 (2004)
[arXiv:astro-ph/0411209]; \\
%%CITATION = IMPAE,A19,5343;%%
%\cite{Meng:2004wg}
%\bibitem{Meng:2004wg}
X.~H.~Meng and P.~Wang,
%``Palatini formulation of the R**(-1) modified gravity with an
%additionally
%squared scalar curvature term,''
Class.\ Quant.\ Grav.\ {\bf 22}, 23 (2005)
[arXiv:gr-qc/0411007]; \\
%%CITATION = CQGRD,22,23;%%
%\cite{Carroll:2004de}
%\bibitem{Carroll:2004de}
S.~M.~Carroll, A.~De Felice, V.~Duvvuri, D.~A.~Easson, M.~Trodden and
M.~S.~Turner,
%``The cosmology of generalized modified gravity models,''
Phys.\ Rev.\ D {\bf 71}, 063513 (2005)
[arXiv:astro-ph/0410031]; \\
%%CITATION = PHRVA,D71,063513;%%
%\cite{Allemandi:2004ca}
%\bibitem{Allemandi:2004ca}
G.~Allemandi, A.~Borowiec and M.~Francaviglia,
%``Accelerated cosmological models in first-order non-linear gravity,''
Phys.\ Rev.\ D {\bf 70}, 043524 (2004)
[arXiv:hep-th/0403264]; \\
%%CITATION = PHRVA,D70,043524;%%
%\cite{Nojiri:2004dw}
%\bibitem{Nojiri:2004dw}
S.~Nojiri,
%``Dark energy and modified gravities,''
TSPU Vestnik {\bf 44N7}, 49 (2004)
[arXiv:hep-th/0407099]; \\
%%CITATION = 00478,44N7,49;%%
%\cite{Brevik:2004sd}
%\bibitem{Brevik:2004sd}
I.~H.~Brevik, S.~Nojiri, S.~D.~Odintsov and L.~Vanzo,
%``Entropy and universality of Cardy-Verlinde formula in dark energy
%universe,''
Phys.\ Rev.\ D {\bf 70}, 043520 (2004)
[arXiv:hep-th/0401073]; \\
%%CITATION = PHRVA,D70,043520;%%
%\cite{Vollick:2003ic}
%\bibitem{Vollick:2003ic}
D.~N.~Vollick,
%``On the viability of the Palatini form of 1/R gravity,''
Class.\ Quant.\ Grav.\ {\bf 21}, 3813 (2004)
[arXiv:gr-qc/0312041]; \\
%%CITATION = CQGRD,21,3813;%%
%\cite{Olmo:2004hj}
%\bibitem{Olmo:2004hj}
G.~J.~Olmo and W.~Komp,
%``Nonlinear gravity theories in the metric and Palatini formalisms,''
arXiv:gr-qc/0403092; \\
%%CITATION = GR-QC/0403092;%%
%\cite{Flanagan:2003iw}
%\bibitem{Flanagan:2003iw}
E.~E.~Flanagan,
%``Higher order gravity theories and scalar tensor theories,''
Class.\ Quant.\ Grav.\ {\bf 21}, 417 (2003)
[arXiv:gr-qc/0309015]; \\
%%CITATION = CQGRD,21,417;%%
%\cite{Nojiri:2003rz}
%\bibitem{Nojiri:2003rz}
S.~Nojiri and S.~D.~Odintsov,
%``Where new gravitational physics comes from: M-theory,''
Phys.\ Lett.\ B {\bf 576}, 5 (2003)
[arXiv:hep-th/0307071].
%%CITATION = PHLTA,B576,5;%%

\bibitem{SOLUTIONS}
%\cite{Leon:2010pu}
%\bibitem{Leon:2010pu}
G.~Leon and E.~N.~Saridakis,
%``Dynamics of anisotropic f(R) cosmology,''
arXiv:1007.3956 [gr-qc]; \\
%%CITATION = ARXIV:1007.3956;%%
%\cite{Carloni:2010ph}
%\bibitem{Carloni:2010ph}
S.~Carloni, R.~Goswami and P.~K.~S.~Dunsby,
%``A new approach to reconstruction methods in $f(R)$ gravity,''
arXiv:1005.1840 [gr-qc]; \\
%%CITATION = ARXIV:1005.1840;%%
%\cite{Farajollahi:2010pn}
%\bibitem{Farajollahi:2010pn}
H.~Farajollahi and F.~Milani,
%``Bouncing Universe and phantom crossing in Modified Gravity and its
%reconstruction,''
Mod.\ Phys.\ Lett.\ A {\bf 25}, 2349 (2010)
[arXiv:1004.3512 [gr-qc]]; \\
%%CITATION = MPLAE,A25,2349;%%
%\cite{Santos:2010tw}
%\bibitem{Santos:2010tw}
J.~Santos, M.~J.~Reboucas and T.~B.~R.~Oliveira,
%``G\'{o}del-type universes in Palatini f(R) gravity,''
Phys.\ Rev.\ D {\bf 81}, 123017 (2010)
[arXiv:1004.2501 [astro-ph.CO]]; \\
%%CITATION = PHRVA,D81,123017;%%
%\cite{Saaidi:2010yr}
%\bibitem{Saaidi:2010yr}
K.~Saaidi, A.~Vaji and A.~Aghamomammadi,
%``Static Spherically Symmetric Solution of (R +- {\mu}^4/R) Gravity,''
Gen.\ Rel.\ Grav.\ {\bf 42}, 2421 (2010)
[arXiv:1001.4149 [gr-qc]]; \\
%%CITATION = GRGVA,42,2421;%%
%\cite{Capozziello:2009jg}
%\bibitem{Capozziello:2009jg}
S.~Capozziello, M.~De laurentis and A.~Stabile,
%``Axially symmetric solutions in f(R)-gravity,''
Class.\ Quant.\ Grav.\ {\bf 27}, 165008 (2010)
[arXiv:0912.5286 [gr-qc]]; \\
%%CITATION = CQGRD,27,165008;%%
%\cite{Nojiri:2009xh}
%\bibitem{Nojiri:2009xh}
S.~Nojiri, S.~D.~Odintsov, A.~Toporensky and P.~Tretyakov,
%``Reconstruction and deceleration-acceleration transitions in modified
%gravity,''
Gen.\ Rel.\ Grav.\ {\bf 42}, 1997 (2010)
[arXiv:0912.2488 [hep-th]]; \\
%%CITATION = GRGVA,42,1997;%%
%\cite{Reijonen:2009hi}
%\bibitem{Reijonen:2009hi}
V.~Reijonen,
%``On white dwarfs and neutron stars in Palatini f(R) gravity,''
arXiv:0912.0825 [gr-qc]; \\
%%CITATION = ARXIV:0912.0825;%%
%\cite{Sharif:2009xa}
%\bibitem{Sharif:2009xa}
M.~Sharif and M.~F.~Shamir,
%``Exact Solutions of Bianchi Types $I$ and $V$ space-times in $f(R)$
%Theory
%of
%Gravity,''
Class.\ Quant.\ Grav.\ {\bf 26}, 235020 (2009)
[arXiv:0910.5787 [gr-qc]]; \\
%%CITATION = CQGRD,26,235020;%%
%\cite{Lobo:2009ip}
%\bibitem{Lobo:2009ip}
F.~S.~N.~Lobo and M.~A.~Oliveira,
%``Wormhole geometries in f(R) modified theories of gravity,''
Phys.\ Rev.\ D {\bf 80}, 104012 (2009)
[arXiv:0909.5539 [gr-qc]]; \\
%%CITATION = PHRVA,D80,104012;%%
%\cite{Faraoni:2009xb}
%\bibitem{Faraoni:2009xb}
V.~Faraoni,
%``Clifton's spherical solution in f(R) vacuo harbours a naked
%singularity,''
Class.\ Quant.\ Grav.\ {\bf 26}, 195013 (2009)
[arXiv:0909.0514 [gr-qc]]; \\
%%CITATION = CQGRD,26,195013;%%
%\cite{Nzioki:2009av}
%\bibitem{Nzioki:2009av}
A.~M.~Nzioki, S.~Carloni, R.~Goswami and P.~K.~S.~Dunsby,
%``A new framework for studying spherically symmetric static solutions
%in
%f(R)
%gravity,''
Phys.\ Rev.\ D {\bf 81}, 084028 (2010)
[arXiv:0908.3333 [gr-qc]]; \\
%%CITATION = PHRVA,D81,084028;%%
%\cite{delaCruzDombriz:2009et}
%\bibitem{delaCruzDombriz:2009et}
A.~de la Cruz-Dombriz, A.~Dobado and A.~L.~Maroto,
%``Black Holes in f(R) theories,''
Phys.\ Rev.\ D {\bf 80}, 124011 (2009)
[arXiv:0907.3872 [gr-qc]]; \\
%%CITATION = PHRVA,D80,124011;%%
%\cite{Goheer:2009ss}
%\bibitem{Goheer:2009ss}
N.~Goheer, J.~Larena and P.~K.~S.~Dunsby,
%``Power-law cosmic expansion in f(R) gravity models,''
Phys.\ Rev.\ D {\bf 80}, 061301 (2009)
[arXiv:0906.3860 [gr-qc]]; \\
%%CITATION = PHRVA,D80,061301;%%
%\cite{Figueiro:2009mm}
%\bibitem{Figueiro:2009mm}
M.~F.~Figueiro and A.~Saa,
%``Anisotropic singularities in modified gravity models,''
Phys.\ Rev.\ D {\bf 80}, 063504 (2009)
[arXiv:0906.2588 [gr-qc]]; \\
%%CITATION = PHRVA,D80,063504;%%
%\cite{Momeni:2009tk}
%\bibitem{Momeni:2009tk}
D.~Momeni,
%``A note on constant curvature solutions in cylindrically symmetric
%metric
%$f(R)$ Gravity,''
Int.\ J.\ Mod.\ Phys.\ D {\bf 18}, 1719 (2009)
[arXiv:0903.0067 [gr-qc]]; \\
%%CITATION = IMPAE,D18,1719;%%
%\cite{Cognola:2008zp}
%\bibitem{Cognola:2008zp}
G.~Cognola, E.~Elizalde, S.~D.~Odintsov, P.~Tretyakov and S.~Zerbini,
%``Initial and final de Sitter universes from modified $f(R)$
%gravity,''
Phys.\ Rev.\ D {\bf 79}, 044001 (2009)
[arXiv:0810.4989 [gr-qc]]; \\
%%CITATION = PHRVA,D79,044001;%%
%\cite{Clifton:2008bn}
%\bibitem{Clifton:2008bn}
T.~Clifton,
%``Higher Powers in Gravitation,''
Phys.\ Rev.\ D {\bf 78}, 083501 (2008)
[arXiv:0807.4682 [gr-qc]]; \\
%%CITATION = PHRVA,D78,083501;%%
%\cite{Oyaizu:2008tb}
%\bibitem{Oyaizu:2008tb}
H.~Oyaizu, M.~Lima and W.~Hu,
%``Non-linear evolution of f(R) cosmologies II: power spectrum,''
Phys.\ Rev.\ D {\bf 78}, 123524 (2008)
[arXiv:0807.2462 [astro-ph]]; \\
%%CITATION = PHRVA,D78,123524;%%
%\cite{Hollenstein:2008hp}
%\bibitem{Hollenstein:2008hp}
L.~Hollenstein and F.~S.~N.~Lobo,
%``Exact solutions of f(R) gravity coupled to nonlinear
%electrodynamics,''
Phys.\ Rev.\ D {\bf 78}, 124007 (2008)
[arXiv:0807.2325 [gr-qc]]; \\
%%CITATION = PHRVA,D78,124007;%%
%\cite{Pun:2008ae}
%\bibitem{Pun:2008ae}
C.~S.~J.~Pun, Z.~Kovacs and T.~Harko,
%``Thin accretion disks in f(R) modified gravity models,''
Phys.\ Rev.\ D {\bf 78}, 024043 (2008)
[arXiv:0806.0679 [gr-qc]]; \\
%%CITATION = PHRVA,D78,024043;%%
%\cite{Goswami:2008fs}
%\bibitem{Goswami:2008fs}
R.~Goswami, N.~Goheer and P.~K.~S.~Dunsby,
%``The Existence of Einstein Static Universes and their Stability in
%Fourth
%order Theories of Gravity,''
Phys.\ Rev.\ D {\bf 78}, 044011 (2008)
[arXiv:0804.3528 [gr-qc]]; \\
%%CITATION = PHRVA,D78,044011;%%
%\cite{Vollick:2007fh}
%\bibitem{Vollick:2007fh}
D.~N.~Vollick,
%``Noether Charge and Black Hole Entropy in Modified Theories of
%Gravity,''
Phys.\ Rev.\ D {\bf 76}, 124001 (2007)
[arXiv:0710.1859 [gr-qc]]; \\
%%CITATION = PHRVA,D76,124001;%%
%\cite{LanahanTremblay:2007sg}
%\bibitem{LanahanTremblay:2007sg}
N.~Lanahan-Tremblay and V.~Faraoni,
%``The Cauchy problem of f(R) gravity,''
Class.\ Quant.\ Grav.\ {\bf 24}, 5667 (2007)
[arXiv:0709.4414 [gr-qc]]; \\
%%CITATION = CQGRD,24,5667;%%
%\cite{Capozziello:2007id}
%\bibitem{Capozziello:2007id}
S.~Capozziello, A.~Stabile and A.~Troisi,
%``Spherical symmetry in $f(R)-$gravity,''
Class.\ Quant.\ Grav.\ {\bf 25}, 085004 (2008)
[arXiv:0709.0891 [gr-qc]]; \\
%%CITATION = CQGRD,25,085004;%%
%\cite{Boehmer:2007tr}
%\bibitem{Boehmer:2007tr}
C.~G.~Boehmer, L.~Hollenstein and F.~S.~N.~Lobo,
%``Stability of the Einstein static universe in f(R) gravity,''
Phys.\ Rev.\ D {\bf 76}, 084005 (2007)
[arXiv:0706.1663 [gr-qc]]; \\
%%CITATION = PHRVA,D76,084005;%%
%\cite{Henttunen:2007bz}
%\bibitem{Henttunen:2007bz}
K.~Henttunen, T.~Multamaki and I.~Vilja,
%``Stellar configurations in f(R) theories of gravity,''
Phys.\ Rev.\ D {\bf 77}, 024040 (2008)
[arXiv:0705.2683 [astro-ph]]; \\
%%CITATION = PHRVA,D77,024040;%%
%\cite{Kainulainen:2007bt}
%\bibitem{Kainulainen:2007bt}
K.~Kainulainen, J.~Piilonen, V.~Reijonen and D.~Sunhede,
%``Spherically symmetric space-times in f(R) gravity theories,''
Phys.\ Rev.\ D {\bf 76}, 024020 (2007)
[arXiv:0704.2729 [gr-qc]]; \\
%%CITATION = PHRVA,D76,024020;%%
%\cite{Clifton:2007ih}
%\bibitem{Clifton:2007ih}
T.~Clifton,
%``Exact Friedmann solutions in higher-order gravity theories,''
Class.\ Quant.\ Grav.\ {\bf 24}, 5073 (2007)
[arXiv:gr-qc/0703126]; \\
%%CITATION = CQGRD,24,5073;%%
%\cite{Seifert:2007fr}
%\bibitem{Seifert:2007fr}
M.~D.~Seifert,
%``Stability of spherically symmetric solutions in modified theories of
%gravity,''
Phys.\ Rev.\ D {\bf 76}, 064002 (2007)
[arXiv:gr-qc/0703060]; \\
%%CITATION = PHRVA,D76,064002;%%
%\cite{Sawicki:2007tf}
%\bibitem{Sawicki:2007tf}
I.~Sawicki and W.~Hu,
%``Stability of Cosmological Solution in f(R) Models of Gravity,''
Phys.\ Rev.\ D {\bf 75}, 127502 (2007)
[arXiv:astro-ph/0702278]; \\
%%CITATION = PHRVA,D75,127502;%%
%\cite{Cognola:2007vq}
%\bibitem{Cognola:2007vq}
G.~Cognola, M.~Gastaldi and S.~Zerbini,
%``On the Stability of a class of Modified Gravitational Models,''
Int.\ J.\ Theor.\ Phys.\ {\bf 47}, 898 (2008)
[arXiv:gr-qc/0701138]; \\
%%CITATION = IJTPB,47,898;%%
%\cite{Bazeia:2007jj}
%\bibitem{Bazeia:2007jj}
D.~Bazeia, B.~Carneiro da Cunha, R.~Menezes and A.~Y.~Petrov,
%``Perturbative aspects and conformal solutions of F(R) gravity,''
Phys.\ Lett.\ B {\bf 649}, 445 (2007)
[arXiv:hep-th/0701106]; \\
%%CITATION = PHLTA,B649,445;%%
%\cite{Baghram:2007df}
%\bibitem{Baghram:2007df}
S.~Baghram, M.~Farhang and S.~Rahvar,
%``Modified gravity with $f(R) = \sqrt{R^2 - R_0^2}$,''
Phys.\ Rev.\ D {\bf 75}, 044024 (2007)
[arXiv:astro-ph/0701013]; \\
%%CITATION = PHRVA,D75,044024;%%
%\cite{Multamaki:2006ym}
%\bibitem{Multamaki:2006ym}
T.~Multamaki and I.~Vilja,
%``Static spherically symmetric perfect fluid solutions in $f(R)$
%theories
%of
%gravity,''
Phys.\ Rev.\ D {\bf 76}, 064021 (2007)
[arXiv:astro-ph/0612775]; \\
%%CITATION = PHRVA,D76,064021;%%
%\cite{Leach:2006br}
%\bibitem{Leach:2006br}
J.~A.~Leach, S.~Carloni and P.~K.~S.~Dunsby,
%``Shear dynamics in Bianchi I cosmologies with R**n-gravity,''
Class.\ Quant.\ Grav.\ {\bf 23}, 4915 (2006)
[arXiv:gr-qc/0603012]; \\
%%CITATION = CQGRD,23,4915;%%
%\cite{Clifton:2006kc}
%\bibitem{Clifton:2006kc}
T.~Clifton and J.~D.~Barrow,
%``Further exact cosmological solutions to higher-order gravity
%theories,''
Class.\ Quant.\ Grav.\ {\bf 23}, 2951 (2006)
[arXiv:gr-qc/0601118]; \\
%%CITATION = CQGRD,23,2951;%%
%\cite{Faraoni:2005vk}
%\bibitem{Faraoni:2005vk}
V.~Faraoni,
%``The stability of modified gravity models,''
Phys.\ Rev.\ D {\bf 72}, 124005 (2005)
[arXiv:gr-qc/0511094]; \\
%%CITATION = PHRVA,D72,124005;%%
%\cite{Clifton:2005aj}
%\bibitem{Clifton:2005aj}
T.~Clifton and J.~D.~Barrow,
%``The Power of General Relativity,''
Phys.\ Rev.\ D {\bf 72}, 103005 (2005)
[arXiv:gr-qc/0509059]; \\
%%CITATION = PHRVA,D72,103005;%%
%\cite{Furey:2004rq}
%\bibitem{Furey:2004rq}
N.~Furey and A.~DeBenedictis,
%``Wormhole throats in R**m gravity,''
Class.\ Quant.\ Grav.\ {\bf 22}, 313 (2005)
[arXiv:gr-qc/0410088]; \\
%%CITATION = CQGRD,22,313;%%
%\cite{Carloni:2004kp}
%\bibitem{Carloni:2004kp}
S.~Carloni, P.~K.~S.~Dunsby, S.~Capozziello and A.~Troisi,
%``Cosmological dynamics of R^n gravity,''
Class.\ Quant.\ Grav.\ {\bf 22}, 4839 (2005)
[arXiv:gr-qc/0410046]; \\
%%CITATION = CQGRD,22,4839;%%
%\cite{Gunther:2004ht}
%\bibitem{Gunther:2004ht}
U.~Gunther, A.~Zhuk, V.~B.~Bezerra and C.~Romero,
%``AdS and stabilized extra dimensions in multidimensional
%gravitational
%models with nonlinear scalar curvature terms R**-1 and R**4,''
Class.\ Quant.\ Grav.\ {\bf 22}, 3135 (2005)
[arXiv:hep-th/0409112]; \\
%%CITATION = CQGRD,22,3135;%%
%\cite{Kleinert:2000rt}
%\bibitem{Kleinert:2000rt}
H.~Kleinert and H.~J.~Schmidt,
%``Cosmology with Curvature-Saturated Gravitational Lagrangian
%R/\sqrt{1
%+
%l^4
%R^2},''
Gen.\ Rel.\ Grav.\ {\bf 34}, 1295 (2002)
[arXiv:gr-qc/0006074].
%%CITATION = GRGVA,34,1295;%%

\bibitem{TESTS}
%\cite{Capozziello:2010iy}
%\bibitem{Capozziello:2010iy}
S.~Capozziello, R.~Cianci, M.~De Laurentis and S.~Vignolo,
%``Testing metric-affine f(R)-gravity by relic scalar gravitational
%waves,''
arXiv:1007.3670 [gr-qc]; \\
%%CITATION = ARXIV:1007.3670;%%
%\cite{Saaidi:2010js}
%\bibitem{Saaidi:2010js}
K.~Saaidi and A.~Aghamohammadi,
%``Solar System Constraints on a Viable $f(R)$) Gravity by Chameleon
%Mechanism,''
arXiv:1006.1842 [gr-qc]; \\
%%CITATION = ARXIV:1006.1842;%%
%\cite{Fu:2010zza}
%\bibitem{Fu:2010zza}
X.~Fu, P.~Wu and H.~Yu,
%``The Growth Factor Of Matter Perturbations In F(R) Gravity,''
Eur.\ Phys.\ J.\ C {\bf 68}, 271 (2010); \\
%%CITATION = EPHJA,C68,271;%%
%\cite{Jain:2010ka}
%\bibitem{Jain:2010ka}
B.~Jain and J.~Khoury,
%``Cosmological Tests of Gravity,''
Annals Phys.\ {\bf 325}, 1479 (2010)
[arXiv:1004.3294 [astro-ph.CO]]; \\
%%CITATION = APNYA,325,1479;%%
%\cite{Lombriser:2010mp}
%\bibitem{Lombriser:2010mp}
L.~Lombriser, A.~Slosar, U.~Seljak and W.~Hu,
%``Constraints on f(R) gravity from probing the large-scale
%structure,''
arXiv:1003.3009 [astro-ph.CO]; \\
%%CITATION = ARXIV:1003.3009;%%
%\cite{Cui:2010wb}
%\bibitem{Cui:2010wb}
W.~Cui, P.~Zhang and X.~Yang,
%``Nonlinearities in modified gravity cosmology I: signatures of
%modified
%gravity in the nonlinear matter power spectrum,''
Phys.\ Rev.\ D {\bf 81}, 103528 (2010)
[arXiv:1001.5184 [astro-ph.CO]]; \\
%%CITATION = PHRVA,D81,103528;%%
%\cite{Narikawa:2009ux}
%\bibitem{Narikawa:2009ux}
T.~Narikawa and K.~Yamamoto,
%``Characterizing the linear growth rate of cosmological density
%perturbations
%in an f(R) model,''
Phys.\ Rev.\ D {\bf 81}, 043528 (2010)
[Erratum-ibid.\ D {\bf 81}, 129903 (2010)]
[arXiv:0912.1445 [astro-ph.CO]]; \\
%%CITATION = PHRVA,D81,043528;%%
%\cite{Schmidt:2009am}
%\bibitem{Schmidt:2009am}
F.~Schmidt, A.~Vikhlinin and W.~Hu,
%``Cluster Constraints on f(R) Gravity,''
Phys.\ Rev.\ D {\bf 80}, 083505 (2009)
[arXiv:0908.2457 [astro-ph.CO]]; \\
%%CITATION = PHRVA,D80,083505;%%
%\cite{Bisabr:2009ee}
%\bibitem{Bisabr:2009ee}
Y.~Bisabr,
%``Solar System Constraints on a Cosmologically Viable $f(R)$ Theory,''
Phys.\ Lett.\ B {\bf 683}, 96 (2010)
[arXiv:0907.3838 [gr-qc]]; \\
%%CITATION = PHLTA,B683,96;%%
%\cite{Nesseris:2009jf}
%\bibitem{Nesseris:2009jf}
S.~Nesseris and A.~Mazumdar,
%``Newton's constant in f(R,R_{\mu\nu}R^{\mu\nu},\Box R) theories of
%gravity
%and constraints from BBN,''
Phys.\ Rev.\ D {\bf 79}, 104006 (2009)
[arXiv:0902.1185 [astro-ph.CO]]; \\
%%CITATION = PHRVA,D79,104006;%%
%\cite{Bellucci:2008jt}
%\bibitem{Bellucci:2008jt}
S.~Bellucci, S.~Capozziello, M.~De Laurentis and V.~Faraoni,
%``Position and frequency shifts induced by massive modes of the
%gravitational
%wave background in alternative gravity,''
Phys.\ Rev.\ D {\bf 79}, 104004 (2009)
[arXiv:0812.1348 [gr-qc]]; \\
%%CITATION = PHRVA,D79,104004;%%
%\cite{Schmidt:2008tn}
%\bibitem{Schmidt:2008tn}
F.~Schmidt, M.~V.~Lima, H.~Oyaizu and W.~Hu,
%``Non-linear Evolution of f(R) Cosmologies III: Halo Statistics,''
Phys.\ Rev.\ D {\bf 79}, 083518 (2009)
[arXiv:0812.0545 [astro-ph]]; \\
%%CITATION = PHRVA,D79,083518;%%
%\cite{Borisov:2008xn}
%\bibitem{Borisov:2008xn}
A.~Borisov and B.~Jain,
%``Three-Point Correlations in f(R) Models of Gravity,''
Phys.\ Rev.\ D {\bf 79}, 103506 (2009)
[arXiv:0812.0013 [astro-ph]]; \\
%%CITATION = PHRVA,D79,103506;%%
%\cite{Lambiase:2008zz}
%\bibitem{Lambiase:2008zz}
G.~Lambiase, S.~Mohanty and G.~Scarpetta,
%``Magnetic field amplification in f(R) theories of gravity,''
JCAP {\bf 0807}, 019 (2008); \\
%%CITATION = JCAPA,0807,019;%%
%\cite{Capozziello:2008fn}
%\bibitem{Capozziello:2008fn}
S.~Capozziello, M.~De Laurentis, S.~Nojiri and S.~D.~Odintsov,
%``$f(R)$ gravity constrained by PPN parameters and stochastic
%background
%of
%gravitational waves,''
Gen.\ Rel.\ Grav.\ {\bf 41}, 2313 (2009)
[arXiv:0808.1335 [hep-th]]; arXiv:0911.2139;\\
%%CITATION = GRGVA,41,2313;%%
%\cite{Dev:2008rx}
%\bibitem{Dev:2008rx}
A.~Dev, D.~Jain, S.~Jhingan, S.~Nojiri, M.~Sami and I.~Thongkool,
%``Delicate f(R) gravity models with disappearing cosmological constant
%and
%observational constraints on the model parameters,''
Phys.\ Rev.\ D {\bf 78}, 083515 (2008)
[arXiv:0807.3445 [hep-th]]; \\
%%CITATION = PHRVA,D78,083515;%%
%\cite{Kang:2008zi}
%\bibitem{Kang:2008zi}
J.~U.~Kang and G.~Panotopoulos,
%``Big-Bang Nucleosynthesis and neutralino dark matter in modified
%gravity,''
Phys.\ Lett.\ B {\bf 677}, 6 (2009)
[arXiv:0806.1493 [astro-ph]]; \\
%%CITATION = PHLTA,B677,6;%%
%\cite{Carvalho:2008am}
%\bibitem{Carvalho:2008am}
F.~C.~Carvalho, E.~M.~Santos, J.~S.~Alcaniz and J.~Santos,
%``Cosmological Constraints from Hubble Parameter on f(R)
%Cosmologies,''
JCAP {\bf 0809}, 008 (2008)
[arXiv:0804.2878 [astro-ph]]; \\
%%CITATION = JCAPA,0809,008;%%
%\cite{Capozziello:2008qc}
%\bibitem{Capozziello:2008qc}
S.~Capozziello, V.~F.~Cardone and V.~Salzano,
%``Cosmography of f(R) gravity,''
Phys.\ Rev.\ D {\bf 78}, 063504 (2008)
[arXiv:0802.1583 [astro-ph]]; \\
%%CITATION = PHRVA,D78,063504;%%
%\cite{Iorio:2007rk}
%\bibitem{Iorio:2007rk}
L.~Iorio and M.~L.~Ruggiero,
%``Solar System tests of some models of modified gravity proposed to
%explain
%galactic rotation curves without dark matter,''
Schol.\ Res.\ Exch.\ {\bf 2008}, 968393 (2008)
[arXiv:0711.0256 [gr-qc]]; \\
%%CITATION = 00698,2008,968393;%%
%\cite{Faraoni:2007sn}
%\bibitem{Faraoni:2007sn}
V.~Faraoni,
%``A viability criterion for modified gravity with an extra force,''
Phys.\ Rev.\ D {\bf 76}, 127501 (2007)
[arXiv:0710.1291 [gr-qc]]; \\
%%CITATION = PHRVA,D76,127501;%%
%\cite{Multamaki:2007jk}
%\bibitem{Multamaki:2007jk}
T.~Multamaki and I.~Vilja,
%``Constraining Newtonian stellar configurations in f(R) theories of
%gravity,''
Phys.\ Lett.\ B {\bf 659}, 843 (2008)
[arXiv:0709.3422 [astro-ph]]; \\
%%CITATION = PHLTA,B659,843;%%
%\cite{Ananda:2007xh}
%\bibitem{Ananda:2007xh}
K.~N.~Ananda, S.~Carloni and P.~K.~S.~Dunsby,
%``The evolution of cosmological gravitational waves in f(R) gravity,''
Phys.\ Rev.\ D {\bf 77}, 024033 (2008)
[arXiv:0708.2258 [gr-qc]]; \\
%%CITATION = PHRVA,D77,024033;%%
%\cite{Hu:2007pj}
%\bibitem{Hu:2007pj}
W.~Hu and I.~Sawicki,
%``A Parameterized Post-Friedmann Framework for Modified Gravity,''
Phys.\ Rev.\ D {\bf 76}, 104043 (2007)
[arXiv:0708.1190 [astro-ph]]; \\
%%CITATION = PHRVA,D76,104043;%%
%\cite{Capozziello:2007ms}
%\bibitem{Capozziello:2007ms}
S.~Capozziello, A.~Stabile and A.~Troisi,
%``The Newtonian Limit of F(R) gravity,''
Phys.\ Rev.\ D {\bf 76}, 104019 (2007)
[arXiv:0708.0723 [gr-qc]]; \\
%%CITATION = PHRVA,D76,104019;%%
%\cite{Song:2007da}
%\bibitem{Song:2007da}
Y.~S.~Song, H.~Peiris and W.~Hu,
%``Cosmological Constraints on f(R) Acceleration Models,''
Phys.\ Rev.\ D {\bf 76}, 063517 (2007)
[arXiv:0706.2399 [astro-ph]]; \\
%%CITATION = PHRVA,D76,063517;%%
%\cite{Fay:2007uy}
%\bibitem{Fay:2007uy}
S.~Fay, S.~Nesseris and L.~Perivolaropoulos,
%``Can $f(R)$ modified gravity theories mimic a $\Lambda$CDM
%cosmology?,''
Phys.\ Rev.\ D {\bf 76}, 063504 (2007)
[arXiv:gr-qc/0703006]; \\
%%CITATION = PHRVA,D76,063504;%%
%\cite{Zhang:2007ne}
%\bibitem{Zhang:2007ne}
P.~J.~Zhang,
%``The behavior of $f(R)$ gravity in the solar system, galaxies and
%clusters,''
Phys.\ Rev.\ D {\bf 76}, 024007 (2007)
[arXiv:astro-ph/0701662]; \\
%%CITATION = PHRVA,D76,024007;%%
%\cite{Fay:2007gg}
%\bibitem{Fay:2007gg}
S.~Fay, R.~Tavakol and S.~Tsujikawa,
%``f(R) gravity theories in Palatini formalism: cosmological dynamics
%and
%observational constraints,''
Phys.\ Rev.\ D {\bf 75}, 063509 (2007)
[arXiv:astro-ph/0701479]; \\
%%CITATION = PHRVA,D75,063509;%%
%\cite{Amendola:2006we}
%\bibitem{Amendola:2006we}
L.~Amendola, R.~Gannouji, D.~Polarski and S.~Tsujikawa,
%``Conditions for the cosmological viability of f(R) dark energy
%models,''
Phys.\ Rev.\ D {\bf 75}, 083504 (2007)
[arXiv:gr-qc/0612180]; \\
%%CITATION = PHRVA,D75,083504;%%
%\cite{Olmo:2006zu}
%\bibitem{Olmo:2006zu}
G.~J.~Olmo,
%``Violation of the equivalence principle in modified theories of
%gravity,''
Phys.\ Rev.\ Lett.\ {\bf 98}, 061101 (2007)
[arXiv:gr-qc/0612002]; \\
%%CITATION = PRLTA,98,061101;%%
%\cite{Bustelo:2006ms}
%\bibitem{Bustelo:2006ms}
A.~J.~Bustelo and D.~E.~Barraco,
%``Equilibrium hydrostatic equation and Newtonian limit of the singular
%f(R)
%gravity,''
Class.\ Quant.\ Grav.\ {\bf 24}, 2333 (2007)
[arXiv:gr-qc/0611149]; \\
%%CITATION = CQGRD,24,2333;%%
%\cite{Jin:2006if}
%\bibitem{Jin:2006if}
X.~H.~Jin, D.~J.~Liu and X.~Z.~Li,
%``Solar System tests disfavor $f(R)$ gravities,''
arXiv:astro-ph/0610854; \\
%%CITATION = ASTRO-PH/0610854;%%
%\cite{Poplawski:2006ew}
%\bibitem{Poplawski:2006ew}
N.~J.~Poplawski,
%``The cosmic snap parameter in f(R) gravity,''
Class.\ Quant.\ Grav.\ {\bf 24}, 3013 (2007)
[arXiv:gr-qc/0610133]; \\
%%CITATION = CQGRD,24,3013;%%
%\cite{Erickcek:2006vf}
%\bibitem{Erickcek:2006vf}
A.~L.~Erickcek, T.~L.~Smith and M.~Kamionkowski,
%``Solar system tests do rule out 1/R gravity,''
Phys.\ Rev.\ D {\bf 74}, 121501 (2006)
[arXiv:astro-ph/0610483]; \\
%%CITATION = PHRVA,D74,121501;%%
%\cite{Li:2006ag}
%\bibitem{Li:2006ag}
B.~Li, K.~C.~Chan and M.~C.~Chu,
%``Constraints on f(R) Cosmology in the Palatini Formalism,''
Phys.\ Rev.\ D {\bf 76}, 024002 (2007)
[arXiv:astro-ph/0610794]; \\
%%CITATION = PHRVA,D76,024002;%%
%\cite{Allemandi:2006bm}
%\bibitem{Allemandi:2006bm}
G.~Allemandi and M.~L.~Ruggiero,
%``Constraining Extended Theories of Gravity using Solar System
%Tests,''
Gen.\ Rel.\ Grav.\ {\bf 39}, 1381 (2007)
[arXiv:astro-ph/0610661]; \\
%%CITATION = GRGVA,39,1381;%%
%\cite{Song:2006ej}
%\bibitem{Song:2006ej}
Y.~S.~Song, W.~Hu and I.~Sawicki,
%``The large scale structure of f(R) gravity,''
Phys.\ Rev.\ D {\bf 75}, 044004 (2007)
[arXiv:astro-ph/0610532]; \\
%%CITATION = PHRVA,D75,044004;%%
%\cite{Mendoza:2006hs}
%\bibitem{Mendoza:2006hs}
S.~Mendoza and Y.~M.~Rosas-Guevara,
%``Gravitational waves and lensing of the metric theory proposed by
%Sobouti,''
Astron.\ Astrophys.\ {\bf 472}, 367 (2007)
[arXiv:astro-ph/0610390]; \\
%%CITATION = AAEJA,472,367;%%
%\cite{Lambiase:2006dq}
%\bibitem{Lambiase:2006dq}
G.~Lambiase and G.~Scarpetta,
%``Baryogenesis in $f(R)$ theories of gravity,''
Phys.\ Rev.\ D {\bf 74}, 087504 (2006)
[arXiv:astro-ph/0610367]; \\
%%CITATION = PHRVA,D74,087504;%%
%\cite{Huterer:2006mva}
%\bibitem{Huterer:2006mva}
D.~Huterer and E.~V.~Linder,
%``Separating dark physics from physical darkness: Minimalist modified
%gravity vs. dark energy,''
Phys.\ Rev.\ D {\bf 75}, 023519 (2007)
[arXiv:astro-ph/0608681]; \\
%%CITATION = PHRVA,D75,023519;%%
%\cite{Brookfield:2006mq}
%\bibitem{Brookfield:2006mq}
A.~W.~Brookfield, C.~van de Bruck and L.~M.~H.~Hall,
%``Viability of $f(R)$ theories with additional powers of curvature,''
Phys.\ Rev.\ D {\bf 74}, 064028 (2006)
[arXiv:hep-th/0608015]; \\
%%CITATION = PHRVA,D74,064028;%%
%\cite{Ruggiero:2006qv}
%\bibitem{Ruggiero:2006qv}
M.~L.~Ruggiero and L.~Iorio,
%``Solar System planetary orbital motions and $f(R)$ Theories of
%Gravity,''
JCAP {\bf 0701}, 010 (2007)
[arXiv:gr-qc/0607093]; \\
%%CITATION = JCAPA,0701,010;%%
%\cite{Faraoni:2006hx}
%\bibitem{Faraoni:2006hx}
V.~Faraoni,
%``Solar system experiments do not yet veto modified gravity models,''
Phys.\ Rev.\ D {\bf 74}, 023529 (2006)
[arXiv:gr-qc/0607016]; \\
%%CITATION = PHRVA,D74,023529;%%
%\cite{Mena:2005ta}
%\bibitem{Mena:2005ta}
O.~Mena, J.~Santiago and J.~Weller,
%``Constraining inverse curvature gravity with supernovae,''
Phys.\ Rev.\ Lett.\ {\bf 96}, 041103 (2006)
[arXiv:astro-ph/0510453].
%%CITATION = PRLTA,96,041103;%%

%%%%%%%%%%%%%%% Until introduction

%\cite{Cognola:2005de}
\bibitem{Cognola:2005de}
G.~Cognola, E.~Elizalde, S.~Nojiri, S.~D.~Odintsov and
S.~Zerbini,
%``One-loop f(R) gravity in de Sitter universe,''
JCAP {\bf 0502}, 010 (2005)
[arXiv:hep-th/0501096].
%%CITATION = JCAPA,0502,010;%%

%\cite{Nojiri:2009xw}
\bibitem{Nojiri:2009xw}
S.~Nojiri and S.~D.~Odintsov,
%``Non-singular modified gravity: the unification of the
%inflation,
%dark
%energy and dark mater,''
AIP Conf.\ Proc.\ {\bf 1241}, 1094 (2010)
[arXiv:0910.1464 [hep-th]].
%%CITATION = APCPC,1241,1094;%%

%\cite{DeFelice:2010aj}
\bibitem{DeFelice:2010aj}
A.~De Felice and S.~Tsujikawa,
%``f(R) theories,''
Living Rev.\ Rel.\ {\bf 13}, 3 (2010)
[arXiv:1002.4928 [gr-qc]].
%%CITATION = 00222,13,3;%%

%\cite{Carloni:2007yv}
\bibitem{Carloni:2007yv}
S.~Carloni, P.~K.~S.~Dunsby and A.~Troisi,
%``The evolution of density perturbations in $f(R)$ gravity,''
Phys.\ Rev.\ D {\bf 77}, 024024 (2008)
[arXiv:0707.0106 [gr-qc]].
%%CITATION = PHRVA,D77,024024;%%

%\cite{Carloni:2009gp}
\bibitem{Carloni:2009gp}
S.~Carloni, E.~Elizalde and S.~Odintsov,
%``Conformal Transformations in Cosmology of Modified Gravity: the
%Covariant
%Approach Perspective,''
Gen.\ Rel.\ Grav.\ {\bf 42}, 1667 (2010)
[arXiv:0907.3941 [gr-qc]].
%%CITATION = GRGVA,42,1667;%%

%\cite{Hu:2007nk}
\bibitem{Hu:2007nk}
W.~Hu and I.~Sawicki,
%``Models of f(R) Cosmic Acceleration that Evade Solar-System
%Tests,''
Phys.\ Rev.\ D {\bf 76}, 064004 (2007)
[arXiv:0705.1158 [astro-ph]].
%%CITATION = PHRVA,D76,064004;%%

%\cite{Capozziello:2005mj}
\bibitem{Capozziello:2005mj}
S.~Capozziello, S.~Nojiri and S.~D.~Odintsov,
%``Dark energy: The equation of state description versus
%scalar-tensor
%or
%modified gravity,''
Phys.\ Lett.\ B {\bf 634}, 93 (2006)
[arXiv:hep-th/0512118].
%%CITATION = PHLTA,B634,93;%%

%\cite{Caldwell:1999ew}
\bibitem{Caldwell:1999ew}
R.~R.~Caldwell,
%``A Phantom Menace?,''
Phys.\ Lett.\ B {\bf 545}, 23 (2002)
[arXiv:astro-ph/9908168].
%%CITATION = PHLTA,B545,23;%%

%\cite{Capozziello:2006dj}
\bibitem{Capozziello:2006dj}
S.~Capozziello, S.~Nojiri, S.~D.~Odintsov and A.~Troisi,
%``Cosmological viability of f(R)-gravity as an ideal fluid and
%its
%compatibility with a matter dominated phase,''
Phys.\ Lett.\ B {\bf 639}, 135 (2006)
[arXiv:astro-ph/0604431].
%%CITATION = PHRVA,D74,086005;%%

%\cite{Dolgov:2003px}
\bibitem{Dolgov:2003px}
A.~D.~Dolgov and M.~Kawasaki,
%``Can modified gravity explain accelerated cosmic expansion?,''
Phys.\ Lett.\ B {\bf 573}, 1 (2003)
[arXiv:astro-ph/0307285].
%%CITATION = PHLTA,B573,1;%%

%\cite{Capozziello:2002rd}
\bibitem{Capozziello:2002rd}
S.~Capozziello,
%``Curvature Quintessence,''
Int.\ J.\ Mod.\ Phys.\ D {\bf 11}, 483 (2002)
[arXiv:gr-qc/0201033].
%%CITATION = IMPAE,D11,483;%%

%\cite{Carroll:2003wy}
\bibitem{Carroll:2003wy}
S.~M.~Carroll, V.~Duvvuri, M.~Trodden and M.~S.~Turner,
%``Is Cosmic Speed-Up Due to New Gravitational Physics?,''
Phys.\ Rev.\ D {\bf 70}, 043528 (2004)
[arXiv:astro-ph/0306438].
%%CITATION = PHRVA,D70,043528;%%

%\cite{Capozziello:2003tk}
\bibitem{Capozziello:2003tk}
S.~Capozziello, S.~Carloni and A.~Troisi,
%``Quintessence without scalar fields,''
Recent Res.\ Dev.\ Astron.\ Astrophys.\ {\bf 1}, 625 (2003)
[arXiv:astro-ph/0303041].
%%CITATION = 00638,1,625;%%

%\cite{Capozziello:2004sm}
\bibitem{Capozziello:2004sm}
S.~Capozziello,
%``Newtonian limit of extended theories of gravity,''
arXiv:gr-qc/0412088; \\
%%CITATION = GR-QC/0412088;%%
%\cite{Meng:2003sx}
%\bibitem{Meng:2003sx}
X.~H.~Meng and P.~Wang,
%``Gravitational potential in Palatini formulation of modified
%gravity,''
Gen.\ Rel.\ Grav.\ {\bf 36}, 1947 (2004)
[arXiv:gr-qc/0311019]; \\
%%CITATION = GRGVA,36,1947;%%
%\cite{Dominguez:2004ds}
%\bibitem{Dominguez:2004ds}
A.~E.~Dominguez and D.~E.~Barraco,
%``Newtonian limit of the singular $f(R)$ gravity in the Palatini
%formalism,''
Phys.\ Rev.\ D {\bf 70}, 043505 (2004)
[arXiv:gr-qc/0408069]; \\
%%CITATION = PHRVA,D70,043505;%%
%\cite{Allemandi:2005tg}
%\bibitem{Allemandi:2005tg}
G.~Allemandi, M.~Francaviglia, M.~L.~Ruggiero and A.~Tartaglia,
%``Post-Newtonian parameters from alternative theories of
%gravity,''
Gen.\ Rel.\ Grav.\ {\bf 37}, 1891 (2005)
[arXiv:gr-qc/0506123]; \\
%%CITATION = GRGVA,37,1891;%%
%\cite{Koivisto:2005yk}
%\bibitem{Koivisto:2005yk}
T.~Koivisto,
%``Covariant conservation of energy momentum in modified
%gravities,''
Class.\ Quant.\ Grav.\ {\bf 23}, 4289 (2006)
[arXiv:gr-qc/0505128]; \\
%%CITATION = CQGRD,23,4289;%%
%\cite{Olmo:2005jd}
%\bibitem{Olmo:2005jd}
G.~J.~Olmo,
%``Post-Newtonian constraints on $f(R)$ cosmologies in metric and
%Palatini
%formalism,''
Phys.\ Rev.\ D {\bf 72}, 083505 (2005); \\
%%CITATION = PHRVA,D72,083505;%%
%\cite{Navarro:2005gh}
%\bibitem{Navarro:2005gh}
I.~Navarro and K.~Van Acoleyen,
%``On the Newtonian limit of Generalized Modified Gravity
%Models,''
Phys.\ Lett.\ B {\bf 622}, 1 (2005)
[arXiv:gr-qc/0506096]; \\
%%CITATION = PHLTA,B622,1;%%
%\cite{Cembranos:2005fi}
%\bibitem{Cembranos:2005fi}
J.~A.~R.~Cembranos,
%``The Newtonian limit at intermediate energies,''
Phys.\ Rev.\ D {\bf 73}, 064029 (2006)
[arXiv:gr-qc/0507039].
%%CITATION = PHRVA,D73,064029;%%
%\cite{Sotiriou:2005xe}
%\bibitem{Sotiriou:2005xe}
T.~P.~Sotiriou,
%``The nearly Newtonian regime in Non-Linear Theories of
%Gravity,''
Gen.\ Rel.\ Grav.\ {\bf 38}, 1407 (2006)
[arXiv:gr-qc/0507027]; \\
%%CITATION = GRGVA,38,1407;%%
%\cite{Amarzguioui:2005zq}
%\bibitem{Amarzguioui:2005zq}
M.~Amarzguioui, O.~Elgaroy, D.~F.~Mota and T.~Multamaki,
%``Cosmological constraints on f(R) gravity theories within the
%Palatini
%approach,''
Astron.\ Astrophys.\ {\bf 454}, 707 (2006)
[arXiv:astro-ph/0510519]; \\
%%CITATION = AAEJA,454,707;%%
%\cite{Shao:2005wt}
%\bibitem{Shao:2005wt}
C.~G.~Shao, R.~G.~Cai, B.~Wang and R.~K.~Su,
%``An Alternative explanation of the conflict between 1/R gravity
%and
%solar
%system tests,''
Phys.\ Lett.\ B {\bf 633}, 164 (2006)
[arXiv:gr-qc/0511034]; \\
%%CITATION = PHLTA,B633,164;%%
%\cite{Bronnikov:2006jy}
%\bibitem{Bronnikov:2006jy}
K.~A.~Bronnikov, S.~A.~Kononogov and V.~N.~Melnikov,
%``Brane world corrections to Newton's law,''
Gen.\ Rel.\ Grav.\ {\bf 38}, 1215 (2006)
[arXiv:gr-qc/0601114]; \\
%%CITATION = GRGVA,38,1215;%%
%\cite{Atazadeh:2006re}
%\bibitem{Atazadeh:2006re}
K.~Atazadeh and H.~R.~Sepangi,
%``Accelerated expansion in modified gravity with a Yukawa-like
%term,''
Int.\ J.\ Mod.\ Phys.\ D {\bf 16}, 687 (2007)
[arXiv:gr-qc/0602028].
%%CITATION = IMPAE,D16,687;%%

%\cite{Soussa:2003re}
\bibitem{Soussa:2003re}
M.~E.~Soussa and R.~P.~Woodard,
%``The Force of Gravity from a Lagrangian containing Inverse
%Powers
%of
%the
%Ricci Scalar,''
Gen.\ Rel.\ Grav.\ {\bf 36}, 855 (2004)
[arXiv:astro-ph/0308114].
%%CITATION = GRGVA,36,855;%%

%\cite{Woodard:2006nt}
\bibitem{Woodard:2006nt}
R.~P.~Woodard,
%``Avoiding dark energy with 1/R modifications of gravity,''
Lect.\ Notes Phys.\ {\bf 720}, 403 (2007)
[arXiv:astro-ph/0601672].
%%CITATION = LNPHA,720,403;%%

%\cite{Nojiri:2003wx}
\bibitem{Nojiri:2003wx}
S.~Nojiri and S.~D.~Odintsov,
%``The minimal curvature of the universe in modified gravity and
%conformal
%anomaly resolution of the instabilities,''
Mod.\ Phys.\ Lett.\ A {\bf 19}, 627 (2004)
[arXiv:hep-th/0310045].
%%CITATION = MPLAE,A19,627;%%

%\cite{Chiba:2003ir}
\bibitem{Chiba:2003ir}
T.~Chiba,
%``1/R gravity and scalar-tensor gravity,''
Phys.\ Lett.\ B {\bf 575}, 1 (2003)
[arXiv:astro-ph/0307338].
%%CITATION = PHLTA,B575,1;%%

%\cite{Nojiri:2003ni}
\bibitem{Nojiri:2003ni}
S.~Nojiri and S.~D.~Odintsov,
%``Modified gravity with ln R terms and cosmic acceleration,''
Gen.\ Rel.\ Grav.\ {\bf 36}, 1765 (2004)
[arXiv:hep-th/0308176].
%%CITATION = GRGVA,36,1765;%%

%\cite{Meng:2003en}
\bibitem{Meng:2003en}
X.~H.~Meng and P.~Wang,
%``Palatini formation of modified gravity with ln(R) terms,''
Phys.\ Lett.\ B {\bf 584}, 1 (2004)
[arXiv:hep-th/0309062].
%%CITATION = PHLTA,B584,1;%%

%\cite{Nojiri:2007as}
\bibitem{Nojiri:2007as}
S.~Nojiri and S.~D.~Odintsov,
%``Unifying inflation with LambdaCDM epoch in modified f(R)
%gravity
%consistent
%with Solar System tests,''
Phys.\ Lett.\ B {\bf 657}, 238 (2007)
[arXiv:0707.1941 [hep-th]].
%%CITATION = PHLTA,B657,238;%%

%\cite{Nojiri:2007cq}
\bibitem{Nojiri:2007cq}
S.~Nojiri and S.~D.~Odintsov,
%``Modified f(R) gravity unifying R^m inflation with $\Lambda$CDM
%epoch,''
Phys.\ Rev.\ D {\bf 77}, 026007 (2008)
[arXiv:0710.1738 [hep-th]].
%%CITATION = PHRVA,D77,026007;%%

%\cite{Cognola:2007zu}
\bibitem{Cognola:2007zu}
G.~Cognola, E.~Elizalde, S.~Nojiri, S.~D.~Odintsov, L.~Sebastiani
and S.~Zerbini,
%``A class of viable modified $f(R)$ gravities describing
%inflation
%and
%the
%onset of accelerated expansion,''
Phys.\ Rev.\ D {\bf 77}, 046009 (2008)
[arXiv:0712.4017 [hep-th]].
%%CITATION = PHRVA,D77,046009;%%

\bibitem{ref5}
%\cite{McInnes:2001zw}
%\bibitem{McInnes:2001zw}
B.~McInnes,
%``The dS/CFT correspondence and the big smash,''
JHEP {\bf 0208} (2002) 029
[arXiv:hep-th/0112066]; \\
%%CITATION = JHEPA,0208,029;%%
%\cite{Nojiri:2003vn}
%\bibitem{Nojiri:2003vn}
S.~Nojiri and S.~D.~Odintsov,
%``Quantum deSitter cosmology and phantom matter,''
Phys.\ Lett.\ B {\bf 562}, 147 (2003)
[arXiv:hep-th/0303117]; \\
%%CITATION = PHLTA,B562,147;%%
%\cite{Nojiri:2003ag}
%\bibitem{Nojiri:2003ag}
S.~Nojiri and S.~D.~Odintsov,
%``Effective equation of state and energy conditions in phantom /
%tachyon
%inflationary cosmology perturbed by quantum effects,''
Phys.\ Lett.\ B {\bf 571}, 1 (2003)
[arXiv:hep-th/0306212]; \\
%%CITATION = PHLTA,B571,1;%%
%\cite{Faraoni:2001tq}
%\bibitem{Faraoni:2001tq}
V.~Faraoni,
%``Superquintessence,''
Int.\ J.\ Mod.\ Phys.\ D {\bf 11}, 471 (2002)
[arXiv:astro-ph/0110067]; \\
%%CITATION = IMPAE,D11,471;%%
%\cite{GonzalezDiaz:2003rf}
%\bibitem{GonzalezDiaz:2003rf}
P.~F.~Gonzalez-Diaz,
%``K-essential phantom energy: Doomsday around the corner?,''
Phys.\ Lett.\ B {\bf 586}, 1 (2004)
[arXiv:astro-ph/0312579]; \\
%%CITATION = PHLTA,B586,1;%%
%\cite{GonzalezDiaz:2004as}
%\bibitem{GonzalezDiaz:2004as}
P.~F.~Gonzalez-Diaz,
%``On tachyon and sub-quantum phantom cosmologies,''
TSPU Vestnik {\bf 44N7}, 36 (2004)
[arXiv:hep-th/0408225]; \\
%%CITATION = 00478,44N7,36;%%
%\cite{McInnes:2005vp}
%\bibitem{McInnes:2005vp}
B.~McInnes,
%``The phantom divide in string gas cosmology,''
Nucl.\ Phys.\ B {\bf 718}, 55 (2005)
[arXiv:hep-th/0502209]; \\
%%CITATION = NUPHA,B718,55;%%
%\cite{Singh:2003vx}
%\bibitem{Singh:2003vx}
P.~Singh, M.~Sami and N.~Dadhich,
%``Cosmological dynamics of phantom field,''
Phys.\ Rev.\ D {\bf 68}, 023522 (2003)
[arXiv:hep-th/0305110]; \\
%%CITATION = PHRVA,D68,023522;%%
%\cite{Csaki:2004ha}
%\bibitem{Csaki:2004ha}
C.~Csaki, N.~Kaloper and J.~Terning,
%``Exorcising w < -1,''
Annals Phys.\ {\bf 317}, 410 (2005)
[arXiv:astro-ph/0409596].
%%CITATION = APNYA,317,410;%%
%\cite{Wu:2004ex}
%\bibitem{Wu:2004ex}
P.~X.~Wu and H.~W.~Yu,
%``Avoidance of Big Rip In Phantom Cosmology by Gravitational
%Back
%Reaction,''
Nucl.\ Phys.\ B {\bf 727}, 355 (2005)
[arXiv:astro-ph/0407424].
%%CITATION = NUPHA,B727,355;%%
%\cite{Nesseris:2004uj}
%\bibitem{Nesseris:2004uj}
S.~Nesseris and L.~Perivolaropoulos,
%``The fate of bound systems in phantom and quintessence
%cosmologies,''
Phys.\ Rev.\ D {\bf 70}, 123529 (2004)
[arXiv:astro-ph/0410309].
%%CITATION = PHRVA,D70,123529;%%
%\cite{Sami:2003xv}
%\bibitem{Sami:2003xv}
M.~Sami and A.~Toporensky,
%``Phantom Field and the Fate of Universe,''
Mod.\ Phys.\ Lett.\ A {\bf 19}, 1509 (2004)
[arXiv:gr-qc/0312009]; \\
%%CITATION = MPLAE,A19,1509;%%
%\cite{Stefancic:2003rc}
%\bibitem{Stefancic:2003rc}
H.~Stefancic,
%``Generalized phantom energy,''
Phys.\ Lett.\ B {\bf 586}, 5 (2004)
[arXiv:astro-ph/0310904]; \\
%%CITATION = PHLTA,B586,5;%%
%\cite{Chimento:2003qy}
%\bibitem{Chimento:2003qy}
L.~P.~Chimento and R.~Lazkoz,
%``On the link between phantom and standard cosmologies,''
Phys.\ Rev.\ Lett.\ {\bf 91}, 211301 (2003)
[arXiv:gr-qc/0307111]; \\
%%CITATION = PRLTA,91,211301;%%
%``On big rip singularities,''
Mod.\ Phys.\ Lett.\ A {\bf 19}, 2479 (2004)
[arXiv:gr-qc/0405020]; \\
%%CITATION = MPLAE,A19,2479;%%
%\cite{Hao:2004ky}
%\bibitem{Hao:2004ky}
J.~G.~Hao and X.~Z.~Li,
%``Generalized quartessence cosmic dynamics: Phantom or quintessence
%with
%de
%Sitter attractor,''
Phys.\ Lett.\ B {\bf 606}, 7 (2005)
[arXiv:astro-ph/0404154]; \\
%%CITATION = PHLTA,B606,7;%%
%\cite{Babichev:2004qp}
%\bibitem{Babichev:2004qp}
E.~Babichev, V.~Dokuchaev and Yu.~Eroshenko,
%``Dark energy cosmology with generalized linear equation of
%state,''
Class.\ Quant.\ Grav.\ {\bf 22}, 143 (2005)
[arXiv:astro-ph/0407190]; \\
%%CITATION = CQGRD,22,143;%%
%\cite{Zhang:2005eg}
%\bibitem{Zhang:2005eg}
X.~F.~Zhang, H.~Li, Y.~S.~Piao and X.~M.~Zhang,
%``Two-field models of dark energy with equation of state across
%-1,''
Mod.\ Phys.\ Lett.\ A {\bf 21}, 231 (2006)
[arXiv:astro-ph/0501652]; \\
%%CITATION = MPLAE,A21,231;%%
%\cite{Elizalde:2005ju}
%\bibitem{Elizalde:2005ju}
E.~Elizalde, S.~Nojiri, S.~D.~Odintsov and P.~Wang,
%``Dark energy: Vacuum fluctuations, the effective phantom phase,
%and
%holography,''
Phys.\ Rev.\ D {\bf 71}, 103504 (2005)
[arXiv:hep-th/0502082]; \\
%%CITATION = PHRVA,D71,103504;%%
%\cite{Dabrowski:2004hx}
%\bibitem{Dabrowski:2004hx}
M.~P.~Dabrowski and T.~Stachowiak,
%``Phantom Friedmann cosmologies and higher-order characteristics of
%expansion,''
Annals Phys.\ {\bf 321}, 771 (2006)
[arXiv:hep-th/0411199]; \\
%%CITATION = APNYA,321,771;%%
%\cite{Lobo:2005us}
%\bibitem{Lobo:2005us}
F.~S.~N.~Lobo,
%``Phantom energy traversable wormholes,''
Phys.\ Rev.\ D {\bf 71}, 084011 (2005)
[arXiv:gr-qc/0502099]; \\
%%CITATION = PHRVA,D71,084011;%%
%\cite{Cai:2005ie}
%\bibitem{Cai:2005ie}
R.~G.~Cai, H.~S.~Zhang and A.~Wang,
%``Crossing w = -1 in Gauss-Bonnet brane world with induced
%gravity,''
Commun.\ Theor.\ Phys.\ {\bf 44}, 948 (2005)
[arXiv:hep-th/0505186]; \\
%%CITATION = CTPMD,44,948;%%
%\cite{Aref'eva:2004vw}
%\bibitem{Aref'eva:2004vw}
I.~Y.~Aref'eva, A.~S.~Koshelev and S.~Y.~Vernov,
%``Exactly Solvable SFT Inspired Phantom Model,''
Theor.\ Math.\ Phys.\ {\bf 148}, 895 (2006)
[Teor.\ Mat.\ Fiz.\ {\bf 148}, 23 (2006)]
[arXiv:astro-ph/0412619]; \\
%%CITATION = TMFZA,148,23;%%
%\cite{Aref'eva:2005fu}
%\bibitem{Aref'eva:2005fu}
%``Crossing of the w = -1 barrier by D3-brane dark energy model,''
Phys.\ Rev.\ D {\bf 72}, 064017 (2005)
[arXiv:astro-ph/0507067]; \\
%%CITATION = PHRVA,D72,064017;%%
%\cite{Lu:2005qy}
%\bibitem{Lu:2005qy}
H.~Q.~Lu, Z.~G.~Huang and W.~Fang,
%``Quantum and classical cosmology with Born-Infeld scalar field,''
arXiv:hep-th/0504038; \\
%%CITATION = HEP-TH/0504038;%%
%\cite{Godlowski:2005tw}
%\bibitem{Godlowski:2005tw}
W.~Godlowski and M.~Szydlowski,
%``Which cosmological model with dark energy - phantom or
%LambdaCDM,''
Phys.\ Lett.\ B {\bf 623}, 10 (2005)
[arXiv:astro-ph/0507322]; \\
%%CITATION = PHLTA,B623,10;%%
%\cite{Sola:2005et}
%\bibitem{Sola:2005et}
J.~Sola and H.~Stefancic,
%``Effective equation of state for dark energy: mimicking
%quintessence
%and
%phantom energy through a variable Lambda,''
Phys.\ Lett.\ B {\bf 624}, 147 (2005)
[arXiv:astro-ph/0505133]; \\
%%CITATION = PHLTA,B624,147;%%
%\cite{Guberina:2005mp}
%\bibitem{Guberina:2005mp}
B.~Guberina, R.~Horvat and H.~Nikolic,
%``Generalized holographic dark energy and the IR cutoff problem,''
Phys.\ Rev.\ D {\bf 72}, 125011 (2005)
[arXiv:astro-ph/0507666]; \\
%%CITATION = PHRVA,D72,125011;%%
%\cite{Dabrowski:2006dd}
%\bibitem{Dabrowski:2006dd}
M.~P.~Dabrowski, C.~Kiefer and B.~Sandhofer,
%``Quantum phantom cosmology,''
Phys.\ Rev.\ D {\bf 74}, 044022 (2006)
[arXiv:hep-th/0605229]; \\
%%CITATION = PHRVA,D74,044022;%%
%\cite{Barbaoza:2006hf}
%\bibitem{Barbaoza:2006hf}
E.~M.~Barbaoza and N.~A.~Lemos,
%``Does the big rip survive quantization?,''
arXiv:gr-qc/0606084.

%\cite{Nojiri:2005sx}
\bibitem{Nojiri:2005sx}
S.~Nojiri, S.~D.~Odintsov and S.~Tsujikawa,
%``Properties of singularities in (phantom) dark energy
%universe,''
Phys.\ Rev.\ D {\bf 71}, 063004 (2005)
[arXiv:hep-th/0501025].
%%CITATION = PHRVA,D71,063004;%%

\bibitem{barrow}
%\cite{Barrow:2004xh}
%\bibitem{Barrow:2004xh}
J.~D.~Barrow,
%``Sudden Future Singularities,''
Class.\ Quant.\ Grav.\ {\bf 21}, L79 (2004)
[arXiv:gr-qc/0403084]; \\
%%CITATION = CQGRD,21,L79;%%
S.~Cotsakis and I.~Klaoudatou,
%``Future singularities of isotropic cosmologies,''
J.\ Geom.\ Phys.\ {\bf 55}, 306 (2005)
[arXiv:gr-qc/0409022]; \\
%%CITATION = JGPHE,55,306;%%
S.~Nojiri and S.~D.~Odintsov,
%``Quantum escape of sudden future singularity,''
Phys.\ Lett.\ B {\bf 595}, 1 (2004)
[arXiv:hep-th/0405078]; \\
%%CITATION = PHLTA,B595,1;%%
%``The final state and thermodynamics of dark energy universe,''
%Phys.\ Rev.\ D {\bf 70}, 103522 (2004)
%[arXiv:hep-th/0408170]; \\
%%CITATION = PHRVA,D70,103522;%%
%\cite{Barrow:2004hk}
%\bibitem{Barrow:2004hk}
J.~D.~Barrow,
%``More general sudden singularities,''
Class.\ Quant.\ Grav.\ {\bf 21}, 5619 (2004)
[arXiv:gr-qc/0409062]; \\
%%CITATION = CQGRD,21,5619; \\
%\cite{Sahni:2002dx}
%\bibitem{Sahni:2002dx}
V.~Sahni and Y.~Shtanov,
%``Braneworld models of dark energy,''
JCAP {\bf 0311}, 014 (2003)
[arXiv:astro-ph/0202346]; \\
%%CITATION = JCAPA,0311,014;%%
%\cite{Lake:2004fu}
%\bibitem{Lake:2004fu}
K.~Lake,
%``Sudden future singularities in FLRW cosmologies,''
Class.\ Quant.\ Grav.\ {\bf 21}, L129 (2004)
[arXiv:gr-qc/0407107]; \\
%%CITATION = CQGRD,21,L129;%%
J.~D.~Barrow and C.~G.~Tsagas,
%``New Isotropic and Anisotropic Sudden Singularities,''
Class.\ Quant.\ Grav.\ {\bf 22}, 1563 (2005)
[arXiv:gr-qc/0411045]; \\
%%CITATION = CQGRD,22,1563;%%
M.~P.~Dabrowski,
%``Inhomogenized sudden future singularities,''
Phys.\ Rev.\ D {\bf 71}, 103505 (2005)
[arXiv:gr-qc/0410033]; \\
%%CITATION = PHRVA,D71,103505;%%
Phys. Lett. B {\bf 625}, 184 (2005);
A.~Balcerzak and M.~P.~Dabrowski,
%``Strings at future singularities,''
Phys.\ Rev.\ D {\bf 73}, 101301 (2006)
[arXiv:hep-th/0604034]; \\
%%CITATION = PHRVA,D73,101301;%%
L.~Fernandez-Jambrina and R.~Lazkoz,
%``Geodesic behavior of sudden future singularities,''
Phys.\ Rev.\ D {\bf 70}, 121503 (2004)
[arXiv:gr-qc/0410124];
%%CITATION = PHRVA,D70,121503;%%
%``Classification of cosmological milestones,''
Phys.\ Rev.\ D {\bf 74}, 064030 (2006)
[arXiv:gr-qc/0607073];
%%CITATION = PHRVA,D74,064030;%%
%``Singular fate of the universe in modified theories of gravity,''
arXiv:0805.2284 [gr-qc]; \\
%%CITATION = ARXIV:0805.2284;%%
P.~Tretyakov, A.~Toporensky, Y.~Shtanov and V.~Sahni,
%``Quantum effects, soft singularities and the fate of the universe
%in
%a
%braneworld cosmology,''
Class.\ Quant.\ Grav.\ {\bf 23}, 3259 (2006)
[arXiv:gr-qc/0510104]; \\
%%CITATION = CQGRD,23,3259;%%
H.~Stefancic,
%``'Expansion' around the vacuum equation of state: Sudden future
%singularities and asymptotic behavior,''
Phys.\ Rev.\ D {\bf 71}, 084024 (2005)
[arXiv:astro-ph/0411630]; \\
%%CITATION = PHRVA,D71,084024;%%
A.~V.~Yurov, A.~V.~Astashenok and P.~F.~Gonzalez-Diaz,
%``Astronomical bounds on future big freeze singularity,''
Grav.\ Cosmol.\ {\bf 14}, 205 (2008)
[arXiv:0705.4108 [astro-ph]]; \\
%%CITATION = GRCOF,14,205;%%
I.~Brevik and O.~Gorbunova,
%``Viscous Dark Cosmology with Account of Quantum Effects,''
Eur.\ Phys.\ J.\ C {\bf 56}, 425 (2008)
[arXiv:0806.1399 [gr-qc]]; \\
%%CITATION = EPHJA,C56,425;%%
M.~Bouhmadi-Lopez, P.~F.~Gonzalez-Diaz and P.~Martin-Moruno,
%``Worse than a big rip?,''
Phys.\ Lett.\ B {\bf 659}, 1 (2008)
[arXiv:gr-qc/0612135];
%%CITATION = PHLTA,B659,1;%%
%``On the generalised Chaplygin gas: worse than a big rip or quieter than 
%a
%sudden singularity?,''
arXiv:0707.2390 [gr-qc]; \\
%%CITATION = ARXIV:0707.2390;%%
%\cite{Sami:2006wj}
%\bibitem{Sami:2006wj}
M.~Sami, P.~Singh and S.~Tsujikawa,
%``Avoidance of future singularities in loop quantum cosmology,''
Phys.\ Rev.\ D {\bf 74}, 043514 (2006)
[arXiv:gr-qc/0605113]; \\
%%CITATION = PHRVA,D74,043514;%%
C.~Cattoen and M.~Visser,
%``Necessary and sufficient conditions for big bangs, bounces,
%crunches,
%rips, sudden singularities, and extremality events,''
Class.\ Quant.\ Grav.\ {\bf 22}, 4913 (2005)
[arXiv:gr-qc/0508045]; \\
%%CITATION = CQGRD,22,4913;%%
J.~D.~Barrow and S.~Z.~W.~Lip,
%``The Classical Stability of Sudden and Big Rip Singularities,''
arXiv:0901.1626 [gr-qc]; \\
%%CITATION = ARXIV:0901.1626;%%
%\cite{BouhmadiLopez:2009jk}
%\bibitem{BouhmadiLopez:2009jk}
M.~Bouhmadi-Lopez, Y.~Tavakoli and P.~V.~Moniz,
%``Appeasing the Phantom Menace?,''
arXiv:0911.1428 [gr-qc].
%%CITATION = ARXIV:0911.1428;%%

%\cite{Nojiri:2008fk}
\bibitem{Nojiri:2008fk}
S.~Nojiri and S.~D.~Odintsov,
%``The future evolution and finite-time singularities in
%$F(R)$-gravity
%unifying the inflation and cosmic acceleration,''
Phys.\ Rev.\ D {\bf 78}, 046006 (2008)
[arXiv:0804.3519 [hep-th]].
%%CITATION = PHRVA,D78,046006;%%

%\cite{Abdalla:2004sw}
\bibitem{Abdalla:2004sw}
M.~C.~B.~Abdalla, S.~Nojiri and S.~D.~Odintsov,
%``Consistent modified gravity: Dark energy, acceleration and the
%absence
%of
%cosmic doomsday,''
Class.\ Quant.\ Grav.\ {\bf 22}, L35 (2005)
[arXiv:hep-th/0409177].
%%CITATION = CQGRD,22,L35;%%%\cite{Nojiri:2004pf}

\bibitem{others}
%\cite{Appleby:2007vb}
%\bibitem{Appleby:2007vb}
S.~A.~Appleby and R.~A.~Battye,
%``Do consistent $F(R)$ models mimic General Relativity plus
%$\Lambda$?,''
Phys.\ Lett.\ B {\bf 654}, 7 (2007)
[arXiv:0705.3199 [astro-ph]]; \\
%%CITATION = PHLTA,B654,7;%%
%\cite{Starobinsky:2007hu}
%\bibitem{Starobinsky:2007hu}
A.~A.~Starobinsky,
%``Disappearing cosmological constant in f(R) gravity,''
JETP Lett.\ {\bf 86}, 157 (2007)
[arXiv:0706.2041 [astro-ph]]; \\
%%CITATION = JTPLA,86,157;%%
%\cite{Nojiri:2007jr}
%\bibitem{Nojiri:2007jr}
S.~Nojiri and S.~D.~Odintsov,
%``Newton law corrections and instabilities in $f(R)$ gravity
%with
%the
%effective cosmological constant epoch,''
Phys.\ Lett.\ B {\bf 652}, 343 (2007)
[arXiv:0706.1378 [hep-th]]; \\
%%CITATION = PHLTA,B652,343;%%
%\cite{Pogosian:2007sw}
%\bibitem{Pogosian:2007sw}
L.~Pogosian and A.~Silvestri,
%``The pattern of growth in viable f(R) cosmologies,''
Phys.\ Rev.\ D {\bf 77}, 023503 (2008)
[Erratum-ibid.\ D {\bf 81}, 049901 (2010)]
[arXiv:0709.0296 [astro-ph]]; \\
%%CITATION = PHRVA,D77,023503;%%
%\cite{Capozziello:2007eu}
%\bibitem{Capozziello:2007eu}
S.~Capozziello and S.~Tsujikawa,
%``Solar system and equivalence principle constraints on $f(R)$ gravity by
%chameleon approach,''
Phys.\ Rev.\ D {\bf 77}, 107501 (2008)
[arXiv:0712.2268 [gr-qc]].
%%CITATION = PHRVA,D77,107501;%%

%\cite{Bamba:2008ut}
\bibitem{Bamba:2008ut}
K.~Bamba, S.~Nojiri and S.~D.~Odintsov,
%``The universe future in modified gravity theories:
%approaching
%the
%finite-time future singularity,''
JCAP {\bf 0810}, 045 (2008)
[arXiv:0807.2575 [hep-th]].
%%CITATION = JCAPA,0810,045;%%

%\cite{Capozziello:2009hc}
\bibitem{Capozziello:2009hc}
S.~Capozziello, M.~De Laurentis, S.~Nojiri and S.~D.~Odintsov,
%``Classifying and avoiding singularities in the alternative
%gravity
%dark
%energy models,''
Phys.\ Rev.\ D {\bf 79}, 124007 (2009)
[arXiv:0903.2753 [hep-th]].
%%CITATION = PHRVA,D79,124007;%%

%\cite{Bamba:2009uf}
\bibitem{Bamba:2009uf}
K.~Bamba, S.~D.~Odintsov, L.~Sebastiani and S.~Zerbini,
%``Finite-time future singularities in modified Gauss-Bonnet
%and
%$\mathcal{F}(R,G)$ gravity and singularity avoidance,''
Eur.\ Phys.\ J.\ C {\bf 67}, 295 (2010)
[arXiv:0911.4390 [hep-th]].
%%CITATION = EPHJA,C67,295;%%

%\cite{Kobayashi:2008wc}
\bibitem{Kobayashi:2008wc}
T.~Kobayashi and K.~i.~Maeda,
%``Can higher curvature corrections cure the singularity
%problem
%in
%f(R)
%gravity?,''
Phys.\ Rev.\ D {\bf 79}, 024009 (2009)
[arXiv:0810.5664 [astro-ph]].
%%CITATION = PHRVA,D79,024009;%%

%\cite{Sami:2009jx}
\bibitem{Sami:2009jx}
M.~Sami,
%``A primer on problems and prospects of dark energy,''
arXiv:0904.3445 [hep-th].
%%CITATION = ARXIV:0904.3445;%%

%\cite{Thongkool:2009js}
\bibitem{Thongkool:2009js}
I.~Thongkool, M.~Sami, R.~Gannouji and S.~Jhingan,
%``The generosity of $f(R)$ gravity models with disappearing
%cosmological
%constant,''
Phys.\ Rev.\ D {\bf 80}, 043523 (2009)
[arXiv:0906.2460 [hep-th]].
%%CITATION = PHRVA,D80,043523;%%

%\cite{Babichev:2009fi}
\bibitem{Babichev:2009fi}
E.~Babichev and D.~Langlois,
%``Relativistic stars in f(R) and scalar-tensor theories,''
Phys.\ Rev.\ D {\bf 81}, 124051 (2010)
[arXiv:0911.1297 [gr-qc]].
%%CITATION = PHRVA,D81,124051;%%

%\cite{Appleby:2010dx}
\bibitem{Appleby:2010dx}
S.~A.~Appleby and J.~Weller,
%``Parameterizing scalar-tensor theories for cosmological
%probes,''
arXiv:1008.2693 [astro-ph.CO].
%%CITATION = ARXIV:1008.2693;%%

%\cite{Nojiri:2009pf}
\bibitem{Nojiri:2009pf}
S.~Nojiri and S.~D.~Odintsov,
%``Is the future universe singular: Dark Matter versus
%modified
%gravity?,''
Phys.\ Lett.\ B {\bf 686}, 44 (2010)
[arXiv:0911.2781 [hep-th]].
%%CITATION = PHLTA,B686,44;%%

\bibitem{Yang}
%\cite{Bamba:2010ws}
%\bibitem{Bamba:2010ws}
K.~Bamba, C.~Q.~Geng and C.~C.~Lee,
%``Cosmological evolution in exponential gravity,''
JCAP {\bf 1008}, 021 (2010)
[arXiv:1005.4574 [astro-ph.CO]]; \\
%%CITATION = JCAPA,1008,021;%%
%\cite{Linder:2009jz}
%\bibitem{Linder:2009jz}
E.~V.~Linder,
%``Exponential Gravity,''
Phys.\ Rev.\ D {\bf 80}, 123528 (2009)
[arXiv:0905.2962 [astro-ph.CO]]; \\
%%CITATION = PHRVA,D80,123528;%%
%\cite{Yang:2010xq}
%\bibitem{Yang:2010xq}
L.~Yang, C.~C.~Lee, L.~W.~Luo and C.~Q.~Geng,
%``Observational Constraints on Exponential Gravity,''
arXiv:1010.2058 [astro-ph.CO].
%%CITATION = ARXIV:1010.2058;%%

%\cite{Nojiri:2010ny}
\bibitem{Nojiri:2010ny}
S.~Nojiri and S.~D.~Odintsov,
%``Non-singular modified gravity unifying inflation with
%late-time
%acceleration and universality of viscous ratio bound in F(R)
%theory,''
arXiv:1008.4275 [hep-th].
%%CITATION = ARXIV:1008.4275;%%

%\cite{Khoury:2003rn}
\bibitem{Khoury:2003rn}
J.~Khoury and A.~Weltman,
%``Chameleon Cosmology,''
Phys.\ Rev.\ D {\bf 69}, 044026 (2004)
[arXiv:astro-ph/0309411].
%%CITATION = PHRVA,D69,044026;%%

%\cite{Brax:2004qh}
\bibitem{Brax:2004qh}
P.~Brax, C.~van de Bruck, A.~C.~Davis, J.~Khoury and A.~Weltman,
%``Detecting dark energy in orbit: The cosmological chameleon,''
Phys.\ Rev.\ D {\bf 70} (2004) 123518
[arXiv:astro-ph/0408415].
%%CITATION = PHRVA,D70,123518;%%

%\cite{Nojiri:2005jg}
\bibitem{Nojiri:2005jg}
S.~Nojiri and S.~D.~Odintsov,
%``Modified Gauss-Bonnet theory as gravitational alternative for
%dark
%energy,''
Phys.\ Lett.\ B {\bf 631}, 1 (2005)
[arXiv:hep-th/0508049].
%%CITATION = PHLTA,B631,1;%%

%\cite{Nojiri:2005am}
\bibitem{Nojiri:2005am}
S.~Nojiri, S.~D.~Odintsov and O.~G.~Gorbunova,
%``Dark energy problem: From phantom theory to modified
%Gauss-Bonnet
%gravity,''
J.\ Phys.\ A {\bf 39}, 6627 (2006)
[arXiv:hep-th/0510183].
%%CITATION = JPAGB,A39,6627;%%

%\cite{Cognola:2006eg}
\bibitem{Cognola:2006eg}
G.~Cognola, E.~Elizalde, S.~Nojiri, S.~D.~Odintsov and
S.~Zerbini,
%``Dark energy in modified Gauss-Bonnet gravity: Late-time
%acceleration
%and
%the hierarchy problem,''
Phys.\ Rev.\ D {\bf 73}, 084007 (2006)
[arXiv:hep-th/0601008].
%%CITATION = PHRVA,D73,084007;%%
%\cite{Nojiri:2005vv}

\bibitem{Nojiri:2005vv}
S.~Nojiri, S.~D.~Odintsov and M.~Sasaki,
%``Gauss-Bonnet dark energy,''
Phys.\ Rev.\ D {\bf 71}, 123509 (2005)
[arXiv:hep-th/0504052].
%%CITATION = PHRVA,D71,123509;%%

%\cite{Sami:2005zc}
\bibitem{Sami:2005zc}
M.~Sami, A.~Toporensky, P.~V.~Tretjakov and S.~Tsujikawa,
%``The fate of (phantom) dark energy universe with string
%curvature
%corrections,''
Phys.\ Lett.\ B {\bf 619}, 193 (2005)
[arXiv:hep-th/0504154].
%%CITATION = PHLTA,B619,193;%%

%\cite{Calcagni:2005im}
\bibitem{Calcagni:2005im}
G.~Calcagni, S.~Tsujikawa and M.~Sami,
%``Dark energy and cosmological solutions in second-order string
%gravity,''
Class.\ Quant.\ Grav.\ {\bf 22}, 3977 (2005)
[arXiv:hep-th/0505193].
%%CITATION = CQGRD,22,3977;%%

\bibitem{1GB}
%\cite{Sadjadi:2010kp}
%\bibitem{Sadjadi:2010kp}
H.~M.~Sadjadi,
%``Cosmological entropy, and thermodynamics second law in $F(R,G)$
%gravity,''
arXiv:1009.2941 [gr-qc]; \\
%%CITATION = ARXIV:1009.2941;%%
%\cite{Myrzakulov:2010gt}
%\bibitem{Myrzakulov:2010gt}
R.~Myrzakulov, D.~Saez-Gomez and A.~Tureanu,
%``On the $\Lambda$CDM Universe in $f(G)$ gravity,''
arXiv:1009.0902 [gr-qc]; \\
%%CITATION = ARXIV:1009.0902;%%
%\cite{Sebastiani:2010nb}
%\bibitem{Sebastiani:2010nb}
L.~Sebastiani,
%``Finite-time singularities in modified $\mathcal{F}(R,G)$-gravity and
%singularity avoidance,''
arXiv:1008.3041 [gr-qc]; \\
%%CITATION = ARXIV:1008.3041;%%
%\cite{Middleton:2010bv}
%\bibitem{Middleton:2010bv}
J.~Middleton,
%``On The Existence Of Anisotropic Cosmological Models In Higher-Order
%Theories Of Gravity,''
arXiv:1007.4669 [gr-qc]; \\
%%CITATION = ARXIV:1007.4669;%%
%\cite{DeFelice:2010hg}
%\bibitem{DeFelice:2010hg}
A.~De Felice and T.~Tanaka,
%``Inevitable ghost and the degrees of freedom in f(R,G) gravity,''
Prog.\ Theor.\ Phys.\ {\bf 124}, 503 (2010)
[arXiv:1006.4399 [astro-ph.CO]]; \\
%%CITATION = PTPKA,124,503;%%
%\cite{Gorbunova:2010sf}
%\bibitem{Gorbunova:2010sf}
O.~Gorbunova and L.~Sebastiani,
%``Viscous Fluids and Gauss-Bonnet Modified Gravity,''
arXiv:1004.1505 [gr-qc]; \\
%%CITATION = ARXIV:1004.1505;%%
%\cite{Moldenhauer:2010zz}
%\bibitem{Moldenhauer:2010zz}
J.~Moldenhauer, M.~Ishak, J.~Thompson and D.~A.~Easson,
%``Supernova, Baryon Acoustic Oscillations, And Cmb Surface Distance
%Constraints On F (G) Higher Order Gravity Models,''
Phys.\ Rev.\ D {\bf 81}, 063514 (2010)
[arXiv:1004.2459 [astro-ph.CO]]; \\
%%CITATION = PHRVA,D81,063514;%%
%\cite{Saltas:2010ga}
%\bibitem{Saltas:2010ga}
I.~D.~Saltas and M.~Hindmarsh,
%``The dynamical equivalence of modified gravity revisited,''
arXiv:1002.1710 [gr-qc]; \\
%%CITATION = ARXIV:1002.1710;%%
%\cite{Elizalde:2010jx}
%\bibitem{Elizalde:2010jx}
E.~Elizalde, R.~Myrzakulov, V.~V.~Obukhov and D.~Saez-Gomez,
%``LambdaCDM epoch reconstruction from F(R,G) and modified Gauss-Bonnet
%gravities,''
Class.\ Quant.\ Grav.\ {\bf 27}, 095007 (2010)
[arXiv:1001.3636 [gr-qc]]; \\
%%CITATION = CQGRD,27,095007;%%
%\cite{DiCriscienzo:2009pb}
%\bibitem{DiCriscienzo:2009pb}
R.~Di Criscienzo and S.~Zerbini,
%``Functional Determinants in Higher Derivative Lagrangian Theories,''
J.\ Math.\ Phys.\ {\bf 50}, 103517 (2009)
[Erratum-ibid.\ {\bf 51}, 059901 (2010)]
[arXiv:0907.4265 [hep-th]]; \\
%%CITATION = JMAPA,50,103517;%%
%\cite{DeFelice:2009aj}
%\bibitem{DeFelice:2009aj}
A.~De Felice and S.~Tsujikawa,
%``Solar system constraints on f(G) gravity models,''
Phys.\ Rev.\ D {\bf 80}, 063516 (2009)
[arXiv:0907.1830 [hep-th]]; \\
%%CITATION = PHRVA,D80,063516;%%
%\cite{Goheer:2009qh}
%\bibitem{Goheer:2009qh}
N.~Goheer, R.~Goswami, P.~K.~S.~Dunsby and K.~Ananda,
%``On the co-existence of matter dominated and accelerating solutions
%in
%f(G)-gravity,''
Phys.\ Rev.\ D {\bf 79}, 121301 (2009)
[arXiv:0904.2559 [gr-qc]]; \\
%%CITATION = PHRVA,D79,121301;%%
%\cite{Zhou:2009cy}
%\bibitem{Zhou:2009cy}
S.~Y.~Zhou, E.~J.~Copeland and P.~M.~Saffin,
%``Cosmological Constraints on $f(G)$ Dark Energy Models,''
JCAP {\bf 0907}, 009 (2009)
[arXiv:0903.4610 [gr-qc]]; \\
%%CITATION = JCAPA,0907,009;%%
%\cite{Uddin:2009wp}
%\bibitem{Uddin:2009wp}
K.~Uddin, J.~E.~Lidsey and R.~Tavakol,
%``Cosmological scaling solutions in generalised Gauss-Bonnet gravity
%theories,''
Gen.\ Rel.\ Grav.\ {\bf 41}, 2725 (2009)
[arXiv:0903.0270 [gr-qc]]; \\
%%CITATION = GRGVA,41,2725;%%
%\cite{Bohmer:2009fc}
%\bibitem{Bohmer:2009fc}
C.~G.~Boehmer and F.~S.~N.~Lobo,
%``Stability of the Einstein static universe in modified Gauss-Bonnet
%gravity,''
Phys.\ Rev.\ D {\bf 79}, 067504 (2009)
[arXiv:0902.2982 [gr-qc]]; \\
%%CITATION = PHRVA,D79,067504;%%
%\cite{Alimohammadi:2008fq}
%\bibitem{Alimohammadi:2008fq}
M.~Alimohammadi and A.~Ghalee,
%``Remarks on generalized Gauss-Bonnet dark energy,''
Phys.\ Rev.\ D {\bf 79}, 063006 (2009)
[arXiv:0811.1286 [gr-qc]]; \\
%%CITATION = PHRVA,D79,063006;%%
%\cite{Davis:2007ta}
%\bibitem{Davis:2007ta}
S.~C.~Davis,
%``Solar System Constraints on f(G) Dark Energy,''
arXiv:0709.4453 [hep-th]; \\
%%CITATION = ARXIV:0709.4453;%%
%\cite{Li:2007jm}
%\bibitem{Li:2007jm}
B.~Li, J.~D.~Barrow and D.~F.~Mota,
%``The Cosmology of Modified Gauss-Bonnet Gravity,''
Phys.\ Rev.\ D {\bf 76}, 044027 (2007)
[arXiv:0705.3795 [gr-qc]]; \\
%%CITATION = PHRVA,D76,044027;%%
%\cite{Brevik:2006nh}
%\bibitem{Brevik:2006nh}
I.~H.~Brevik and J.~Quiroga,
%``Vanishing cosmological constant in modified Gauss-Bonnet gravity
%with
%conformal anomaly,''
Int.\ J.\ Mod.\ Phys.\ D {\bf 16}, 817 (2007)
[arXiv:gr-qc/0610044].
%%CITATION = IMPAE,D16,817;%%

%\cite{Boulware:1986dr}
\bibitem{Boulware:1986dr}
D.~G.~Boulware and S.~Deser,
%``Effective Gravity Theories With Dilatons,''
Phys.\ Lett.\ B {\bf 175}, 409 (1986).
%%CITATION = PHLTA,B175,409;%%

%\cite{Nojiri:2006je}
\bibitem{Nojiri:2006je}
S.~Nojiri, S.~D.~Odintsov and M.~Sami,
%``Dark energy cosmology from higher-order, string-inspired
%gravity
%and
%its
%reconstruction,''
Phys.\ Rev.\ D {\bf 74}, 046004 (2006)
[arXiv:hep-th/0605039].
%%CITATION = PHRVA,D74,046004;%%

\bibitem{sGB}
%\cite{Satoh:2010ep}
%\bibitem{Satoh:2010ep}
M.~Satoh,
%``Slow-roll Inflation with the Gauss-Bonnet and Chern-Simons
%Corrections,''
arXiv:1008.2724 [astro-ph.CO]; \\
%%CITATION = ARXIV:1008.2724;%%
%\cite{Guo:2010jr}
%\bibitem{Guo:2010jr}
Z.~K.~Guo and D.~J.~Schwarz,
%``Slow-roll inflation with a Gauss-Bonnet correction,''
Phys.\ Rev.\ D {\bf 81}, 123520 (2010)
[arXiv:1001.1897 [hep-th]]; \\
%%CITATION = PHRVA,D81,123520;%%
%\cite{Nozari:2009ay}
%\bibitem{Nozari:2009ay}
K.~Nozari, T.~Azizi and M.~R.~Setare,
%``Phantom-Like Behavior of a DGP-Inspired Scalar-Gauss-Bonnet
%Gravity,''
JCAP {\bf 0910}, 022 (2009)
[arXiv:0910.0611 [gr-qc]]; \\
%%CITATION = JCAPA,0910,022;%%
%\cite{Guo:2009uk}
%\bibitem{Guo:2009uk}
Z.~K.~Guo and D.~J.~Schwarz,
%``Power spectra from an inflaton coupled to the Gauss-Bonnet term,''
Phys.\ Rev.\ D {\bf 80}, 063523 (2009)
[arXiv:0907.0427 [hep-th]]; \\
%%CITATION = PHRVA,D80,063523;%%
%\cite{Sadeghi:2009zza}
%\bibitem{Sadeghi:2009zza}
J.~Sadeghi, M.~R.~Setare, A.~Banijamali and F.~Milani,
%``Crossing of the phantom divide using tachyon-Gauss-Bonnet gravity,''
Phys.\ Rev.\ D {\bf 79}, 123003 (2009); \\
%%CITATION = PHRVA,D79,123003;%%
%\cite{Koivisto:2009sd}
%\bibitem{Koivisto:2009sd}
T.~S.~Koivisto, D.~F.~Mota and C.~Pitrou,
%``Inflation from N-Forms and its stability,''
JHEP {\bf 0909}, 092 (2009)
[arXiv:0903.4158 [astro-ph.CO]]; \\
%%CITATION = JHEPA,0909,092;%%
%\cite{Gurses:2008zz}
%\bibitem{Gurses:2008zz}
M.~Gurses,
%``Some solutions of the Gauss-Bonnet gravity with scalar field in four
%dimensions,''
Gen.\ Rel.\ Grav.\ {\bf 40}, 1825 (2008); \\
%%CITATION = GRGVA,40,1825;%%
%\cite{Setare:2008hm}
%\bibitem{Setare:2008hm}
M.~R.~Setare and E.~N.~Saridakis,
%``Correspondence between Holographic and Gauss-Bonnet dark energy
%models,''
Phys.\ Lett.\ B {\bf 670}, 1 (2008)
[arXiv:0810.3296 [hep-th]]; \\
%%CITATION = PHLTA,B670,1;%%
%\cite{Paul:2008id}
%\bibitem{Paul:2008id}
B.~C.~Paul and S.~Ghose,
%``Emergent Universe Scenario in the Einstein-Gauss-Bonnet Gravity with
%Dilaton,''
Gen.\ Rel.\ Grav.\ {\bf 42}, 795 (2010)
[arXiv:0809.4131 [hep-th]]; \\
%%CITATION = GRGVA,42,795;%%
%\cite{Satoh:2008ck}
%\bibitem{Satoh:2008ck}
M.~Satoh and J.~Soda,
%``Higher Curvature Corrections to Primordial Fluctuations in Slow-roll
%Inflation,''
JCAP {\bf 0809}, 019 (2008)
[arXiv:0806.4594 [astro-ph]]; \\
%%CITATION = JCAPA,0809,019;%%
%\cite{Cai:2008ht}
%\bibitem{Cai:2008ht}
R.~G.~Cai, B.~Hu and S.~Koh,
%``Gauss-Bonnet Term on Vacuum Decay,''
Phys.\ Lett.\ B {\bf 671}, 181 (2009)
[arXiv:0806.2508 [hep-th]]; \\
%%CITATION = PHLTA,B671,181;%%
%\cite{Bazeia:2008ay}
%\bibitem{Bazeia:2008ay}
D.~Bazeia, R.~Menezes and A.~Y.~Petrov,
%``Liouville-like solutions in dilaton gravity with Gauss-Bonnet
%modifications,''
Eur.\ Phys.\ J.\ C {\bf 58}, 171 (2008)
[arXiv:0806.2299 [hep-th]]; \\
%%CITATION = EPHJA,C58,171;%%
%\cite{Chingangbam:2007yt}
%\bibitem{Chingangbam:2007yt}
R.~Chingangbam, M.~Sami, P.~V.~Tretyakov and A.~V.~Toporensky,
%``A Note on the Viability of Gauss-Bonnet Cosmology,''
Phys.\ Lett.\ B {\bf 661}, 162 (2008)
[arXiv:0711.2122 [hep-th]]; \\
%%CITATION = PHLTA,B661,162;%%
%\cite{Elizalde:2007kb}
%\bibitem{Elizalde:2007kb}
E.~Elizalde, J.~Q.~Hurtado and H.~I.~Arcos,
%``De Sitter cosmology from Gauss-Bonnet dark energy with quantum
%effects,''
Int.\ J.\ Mod.\ Phys.\ D {\bf 17}, 2159 (2008)
[arXiv:0708.0591 [gr-qc]]; \\
%%CITATION = IMPAE,D17,2159;%%
%\cite{Bamba:2007ef}
%\bibitem{Bamba:2007ef}
K.~Bamba, Z.~K.~Guo and N.~Ohta,
%``Accelerating Cosmologies in the Einstein-Gauss-Bonnet Theory with
%Dilaton,''
Prog.\ Theor.\ Phys.\ {\bf 118}, 879 (2007)
[arXiv:0707.4334 [hep-th]]; \\
%%CITATION = PTPKA,118,879;%%
%\cite{Nojiri:2007te}
%\bibitem{Nojiri:2007te}
S.~Nojiri, S.~D.~Odintsov and P.~V.~Tretyakov,
%``Dark energy from modified F(R)-scalar-Gauss-Bonnet gravity,''
Phys.\ Lett.\ B {\bf 651}, 224 (2007)
[arXiv:0704.2520 [hep-th]]; \\
%%CITATION = PHLTA,B651,224;%%
%\cite{Leith:2007bu}
%\bibitem{Leith:2007bu}
B.~M.~Leith and I.~P.~Neupane,
%``Gauss-Bonnet cosmologies: crossing the phantom divide and the
%transition
%from matter dominance to dark energy,''
JCAP {\bf 0705}, 019 (2007)
[arXiv:hep-th/0702002]; \\
%%CITATION = JCAPA,0705,019;%%
%\cite{Cognola:2006sp}
%\bibitem{Cognola:2006sp}
G.~Cognola, E.~Elizalde, S.~Nojiri, S.~Odintsov and S.~Zerbini,
%``String-inspired Gauss-Bonnet gravity reconstructed from the universe
%expansion history and yielding the transition from matter dominance to 
%dark
%energy,''
Phys.\ Rev.\ D {\bf 75}, 086002 (2007)
[arXiv:hep-th/0611198]; \\
%%CITATION = PHRVA,D75,086002;%%
%\cite{Sotiriou:2006pq}
%\bibitem{Sotiriou:2006pq}
T.~P.~Sotiriou and E.~Barausse,
%``Post-Newtonian expansion for Gauss-Bonnet Gravity,''
Phys.\ Rev.\ D {\bf 75}, 084007 (2007)
[arXiv:gr-qc/0612065]; \\
%%CITATION = PHRVA,D75,084007;%%
%\cite{Koivisto:2006ai}
%\bibitem{Koivisto:2006ai}
T.~Koivisto and D.~F.~Mota,
%``Gauss-Bonnet quintessence: Background evolution, large scale
%structure
%and
%cosmological constraints,''
Phys.\ Rev.\ D {\bf 75}, 023518 (2007)
[arXiv:hep-th/0609155]; \\
%%CITATION = PHRVA,D75,023518;%%
%\cite{Tsujikawa:2006ph}
%\bibitem{Tsujikawa:2006ph}
S.~Tsujikawa and M.~Sami,
%``String-inspired cosmology: Late time transition from scaling matter
%era
%to
%dark energy universe caused by a Gauss-Bonnet coupling,''
JCAP {\bf 0701}, 006 (2007)
[arXiv:hep-th/0608178]; \\
%%CITATION = JCAPA,0701,006;%%
%\cite{Sanyal:2006wi}
%\bibitem{Sanyal:2006wi}
A.~K.~Sanyal,
%``If Gauss-Bonnet interaction plays the role of dark energy,''
Phys.\ Lett.\ B {\bf 645}, 1 (2007)
[arXiv:astro-ph/0608104]; \\
%%CITATION = PHLTA,B645,1;%%
%\cite{Koivisto:2006xf}
%\bibitem{Koivisto:2006xf}
T.~Koivisto and D.~F.~Mota,
%``Cosmology and astrophysical constraints of Gauss-Bonnet dark
%energy,''
Phys.\ Lett.\ B {\bf 644}, 104 (2007)
[arXiv:astro-ph/0606078]; \\
%%CITATION = PHLTA,B644,104;%%
%\cite{Carter:2005fu}
%\bibitem{Carter:2005fu}
B.~M.~N.~Carter and I.~P.~Neupane,
%``Towards inflation and dark energy cosmologies from modified
%Gauss-Bonnet
%theory,''
JCAP {\bf 0606}, 004 (2006)
[arXiv:hep-th/0512262]; \\
%%CITATION = JCAPA,0606,004;%%
%\cite{Carter:2005sd}
%\bibitem{Carter:2005sd}
B.~M.~N.~Carter and I.~P.~Neupane,
%``Dynamical relaxation of dark energy: Solution to both the inflation
%and
%the cosmological constant,''
Phys.\ Lett.\ B {\bf 638}, 94 (2006)
[arXiv:hep-th/0510109]. 
%%CITATION = PHLTA,B638,94;%%

%\cite{Elizalde:2007pi}
\bibitem{Elizalde:2007pi}
E.~Elizalde, S.~Jhingan, S.~Nojiri, S.~D.~Odintsov, M.~Sami and
I.~Thongkool,
%``Dark energy generated from a (super)string effective action
%with
%higher
%order curvature corrections and a dynamical dilaton,''
Eur.\ Phys.\ J.\ C {\bf 53}, 447 (2008)
[arXiv:0705.1211 [hep-th]].
%%CITATION = EPHJA,C53,447;%%

%\cite{Deser:2007jk}
\bibitem{Deser:2007jk}
S.~Deser and R.~P.~Woodard,
%``Nonlocal Cosmology,''
Phys.\ Rev.\ Lett.\ {\bf 99}, 111301 (2007)
[arXiv:0706.2151 [astro-ph]]; \\
%%CITATION = PRLTA,99,111301;%%
%\cite{Deffayet:2009ca}
%\bibitem{Deffayet:2009ca}
C.~Deffayet and R.~P.~Woodard,
%``Reconstructing the Distortion Function for Nonlocal Cosmology,''
JCAP {\bf 0908}, 023 (2009)
[arXiv:0904.0961 [gr-qc]].
%%CITATION = JCAPA,0908,023;%%

%\cite{Nojiri:2007uq}
\bibitem{Nojiri:2007uq}
S.~Nojiri and S.~D.~Odintsov,
%``Modified non-local-F(R) gravity as the key for the inflation
%and
%dark
%energy,''
Phys.\ Lett.\ B {\bf 659}, 821 (2008)
[arXiv:0708.0924 [hep-th]].
%%CITATION = PHLTA,B659,821;%%

%\cite{Jhingan:2008ym}
\bibitem{Jhingan:2008ym}
S.~Jhingan, S.~Nojiri, S.~D.~Odintsov, M.~Sami, I.~Thongkool and
S.~Zerbini,
%``Phantom and non-phantom dark energy: The cosmological
%relevance
%of
%non-locally corrected gravity,''
Phys.\ Lett.\ B {\bf 663}, 424 (2008)
[arXiv:0803.2613 [hep-th]].
%%CITATION = PHLTA,B663,424;%%

\bibitem{nonlocal}
%\cite{Nojiri:2010pw}
%\bibitem{Nojiri:2010pw}
S.~Nojiri, S.~D.~Odintsov, M.~Sasaki and Y.~l.~Zhang,
%``Screening of cosmological constant in non-local gravity,''
arXiv:1010.5375 [gr-qc]; \\
%%CITATION = ARXIV:1010.5375;%%
%\cite{Koivisto:2008xfa}
%\bibitem{Koivisto:2008xfa}
T.~Koivisto,
%``Dynamics of Nonlocal Cosmology,''
Phys.\ Rev.\ D {\bf 77}, 123513 (2008)
[arXiv:0803.3399 [gr-qc]]; \\
%%CITATION = PHRVA,D77,123513;%%
%\cite{Koivisto:2008dh}
%\bibitem{Koivisto:2008dh}
T.~S.~Koivisto,
%``Newtonian limit of nonlocal cosmology,''
Phys.\ Rev.\ D {\bf 78}, 123505 (2008)
[arXiv:0807.3778 [gr-qc]]; \\
%%CITATION = PHRVA,D78,123505;%%
%\cite{Koshelev:2008ie}
%\bibitem{Koshelev:2008ie}
N.~A.~Koshelev,
%``Comments on scalar-tensor representation of nonlocally corrected
%gravity,''
Grav.\ Cosmol.\ {\bf 15}, 220 (2009)
[arXiv:0809.4927 [gr-qc]]; \\
%%CITATION = GRCOF,15,220;%%
%\cite{Nesseris:2009jf}
%\bibitem{Nesseris:2009jf}
S.~Nesseris and A.~Mazumdar,
%``Newton's constant in f(R,R_{\mu\nu}R^{\mu\nu},\Box R) theories of
%gravity
%and constraints from BBN,''
Phys.\ Rev.\ D {\bf 79}, 104006 (2009)
[arXiv:0902.1185 [astro-ph.CO]]; \\
%%CITATION = PHRVA,D79,104006;%%
%\cite{Bronnikov:2009az}
%\bibitem{Bronnikov:2009az}
K.~A.~Bronnikov and E.~Elizalde,
%``Spherical systems in models of nonlocally corrected gravity,''
Phys.\ Rev.\ D {\bf 81}, 044032 (2010)
[arXiv:0910.3929 [hep-th]]; \\
%%CITATION = PHRVA,D81,044032;%%
%\cite{Calcagni:2010ab}
%\bibitem{Calcagni:2010ab}
G.~Calcagni and G.~Nardelli,
%``Non-local gravity and the diffusion equation,''
arXiv:1004.5144 [hep-th].
%%CITATION = ARXIV:1004.5144;%%

%\cite{Capozziello:2008gu}
\bibitem{Capozziello:2008gu}
S.~Capozziello, E.~Elizalde, S.~Nojiri and S.~D.~Odintsov,
%``Accelerating cosmologies from non-local higher-derivative
%gravity,''
Phys.\ Lett.\ B {\bf 671} (2009) 193
[arXiv:0809.1535 [hep-th]].
%%CITATION = PHLTA,B671,193;%%

%\cite{Cognola:2009jx}
\bibitem{Cognola:2009jx}
G.~Cognola, E.~Elizalde, S.~Nojiri, S.~D.~Odintsov and
S.~Zerbini,
%``One-loop effective action for non-local modified Gauss-Bonnet
%gravity
%in
%de
%Sitter space,''
Eur.\ Phys.\ J.\ C {\bf 64}, 483 (2009)
[arXiv:0905.0543 [gr-qc]].
%%CITATION = EPHJA,C64,483;%%

\bibitem{Nojiri:2004bi}
S.~Nojiri and S.~D.~Odintsov,
%``Gravity assisted dark energy dominance and cosmic
%acceleration,''
Phys.\ Lett.\ B {\bf 599}, 137 (2004)
[arXiv:astro-ph/0403622].
%%CITATION = PHLTA,B599,137;%%

%\cite{Allemandi:2005qs}
\bibitem{Allemandi:2005qs}
G.~Allemandi, A.~Borowiec, M.~Francaviglia and S.~D.~Odintsov,
%``Dark energy dominance and cosmic acceleration in first order
%formalism,''
Phys.\ Rev.\ D {\bf 72}, 063505 (2005)
[arXiv:gr-qc/0504057].
%%CITATION = PHRVA,D72,063505;%%

%\cite{Capozziello:1999xt}
\bibitem{Capozziello:1999xt}
S.~Capozziello, G.~Lambiase and H.~J.~Schmidt,
%``Nonminimal derivative couplings and inflation in generalized
%theories
%of
%gravity,''
Annalen Phys.\ {\bf 9}, 39 (2000)
[arXiv:gr-qc/9906051].
%%CITATION = ANPYA,9,39;%%

\bibitem{nonlinear}
%\cite{Harko:2008qz}
%\bibitem{Harko:2008qz}
T.~Harko,
%``Modified gravity with arbitrary coupling between matter and
%geometry,''
Phys.\ Lett.\ B {\bf 669}, 376 (2008)
[arXiv:0810.0742 [gr-qc]]; \\
%%CITATION = PHLTA,B669,376;%%
%\cite{Harko:2010zi}
%\bibitem{Harko:2010zi}
T.~Harko,
%``The matter Lagrangian and the energy-momentum tensor in modified
%gravity
%with non-minimal coupling between matter and geometry,''
Phys.\ Rev.\ D {\bf 81}, 044021 (2010)
[arXiv:1001.5349 [gr-qc]]; \\
%%CITATION = PHRVA,D81,044021;%%
%\cite{Granda:2010hb}
%\bibitem{Granda:2010hb}
L.~N.~Granda and W.~Cardona,
%``General Non-minimal Kinetic coupling to gravity,''
JCAP {\bf 1007}, 021 (2010)
[arXiv:1005.2716 [hep-th]]; \\
%%CITATION = JCAPA,1007,021;%%
%\cite{Harko:2010mv}
%\bibitem{Harko:2010mv}
T.~Harko and F.~S.~N.~Lobo,
%``$f\left(R,L_m\right)$ gravity,''
arXiv:1008.4193 [gr-qc]; \\
%%CITATION = ARXIV:1008.4193;%%
%\cite{Harko:2011kv}
%\bibitem{Harko:2011kv}
T.~Harko, F.~S.~N.~Lobo, S.~Nojiri and S.~D.~Odintsov,
%``$f(R,T)$ gravity,''
arXiv:1104.2669 [gr-qc].
%%CITATION = ARXIV:1104.2669;%%




\bibitem{extra}
%\cite{Faraoni:2009rk}
%\bibitem{Faraoni:2009rk}
V.~Faraoni,
%``The Lagrangian description of perfect fluids and modified gravity
%with
%an
%extra force,''
Phys.\ Rev.\ D {\bf 80}, 124040 (2009)
[arXiv:0912.1249 [astro-ph.GA]]; \\
%%CITATION = PHRVA,D80,124040;%%
%\cite{Bertolami:2007gv}
%\bibitem{Bertolami:2007gv}
O.~Bertolami, C.~G.~Boehmer, T.~Harko and F.~S.~N.~Lobo,
%``Extra force in $f(R)$ modified theories of gravity,''
Phys.\ Rev.\ D {\bf 75}, 104016 (2007)
[arXiv:0704.1733 [gr-qc]].
%%CITATION = PHRVA,D75,104016;%%

\bibitem{nonlinear1}
%\cite{Harko:2010hw}
%\bibitem{Harko:2010hw}
T.~Harko and F.~S.~N.~Lobo,
%``Palatini formulation of modified gravity with a nonminimal
%curvature-matter
%coupling,''
arXiv:1007.4415 [gr-qc]; \\
%%CITATION = ARXIV:1007.4415;%%
%\cite{Deffayet:2010qz}
%\bibitem{Deffayet:2010qz}
C.~Deffayet, O.~Pujolas, I.~Sawicki and A.~Vikman,
%``Imperfect Dark Energy from Kinetic Gravity Braiding,''
arXiv:1008.0048 [hep-th]; \\
%%CITATION = ARXIV:1008.0048;%%
%\cite{Mohseni:2009ns}
%\bibitem{Mohseni:2009ns}
M.~Mohseni,
%``Non-geodesic motion in $f({\mathcal G})$ gravity with non-minimal
%coupling,''
Phys.\ Lett.\ B {\bf 682}, 89 (2009)
[arXiv:0911.2754 [hep-th]]; \\
%%CITATION = PHLTA,B682,89;%%
%\cite{Granda:2009fh}
%\bibitem{Granda:2009fh}
L.~N.~Granda,
%``Non-minimal Kinetic coupling to gravity and accelerated
%expansion,''
JCAP {\bf 1007}, 006 (2010)
[arXiv:0911.3702 [hep-th]]; \\
%%CITATION = JCAPA,1007,006;%%
%\cite{Gao:2010vr}
%\bibitem{Gao:2010vr}
C.~Gao,
%``When scalar field is kinetically coupled to the Einstein
%tensor,''
JCAP {\bf 1006}, 023 (2010)
[arXiv:1002.4035 [gr-qc]]; \\
%%CITATION = JCAPA,1006,023;%%
%\cite{Cognola:2009za}
%\bibitem{Cognola:2009za}
G.~Cognola, E.~Elizalde, S.~Nojiri and S.~D.~Odintsov,
%``Vacuum energy fluctuations, the induced cosmological constant and
%cosmological reconstruction in non-minimal modified gravity
%models,''
Open Astron.\ J.\ {\bf 3}, 20 (2010)
[arXiv:0909.2747 [gr-qc]]; \\
%%CITATION = 00695,3,20;%%
%\cite{Sadeghi:2009pu}
%\bibitem{Sadeghi:2009pu}
J.~Sadeghi, M.~R.~Setare and A.~Banijamali,
%``Non-minimal Maxwell-Modified Gauss-Bonnet Cosmologies: Inflation
%and
%Dark
%Energy,''
Eur.\ Phys.\ J.\ C {\bf 64}, 433 (2009)
[arXiv:0906.0713 [hep-th]]; \\
%%CITATION = EPHJA,C64,433;%%
%\cite{Borowiec:2008js}
%\bibitem{Borowiec:2008js}
A.~Borowiec,
%``From Dark Energy to Dark Matter via Non-Minimal Coupling,''
arXiv:0812.4383 [gr-qc]; \\
%%CITATION = ARXIV:0812.4383;%%
%\cite{Nesseris:2008mq}
%\bibitem{Nesseris:2008mq}
S.~Nesseris,
%``Matter density perturbations in modified gravity models with
%arbitrary
%coupling between matter and geometry,''
Phys.\ Rev.\ D {\bf 79}, 044015 (2009)
[arXiv:0811.4292 [astro-ph]]; \\
%%CITATION = PHRVA,D79,044015;%%
%\cite{Koivisto:2008ak}
%\bibitem{Koivisto:2008ak}
T.~S.~Koivisto,
%``Disformal quintessence,''
arXiv:0811.1957 [astro-ph]; \\
%%CITATION = ARXIV:0811.1957;%%
%\cite{Puetzfeld:2008xu}
%\bibitem{Puetzfeld:2008xu}
D.~Puetzfeld and Y.~N.~Obukhov,
%``On the motion of test bodies in theories with non-minimal
%coupling,''
Phys.\ Rev.\ D {\bf 78}, 121501 (2008)
[arXiv:0811.0913 [astro-ph]]; \\
%%CITATION = PHRVA,D78,121501;%%
%\cite{Sotiriou:2008it}
%\bibitem{Sotiriou:2008it}
T.~P.~Sotiriou and V.~Faraoni,
%``Modified gravity with R-matter couplings and (non-)geodesic
%motion,''
Class.\ Quant.\ Grav.\ {\bf 25}, 205002 (2008)
[arXiv:0805.1249 [gr-qc]]; \\
%%CITATION = CQGRD,25,205002;%%
%\cite{Sotiriou:2008dh}
%\bibitem{Sotiriou:2008dh}
T.~P.~Sotiriou,
%``The viability of theories with matter coupled to the Ricci
%scalar,''
Phys.\ Lett.\ B {\bf 664}, 225 (2008)
[arXiv:0805.1160 [gr-qc]]; \\
%%CITATION = PHLTA,B664,225;%%
%\cite{Barr:2006mp}
%\bibitem{Barr:2006mp}
S.~M.~Barr, S.~P.~Ng and R.~J.~Scherrer,
%``Classical cancellation of the cosmological constant
%re-considered,''
Phys.\ Rev.\ D {\bf 73}, 063530 (2006)
[arXiv:hep-ph/0601053]; \\
%%CITATION = PHRVA,D73,063530;%%
%\cite{Inagaki:2005qp}
%\bibitem{Inagaki:2005qp}
T.~Inagaki, S.~Nojiri and S.~D.~Odintsov,
%``The one-loop effective action in phi**4 theory coupled
%non-linearly
%with
%curvature power and dynamical origin of cosmological constant,''
JCAP {\bf 0506}, 010 (2005)
[arXiv:gr-qc/0504054]; \\
%%CITATION = JCAPA,0506,010;%%
%\cite{Bertolami:2008zh}
%\bibitem{Bertolami:2008zh}
O.~Bertolami, J.~Paramos, T.~Harko and F.~S.~N.~Lobo,
%``Non-minimal curvature-matter couplings in modified gravity,''
arXiv:0811.2876 [gr-qc]; \\
%%CITATION = ARXIV:0811.2876;%%
%\cite{Bertolami:2008im}
%\bibitem{Bertolami:2008im}
O.~Bertolami and J.~Paramos,
%``On the non-trivial gravitational coupling to matter,''
Class.\ Quant.\ Grav.\ {\bf 25}, 245017 (2008)
[arXiv:0805.1241 [gr-qc]].
%%CITATION = CQGRD,25,245017;%%

%\cite{Bamba:2008ja}
\bibitem{Bamba:2008ja}
K.~Bamba and S.~D.~Odintsov,
%``Inflation and late-time cosmic acceleration in non-minimal
%Maxwell-$F(R)$
%gravity and the generation of large-scale magnetic fields,''
JCAP {\bf 0804}, 024 (2008)
[arXiv:0801.0954 [astro-ph]].
%%CITATION = JCAPA,0804,024;%%

\bibitem{bal}
%\cite{Balakin:2009rg}
%\bibitem{Balakin:2009rg}
A.~B.~Balakin and W.~T.~Ni,
%``Non-minimal coupling of photons and axions,''
Class.\ Quant.\ Grav.\ {\bf 27}, 055003 (2010)
[arXiv:0911.2946 [gr-qc]]; \\
%%CITATION = CQGRD,27,055003;%%
%\cite{MosqueraCuesta:2009tf}
%\bibitem{MosqueraCuesta:2009tf}
H.~J.~Mosquera Cuesta and G.~Lambiase,
%``Primordial magnetic fields and gravitational baryogenesis in
%nonlinear
%electrodynamics,''
Phys.\ Rev.\ D {\bf 80}, 023013 (2009)
[arXiv:0907.3678 [astro-ph.CO]]; \\
%%CITATION = PHRVA,D80,023013;%%
%\cite{Balakin:2007am}
%\bibitem{Balakin:2007am}
A.~B.~Balakin, V.~V.~Bochkarev and J.~P.~S.~Lemos,
%``Non-minimal coupling for the gravitational and electromagnetic
%fields:
%black hole solutions and solitons,''
Phys.\ Rev.\ D {\bf 77}, 084013 (2008)
[arXiv:0712.4066 [gr-qc]]; \\
%%CITATION = PHRVA,D77,084013;%%
%\cite{Balakin:2005fu}
%\bibitem{Balakin:2005fu}
A.~B.~Balakin and J.~P.~S.~Lemos,
%``Non-minimal coupling for the gravitational and electromagnetic
%fields:
%A
%general system of equations,''
Class.\ Quant.\ Grav.\ {\bf 22}, 1867 (2005)
[arXiv:gr-qc/0503076]; \\
%%CITATION = CQGRD,22,1867;%%
%\cite{Jimenez:2009dt}
%\bibitem{Jimenez:2009dt}
J.~B.~Jimenez and A.~L.~Maroto,
%``The electromagnetic dark sector,''
Phys.\ Lett.\ B {\bf 686}, 175 (2010)
[arXiv:0903.4672 [astro-ph.CO]].
%%CITATION = PHLTA,B686,175;%%

%\cite{Bamba:2008xa}
\bibitem{Bamba:2008xa}
K.~Bamba, S.~Nojiri and S.~D.~Odintsov,
%``Inflationary cosmology and the late-time accelerated expansion
%of
%the
%universe in non-minimal Yang-Mills-$F(R)$ gravity and
%non-minimal
%vector-$F(R)$ gravity,''
Phys.\ Rev.\ D {\bf 77}, 123532 (2008)
[arXiv:0803.3384 [hep-th]].
%%CITATION = PHRVA,D77,123532;%%

\bibitem{bal1}
%\cite{Zhao:2009wy}
%\bibitem{Zhao:2009wy}
W.~Zhao, Y.~Zhang and M.~L.~Tong,
%``Quantum Yang-Mills Condensate Dark Energy Models,''
arXiv:0909.3874 [astro-ph.CO]; \\
%%CITATION = ARXIV:0909.3874;%%
%\cite{Himmetoglu:2009qi}
%\bibitem{Himmetoglu:2009qi}
B.~Himmetoglu, C.~R.~Contaldi and M.~Peloso,
%``Ghost instabilities of cosmological models with vector fields
%nonminimally
%coupled to the curvature,''
Phys.\ Rev.\ D {\bf 80}, 123530 (2009)
[arXiv:0909.3524 [astro-ph.CO]]; \\
%%CITATION = PHRVA,D80,123530;%%
%\cite{Balakin:2008cx}
%\bibitem{Balakin:2008cx}
A.~B.~Balakin, H.~Dehnen and A.~E.~Zayats,
%``Effective metrics in the non-minimal Einstein-Yang-Mills-Higgs
%theory,''
Annals Phys.\ {\bf 323}, 2183 (2008)
[arXiv:0804.2196 [gr-qc]]; \\
%%CITATION = APNYA,323,2183;%%
%\cite{Balakin:2007mp}
%\bibitem{Balakin:2007mp}
A.~B.~Balakin, H.~Dehnen and A.~E.~Zayats,
%``Non-minimal isotropic cosmological model with Yang-Mills and
%Higgs
%fields,''
Int.\ J.\ Mod.\ Phys.\ D {\bf 17}, 1255 (2008)
[arXiv:0710.4992 [gr-qc]]; \\
%%CITATION = IMPAE,D17,1255;%%
%\cite{Elizalde:2003ku}
%\bibitem{Elizalde:2003ku}
E.~Elizalde, J.~E.~Lidsey, S.~Nojiri and S.~D.~Odintsov,
%``Born-Infeld quantum condensate as dark energy in the universe,''
Phys.\ Lett.\ B {\bf 574}, 1 (2003)
[arXiv:hep-th/0307177].
%%CITATION = PHLTA,B574,1;%%

%\cite{Carloni:2010nx}
\bibitem{Carloni:2010nx}
%\cite{Chaichian:2010yi}
%\bibitem{Chaichian:2010yi}
M.~Chaichian, S.~Nojiri, S.~D.~Odintsov, M.~Oksanen and A.~Tureanu,
%``Modified F(R) Horava-Lifshitz gravity: a way to accelerating FRW
%cosmology,''
Class.\ Quant.\ Grav.\ {\bf 27}, 185021 (2010)
[arXiv:1001.4102 [hep-th]]; \\
%%CITATION = CQGRD,27,185021;%%
%\cite{Carloni:2010nx}
%\bibitem{Carloni:2010nx}
S.~Carloni, M.~Chaichian, S.~Nojiri, S.~D.~Odintsov, M.~Oksanen
and A.~Tureanu,
%``Modified first-order Horava-Lifshitz gravity: Hamiltonian
%analysis
%of
%the
%general theory and accelerating FRW cosmology in power-law F(R)
%model,''
arXiv:1003.3925 [hep-th].
%%CITATION = ARXIV:1003.3925;%%

%\cite{Arnowitt:1962hi}
\bibitem{Arnowitt:1962hi}
R.~L.~Arnowitt, S.~Deser and C.~W.~Misner,
%``The dynamics of general relativity,''
arXiv:gr-qc/0405109, originally
``Gravitation: An Introduction to Current Research'',
L. Witten ed., Wiley, New York, 1962.
%%CITATION = GR-QC/0405109;%%

%\cite{Gao:2009er}
\bibitem{Gao:2009er}
C.~Gao,
%``Modified gravity in Arnowitt-Deser-Misner formalism,''
arXiv:0905.0310 [astro-ph.CO].
%%CITATION = ARXIV:0905.0310;%%

\bibitem{Wald:1984}
R.~M.~Wald, \emph{General Relativity},
University of Chicago Press, 1984, Chicago and London.

\bibitem{Gourgoulhon:2007}
\'E.~Gourgoulhon,
%``3+1 Formalism and Bases of Numerical Relativity,''
arXiv:gr-qc/0703035.

%\cite{Horava:2009uw}
\bibitem{Horava:2009uw}
P.~Ho\v{r}ava,
%``Quantum Gravity at a Lifshitz Point,''
Phys.\ Rev.\ D {\bf 79}, 084008 (2009)
[arXiv:0901.3775 [hep-th]].
%%CITATION = PHRVA,D79,084008;%%

%\cite{Kluson:2009xx}
\bibitem{Kluson:2009xx}
J.~Kluso\v{n},
%``New Models of f(R) Theories of Gravity,''
arXiv:0910.5852 [hep-th]; \\
%%CITATION = ARXIV:0910.5852;%%
%\cite{Kluson:2009rk}
%\bibitem{Kluson:2009rk}
J.~Kluson,
%``Horava-Lifshitz f(R) Gravity,''
JHEP {\bf 0911}, 078 (2009)
[arXiv:0907.3566 [hep-th]].
%%CITATION = JHEPA,0911,078;%%

%\cite{Henneaux:2009zb}
\bibitem{Henneaux:2009zb}
M.~Henneaux, A.~Kleinschmidt and G.~L.~Gomez,
%``A dynamical inconsistency of Horava gravity,''
arXiv:0912.0399 [hep-th].
%%CITATION = ARXIV:0912.0399;%%

\bibitem{Li:2009}
M.~Li and Y.~Pang,
%``A Trouble with Ho\v{r}ava-Lifshitz Gravity,''
JHEP {\bf 0908}, 015 (2009)
[arXiv:0905.2751 [hep-th]].

%\cite{Mukohyama:2009mz}
\bibitem{Mukohyama:2009mz}
S.~Mukohyama,
%``Dark matter as integration constant in Horava-Lifshitz
%gravity,''
Phys.\ Rev.\ D {\bf 80}, 064005 (2009)
[arXiv:0905.3563 [hep-th]].
%%CITATION = PHRVA,D80,064005;%%

%\cite{Elizalde:2010ep}
\bibitem{Elizalde:2010ep}
E.~Elizalde, S.~Nojiri, S.~D.~Odintsov and D.~Saez-Gomez,
%``Unifying inflation with dark energy in modified F(R)
%Horava-Lifshitz
%gravity,''
arXiv:1006.3387 [hep-th].
%%CITATION = ARXIV:1006.3387;%%

%\cite{Takahashi:2009wc}
\bibitem{Takahashi:2009wc}
T.~Takahashi and J.~Soda,
%``Chiral Primordial Gravitational Waves from a Lifshitz
%Point,''
Phys.\ Rev.\ Lett.\ {\bf 102}, 231301 (2009)
[arXiv:0904.0554 [hep-th]].
%%CITATION = PRLTA,102,231301;%%

%\cite{Kiritsis:2009sh}
\bibitem{Kiritsis:2009sh}
E.~Kiritsis and G.~Kofinas,
%``Horava-Lifshitz Cosmology,''
Nucl.\ Phys.\ B {\bf 821}, 467 (2009)
[arXiv:0904.1334 [hep-th]].
%%CITATION = NUPHA,B821,467;%%

%\cite{Brandenberger:2009yt}
\bibitem{Brandenberger:2009yt}
R.~Brandenberger,
%``Matter Bounce in Horava-Lifshitz Cosmology,''
Phys.\ Rev.\ D {\bf 80}, 043516 (2009)
[arXiv:0904.2835 [hep-th]].
%%CITATION = PHRVA,D80,043516;%%

%\cite{Mukohyama:2009zs}
\bibitem{Mukohyama:2009zs}
S.~Mukohyama, K.~Nakayama, F.~Takahashi and S.~Yokoyama,
%``Phenomenological Aspects of Horava-Lifshitz Cosmology,''
Phys.\ Lett.\ B {\bf 679}, 6 (2009)
[arXiv:0905.0055 [hep-th]].
%%CITATION = PHLTA,B679,6;%%

%\cite{Sotiriou:2009bx}
\bibitem{Sotiriou:2009bx}
T.~P.~Sotiriou, M.~Visser and S.~Weinfurtner,
%``Quantum gravity without Lorentz invariance,''
JHEP {\bf 0910}, 033 (2009)
[arXiv:0905.2798 [hep-th]].
%%CITATION = JHEPA,0910,033;%%

%\cite{Saridakis:2009bv}
\bibitem{Saridakis:2009bv}
E.~N.~Saridakis,
%``Horava-Lifshitz Dark Energy,''
Eur.\ Phys.\ J.\ C {\bf 67}, 229 (2010)
[arXiv:0905.3532 [hep-th]].
%%CITATION = EPHJA,C67,229;%%

%\cite{Minamitsuji:2009ii}
\bibitem{Minamitsuji:2009ii}
M.~Minamitsuji,
%``Classification of cosmology with arbitrary matter in the
%Ho\v{r}ava-Lifshitz theory,''
Phys.\ Lett.\ B {\bf 684}, 194 (2010)
[arXiv:0905.3892 [astro-ph.CO]].
%%CITATION = PHLTA,B684,194;%%

%\cite{Calcagni:2009qw}
\bibitem{Calcagni:2009qw}
G.~Calcagni,
%``Detailed balance in Horava-Lifshitz gravity,''
Phys.\ Rev.\ D {\bf 81}, 044006 (2010)
[arXiv:0905.3740 [hep-th]].
%%CITATION = PHRVA,D81,044006;%%

%\cite{Wang:2009rw}
\bibitem{Wang:2009rw}
A.~Wang and Y.~Wu,
%``Thermodynamics and classification of cosmological models in
%the
%Horava-Lifshitz theory of gravity,''
JCAP {\bf 0907}, 012 (2009)
[arXiv:0905.4117 [hep-th]].
%%CITATION = JCAPA,0907,012;%%

\bibitem{Park:2009zra}
M.~i.~Park,
%``The Black Hole and Cosmological Solutions in IR modified
%Horava
%Gravity,''
JHEP {\bf 0909}, 123 (2009)
[arXiv:0905.4480 [hep-th]].
%%CITATION = JHEPA,0909,123;%%

%\cite{Nojiri:2009th}
\bibitem{Nojiri:2009th}
S.~Nojiri and S.~D.~Odintsov,
%``Covariant Horava-like renormalizable gravity and its FRW
%cosmology,''
Phys.\ Rev.\ D {\bf 81}, 043001 (2010)
[arXiv:0905.4213 [hep-th]].
%%CITATION = PHRVA,D81,043001;%%

%\cite{Jamil:2009sq}
\bibitem{Jamil:2009sq}
M.~Jamil, E.~N.~Saridakis and M.~R.~Setare,
%``Holographic dark energy with varying gravitational
%constant,''
Phys.\ Lett.\ B {\bf 679}, 172 (2009)
[arXiv:0906.2847 [hep-th]].
%%CITATION = PHLTA,B679,172;%%

%\cite{Bogdanos:2009uj}
\bibitem{Bogdanos:2009uj}
C.~Bogdanos and E.~N.~Saridakis,
%``Perturbative instabilities in Horava gravity,''
Class.\ Quant.\ Grav.\ {\bf 27}, 075005 (2010)
[arXiv:0907.1636 [hep-th]].
%%CITATION = CQGRD,27,075005;%%

%\cite{Boehmer:2009yz}
\bibitem{Boehmer:2009yz}
C.~G.~Boehmer and F.~S.~N.~Lobo,
%``Stability of the Einstein static universe in IR modified
%Ho\v{r}ava
%gravity,''
arXiv:0909.3986 [gr-qc].
%%CITATION = ARXIV:0909.3986;%%

%\cite{Bakas:2009ku}
\bibitem{Bakas:2009ku}
I.~Bakas, F.~Bourliot, D.~Lust and M.~Petropoulos,
%``Mixmaster universe in Horava-Lifshitz gravity,''
Class.\ Quant.\ Grav.\ {\bf 27}, 045013 (2010)
[arXiv:0911.2665 [hep-th]].
%%CITATION = CQGRD,27,045013;%%

%\cite{Calcagni:2009ar}
\bibitem{Calcagni:2009ar}
G.~Calcagni,
%``Cosmology of the Lifshitz universe,''
JHEP {\bf 0909}, 112 (2009)
[arXiv:0904.0829 [hep-th]].
%%CITATION = JHEPA,0909,112;%%

%\cite{Carloni:2009jc}
\bibitem{Carloni:2009jc}
S.~Carloni, E.~Elizalde and P.~J.~Silva,
%``An analysis of the phase space of Horava-Lifshitz
%cosmologies,''
Class.\ Quant.\ Grav.\ {\bf 27}, 045004 (2010)
[arXiv:0909.2219 [hep-th]].
%%CITATION = CQGRD,27,045004;%%

%\cite{Gao:2009wn}
\bibitem{Gao:2009wn}
X.~Gao, Y.~Wang, W.~Xue and R.~Brandenberger,
%``Fluctuations in a Ho\v{r}ava-Lifshitz Bouncing Cosmology,''
JCAP {\bf 1002}, 020 (2010)
[arXiv:0911.3196 [hep-th]].
%%CITATION = JCAPA,1002,020;%%

%\cite{Myung:2009if}
\bibitem{Myung:2009if}
Y.~S.~Myung, Y.~W.~Kim, W.~S.~Son and Y.~J.~Park,
%``Chaotic universe in the z=2 Hovava-Lifshitz gravity,''
arXiv:0911.2525 [gr-qc].
%%CITATION = ARXIV:0911.2525;%%

%\cite{Son:2010qh}
\bibitem{Son:2010qh}
E.~J.~Son and W.~Kim,
%``Smooth cosmological phase transition in the Horava-Lifshitz
%gravity,''
arXiv:1003.3055 [hep-th].
%%CITATION = ARXIV:1003.3055;%%

%\cite{Wang:2010mw}
\bibitem{Wang:2010mw}
A.~Wang,
%``$f(R)$ theory and geometric origin of the dark sector in
%Horava-Lifshitz
%gravity,''
arXiv:1003.5152 [hep-th].
%%CITATION = ARXIV:1003.5152;%%

%\cite{Ali:2010sv}
\bibitem{Ali:2010sv}
A.~Ali, S.~Dutta, E.~N.~Saridakis and A.~A.~Sen,
%``Horava-Lifshitz cosmology with generalized Chaplygin gas,''
arXiv:1004.2474 [astro-ph.CO].
%%CITATION = ARXIV:1004.2474;%%


%\cite{Gong:2010xp}
\bibitem{Gong:2010xp}
J.~O.~Gong, S.~Koh and M.~Sasaki,
%``A complete analysis of linear cosmological perturbations in
%Ho\v{r}ava-Lifshitz gravity,''
Phys.\ Rev.\ D {\bf 81}, 084053 (2010)
[arXiv:1002.1429 [hep-th]].
%%CITATION = PHRVA,D81,084053;%%

%\cite{Elizalde:2009gx}
\bibitem{Elizalde:2009gx}
E.~Elizalde and D.~Saez-Gomez,
%``F(R) cosmology in presence of a phantom fluid and its
%scalar-tensor
%counterpart:towards a unified precision model of the universe
%evolution,''
Phys.\ Rev.\ D {\bf 80}, 044030 (2009)
[arXiv:0903.2732 [hep-th]].
%%CITATION = PHRVA,D80,044030;%%
%E.~Elizalde, D.~S\'aez-G\'omez, Phys. Rev. D \textbf{80}
%044030 (2009) [arxiv:0903.2732]

%\cite{Nojiri:2010tv}
\bibitem{Nojiri:2010tv}
S.~Nojiri and S.~D.~Odintsov,
%``A proposal for covariant renormalizable field theory of
%gravity,''
Phys.\ Lett.\ B {\bf 691}, 60 (2010)
[arXiv:1004.3613 [hep-th]].
%%CITATION = PHLTA,B691,60;%%

%\cite{Nojiri:2010kx}
\bibitem{Nojiri:2010kx}
S.~Nojiri and S.~D.~Odintsov,
%``Covariant power-counting renormalizable gravity: Lorentz
%symmetry
%breaking
%and accelerating early-time FRW universe,''
arXiv:1007.4856 [hep-th].
%%CITATION = ARXIV:1007.4856;%%

%\cite{Kluson:2011rs}
\bibitem{Kluson:2011rs}
J.~Kluson, S.~Nojiri and S.~D.~Odintsov,
%``Covariant Lagrange multiplier constrained higher derivative gravity with
%scalar projectors,''
arXiv:1104.4286 [hep-th].
%%CITATION = ARXIV:1104.4286;%%





\bibitem{masud}
M.~Chaichian and A.~Demichev,
{\it Path Integrals in Physics}, Vol. I and II,
Institute of Physics Publishing, 2001, Bristol
and Philadelphia; \\
M.~Henneaux and C.~Teitelboim, {\it Quantization of Gauge Systems},
Princeton University Press, 1994, Princeton, New
Jersey.

%\cite{Nojiri:2005sr}
\bibitem{Nojiri:2005sr}
S.~Nojiri and S.~D.~Odintsov,
%``Inhomogeneous equation of state of the universe: Phantom era,
%future
%singularity and crossing the phantom barrier,''
Phys.\ Rev.\ D {\bf 72}, 023003 (2005)
[arXiv:hep-th/0505215]; \\
%%CITATION = PHRVA,D72,023003;%%
%\cite{Capozziello:2005pa}
%\bibitem{Capozziello:2005pa}
S.~Capozziello, V.~F.~Cardone, E.~Elizalde, S.~Nojiri and S.~D.~Odintsov,
%``Observational constraints on dark energy with generalized equations of
%state,''
Phys.\ Rev.\ D {\bf 73}, 043512 (2006)
[arXiv:astro-ph/0508350].
%%CITATION = PHRVA,D73,043512;%%

\bibitem{Nojiri:2004pf}
S.~Nojiri and S.~D.~Odintsov,
%``The final state and thermodynamics of dark energy universe,''
Phys.\ Rev.\ D {\bf 70}, 103522 (2004)
[arXiv:hep-th/0408170].
%%CITATION = PHRVA,D70,103522;%%

%%%%%%%%%%%% Until chapter 2


%\cite{Nojiri:2006gh}
\bibitem{Nojiri:2006gh}
S.~Nojiri and S.~D.~Odintsov,
%``Modified f(R) gravity consistent with realistic cosmology:
%From
%matter
%dominated epoch to dark energy universe,''
Phys.\ Rev.\ D {\bf 74}, 086005 (2006)
[arXiv:hep-th/0608008].
%%CITATION = PHLTA,B639,135;%%

%\cite{Nojiri:2006be}
\bibitem{Nojiri:2006be}
S.~Nojiri and S.~D.~Odintsov,
%``Modified gravity and its reconstruction from the universe
%expansion
%history,''
J.\ Phys.\ Conf.\ Ser.\ {\bf 66}, 012005 (2007)
[arXiv:hep-th/0611071];
hep-th/0610164;
%%CITATION = 00462,66,012005;%%


%\cite{Nojiri:2009uu}
\bibitem{Nojiri:2009uu}
S.~Nojiri and S.~D.~Odintsov,
%``Singularity of spherically-symmetric space-time in
%quintessence/phantom
%dark
%energy universe,''
Phys.\ Lett.\ B {\bf 676}, 94 (2009)
[arXiv:0903.5231 [hep-th]]; \\
%%CITATION = PHLTA,B676,94;%%
%\cite{Nojiri:2006jy}
%\bibitem{Nojiri:2006jy}
S.~Nojiri, S.~D.~Odintsov and H.~Stefancic,
%``On the way from matter-dominated era to dark energy universe,''
Phys.\ Rev.\ D {\bf 74}, 086009 (2006)
[arXiv:hep-th/0608168].
%%CITATION = PHRVA,D74,086009;%%

%\cite{Nojiri:2005pu}
\bibitem{Nojiri:2005pu}
S.~Nojiri and S.~D.~Odintsov,
%``Unifying phantom inflation with late-time acceleration: Scalar
%phantom-non-phantom transition model and generalized holographic
%dark
%energy,''
Gen.\ Rel.\ Grav.\ {\bf 38}, 1285 (2006)
[arXiv:hep-th/0506212].
%%CITATION = GRGVA,38,1285;%%

%\cite{Capozziello:2005tf}
\bibitem{Capozziello:2005tf}
S.~Capozziello, S.~Nojiri and S.~D.~Odintsov,
%``Unified phantom cosmology: inflation, dark energy and dark
%matter
%under
%the
%same standard,''
Phys.\ Lett.\ B {\bf 632}, 597 (2006)
[arXiv:hep-th/0507182].
%%CITATION = PHLTA,B632,597;%%

%\cite{Vikman:2004dc}
\bibitem{Vikman:2004dc}
A.~Vikman,
%``Can dark energy evolve to the phantom?,''
Phys.\ Rev.\  D {\bf 71}, 023515 (2005)
[arXiv:astro-ph/0407107].
%%CITATION = PHRVA,D71,023515;%%


%\cite{Komatsu:2008hk}
\bibitem{Komatsu:2008hk}
E.~Komatsu {\it et al.} [WMAP Collaboration],
%``Five-Year Wilkinson Microwave Anisotropy Probe
%(WMAP\altaffilmark
%1
%)
%Observations:Cosmological Interpretation,''
Astrophys.\ J.\ Suppl.\ {\bf 180}, 330 (2009)
[arXiv:0803.0547 [astro-ph]].
%%CITATION = APJSA,180,330;%%


%\cite{Spergel:2003cb}
\bibitem{Spergel:2003cb}
D.~N.~Spergel {\it et al.} [WMAP Collaboration],
%``First Year Wilkinson Microwave Anisotropy Probe (WMAP)
%Observations:
%Determination of Cosmological Parameters,''
Astrophys.\ J.\ Suppl.\ {\bf 148}, 175 (2003)
[arXiv:astro-ph/0302209].
%%CITATION = APJSA,148,175;%%

%\cite{Peiris:2003ff}
\bibitem{Peiris:2003ff}
H.~V.~Peiris {\it et al.} [WMAP Collaboration],
%``First year Wilkinson Microwave Anisotropy Probe (WMAP)
%observations:
%Implications for inflation,''
Astrophys.\ J.\ Suppl.\ {\bf 148}, 213 (2003)
[arXiv:astro-ph/0302225].
%%CITATION = APJSA,148,213;%%

%\cite{Spergel:2006hy}
\bibitem{Spergel:2006hy}
D.~N.~Spergel {\it et al.} [WMAP Collaboration],
%``Wilkinson Microwave Anisotropy Probe (WMAP) three year
%results:
%Implications for cosmology,''
Astrophys.\ J.\ Suppl.\ {\bf 170}, 377 (2007)
[arXiv:astro-ph/0603449].
%%CITATION = APJSA,170,377;%%

%\cite{Brans:1961sx}
\bibitem{Brans:1961sx}
C.~Brans and R.~H.~Dicke,
%``Mach's principle and a relativistic theory of gravitation,''
Phys.\ Rev.\ {\bf 124}, 925 (1961).
%%CITATION = PHRVA,124,925;%%

%\cite{Elizalde:2004mq}
\bibitem{Elizalde:2004mq}
E.~Elizalde, S.~Nojiri and S.~D.~Odintsov,
%``Late-time cosmology in (phantom) scalar-tensor theory: Dark
%energy
%and
%the
%cosmic speed-up,''
Phys.\ Rev.\ D {\bf 70}, 043539 (2004)
[arXiv:hep-th/0405034].
%%CITATION = PHRVA,D70,043539;%%

\bibitem{yoshioka}
S.~Nojiri, H.~Yoshioka, unpublished.

%\cite{Elizalde:2008yf}
\bibitem{Elizalde:2008yf}
E.~Elizalde, S.~Nojiri, S.~D.~Odintsov, D.~Saez-Gomez and
V.~Faraoni,
%``Reconstructing the universe history, from inflation to
%acceleration,
%with
%phantom and canonical scalar fields,''
Phys.\ Rev.\ D {\bf 77} (2008) 106005
[arXiv:0803.1311 [hep-th]].
%%CITATION = PHRVA,D77,106005;%%

%\cite{Matsumoto:2010uv}
\bibitem{Matsumoto:2010uv}
J.~Matsumoto and S.~Nojiri,
%``Reconstruction of k-essence model,''
Phys.\ Lett.\ B {\bf 687}, 236 (2010)
[arXiv:1001.0220 [hep-th]].
%%CITATION = PHLTA,B687,236;%%

%\cite{Bamba:2008hq}
\bibitem{Bamba:2008hq}
K.~Bamba, C.~Q.~Geng, S.~Nojiri and S.~D.~Odintsov,
%``Crossing of the phantom divide in modified gravity,''
Phys.\ Rev.\ D {\bf 79}, 083014 (2009)
[arXiv:0810.4296 [hep-th]].
%%CITATION = PHRVA,D79,083014;%%

%\cite{Capozziello:2010uv}
\bibitem{Capozziello:2010uv}
S.~Capozziello, J.~Matsumoto, S.~Nojiri and S.~D.~Odintsov,
%``Dark energy from modified gravity with Lagrange multipliers,''
Phys.\ Lett.\ B {\bf 693}, 198 (2010)
[arXiv:1004.3691 [hep-th]].
%%CITATION = PHLTA,B693,198;%%

\bibitem{vikman}
%\cite{Lim:2010yk}
%\bibitem{Lim:2010yk}
E.~A.~Lim, I.~Sawicki and A.~Vikman,
%``Dust of Dark Energy,''
JCAP {\bf 1005}, 012 (2010)
[arXiv:1003.5751 [astro-ph.CO]]; \\
%%CITATION = JCAPA,1005,012;%%
%\cite{Gao:2010gj}
%\bibitem{Gao:2010gj}
C.~Gao, Y.~Gong, X.~Wang and X.~Chen,
%``Cosmological models with Lagrange Multiplier Field,''
arXiv:1003.6056 [astro-ph.CO]; \\
%%CITATION = ARXIV:1003.6056;%%
%\cite{Cai:2010zm}
%\bibitem{Cai:2010zm}
Y.~F.~Cai and E.~N.~Saridakis,
%``Cyclic cosmology from Lagrange-multiplier modified gravity,''
arXiv:1007.3204 [astro-ph.CO]; \\
%%CITATION = ARXIV:1007.3204;%%
%\cite{Kluson:2010af}
%\bibitem{Kluson:2010af}
J.~Kluson,
%``Hamiltonian Analysis of Lagrange Multiplier Modified Gravity,''
arXiv:1009.6067 [hep-th].
%%CITATION = ARXIV:1009.6067;%%

%\cite{Nojiri:2009kx}
\bibitem{Nojiri:2009kx}
S.~Nojiri, S.~D.~Odintsov and D.~Saez-Gomez,
%``Cosmological reconstruction of realistic modified F(R)
%gravities,''
Phys.\ Lett.\ B {\bf 681}, 74 (2009)
[arXiv:0908.1269 [hep-th]].
%%CITATION = PHLTA,B681,74;%%

%\cite{Dunsby:2010wg}
\bibitem{Dunsby:2010wg}
P.~K.~S.~Dunsby, E.~Elizalde, R.~Goswami, S.~Odintsov and
D.~S.~Gomez,
%``On the LCDM Universe in f(R) gravity,''
arXiv:1005.2205 [gr-qc].
%%CITATION = ARXIV:1005.2205;%%

%\cite{Elizalde:2010xq}
\bibitem{Elizalde:2010xq}
E.~Elizalde and A.~J.~Lopez-Revelles,
%``Reconstructing cosmic acceleration from modified and
%non-minimal
%gravity:
%The Yang-Mills case,''
Phys.\ Rev.\ D {\bf 82}, 063504 (2010)
[arXiv:1004.5021 [hep-th]].
%%CITATION = PHRVA,D82,063504;%%

%\cite{Barrow:1990vx}
\bibitem{Barrow:1990vx}
J.~D.~Barrow,
%``GRADUATED INFLATIONARY UNIVERSES,''
Phys.\ Lett.\ B {\bf 235}, 40 (1990).
%%CITATION = PHLTA,B235,40;%%

%\cite{Briscese:2006xu}
\bibitem{Briscese:2006xu}
F.~Briscese, E.~Elizalde, S.~Nojiri and S.~D.~Odintsov,
%``Phantom scalar dark energy as modified gravity: Understanding the 
%origin of
%the big rip singularity,''
Phys.\ Lett.\ B {\bf 646}, 105 (2007)
[arXiv:hep-th/0612220].
%%CITATION = PHLTA,B646,105;%%

\end{thebibliography}
\end{document}